%% file: main_book_format_for_arxiv.tex
\documentclass[11pt,fleqn]{book} %

\usepackage{lineno}
\usepackage{wrapfig}
\usepackage[font={sf,scriptsize}]{caption}
\usepackage{multirow}
\usepackage[utf8]{inputenc}
\usepackage[T1]{fontenc}
\usepackage{natbib}
\usepackage{lipsum}  
\usepackage{amsmath,bm}  
\usepackage{array}
\usepackage{makecell}
\usepackage{booktabs}
\usepackage{tablefootnote}
\usepackage{nicefrac}
\usepackage{todonotes}  
\usepackage{verbatim}

\usepackage{color, colortbl}
\usepackage{xcolor}

\definecolor{shadecolor}{rgb}{0.2,0.9,0.2}
\definecolor{green}{rgb}{0,0.4,0}

\input{commands.tex}
\usepackage{tikz}
\usepackage{pgf}
\usetikzlibrary{arrows,shapes,positioning,shadows,trees,automata}

\tikzset{
  basic/.style  = {draw, text width=5cm, drop shadow, font=\sffamily, rectangle},
  root/.style   = {basic, rounded corners=2pt, thin, align=center, fill=green!50},
  level 2/.style = {basic, rounded corners=2pt, text width=3cm, thin, align=left, fill=blue!20},
  level 3/.style = {basic, rounded corners=2pt, text width=3cm, thin, align=left, fill=green!30},
  level 4/.style = {basic, rounded corners=2pt, text width=3cm, thin, align=left, fill=orange!20}
}

\def\author{
\begin{quote}
Gustau Camps-Valls, Andreas Gerhardus, Urmi Ninad, Gherardo Varando, Georg Martius, Ricardo Vinuesa, Emili Balaguer-Ballester, Emiliano Diaz, Laure Zanna, Jakob Runge
\end{quote}
}

\def\date{\today} %

\input{structure} %

\begin{document}

\begingroup
\thispagestyle{empty}
\begin{tikzpicture}[remember picture,overlay]
\coordinate [below=12cm] (midpoint) at (current page.north);
\node at (current page.north west)
{\begin{tikzpicture}[remember picture,overlay]
\node[anchor=north west,inner sep=0pt] at (0,0) {\includegraphics[width=\paperwidth]{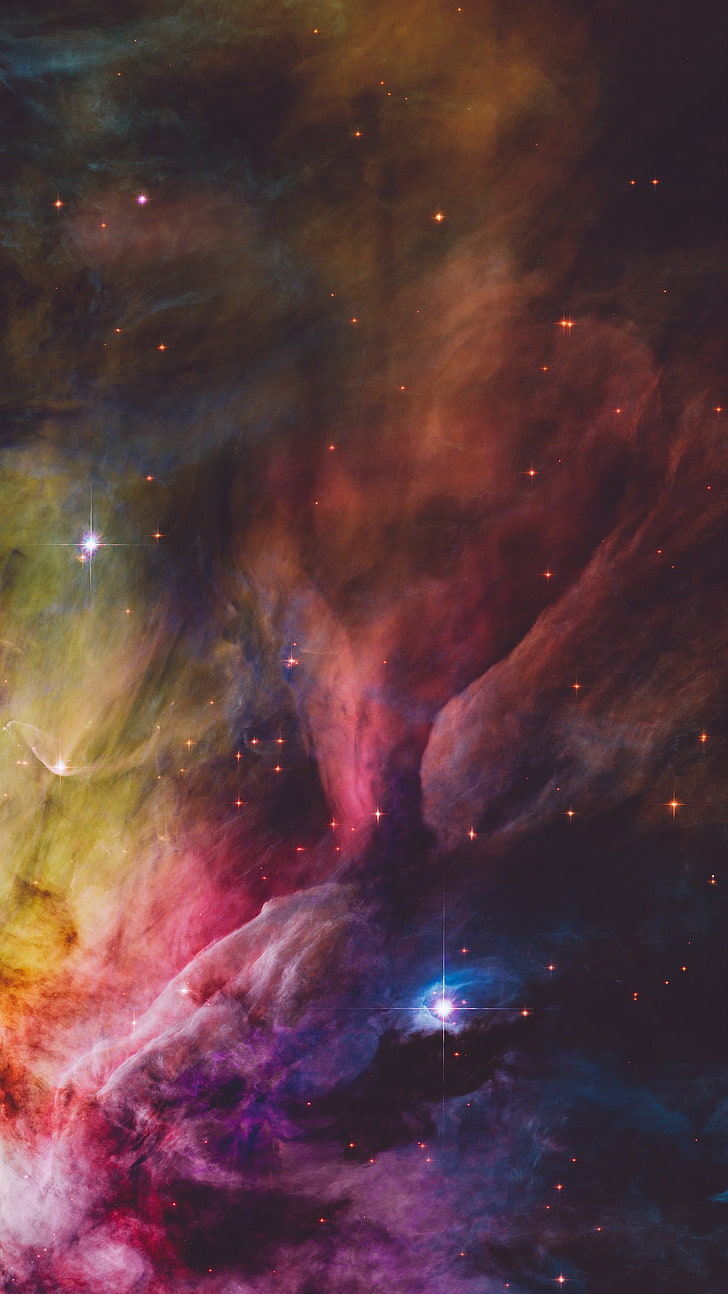}}; %
\draw[anchor=north] (midpoint) node [fill=ocre!30!white,fill opacity=0.6,text opacity=1,inner sep=1cm]{\Huge\centering\bfseries\sffamily\parbox[c][][t]{\paperwidth}{\centering \Huge Discovering Causal Relations \\and Equations from Data\\[15pt] %
{\LARGE A review of methods, challenges and opportunities}\\[20pt] %
{\Large \author\vspace*{0.5cm}
\color{white}\large\date}}}; %
\end{tikzpicture}};
\end{tikzpicture}
\vfill
\endgroup

\newpage
~\vfill
\thispagestyle{empty}

\noindent Copyright \copyright\ 2023 
Gustau Camps-Valls, 
Andreas Gerhardus, 
Urmi Ninad, 
Gherardo Varando, 
Georg Martius, 
Emili Balaguer-Ballester, 
Ricardo Vinuesa, 
Emiliano Diaz,  
Laure Zanna, 
Jakob Runge\\ %

\noindent \textsc{http://isp.uv.es}\\ %

\noindent \textit{First version, \date} %

\include{Cover}

\pagestyle{empty}
\tableofcontents
\pagestyle{fancy}

\addcontentsline{toc}{chapter}{Preface}

\include{00_Abstract_book}

\input{01_Introduction_book}
\input{02_CausalDiscovery_book}
\input{03_LearningPhysics_book}

\input{04_CaseStudies_book}
\input{05_Conclusions_book}
\addcontentsline{toc}{chapter}{Acknowledgements and Contributions}
\input{06_Acks_book}

\input{07_Contributions_book}

\addcontentsline{toc}{chapter}{Bibliography}
\bibliographystyle{abbrvnat}
\bibliography{references_cleaned}

\end{document}

%% file: commands.tex
\newcommand{\ind}{\rotatebox[origin=c]{90}{$\models$}} %
\newcommand{\Real}{\ensuremath{\mathbb R}}        %
\newcommand{\T}{\ensuremath{\top}}                %

\DeclareMathOperator*{\argmin}{arg\,min}   %

\usepackage{pifont}%
\newcommand{\cmark}{\ding{51}}%
\newcommand{\xmark}{\ding{55}}%

\usepackage{xspace}
\makeatletter
\DeclareRobustCommand\onedot{\futurelet\@let@token\@onedot}
\def\@onedot{\ifx\@let@token.\else.\null\fi\xspace}
\makeatother

\newcommand{\eg}{e.g\onedot}
\newcommand{\ie}{i.e\onedot}

\renewcommand{\cite}[1]{\citep{#1}}

\newcommand{\Figref}[1]{Figure~\ref{#1}}  %
\newcommand{\figref}[1]{Fig.~\ref{#1}}    %

\newcommand{\tabref}[1]{Table~\ref{#1}}

\newcommand{\eqnpref}[1]{(Eq.~\ref{#1})} %
\newcommand{\secref}[1]{Sec.~\ref{#1}} %

%% file: structure.tex
\usepackage[top=3cm,bottom=3cm,left=3cm,right=3cm,headsep=10pt,a4paper]{geometry} %

\usepackage{graphicx} %
\graphicspath{{images/}} %

\usepackage{lipsum} %

\usepackage{tikz} %

\usepackage[english]{babel} %

\usepackage{enumitem} %
\setlist{nolistsep} %

\usepackage{booktabs} %

\usepackage{xcolor} %
\definecolor{ocre}{RGB}{243,102,25} %

\usepackage{avant} %
\usepackage{mathptmx} %

\usepackage{microtype} %
\usepackage[utf8]{inputenc} %
\usepackage[T1]{fontenc} %

\usepackage{calc} %
\usepackage{makeidx} %
\makeindex %

\usepackage{titletoc} %

\contentsmargin{0cm} %

\titlecontents{part}[0cm]
{\addvspace{20pt}\centering\large\bfseries}
{}
{}
{}

\titlecontents{chapter}[1.25cm] %
{\addvspace{12pt}\large\sffamily\bfseries} %
{\color{ocre!60}\contentslabel[\Large\thecontentslabel]{1.25cm}\color{ocre}} %
{\color{ocre}}  
{\color{ocre!60}\normalsize\;\titlerule*[.5pc]{.}\;\thecontentspage} %

\titlecontents{section}[1.25cm] %
{\addvspace{3pt}\sffamily\bfseries} %
{\contentslabel[\thecontentslabel]{1.25cm}} %
{}
{\hfill\color{black}\thecontentspage} %
[]

\titlecontents{subsection}[1.25cm] %
{\addvspace{1pt}\sffamily\small} %
{\contentslabel[\thecontentslabel]{1.25cm}} %
{}
{\ \titlerule*[.5pc]{.}\;\thecontentspage} %
[]

\titlecontents{figure}[0em]
{\addvspace{-5pt}\sffamily}
{\thecontentslabel\hspace*{1em}}
{}
{\ \titlerule*[.5pc]{.}\;\thecontentspage}
[]

\titlecontents{table}[0em]
{\addvspace{-5pt}\sffamily}
{\thecontentslabel\hspace*{1em}}
{}
{\ \titlerule*[.5pc]{.}\;\thecontentspage}
[]

\titlecontents{lchapter}[0em] %
{\addvspace{15pt}\large\sffamily\bfseries} %
{\color{ocre}\contentslabel[\Large\thecontentslabel]{1.25cm}\color{ocre}} %
{}  
{\color{ocre}\normalsize\sffamily\bfseries\;\titlerule*[.5pc]{.}\;\thecontentspage} %

\titlecontents{lsection}[0em] %
{\sffamily\small} %
{\contentslabel[\thecontentslabel]{1.25cm}} %
{}
{}

\titlecontents{lsubsection}[.5em] %
{\normalfont\footnotesize\sffamily} %
{}
{}
{}

\usepackage{fancyhdr} %

\fancypagestyle{plain}{\fancyhead{}} %

\makeatletter
\renewcommand{\cleardoublepage}{
\clearpage\ifodd\c@page\else
\hbox{}
\vspace*{\fill}
\thispagestyle{empty}
\newpage
\fi}

\usepackage{amsmath,amsfonts,amssymb,amsthm} %

\newtheoremstyle{ocrenumbox}%
{0pt}%
{0pt}%
{\normalfont}%
{}%
{\small\bf\sffamily\color{ocre}}%
{\;}%
{0.25em}%
{\small\sffamily\color{ocre}\thmname{#1}\nobreakspace\thmnumber{\@ifnotempty{#1}{}\@upn{#2}}%
\thmnote{\nobreakspace\the\thm@notefont\sffamily\bfseries\color{black}---\nobreakspace#3.}} %

\newtheoremstyle{blacknumex}%
{5pt}%
{5pt}%
{\normalfont}%
{} %
{\small\bf\sffamily}%
{\;}%
{0.25em}%
{\small\sffamily{\tiny\ensuremath{\blacksquare}}\nobreakspace\thmname{#1}\nobreakspace\thmnumber{\@ifnotempty{#1}{}\@upn{#2}}%
\thmnote{\nobreakspace\the\thm@notefont\sffamily\bfseries---\nobreakspace#3.}}%

\newtheoremstyle{blacknumbox} %
{0pt}%
{0pt}%
{\normalfont}%
{}%
{\small\bf\sffamily}%
{\;}%
{0.25em}%
{\small\sffamily\thmname{#1}\nobreakspace\thmnumber{\@ifnotempty{#1}{}\@upn{#2}}%
\thmnote{\nobreakspace\the\thm@notefont\sffamily\bfseries---\nobreakspace#3.}}%

\newtheoremstyle{ocrenum}%
{5pt}%
{5pt}%
{\normalfont}%
{}%
{\small\bf\sffamily\color{ocre}}%
{\;}%
{0.25em}%
{\small\sffamily\color{ocre}\thmname{#1}\nobreakspace\thmnumber{\@ifnotempty{#1}{}\@upn{#2}}%
\thmnote{\nobreakspace\the\thm@notefont\sffamily\bfseries\color{black}---\nobreakspace#3.}} %
\makeatother

\newcounter{dummy} 
\numberwithin{dummy}{section}
\theoremstyle{ocrenumbox}
\newtheorem{theoremeT}[dummy]{Theorem}

\newtheorem{exerciseT}{Exercise}[chapter]
\theoremstyle{blacknumex}
\newtheorem{exampleT}{Example}[chapter]
\theoremstyle{blacknumbox}

\newtheorem{definitionT}{Definition}[section]
\newtheorem{corollaryT}[dummy]{Corollary}
\theoremstyle{ocrenum}

\RequirePackage[framemethod=default]{mdframed} %

\newmdenv[skipabove=7pt,
skipbelow=7pt,
backgroundcolor=black!5,
linecolor=ocre,
innerleftmargin=5pt,
innerrightmargin=5pt,
innertopmargin=5pt,
leftmargin=0cm,
rightmargin=0cm,
innerbottommargin=5pt]{tBox}

\newmdenv[skipabove=7pt,
skipbelow=7pt,
rightline=false,
leftline=true,
topline=false,
bottomline=false,
backgroundcolor=ocre!10,
linecolor=ocre,
innerleftmargin=5pt,
innerrightmargin=5pt,
innertopmargin=5pt,
innerbottommargin=5pt,
leftmargin=0cm,
rightmargin=0cm,
linewidth=4pt]{eBox}	

\newmdenv[skipabove=7pt,
skipbelow=7pt,
rightline=false,
leftline=true,
topline=false,
bottomline=false,
linecolor=ocre,
innerleftmargin=5pt,
innerrightmargin=5pt,
innertopmargin=0pt,
leftmargin=0cm,
rightmargin=0cm,
linewidth=4pt,
innerbottommargin=0pt]{dBox}	

\newmdenv[skipabove=7pt,
skipbelow=7pt,
rightline=false,
leftline=true,
topline=false,
bottomline=false,
linecolor=gray,
backgroundcolor=black!5,
innerleftmargin=5pt,
innerrightmargin=5pt,
innertopmargin=5pt,
leftmargin=0cm,
rightmargin=0cm,
linewidth=4pt,
innerbottommargin=5pt]{cBox}

\makeatletter
\renewcommand{\@seccntformat}[1]{\llap{\textcolor{ocre}{\csname the#1\endcsname}\hspace{1em}}}                    
\renewcommand{\section}{\@startsection{section}{1}{\z@}
{-4ex \@plus -1ex \@minus -.4ex}
{1ex \@plus.2ex }
{\normalfont\large\sffamily\bfseries}}
\renewcommand{\subsection}{\@startsection {subsection}{2}{\z@}
{-3ex \@plus -0.1ex \@minus -.4ex}
{0.5ex \@plus.2ex }
{\normalfont\sffamily\bfseries}}
\renewcommand{\subsubsection}{\@startsection {subsubsection}{3}{\z@}
{-2ex \@plus -0.1ex \@minus -.2ex}
{.2ex \@plus.2ex }
{\normalfont\small\sffamily\bfseries}}                        
\renewcommand\paragraph{\@startsection{paragraph}{4}{\z@}
{-2ex \@plus-.2ex \@minus .2ex}
{.1ex}
{\normalfont\small\sffamily\bfseries}}

\newcommand{\@mypartnumtocformat}[2]{%
\setlength\fboxsep{0pt}%
\noindent\colorbox{ocre!20}{\strut\parbox[c][.7cm]{\ecart}{\color{ocre!70}\Large\sffamily\bfseries\centering#1}}\hskip\esp\colorbox{ocre!40}{\strut\parbox[c][.7cm]{\linewidth-\ecart-\esp}{\Large\sffamily\centering#2}}}%
\newcommand{\@myparttocformat}[1]{%
\setlength\fboxsep{0pt}%
\noindent\colorbox{ocre!40}{\strut\parbox[c][.7cm]{\linewidth}{\Large\sffamily\centering#1}}}%
\newlength\esp
\newlength\ecart
\def\@part[#1]#2{%
\ifnum \c@secnumdepth >-2\relax%
\refstepcounter{part}%
\addcontentsline{toc}{part}{\texorpdfstring{\protect\@mypartnumtocformat{\thepart}{#1}}{\partname~\thepart\ ---\ #1}}
\else%
\addcontentsline{toc}{part}{\texorpdfstring{\protect\@myparttocformat{#1}}{#1}}%
\fi%
\startcontents%
\markboth{}{}%
{\thispagestyle{empty}%
\begin{tikzpicture}[remember picture,overlay]%
\node at (current page.north west){\begin{tikzpicture}[remember picture,overlay]%
\fill[ocre!20](0cm,0cm) rectangle (\paperwidth,-\paperheight);
\node[anchor=north] at (4cm,-3.25cm){\color{ocre!40}\fontsize{220}{100}\sffamily\bfseries\@Roman\c@part}; 
\node[anchor=south east] at (\paperwidth-1cm,-\paperheight+1cm){\parbox[t][][t]{8.5cm}{
\printcontents{l}{0}{\setcounter{tocdepth}{1}}%
}};
\node[anchor=north east] at (\paperwidth-1.5cm,-3.25cm){\parbox[t][][t]{15cm}{\strut\raggedleft\color{white}\fontsize{30}{30}\sffamily\bfseries#2}};
\end{tikzpicture}};
\end{tikzpicture}}%
\@endpart}
\def\@spart#1{%
\startcontents%
\phantomsection
{\thispagestyle{empty}%
\begin{tikzpicture}[remember picture,overlay]%
\node at (current page.north west){\begin{tikzpicture}[remember picture,overlay]%
\fill[ocre!20](0cm,0cm) rectangle (\paperwidth,-\paperheight);
\node[anchor=north east] at (\paperwidth-1.5cm,-3.25cm){\parbox[t][][t]{15cm}{\strut\raggedleft\color{white}\fontsize{30}{30}\sffamily\bfseries#1}};
\end{tikzpicture}};
\end{tikzpicture}}
\addcontentsline{toc}{part}{\texorpdfstring{%
\setlength\fboxsep{0pt}%
\noindent\protect\colorbox{ocre!40}{\strut\protect\parbox[c][.7cm]{\linewidth}{\Large\sffamily\protect\centering #1\quad\mbox{}}}}{#1}}%
\@endpart}
\def\@endpart{\vfil\newpage
\if@twoside
\if@openright
\null
\thispagestyle{empty}%
\newpage
\fi
\fi
\if@tempswa
\twocolumn
\fi}

\newif\ifusechapterimage
\usechapterimagefalse  %
\newcommand{\thechapterimage}{}%
\def\@makechapterhead#1{%
{\parindent \z@ \raggedright \normalfont
\ifnum \c@secnumdepth >\m@ne
\if@mainmatter
\begin{tikzpicture}[remember picture,overlay]
\node at (current page.north west)
{\begin{tikzpicture}[remember picture,overlay]
\node[anchor=north west,inner sep=0pt] at (0,0) {\ifusechapterimage\includegraphics[width=\paperwidth]{\thechapterimage}\fi};
\draw[anchor=west] (\Gm@lmargin,-9cm) node [line width=2pt,rounded corners=15pt,draw=ocre,fill=white,fill opacity=0.5,inner sep=15pt]{\strut\makebox[22cm]{}};
\draw[anchor=west] (\Gm@lmargin+.3cm,-9cm) node {\huge\sffamily\bfseries\color{black}\thechapter. #1\strut};
\end{tikzpicture}};
\end{tikzpicture}
\else
\begin{tikzpicture}[remember picture,overlay]
\node at (current page.north west)
{\begin{tikzpicture}[remember picture,overlay]
\node[anchor=north west,inner sep=0pt] at (0,0) {\ifusechapterimage\includegraphics[width=\paperwidth]{\thechapterimage}\fi};
\draw[anchor=west] (\Gm@lmargin,-9cm) node [line width=2pt,rounded corners=15pt,draw=ocre,fill=white,fill opacity=0.5,inner sep=15pt]{\strut\makebox[22cm]{}};
\draw[anchor=west] (\Gm@lmargin+.3cm,-9cm) node {\huge\sffamily\bfseries\color{black}#1\strut};
\end{tikzpicture}};
\end{tikzpicture}
\fi\fi\par\vspace*{270\p@}}}

\def\@makeschapterhead#1{%
\begin{tikzpicture}[remember picture,overlay]
\node at (current page.north west)
{\begin{tikzpicture}[remember picture,overlay]
\node[anchor=north west,inner sep=0pt] at (0,0) {\ifusechapterimage\includegraphics[width=\paperwidth]{\thechapterimage}\fi};
\draw[anchor=west] (\Gm@lmargin,-9cm) node [line width=2pt,rounded corners=15pt,draw=ocre,fill=white,fill opacity=0.5,inner sep=15pt]{\strut\makebox[22cm]{}};
\draw[anchor=west] (\Gm@lmargin+.3cm,-9cm) node {\huge\sffamily\bfseries\color{black}#1\strut};
\end{tikzpicture}};
\end{tikzpicture}
\par\vspace*{270\p@}}
\makeatother

\usepackage{hyperref}
\hypersetup{hidelinks,colorlinks=true,linkcolor=ocre,citecolor = ocre, breaklinks=true,urlcolor= ocre,bookmarksopen=false,pdftitle={Causal and equation discovery for the physical sciences},pdfauthor={Gustau Camps-Valls}}

\usepackage{bookmark}
\bookmarksetup{
open,
numbered,
addtohook={%
\ifnum\bookmarkget{level}=0 %
\bookmarksetup{bold}%
\fi
\ifnum\bookmarkget{level}=-1 %
\bookmarksetup{color=ocre,bold}%
\fi
}
}

%% file: Cover.tex
\hspace{5cm}\vspace{4cm}
\begin{quote}
{\LARGE Discovering Causal Relations and Equations from Data}

\hspace{5cm}\vspace{1cm}

Gustau Camps-Valls$^{1,\dag}$, 
Andreas Gerhardus$^{2,*}$, 
Urmi Ninad$^{3,*}$, 
Gherardo Varando$^{1,*}$, 
Georg Martius$^{4,5}$, 
Emili Balaguer-Ballester$^{6,7}$, 
Ricardo Vinuesa$^{8}$, 
Emiliano Diaz$^{1}$, 
Laure Zanna$^{9}$, 
Jakob Runge$^{2,3}$

\hspace{5cm}\vspace{1cm}

$^1$Universitat de Val\`encia, Val\`encia, Spain

$^2$German Aerospace Center, Jena, Germany

$^3$Technische Universit\"at Berlin, Berlin, Germany

$^4$University of Tübingen, T\"ubingen, Germany

$^5$Max Planck Institute for Intelligent Systems, T\"ubingen, Germany

$^6$Bournemouth University, Bournemouth, UK

$^7$Medical Faculty Mannheim and Heidelberg University, Mannheim, Germany.

$^8$KTH Royal Institute of Technology, Stockholm, Sweden

$^9$New York University, New York, USA

$^*$ These authors contributed equally.

$^\dag$ Corresponding author at Image Processing Laboratory (IPL), E4 building - 4th floor, Parc Cient\'ific Universitat de Val\`encia. C/ Cat. Agustín Escardino Benlloch, 9. 46980 Paterna (València). Spain. E-mail address: \url{gustau.camps@uv.es}

\end{quote}

%% file: 00_Abstract_book.tex
\chapter*{Preface}

\begin{quote}
\small
{\em ``As in Mathematics, so in Natural Philosophy, the Investigation of difficult Things by the Method of Analysis ought ever to precede the Method of Composition. This Analysis consists in making Experiments and Observations, and in drawing general Conclusions from them by Induction, and admitting of no Objections against the Conclusions, but such as are taken from Experiments, or other certain Truths ... By this way of Analysis, we may proceed from Compounds to Ingredients, and from Motions to the Forces producing them; and in general, from Effects to their Causes, and particular Causes to more general ones, till the Argument end in the most general. This is the Method of Analysis''}
\hfill{(Newton, 1718).}
\end{quote}

Physics is a field of science that has traditionally used the scientific method to answer questions about why natural phenomena occur and to make testable models that explain the phenomena. Discovering equations, laws and principles that are invariant, robust and causal explanations of the world has been fundamental in physical sciences throughout the centuries. Discoveries emerge from observing the world and, when possible, performing interventional studies in the system under study. With the advent of big data and the use of data-driven methods, causal and equation discovery fields have grown and made progress in computer science, physics, statistics, philosophy, and many applied fields. All these domains are intertwined and can be used to discover causal relations, physical laws, and equations from observational data. 
This paper reviews the concepts, methods, and relevant works on causal and equation discovery in the broad field of Physics and outlines the most important challenges and promising future lines of research. We also provide a taxonomy for observational causal and equation discovery, point out connections, and showcase a complete set of case studies in Earth and climate sciences, fluid dynamics and mechanics, and the neurosciences. 
This review demonstrates that discovering fundamental laws and causal relations by observing natural phenomena is being revolutionised with the efficient exploitation of observational data, modern machine learning algorithms and the interaction with domain knowledge. Exciting times are ahead with many challenges and opportunities to improve our understanding of complex systems.

%% file: 01_Introduction_book.tex
\chapter{Introduction}\label{sec:introduction}

This paper reviews the recent advances in {\em causal discovery} and {\em equation discovery} from data. Both problems are conundrums for scientists and philosophers of science. After all, Science is about studying, discovering, and understanding the structure and behaviour of the physical and natural world through observation and experimentation. Understanding the system's structure involves performing interventions on the systems to evaluate their responses. However, interventional experiments are often not feasible for economic or ethical reasons, so relying on observations, simulations, and domain knowledge must be exploited. In recent decades, discovering causal relations and underlying governing laws from data have emerged as exciting fields of research that promise advancing science. 

\section{Understanding in the physical sciences} %

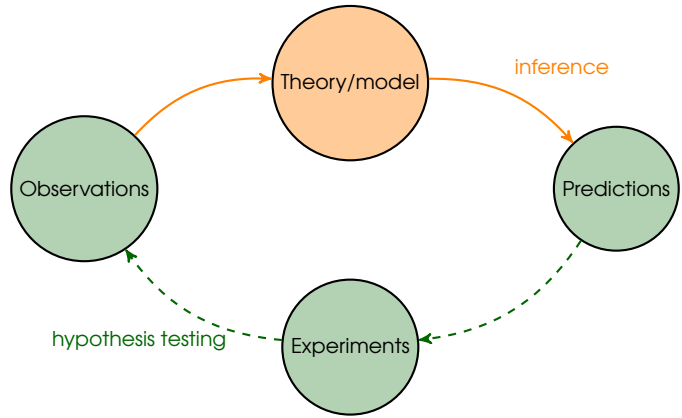
\begin{figure}[t!]
\begin{center}
\begin{tikzpicture}[->,>=stealth',scale=.7,transform shape,node distance=5cm,thick,font=\sffamily]
  \tikzstyle{every state}=[fill=green!30,text=black]
  \node[state] (RET)  [fill=green!30] {Experiments};
  \node[state] (RTM)  [above of=RET,fill=orange!40] {Theory/model};
  \node[state] (OBS)  [right of=RTM,fill=green!30,yshift=-2cm] {Predictions};
  \node[state] (VARS) [left of=RTM,fill=green!30,yshift=-2cm] {Observations};
  \path[solid, bend left=25, thick, color=orange] (VARS) edge node {} (RTM);
  \path[solid, bend left=25, thick, color=orange] (RTM)  edge node[yshift=+0.5cm,xshift=+1cm] {inference} (OBS);
  \path[dashed, bend left=25, thick, color=green] (OBS)   edge node {} (RET);
  \path[dashed, bend left=25, thick, color=green] (RET)   edge node[yshift=-0.5cm,xshift=-1cm] {hypothesis testing} (VARS);
\end{tikzpicture}
\end{center}
\vspace{-0.5cm}
\caption{Standard loop in understanding complex systems following the standard scientific method. Understanding involves experimentation by refining a descriptive mechanistic model. The initial model hypothesis is tested in practice and, through experiments, yields observations that are confronted with the model's predictions. The unexplained processes are then used to improve the model's misspecification and predictions.}
\label{fig:method}
\end{figure}

A pertinent question arises here; what is understanding? Understanding is the ability to comprehend and make sense of processes to gain a deeper knowledge of a system. Understanding involves analysing information, making (eventually causal) connections, and coming to conclusions. Whether the conclusions should be falsifiable has been the subject of active discussion in the Philosophy of Science \citep{popper2005logic, munz2014our}. %
We aim to {\em understand} complex systems by following the {\em scientific method}; make an observation, ask a scientific question, form a hypothesis, theory, model, or explanation of the phenomena, and make predictions, which are ultimately tested and whose results are used to make new hypotheses or predictions %
(see Fig.~\ref{fig:method}).

Understanding involves reasoning and thinking critically about a subject, a system's behaviour, or a problem. Understanding and explaining how a system works %
is more complicated than making  predictions about the system's behaviour. %
The Oracle of Delphi gave accurate predictions of the future and the optimal course of action, but the lack of understanding frequently led to disaster. True understanding is about making (truly accurate) predictions and, more importantly, gaining knowledge of the causal chain. Science generally aims to answer causal questions, infer causal relations, and attain mathematical models (mainly laws and equations) that work well in most situations, explain the system and underlying processes, and are invariant across space and time. Without it, we cannot predict the consequences of our actions ({\em interventions}) or analyse when, where and why things went wrong ({\em counterfactuals})~\citep{pearl2009causality}.

\begin{wrapfigure}{l}{5cm}%
\centerline{\includegraphics[width=5cm]{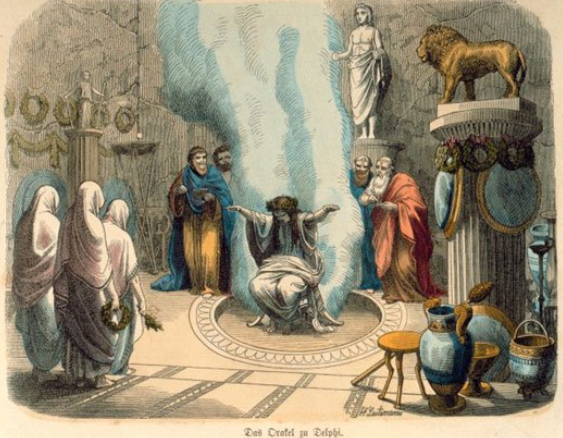}}
\caption{The Oracle of Delphi at inference time.}
\end{wrapfigure}
Yet, what do we want to understand? And how do we generally do it? In physical sciences, one typically analyses phenomena and instantiations of the physical world, uses observations, and refines and tests models. For learning about the system, one aims to (1) characterise its complexity in terms of trajectories, persistence, stability and collapse, bifurcations and viability boundaries \citep{sterman1994learning,kwapien2012physical,salcedo2022persistence,may1972will}, (2) obtain explanatory and causal models of their behaviour \citep{Pearl2000,Peters2017book,runge2019inferring}, and (3) discover and formalise general laws, governing equations, and parameterizations \citep{richardson1996discovery,brunton2016discovering,chen2022automated}. These three components allow us to advance science and technology.
However, in many systems, governing equations and causal relations are (partially) unknown, and recourse to first principles is untenable. Resorting to algorithms that can discover laws, governing equations, and causal relations from data may thus constitute a paradigm shift that promises to accelerate science. %

\section{Scientific discovery}

Before revising the fields, let us take a step back and review the formalism in the logic of science and the key elements, definitions, and challenges we face in addressing such problems.
What are the cornerstones for understanding how science and the scientific method work? The philosophy of science studies science's assumptions, foundations, and implications. It examines the implications of scientific theories and methods for understanding the world. It concerns the fundamental questions at the heart of science, such as the nature of knowledge, the limits of scientific inquiry, and the relationship between science and values.

Scientific discovery is the process or product of successful scientific inquiry. Objects of discovery can be things, events, processes, causes, properties, theories, hypotheses, and their characteristics. Philosophical discussions of scientific discovery vary widely in scope and definition, from the narrowest sense of a ``eureka moment'' to the broadest sense of a ``successful scientific endeavour''. 
The utilisation of data sets to create and test new hypotheses in philosophical discourse has led to a multifaceted and intricate discussion regarding the precise definition and potential misuse of the term ``discovery''.

\begin{wrapfigure}{l}{5cm} %
\centerline{\includegraphics[width=5cm]{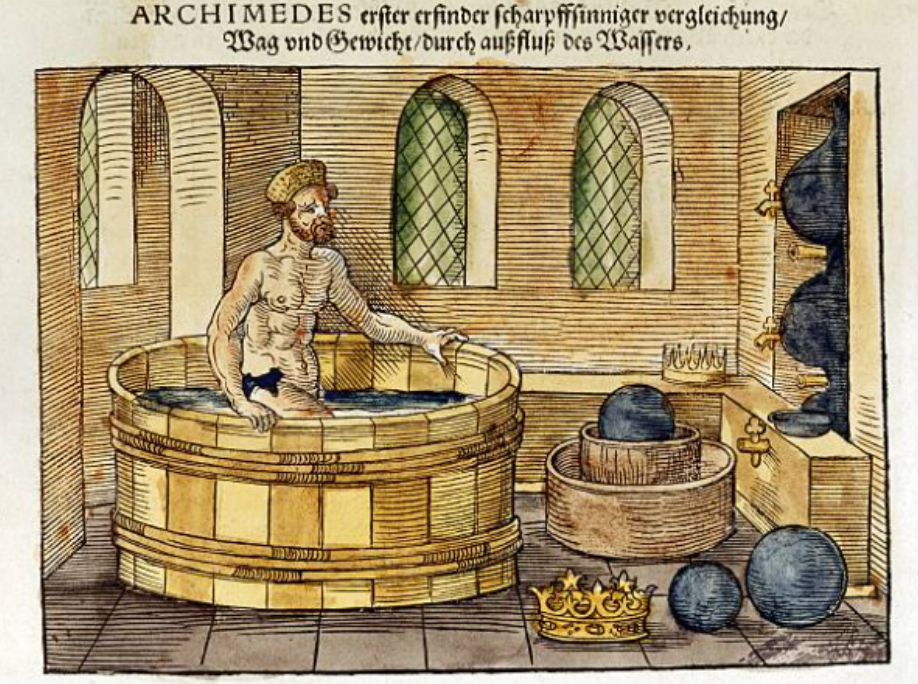}}
\caption{Archimedes (287-212 BC) in his bath. A woodcut by Fl\"ortner (1490-1546).} %
\vspace{-20pt}
\end{wrapfigure}
Human nature aims to discover. Always. Since the Bronze Age\footnote{\href{https://en.wikipedia.org/wiki/Timeline\_of\_scientific\_discoveries}{https://en.wikipedia.org/wiki/Timeline\_of\_scientific\_discoveries
}}. Generations have created and discovered new principles, techniques, and operations through millennia. Right after the Neolithic Revolution, the world stopped except for some remarkable technological advances, like the invention of the water wheel (476-221 BC) and the windmill (ca 644 BC). Romans were amazed by stories of what Archimedes (287-212 BC) had been able to do. But, bold as it may sound, one may claim that modern science was invented between 1572 when Tycho Brahe saw a {\it nova} or new star, and 1704 when Newton published his Opticks~\citep{wootton2015invention}. What happened in that period prepared humanity and scientists for a New Era: a research program endorsed with a scientific method that allowed scrutinising new theories and validating or refuting hypotheses and models of the world, and all that in the light of evidence and observations. 
After fitting many ovoids to observational data, Kepler discovered the laws of planetary motion (1609) and needed four years to discover Mars' orbit was an ellipse. The scientific method was slow but sure. 
Galilei discovered the law of falling bodies (1638) by dropping two cannonballs of different masses from the tower of Pisa and measuring the effect of mass on the fall rate to the ground. 
And in 1662, Boyle discovered the law of ideal gases. Only ten years later, in 1672, Newton discovered that white light is a mixture of distinct coloured rays, and in 1687 he formulated the classical mathematical description of the fundamental force of universal gravitation and the three physical laws of motion. The triumph of Newtonianism marks the end of the beginning of scientific discovery. 

The history of science is a long and complex narrative punctuated by moments of major scientific revolutions \citep{Kuhn1962}. Kuhn identified a general pattern: A discovery is not a simple act but an extended, complex process that culminates in paradigm changes. These events mark a fundamental shift in how science is understood and practised and have greatly impacted scientific progress. 
The first scientific revolution occurred in the 16th and 17th centuries when the Copernican revolution challenged the traditional Ptolemaic view of the universe \citep{Copernicus1543}. 
The next major scientific revolution was the Enlightenment during the 18th century, which saw the emergence of the scientific method and the development of modern physics and chemistry, from formulating the laws of motion to discovering electricity. This period also saw the emergence of scientific societies, which helped to propagate and popularise scientific ideas. The 19th century saw the emergence of the theory of evolution, which revolutionised the field of biology \citep{Darwin1859}. %
This revolution was followed by the rise of modern genetics, which further expanded our understanding of the evolution of life \citep{Watson1953}. The 20th century saw the emergence of the quantum revolution, which revolutionised our understanding of the physical world as it could not be fully explained by classical physics \citep{Heisenberg1925}. 
Revolutions happen only gradually, as it takes time for the scientific community to recognise {\em ``both that something is and what it is''} \citep{Kuhn1962}. Eventually, a new paradigm becomes established, and the strange phenomena become the expected phenomena. 

The idea that there is such a thing as `the Scientific Revolution' and that it took place in the 17th century is thus a fairly recent one \citep{butterfield1965origins}; some have argued that it can be seen as the construction of intellectuals looking back from the 20th century \citep{wootton2015invention}. 
Like the term `Industrial Revolution', the idea of a scientific revolution brings problems of multiplication (how many scientific revolutions?) and periodisation (how often?). Some philosophers of science have argued for continuity, others have sought multiple revolutions: the Darwinian revolution, the Quantum revolution, the DNA revolution, and so on, while others claim that the real Scientific Revolution came in the 19th century when science and technology married. Recently, we are witnessing the so-called 4th Industrial Revolution, which conceptualises rapid change to technology, industries, and societal patterns due to increasing interconnectivity, smart automation and the amalgamation of artificial intelligence and automated machines. Yet, is it only a technological revolution, or can machines discover and explain new science? Are we facing the emergence of machine discovery of science?

\begin{figure}[h!]
\centerline{\includegraphics[height=4.5cm]{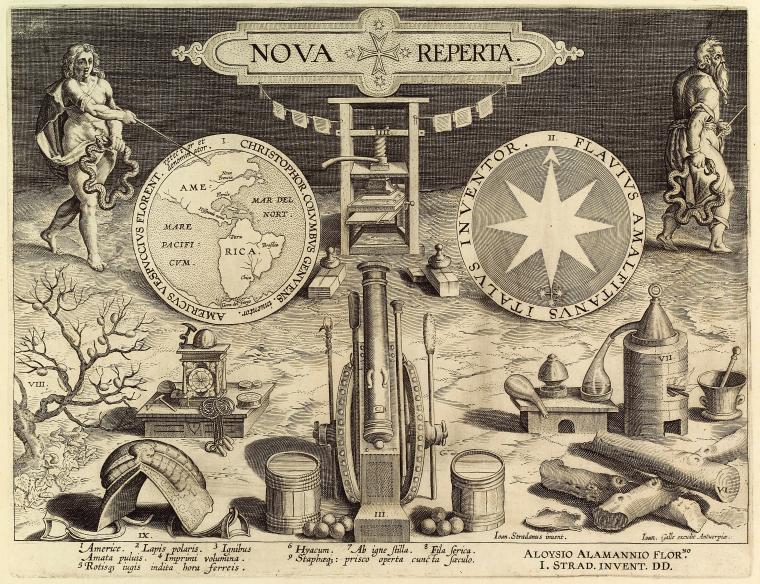}~~~\includegraphics[height=4.5cm]{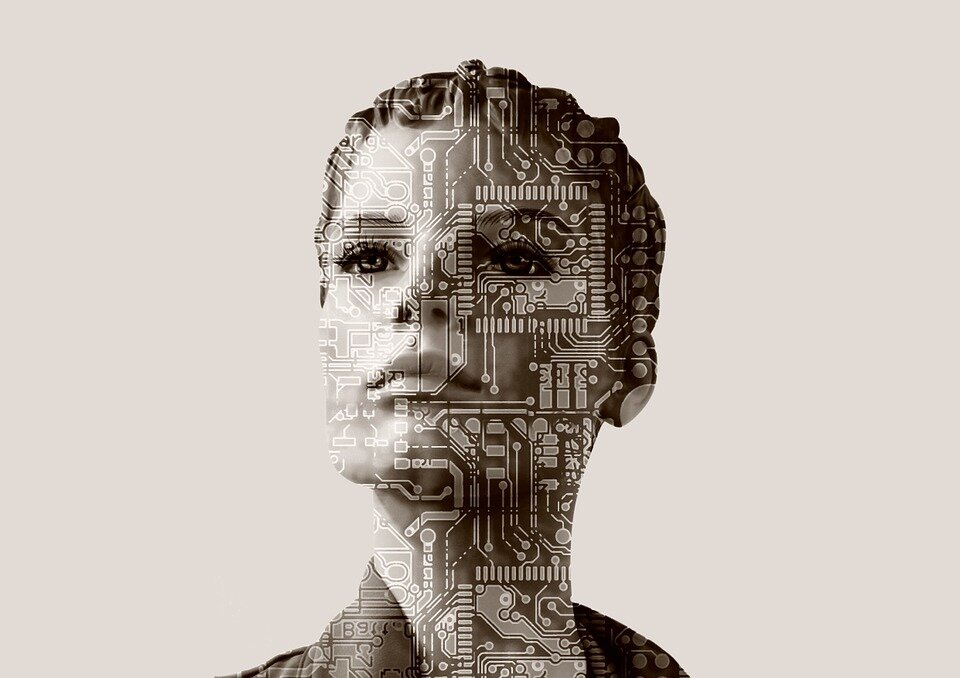}}
\caption{Left: The title page of Johannes Stradanus's New Discoveries ({\em `Nova reperta'}, c.1591) summarises the knowledge that marks off the modern world from the ancient: the discovery of America, the invention of the compass, the printing press, the gunpowder, the clock, silk weaving, distillation and the saddle with stirrups. Right: Can AI start a new scientific revolution? Is AI itself the scientific revolution?}
\end{figure}

\subsection{Elements of scientific discovery}
During the 18th and 19th centuries, the different elements of discovery gradually became separated and discussed in more detail. Discussions concerned the nature of observations and experiments, the act of having insight and the processes of articulating, developing, and testing the novel insight. For Whewell, for example, discovery comprised three elements: the happy thought, the articulation and development of that thought, and the testing or verifying it. %
In contrast to many 20th-century approaches, Whewell's philosophical conception of discovery articulates happy thoughts and integrates the process of verification as an integral part of discovery. Thus, to verify a hypothesis, the scientist needs to show that it accounts for the known facts, that it foretells new, previously unobserved phenomena, and that it can explain and predict phenomena which are explained and predicted by a hypothesis that was obtained through an independent happy thought-cum-colligation \citep{ducasse1951whewell}. Until the late 20th century, most philosophers operated with a narrower notion of discovery than Whewell's. Controversies in the 20th century moved around whether or not the discovery process should include the articulation and development of a novel thought.

\subsection{Elements of knowledge} %

Throughout history, many scientific accounts have described methods of knowledge generation and scientific reasoning without explicitly labelling them as such. These methods include using the senses to gather knowledge, observation and experimentation, analysis and synthesis, induction and deduction, hypotheses, probability, and certainty (Table~\ref{tab:concepts}). By exploring these methods, scientists have generated new knowledge and developed theories about the nature of matter and natural forces. These methods are integral to scientific inquiry and continue to be used today.

In the early modern period, authors such as Bacon and Newton advanced ideas about generating and verifying empirical knowledge and the difficulties that may arise in scientific inquiry. These theories were closely linked to theories about matter and force. However, by the 18th and 19th centuries, authors on scientific method and logic began citing early modern approaches to model proper scientific practice and reasoning. At the same time, the connection between the two was gradually severed. It was common in 20th-century philosophy of science to draw a sharp contrast between those early theories of the scientific method and modern approaches. Yet, recent research in the history of the philosophy of science has shown that the development of the scientific method was a gradual and ongoing process rather than a sudden shift from one approach to another.

Yet, how can we build knowledge? {\em Knowledge} is a well-studied, yet elusive, concept in the philosophy of science \citep{popper2005logic}. Knowledge can be acquired through observation, experimentation, and reasoning and is often expressed through language, symbols, and concepts. In quantitative physical sciences, we encapsulate our domain knowledge in mathematical constructs and equations that describe the system under consideration. Knowledge is built on {\em facts}, that is, {\em observations} about the physical world that are accepted as true and are often the starting point for deeper analysis and understanding. Empiricism offers resorts to perform {\em experiments} that are used to test the validity of facts and to gain further insight into the underlying {\em causes and effects} of phenomena. From there, scientists construct {\em laws}, which are generalisations based on many observations and experiments, the rules describing the behaviour of a system that can be used to predict future outcomes. In the scientific method, {\em hypotheses} are testable explanations for facts and laws based on {\em evidence} but have not been proven true. See Table~\ref{tab:concepts}.

We must use judgement to evaluate the validity of facts, experiments, laws, hypotheses, theories, and evidence. {\em Judgement} involves forming an opinion or deciding based on evidence and knowledge through reasoning and considering our conclusions' potential implications. Judgement is important in all aspects of knowledge, from generating new ideas to applying existing knowledge. It is important to remember that knowledge is not static but an ongoing discovery, analysis, and interpretation process. New theories and hypotheses can be formed as new evidence is discovered, and existing knowledge can be refined or overturned. Formalising new knowledge needs {\em arguments}, which can be inductive, deductive, analogical, and statistical. {\em Inductive} arguments are based on observations and generalisations, while {\em deductive} arguments are based on logical reasoning. {\em Analogical} arguments are based on similarities between two or more cases, while {\em statistical} arguments are based on data and evidence and formally linked to the probability theory. 

\paragraph{Objectivity and subjective conviction.}

Objectivity is the idea that scientific inquiry should be based on factual evidence and logical reasoning, independent of personal bias or opinion.  Objectivity, as a (causal) statement, should be inter-subjectively tested according to Popper \citep{popper2005logic} or be valid, justifiable or verifiable in the sense of Dickerson \citep{dickerson2003kant}. Objectivity can be challenged because some scientific theories can never be fully justifiable or verifiable but testable (i.e. reproducibility). On the contrary, conviction relies on believing the inquiry's findings are true and trustworthy, stating a scientific argument without proofs \citep{popper2005logic}. Conviction implies arguments based on mere association, which are not necessarily causal, verifiable, testable or reproducible and cannot justify a scientific statement \citep{pearl2009causality}.

\paragraph{Universality, falsifiability, and consistency.}

The concepts of universality, falsifiability, and consistency are essential tools for understanding scientific method \citep{popper2005logic} (Table~\ref{tab:concepts}). 
{\em Universality}, the principle of {\em universalism}, is the idea that scientific truths can be applied to all situations. It is a fundamental tenet of scientific inquiry, implying that the same principles can be used to explain phenomena in different contexts. %
This allows scientists to conclude the natural world, not limited to a single instance or context. It also makes experiments more reliable and reproducible, as results can be compared across multiple contexts.
The latter relates to {\em falsifiability} introduced by Popper \citep{popper2005logic}: a scientific hypothesis must be tested and disproved. Falsifiability is important for scientific research, as it allows for the testing and refining of hypotheses. %
The concept of falsifiability is used to evaluate the validity of scientific claims, as it implies that a hypothesis can be disproved if it does not fit the evidence. 
Yet, scientific theories must remain {\em consistent} over time; any change or modification to a scientific theory must be able to be explained by existing evidence. Newton first proposed this concept in the 17th century \citep{newton1833philosophiae}. All three concepts remain an important part of the scientific method.

\paragraph{Empiricism.} 

Empiricism is a philosophical position that asserts that knowledge must be acquired through sensory experience \citep{sellars1956empiricism}. This means knowledge is not innate and can only be gained through observation and experimentation. Empirical evidence is necessary for scientific progress as the only reliable way of assessing the validity of theories and hypotheses. Empiricism is closely associated with the scientific method, as it gathers data and evidence to understand the world comprehensively. This data then form hypotheses and theories explaining why certain phenomena occur worldwide. Empiricism has been challenged most notably by Feyerabend \citep{feyerabend1965problems}, where realism is promoted to enable the proliferation of new and incompatible theories. This way, scientific progress comes through ``theoretical pluralism'' allowing a plurality of incompatible theories, each of which will contribute by competition to maintaining and enhancing the testability, and thus the empirical content, of the others. 

\subsection{Elements of models, governing equations and laws}

\paragraph{Identifiability and equifinality.}

A relevant problem in model development is identifiability, the property by which learning the true values of a model's underlying parameters is theoretically possible from an infinite number of observations \citep{ljung1994global,peters2011identifiability}. The impossibility arises when two more model parameterisations are observationally equivalent, i.e., indistinguishability. This situation is sometimes called equifinality; one can achieve the same result or state description by many potential solutions and model parameterisations. This is a very active field of research in statistics, machine learning and functional analysis, where regularisation helps. Combining domain knowledge with data-driven approaches, for example, may alleviate cases of model misspecification, as the model solution would be confined to a solution subspace with little expressive power and even implausible solutions. 
 
\paragraph{Compressibility, sparsity and compositionally.}

A desirable property of models and governing equations is compositionality, by which theory/models are typically a composition of a small set of elementary functions \citep{udrescu2020ai}. The problem of attaining compositional models is challenging, as misspecification arises when defining model components or facing multidimensional compositions. Compositionality also advocates for sparse models, where combinations of simpler explanations are preferred, directly implementing Occam's razor in the model search. 
Besides, intimately related, we find the desirable property of compressibility by which data/models are compressible if and only if they exhibit a pattern: this way, high-level models are much simpler than their low-level counterparts \citep{dennett1991real,shannon1948mathematical,kolmogorov1963tables}. As for attaining compositional models or achieving the model's sparsity, estimating multivariate information-theoretic estimates is challenging here.

\paragraph{Generalisation, robustness, invariance and extrapolation.}

Models should operate well in unseen but similar situations. The generalisation property (or generalizability) gives confidence in the model's performance, expressive power, and predictions. Models operating in out-of-the-sample regimes tend to extrapolate, thus compromising trustworthiness. Extrapolation is about inferring the unknown from the known, estimating beyond the original observation range, and predicting future data by relying on historical data \citep{scott1993causal}. Since the governing equations and laws of Physics are invariant through space and time, inferred models should be. However, attaining robust models that generalise well and implement properties of invariance is typically challenged by the lack of reliability of the data source, the accuracy of the extrapolation process itself, the complexity of the data, and the potential for errors in the extrapolation.

\begin{table}[p!]
\centering
\tiny
\sf
\caption{Key concepts, definitions, and challenges for understanding complex systems from data. }
\vspace{-0.25cm}
\begin{tabular}{l|p{6cm}|p{6cm}}
\hline
{\bf Concept}  &  {\bf Definition} & {\bf Challenge in causal/equation discovery}\\
\hline

\multicolumn{3}{l}{\em Logic of scientific discovery} \\
\hline
Consistency & theory not leading to logical contradictions; refutes/includes/generalises previous theories;  \citep{godel1931formal,tarski1994introduction,popper2005logic} & impossibility to perform controlled experiments repeated observations, or survey research; cf. hypothesis testing \\
\hline
Deduction  & type of inference where the conclusion follows logically from its premises \citep{johnson1991deduction,popper2005logic} &  deduction breaks under confounders  \\
\hline
Empiricism & principle that only `experience' can decide about the truth or falsity of a factual statement \citep{popper2005logic}; central role of empirical evidence to form ideas, rather than innate ideas or traditions; cf. induction & impossibility to perform experiments or access observations \\
\hline
Explanation & deduction of a statement that describes the event using as premises one or more universal laws, together with certain singular statements (initial conditions) \citep{popper2005logic,pearl2009causality} & no formal principle of causality or universal causation available\\
\hline
Falsifiability & capacity for some proposition, statement, theory, or hypothesis to be proven wrong \citep{popper2005logic} & discovered models and laws can be misspecified, untestable or incomplete\\
\hline
Induction &  method of reasoning where a general principle is derived from a set of observations \citep{popper2005logic,munz2014our,reichenbach1991direction,keynes2013treatise}, cf. generalisation (sample to population), probabilistic inference & impossibility, data scarcity or conditions; unlike in deduction, the truth of the conclusion is only probable\\
\hline
Objectivity & (causal) statement that can be inter-subjectively tested \citep{popper2005logic} or that is valid, justifiable or verifiable \citep{dickerson2003kant} & elusive term, scientific theories can never be fully justifiable or verifiable, but testable, cf. reproducibility \\
\hline
Conviction  & stating a scientific argument without proofs \citep{popper2005logic} & mere association is not causal, verifiable, testable or reproducible and cannot justify a scientific statement   \\
\hline
Universality & universal facts exist and can be progressively discovered, as opposed to relativism \citep{popper2005logic} & model misspecification, reproducibility, theory testability; Universal laws transcend any finite number of their observable instances\\
\hline

\multicolumn{3}{l}{\em Equation discovery: model properties and data challenges} \\
\hline

Compositionality & theory models are typically a composition of a small set of elementary functions \citep{udrescu2020ai} &  misspecification, multidimensional compositions  \\
\hline
Compressibility & data/models are compressible iff exhibit a pattern, high-level models are much simpler than their low-level counterparts \citep{dennett1991real,shannon1948mathematical,kolmogorov1963tables} &  multivariate information-theoretic estimates are difficult, cf. sparsity \\
\hline
Extrapolation & inferring the unknown from the known, estimation beyond the original observation range, predicting future data by relying on historical data  \citep{scott1993causal} & reliability of the data source, the accuracy of the extrapolation process itself, the complexity of the data, and the potential for errors in the extrapolation   \\
\hline
Generalization & model's ability to adapt to new, unseen data drawn from the same/similar distribution as the one used to create the model \citep{vapnik1999overview} &  hard to evaluate, pointless in regime/domain shift, cf. extrapolation  \\
\hline
Identifiability  & learning the true values of a model's underlying parameters is theoretically possible from an infinite number of observations \citep{ljung1994global,peters2011identifiability} & two or more model parameterisations are observationally equivalent, cf. indistinguishability, equifinality \\
\hline
Invariance & probability of a certain event occurring remains the same, regardless of any changes to the underlying system or process \citep{olver1999classical,kallenberg2005probabilistic} & strong assumption in many complex systems  \\
\hline
Robustness & property of models to perform well in noisy environments (presence of outliers, mixed noise distributions, missing data), extrapolation regimes, covariate shift &  hard-to-identify regimes, design model's regularisers, cf. inductive bias, priors, Bayesian inference  \\
\hline
Separability & model can be written as a sum or product of two parts with no variables in common \citep{udrescu2020ai}  & misspecification, oversimplification\\
\hline
Smoothness     & model is continuous and perhaps even analytic in its domain \citep{scholkopf2021toward}  & oversimplification, inappropriate to deal with regime shifts and anomalies \\
\hline
Symmetry  & model exhibits translational, rotational, or scaling symmetry concerning some of its variables \citep{udrescu2020ai} & inappropriate for unstructured data, macroscale-vs-microscale systems  \\
\hline
Uncertainty & epistemic situations involving imperfect or unknown information; it applies to predictions of future events, to physical measurements that are already made, or to the unknown \citep{lindley2013understanding,keynes2013treatise} & partially observable or stochastic environments, yet difficult to disentangle the sources and to quantify it properly\\
\hline
Units & models \& corresponding variables have known physical units \citep{udrescu2020ai} & dealing with unknown factors and confounders  \\
\hline

\multicolumn{3}{l}{\em Causal inference: discovery, cause-effect estimation \& counterfactuals} \\
\hline

Causal discovery & Reconstructing the full causal graph or pairwise causation from data and assumptions \citep{Pearl2018a,Peters2017book} & challenging, limited data, poor assumptions, incomplete knowledge, untestable\\
\hline
Causal effects & Estimating causal effects (of hypothetical interventions) in terms of expectations or full interventional distributions \citep{winship1999estimation,pearl1995causal} & incomplete knowledge, untestable\\
\hline
Confounder  & variable causally associated with both exposure and outcome and not on the causal pathway between $X$ and $Y$ \citep{pearl2009causality,morgan2015counterfactuals,spirtes1995causal} & unobserved processes, hidden confounding, bidirected edges, periodicity, slow-time scales, discrete states\\
\hline
Counterfactuals  &  Attribution of observed events or mediation analysis \citep{pearl1995causal,morgan2015counterfactuals}& incomplete knowledge, untestable\\
\hline
Covariate shift & The system's probability density function changes through time or space \citep{sugiyama2012machine} & hard to detect and correct models for, cf. invariance \\
\hline 
Dependencies & models trained to maximise statistical measures of linear or nonlinear association, or in-distribution prediction \citep{kotz2001correlation} & non-causal, right-for-the-wrong-reasons\\
\hline
Faithfulness & only the variables that are $d$-separated in a DAG will be independent (i.e., all others will be dependent) \citep{pearl2009causality} & requires an understanding of how different variables work together to cause an outcome \\ 
\hline
Missingness &  unobserved quantities or random variables, either missing at random or through particular mechanisms \citep{mcknight2007missing} & potentially leading to selection bias \\
\hline
\end{tabular}
\label{tab:concepts}
\end{table}

\section{Knowledge discovery from data}

\subsection{Discoverability and heuristic strategies}
The questions of what and how phenomena and mechanisms can be discovered have been the subject of intense research and philosophical discussion. In the philosophy of science, {\em discoverability} is the concept that scientific knowledge must be discoverable and verifiable \citep{ducasse1951whewell, popper2005logic, langley2019scientific, langley1987scientific,klahr1999studies}.  This means that hypotheses or theories must be supported by evidence and based on empirical observations and data. Furthermore, scientific knowledge must be available for anyone to see and verify, ensuring that it is not biased or limited to a specific group of people. %

Recent advancements in the philosophy of science have seen a revival of interest in {\em heuristic strategies} to discover knowledge; these strategies are seen as problem-solving activities, whereby a discovery is a solution to a problem. Heuristics-based discovery methodologies are neither completely subjective and intuitive nor algorithmic or formalisable. This view has shifted the scientific researcher from being viewed as a `puzzle solver' to a `problem solver' and `decision maker' in complex, variable, and changing environments \citep{wimsatt2007re}. In this paper, we will review mathematical models that address equation discovery and causal discovery by generally formalising the problems as concrete statistical inference tasks; regression, conditional dependence or density estimation.

\subsection{Modern approaches to data-driven discovery}

Observational discovery relies on modelling. Yet, what types of models? Table \ref{tab:models}, cf. \citep{peters2017elements}, gives a simple categorisation of models from mechanistic/physical models based on first principles and (rigid) equations and laws but with desirable properties of interpretability, invariance and robustness to distribution shifts, to purely statistical (machine learning) models that excel in prediction and are learned from data. In the middle, we have structural causal models, which can answer counterfactual questions but do not necessarily capture physical knowledge. All three models are used in quantitative data-driven science and map to different levels of discovery: %
learning statistical associations in data streams, identifying causal relations between variables, and discovering equations from data. Note the resemblance to the Ladder of Causation proposed by \citet{Pearl2018a} with three rungs: association, intervention, and counterfactual. Discovering causal and physical laws from observations is a paradigm shift in AI and can impact the physical sciences and other disciplines. The fields of discovery of scientific knowledge and causal models of scientific phenomena are intertwined and tightly connected: our scientific endeavour is constantly challenged with causal questions, robust model building, intervention analysis, and hypothesis testing. The fields also share important theoretical challenges, where generalizability, compressibility, robustness, invariance, and extrapolation come into play. %

\begin{table}[h!]
    \centering
    \small
    \caption{Simple taxonomy of models and their level. Figure partly reproduced from \citep{Peters2017book}.}
    \label{tab:models}
    \begin{tabular}{l|c|c|c|c|c}
    \toprule
       {\bf Model/Level}  &  {\bf i.i.d.} & {\bf distribution shifts} & {\bf Counterfactuals} & {\bf Physical insight} & {\bf Data-driven} \\
       \midrule
       3 - Mechanistic  & \cmark & \cmark & \cmark & \cmark & \cmark\\
       \midrule
       2 - Structural causal  & \cmark & \cmark & \xmark & {\bf ?} & {\bf ?}\\
       \midrule
       1 - Statistical/ML  & \cmark & \xmark & \xmark & \xmark & {\bf ?}\\
    \bottomrule
    \end{tabular}
\end{table}

\paragraph{Level 1 -- Learning statistical associations.} %

The most rudimentary approach to building association links %
from multivariate time series data involves computing pairwise Pearson's correlations or mutual information, which capture relationships between variables at lag zero. Networks derived from these measures have found applications in numerous scientific and engineering domains, including climate network analysis \citep{donges2009complex}, financial market network analysis \citep{fiedor2014networks}, and brain network analysis \citep{Johnson2007BRAINSFitMI,Takagi2020PrinciplesOM}. However, mutual information-based associations at lag zero cannot be interpreted directionally since the term ``information'' implies a lack of directionality. Other regions or variables may influence the nodes under investigation, or the association could be due to a common driving process. Lagged association measures are commonly used to account for directional links and quantify the time lag of associations.

Lagged correlation analysis has a long history in climate research \citep{Walker1923} and neuroscience \citep{medaglia2015cognitive,richiardi2013machine}, with the delay at the maximum of the cross-correlation function being used to interpret the delay of the underlying physical mechanism coupling two processes. Other lagged measures of association, such as mutual information \citep{granger1969investigating}, %
have been proposed to determine lags in nonlinear processes. In addition to analysing time lags, the magnitude of the cross-correlation is often used as a measure of the impact of one process on another or as a measure of the strength of an association. This aligns with the statistical interpretation of the square of correlation as the proportion of variance in one process that can be linearly represented by another \citep{von2002statistical, chatfield2013analysis}.

However, relying solely on association measures, even those that account for lags and nonlinearity, cannot uncover directionality, detect the delay of the underlying mechanism, or provide a physically or causally interpretable estimate~\citep{runge2014quantifying}. The widespread use and abuse of association measures in engineering and science throughout the 20th century have impeded the exploration and development of meaningful causation measures and hindered the discovery of new and alternative explanatory laws from data.

\paragraph{Level 2 -- Learning causal relations from observations.} %

A fundamental objective in the scientific enterprise is understanding the causes behind the phenomena we observe \citep{Pearl2000,Peters2017book}. This is particularly challenging in disciplines dealing with complex dynamical systems, where experimental interventions are expensive, unethical, or practically impossible. In some fields (\eg, climate sciences, economics, cardiology, and neurosciences), the current alternative is to rely on computationally expensive simulation experiments. Still, those do not adequately represent all relevant physical processes involved. At the same time, a rapidly increasing amount of time series data is generated by observations and also models. How can we use this wealth of data to gain new insights into our fundamental understanding of these systems?

In recent years, rapid progress has been made in computer science, physics, statistics, philosophy, and applied fields to infer and quantify potential causal dependencies from data without intervening in the systems. Although the truism that correlation does not imply causation holds, the key idea shared by several approaches follows Reichenbach's common cause principle \citep{reichenbach1991direction}: if variables are dependent, then they are either causal to each other (in either direction) or driven by a common driver. To estimate causal relationships among variables, different methods take different, partially strong, assumptions. Granger \citep{granger1969investigating} addressed this question quantitatively using prediction. At the same time, in the last decades, several complementary concepts emerged, from nonlinear dynamics \citep{sugihara2012detecting} based on attractor reconstruction to computer science exploiting statistical independence relations in the data \citep{pearl1995causal,Pearl2000}. More recently, statistics and machine learning research utilised the framework of structural causal models (SCMs) \citep{Peters2017book} for this purpose. Causal inference from data is becoming a mature scientific approach \citep{Pearl2000}. 

Causal inference combines domain knowledge, ML models, and data to learn the underlying system's causal structure and quantify causal effects~\citep{Pearl2009, Pearl2016, Pearl2018a, Spirtes2000,Peters2017book}. 
Causal inference can leverage observational or (interventional) model output data~\citep{Pearl2009} to learn, understand and evaluate the plausibility of specific causal relations among the considered variables. 
Today, many methods and tools are available to address challenges in complex systems~\citep{wagner1999causality,sugihara2012detecting} and many other fields. Causality is pivotal not only for a better academic understanding of processes in science but also for more robust forecasts, attributing the causes of events, and improving the physics embedded in physics models. Many fields of science and engineering are using causal inference/discovery methods, from Earth and climate sciences~\citep{runge2019inferring, PerezSuay19shsic, ebert2017causal, raia2008causality, reitsma2010geoscience, niemeijer2008framing, shepherd2019storyline, goodwell2020debates}, neurosciences \citep{reid19,siddiqi2022causal,stokes2017study}, social sciences~\citep{marini1988causality,russo2010causality}, health and epidemiology \citep{hernan2018c,glass2013causal,hernan2020causal}, or economics \citep{hicks1980causality,leroy2004causality}.

\paragraph{Level 3 -- Equation discovery in physical systems.} 

The scientific enterprise distinctly differs from other intellectual endeavours by relying on formal theories, laws, and models to explain and predict observations and using such observations to construct, revise, and evaluate its formal statements \citep{langley1987scientific,langley2019scientific,klahr1999studies}. Many of these activities have been studied by philosophers of science for over a century. The Logic of Science \citep{popper2005logic} (or justification) aims to characterise how observational data, simulations, and experiments can collectively support or refute laws, models, or theories. 

A common claim was that scientific discovery requires some `creative spark', which cannot be analysed rationally or logically \citep{simon1989scientist}. Popper, Hempel, and many other philosophers of science maintained that the discovery process was inherently irrational and beyond any formal understanding. The key insight came from Simon \citep{popper2005logic,hempel2001philosophy,simon1989scientist}, who proposed that scientific discovery, rather than {\em ``depending on some unknown mystical ability, is a variety of problem-solving that involves searching through a space of problem states generated by applying mental operators and guided by heuristics to make the search tractable.''} 
Such observation established the first heuristic programming methods of hypothesis (model) search to automate the creative process and law discovery. Discovering numeric laws from data has been approached by many authors in the past using grammars, logic rules, propositional bases, entailment, and genetic algorithms, to name a few \citep{falkenhainer1986, kokar1986,zytkow1990,schaffer1990, nordhausen1990, langley2019scientific, moulet1992, gordon1994, murata1994, dzeroski1995, washio1997, bradley2001, koza2001}. Later, \citep{schmidt2009} proposed an automated algorithm to discover Hamiltonians, Lagrangians, and other geometric and momentum conservation laws without prior knowledge of physics, kinematics, or geometry. 
A new field was born; learning explicit mathematical laws from observations, which was often referred to as {\em equation discovery} or {\em data-driven system identification} \citep{simidjievski2020equation}. 

Inspired by earlier work on the DENDRAL system \citep{feigenbaum1971}, which inferred structural models of organic molecules from their mass spectra, the community developed different systems that created models of other scientific phenomena (\eg, \citep{Langley1987}). The field was named computational scientific discovery, and the challenge of automating it has been approached by many researchers since then \citep{evans2010machine,fortunato2018science,Bongard2007,schmidt2011automated,waltz2009automating,king2009robot}. Efforts in this paradigm differ from mainstream work in machine learning by producing  scientific formalisms \citep{langley2002a,langley2002b,langley1987scientific}, ranging from componential models in particle physics \citep{kocabas1991} to reaction pathways in chemistry and to regulatory models in genetics \citep{king2004}. Reviews have been edited by \citet{shrager1990,dzeroski2007,simidjievski2020equation}.

Recently, the field has been approached by scientists in AI, functional analysis and mathematical operators, nonlinear control,  and system identification. Modern approaches that we will review in this paper consider: Automated reverse engineering of nonlinear
dynamical systems \citep{Bongard2007}, sparse-promoting solutions that identify parsimonious models of nonlinear dynamics; \eg relevance vector machine, a sparse Bayesian regression method \citep{tipping2001sparse} or the SINDy method \citep{brunton2016discovering,brunton2022data}, which has been combined with deep neural networks \citep{champion2019data}, reduced-order models \citep{mezic2005spectral} and Koopman operators based on kernel theory and autoencoders \citep{kaiser2020data, kaiser2021data,Kostic2022, Klus2020,Lusch2018}, differentiable networks that can learn the true underlying equation and extrapolate to unseen domains \citep{SahooLampertMartius2018:EQLDiv}, a constrained symbolic regression methodology, named AI Feynman, that enforces desirable constraints in equation learning (compositionality, units, separability, symmetry, smoothness) \citep{udrescu2020ai}, genetic programming to distil laws of physics \citep{SchmidtLipson}, or a transformer-based architecture using massive pretraining can predict formulas from data \citep{Biggio2021:NeSymReS}.

A relevant challenge in this context is the discovery of state variables from experimental/observational data. Almost in all equation and causal discovery algorithms, the variables are given or assumed, which is impossible when trying to understand new or highly complex systems. Some approaches exist in the literature based on the combination of learning compact and expressive feature representations, manifold learning for the determination of the intrinsic dimensionality of the system representation coordinates, and SINDy as a regulariser that enforces dynamic equations in the state space \citep{brunton2022data,chen2022automated,pukrittayakamee2009simultaneous,schutt2017quantum}. These approaches are somewhat related to the discovery of causal relations under noise and latent functions, which is also an active field of research \citep{probaLatZ,autoReg,Diaz22noiseimputation}.

\subsection{AI for scientific discovery}

\begin{wrapfigure}{r}{5cm}
\vspace{-0.5cm}
\begin{mdframed}[backgroundcolor=gray!20] 
\small
{\em ``With the advent of data-driven methods that learn patterns and relations from data, the tedious human endeavour of scientific discovery (laws, equations and causes of phenomena) is being revolutionised... and accelerated in many fields.''}
\end{mdframed}
\vspace{-0.5cm}
\end{wrapfigure}
The debate about AI-based theories of scientific discovery has been ongoing for decades, beginning with whether computers can devise new concepts or merely process the concepts already included in a given computer language. However, the discussion has been revived with the development of new computational tools for data analysis. It is now largely uncontroversial that machine learning tools can aid discovery, though there is still debate about whether they generate new knowledge or merely speed up data processing. Moreover, there is the question of whether data-intensive science fundamentally differs from traditional research and the ethical implications of ``superhuman AI''. Philosophers have also focused on the opacity of machine learning, asking whether we can say that humans and machines are ``co-developers'' of knowledge. Ultimately, the debate about AI-based theories of scientific discovery is still ongoing, with researchers considering both the potential benefits and ethical implications of such tools.

The fields of equation discovery and causal inference that we will review in this paper promise to shed light on the previous fundamental questions: what can an algorithm learn and discover, what can AI explain, and what new science may emerge through a collaborative AI-machine dialogue about science? An integrative approach seems necessary, where domain experts, data and machine work together in a data-driven framework that formulates and answer causal questions and discover new laws \citep{Pearl2000}. At a more general level, it becomes pertinent to ask if machines can start a new scientific revolution and even if AI itself is the ultimate scientific revolution \citep{russell2021human,boden2014creativity,gillies1996artificial}. 

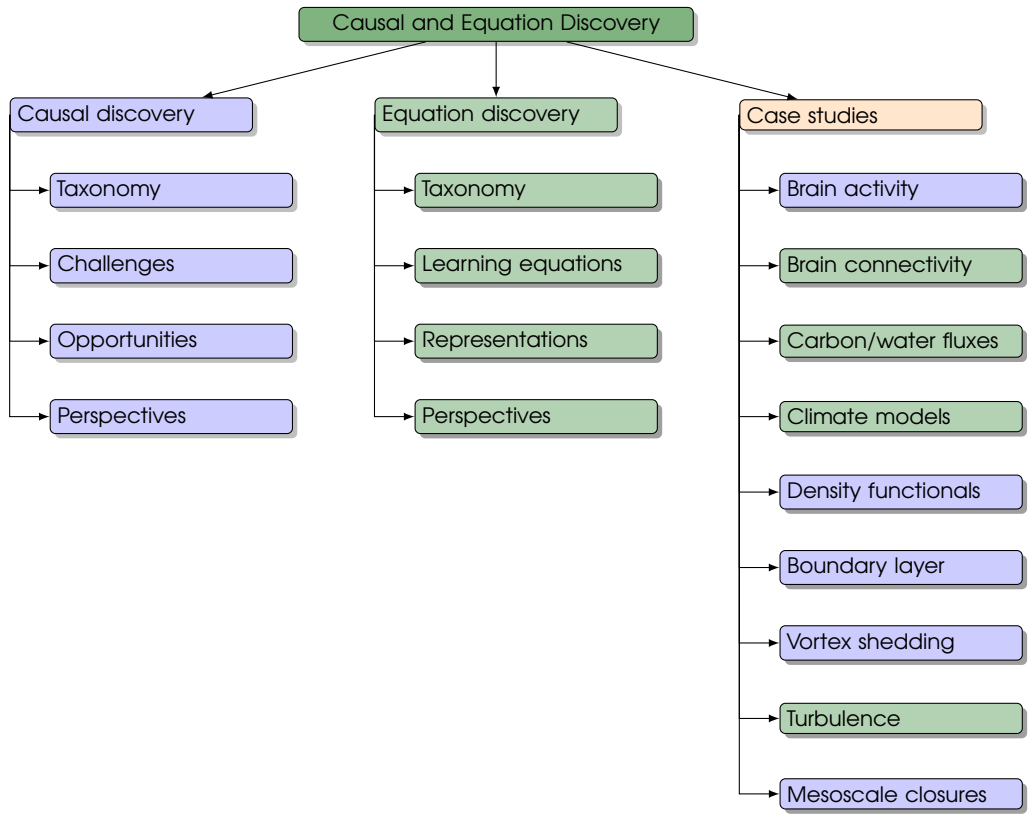
\begin{figure}[t!]
\centering
\scriptsize
\begin{tikzpicture}[scale=0.8, level 1/.style={sibling distance=60mm}, edge from parent/.style={->,draw}, >=latex]

\node[root] {Causal and Equation Discovery}
  child {node[level 2] (c1) {Causal discovery}}
  child {node[level 3] (c2) {Equation discovery}}
  child {node[level 4] (c3) {Case studies}};

\begin{scope}[every node/.style={level 2}] 
\node [below of = c1, xshift=15pt] (c11) {Taxonomy};
\node [below of = c11] (c12) {Challenges};
\node [below of = c12] (c13) {Opportunities};
\node [below of = c13] (c14) {Perspectives};

\node [below of = c2, xshift=15pt, style={level 3}] (c21) {Taxonomy};
\node [below of = c21, style={level 3}] (c22) {Learning equations};
\node [below of = c22, style={level 3}] (c23) {Representations};
\node [below of = c23, style={level 3}] (c24) {Perspectives};

\node [below of = c3, xshift=15pt, style={level 2}] (c31) {Brain activity};
\node [below of = c31, style={level 3}] (c32) {Brain connectivity};
\node [below of = c32, style={level 3}] (c33) {Carbon/water fluxes};
\node [below of = c33, style={level 3}] (c34) {Climate models};
\node [below of = c34, style={level 2}] (c35) {Density functionals};
\node [below of = c35, style={level 2}] (c36) {Boundary layer};
\node [below of = c36, style={level 2}] (c37) {Vortex shedding};
\node [below of = c37, style={level 3}] (c38) {Turbulence};
\node [below of = c38, style={level 2}] (c39) {Mesoscale closures};

\end{scope}

\foreach \value in {1,...,4}
  \draw[->] (c1.180) |- (c1\value.west);

\foreach \value in {1,...,4}
  \draw[->] (c2.180) |- (c2\value.west);

\foreach \value in {1,...,9}
  \draw[->] (c3.180) |- (c3\value.west);
\end{tikzpicture}
    \caption{Roadmap of the article.}
    \label{fig:roadmap}
\end{figure}

\section{Outline}

This paper aims to review the most important concepts, methods, and previous works on causal inference and discovery in the physical sciences. We use statistical learning techniques to discover causal relations, physical laws, and governing equations from data.  %
\secref{sec:causality} and \secref{sec:equationdiscovery} present general frameworks and taxonomies for causal discovery and learning physical laws from data, respectively. Both sections categorise the field, reviewing concepts and methods, their specific characteristics, challenges, and opportunities in the physical sciences. 
\secref{sec:casestudies} provides examples of causal discovery and equation discovery in a wide range of fields of the physical sciences: dynamical systems, neuroscience, classical and quantum systems, fluid mechanics, geosciences and climate sciences. We pay attention to how causality concepts and methods can improve our knowledge of a given physical system from observations. \secref{sec:conclusions} outlines the most promising future lines of research in this area of study at the intersection of machine learning and nonlinear physical processes.

%% file: 02_CausalDiscovery_book.tex
\chapter{Causal discovery in the physical sciences}\label{sec:causality}

Causal discovery, see for example \citep{Spirtes2000, Peters2017book} for extended expositions of the topic, has become increasingly popular in the last years as a tool to discover the underlying causal structure of physical systems \citep{runge2019inferring}. There is an abundance and ever-growing number of methods designed to work under different assumptions and tackle other use cases.
This section reviews several methods for causal discovery, focusing on methods for time series and their potential use in physical sciences. To this end, in \secref{sec:causal-discovery-taxonomy}, we first provide a taxonomy for many available causal discovery methods. \secref{sec:causal-discovery-challenges} discusses causal discovery's challenges in real-world applications. We conclude in \secref{sec:causal-discovery-opportunities} by discussing opportunities for applications of causal discovery in the physical sciences.
We would also like to point to other reviews focusing on causal discovery of time series~\citep{runge2019inferring,assaad2022survey,moraffah2021causal}. In addition, \citet{runge2023causal} provides a shorter accessible summary of methods for causal discovery and causal effect estimation with practical case studies to illustrate typical challenges, such as contemporaneous causation, hidden confounding and non-stationarity.

\section{A taxonomy of causal discovery methods}\label{sec:causal-discovery-taxonomy}

\begin{table}[h!]%
\scriptsize
\begin{center}
\caption{Taxonomy of methods for causal discovery. The entries in parentheses $(\cdot)$ indicate that there are versions of the algorithm %
in which the assumption is relaxed. 
We use the abbreviations `TSG' for time series graph, `summary' for summary graph and `ext. sum. graph' for an extended summary graph.
\label{tab:causal_methods}}
\begin{tabular*}{\textwidth}{p{0.18\textwidth}|p{0.08\textwidth}|p{0.08\textwidth}|p{0.08\textwidth}|p{0.08\textwidth}|p{0.07\textwidth}|p{0.06\textwidth}|p{0.08\textwidth}|p{0.06\textwidth}}
\hline
 \multicolumn{1}{@{}c@{}|}{\bf Method} & \multicolumn{2}{@{}c@{}|}{\bf Target of Inference} &  \multicolumn{1}{@{}c@{}|}{\bf Approach} & \multicolumn{4}{@{}c@{}|}{\bf Process Assumptions} & {\bf Data Assumption} \\%\multicolumn{1}{@{}c@{}}{Data}  \\ 
\hline
 &  Bi-/ Multi- variate  & Graph type  & Indep./ Asymm./ Score & non- /linear & Stoch. / Det.  & Con- temp. &Cycles & Hidden var.\\
\hline
GC \citep{granger1969investigating} &  Bi. & summary & indep. & linear & stoch.  & \xmark & lagged-only & \xmark \\\hline
Mult-GC \citep{geweke1982measurement} &  Multi. & summary &indep. & linear & stoch. & \xmark & lagged-only & \xmark\\\hline
Mult-nonlin-GC \citep{bueso2020explicit}  & Multi. & summary & indep. & nonlin. & stoch. & \xmark & lagged-only & \xmark\\\hline
TE \citep{schreiber2000measuring}  &  Bi. & summary & indep. & nonlin. & stoch. & \xmark & lagged-only & \xmark \\\hline
Multi-TE \citep{barnett2009granger} &   Multi.& summary &indep. & nonlin. & stoch. & \xmark & lagged-only & \xmark \\\hline
CCM \citep{sugihara2012detecting} & Bi. & summary & indep.(?) & nonlin. & det. & \cmark & ? & part\-ially\\\hline
Ext.-CCM \citep{Diaz21rccm}  &  Bi. & summary & indep.(?) & nonlin. & det. & \cmark & ? & part\-ially\\\hline
tsPC \citep{runge2020discovering}  &Multi. & TSG &  indep. &  both & stoch. & \cmark & (\cmark) & \xmark\\\hline
PCMCI \citep{runge2015identifying} &  Multi. & TSG &indep. & both & stoch. & \xmark & lagged-only & \xmark\\\hline
PCMCI$^+$ \citep{runge2020discovering} &  Multi. & TSG &indep. & both & stoch. & \cmark & (\cmark) & \xmark\\\hline
PCGCE \citep{assaad2022discovery} &  Multi. & ext.~sum. graph& indep. & both & stoch. & \cmark & (\cmark) & \xmark\\\hline
FCIGCE \citep{assaad2022discovery} &  Multi. & ext.~sum. graph& indep. & both & stoch. & \cmark & (\cmark) & \cmark\\\hline
tsFCI & Multi. & TSG & indep. & both & stoch. & (\cmark) & (\cmark) & \cmark\\\hline
SVAR-FCI \citep{assaad2022discovery} &  Multi. & TSG &indep. & both & stoch. & \cmark & (\cmark) & \cmark\\\hline
SVAR-GFCI \citep{malinsky2018causal} &  Multi. & TSG & score \& indep. & both & stoch. & \cmark & (\cmark) & \cmark\\\hline
LPCMCI \citep{gerhardus2020high} & Multi. & TSG &indep. & both & stoch. & \cmark & (\cmark) & \cmark\\\hline
(F)GES \citep{Meek1997GraphicalMS,chickering2002optimal,chickering2002learning,ramsey2017million} &   Multi. & summary & score & linear & stoch. & \cmark & lagged-only & \xmark\\\hline
DYNOTEARS \citep{pamfil2020dynotears}  &  Multi & TSG &score & linear & stoch. & \cmark & lagged-only & \xmark\\\hline
IDYNO~\citep{gao22IDYNO}  &  Multi & TSG &score & linear and non-linear & stoch. & \cmark & lagged-only & \xmark\\\hline
NTS-NOTEARS~\citep{sun2023nts}  &  Multi & TSG &score & linear and non-linear & stoch. & \cmark & lagged-only & \xmark\\\hline
TiMiNo \citep{peters2013causal}  &  Multi. & TSG & indep. &both & stoch. & \cmark & lagged-only & \xmark\\\hline
RHINO~\citep{gong2023rhino}  &  Multi. & TSG & indep. &both & stoch. & \cmark & lagged-only & \xmark\\\hline
VARLiNGAM \citep{shimizu2006linear} &  Multi. & TSG & asymm. & linear & stoch. & \cmark & lagged-only & \xmark\\
\hline
\end{tabular*}
\end{center}
\end{table}

In this section, we structure the zoo of existing causal discovery methods to guide method users in finding a method suitable for their application and guide method developers in identifying open challenges. %
To this end, we summarise the central formal aspects of the graphical-model-based causal inference framework in \secref{sec:scms-et-al}. Then, in \secref{sec:axes-of-distinction}, we discuss several characteristics (hereafter referred to as ``axes'') by which methods can be conceptually distinguished. Lastly, in \secref{sec:list-of-causal-discovery-methods}, we present an extensive (but not exhaustive) list of causal discovery methods and characterise these methods according to previously introduced axes.

\subsection{Preliminaries}\label{sec:scms-et-al}
At the heart of the graphical-model-based causal inference framework are \emph{structural causal models (SCMs)}, \eg \citep{bollen1989structural, pearl2009causality, Peters2017book}. An SCM serves as a causal model for the data-generating process and specifies how the system reacts to \emph{interventions}, that is, to idealised experimental manipulations that deliberately hold fixed a subset of the system's variables while not perturbing the system in any other way.

An SCM for a system described by the set of variables $\mathbf{V} = \{V^1, \, \ldots, \, V^n\}$ consists of $n$ so-called \emph{structural assignments}
\begin{equation}\label{eq:scm-basic}
V^i := f^i (pa^i, \epsilon^i) \quad \text{with $1 \leq i \leq n$} \, ,
\end{equation}
together with a product distribution $p_{\epsilon}(\epsilon^1, \, \ldots, \epsilon^n) = p_{\epsilon}^1(\epsilon_1) \cdot \ldots \cdot p_{\epsilon}^n(\epsilon_n)$ of the random variables $\epsilon^i$. Formally, the $f^i$ are measurable functions that depend non-trivially on all of their input arguments, and the $pa^i \subseteq \mathbf{V} \setminus \{V^i\}$ are subsets of the system variables $V^1, \, \ldots, \, V^n$. The functions $f^i$ are interpreted as the causal mechanisms by which the values of the respective variable $V^i$ are determined from the value of $\epsilon^i$ and the values of the variables in $pa^i$. Consequently, the variables in $pa^i$ are referred to as the \emph{causal parents} of $V^i$. The random variables $\epsilon^i$ are interpreted as noise that summarises all factors that are not modelled explicitly, and the factorisation of $p_{\epsilon}(\epsilon^1, \, \ldots, \epsilon^n)$ amounts to the assumption that the $\epsilon^1, \, \ldots, \epsilon^n$ are jointly independent. This joint independence is motivated by the view that any dependence between the noise variables must be due to a causal relationship between them and that such a dependence should then rather be modelled explicitly by enlarging the set $\mathbf{V}$ of system variables.

The \emph{causal graph} of an SCM with system variables $V^1, \, \ldots, \, V^n$ is the directed graph whose vertices are the variables $V^i$ and with a directed edge $V^i \rightarrow V^j$ if and only if $V^i$ is a causal parent of $V^j$, that is, if and only if $V^i \in pa^i$. Consequently, the causal graph of an SCM shows the \emph{qualitative} cause-and-effect relationships as specified by the sets $pa^i$. If the causal graph is acyclic, that is, if the causal graph is a directed acyclic graph (DAG), then the SCM is said to be \emph{acyclic}.

An SCM obtains causal meaning by asserting how the modelled system reacts to interventions. Formally, an intervention on the variable $V^k \in \mathbf{V}$ is a mapping, conventionally denoted as $do(V^k := v^k)$, that maps the original SCM and a number $v_k$ to a new SCM in which the original structural assignment for $V^k$ is replaced by the new structural assignment $V^k := v^k$ and the noise variables $\epsilon^k$ is removed. This new SCM is referred to as an \emph{intervened SCM}, and $do(V^k := v^k)$ is interpreted as an idealised experimental manipulation by which the value of $V^k$ is held fixed at $v_k$ while leaving the system unaltered else. This specification of how the system reacts to interventions is why using the symbol ``$:=$'' instead of ``$=$'' in \eqref{eq:scm-basic} is conventional. On the level of causal graphs, $do(V^k := v^k)$ amounts to removing all edges that point into $V^k$ because, in the intervened SCM, the variable $V^k$ has no causal parents. Interventions on subsets of variables are defined similarly.

In an acyclic SCM, the combination of noise distribution $p_{\epsilon}$ and functions $f^i$ uniquely determines a distribution of the system variables $V^1, \, \ldots, \, V^n$. This distribution is often referred to as the \emph{entailed distribution of the SCM}. The entailed distribution $p(\cdot)$ of the original SCM (that is, of the SCM that models that system without interventions) is often referred to as the \emph{observational distribution}. The entailed distributions of the intervened SCMs are often referred to as \emph{interventional distributions} and are conventionally often denoted as $p(\cdot \vert~ do(V^k :=v^k))$; and similarly for interventions on subsets of system variables.

When using this notation, it is important to keep in mind that $p(\cdot \vert~ do(V^k :=v^k))$ is, in general, not equal to $p(\cdot \vert~ V^k =v_k)$. Indeed, $p(\cdot \vert~ do(V^k :=v^k))$ is the distribution of the \emph{intervened} SCM, whereas $p(\cdot \vert~ V^k =v_k)$ corresponds to \emph{observing} $V^k = v^k$; put differently: Correlation is not equal to causation.

The article \citep{bongers2021foundations} discusses in much detail the more complicated case of cyclic SCMs. As shown there, cyclic SCMs need not entail a unique distribution for the system variables. However, \citep{bongers2021foundations} defines a restricted class of cyclic SCMs, termed \emph{simple SCMs}, that entail a unique distribution and that are closed under interventions. Moreover, acyclic SCMs are a special case of simple SCMs.

In the time series case, which we are predominantly interested in this paper,~\eqref{eq:scm-basic} can be generalised by putting a time index $t$ on $V^i$, $f^i$, $pa^i$ and $\epsilon^i$. The commonly used term \emph{causal stationary} then refers to time-invariance of the qualitative cause-and-effect relationships, that is, to the situation that $pa^i_{t+\Delta t} = \{V^j_{s+\Delta t} ~\vert~ V^j_s \in pa^i_t\}$ for all $t$ and $\Delta t$.

\subsection{Axes for categorising causal discovery methods}\label{sec:axes-of-distinction}
\begin{figure}[h!]
    \centering
    \includegraphics[width=\linewidth]{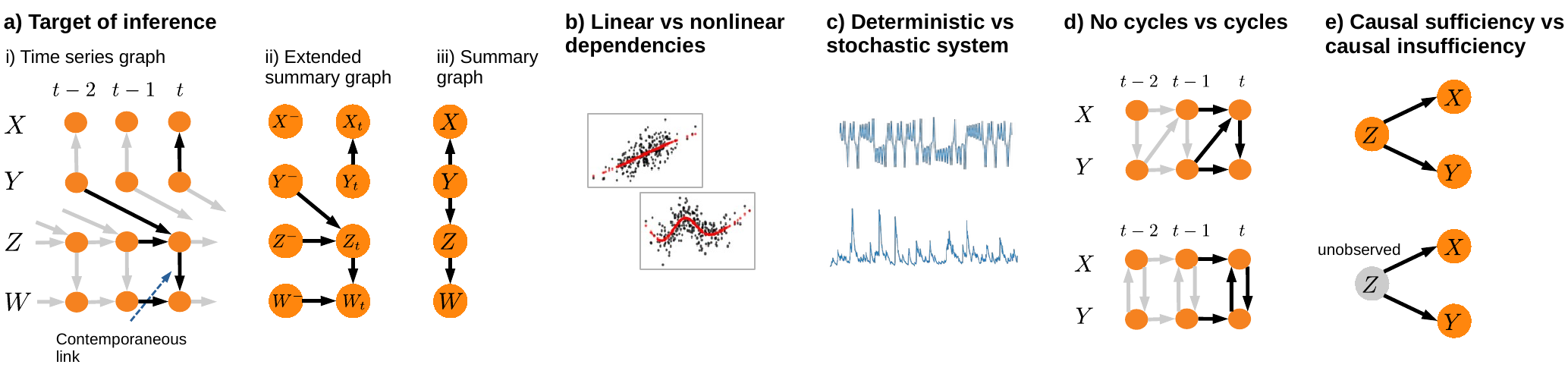}
    \caption{Illustration of some axes along which causal discovery methods are categorised in this review; see references in the main text to the respective subparts of the figure.}
    \label{fig:taxonomy}
\end{figure}

This section introduces and explains several axes for categorising and distinguishing causal discovery methods. While it would be possible to consider more axes yet, the authors believe that the choice of axes presented here is a reasonable compromise between a sufficiently fine-grained categorisation on the one hand and clarity of exposition on the other hand. 
Table~\ref{tab:causal_methods} lists many causal discovery methods and categorises them according to the aforementioned axes. In Figure~\ref{fig:taxonomy}, we graphically illustrate some axes. 

\paragraph{Bivariate vs multivariate causal discovery.}

This axis concerns the number of variables that are being considered. Bivariate causal discovery aims to discover the causal relationship between exactly two variables $X$ and $Y$ (in the non-temporal case) or between exactly two component time series $X^i$ and $X^j$ (in the time series case). Multivariate causal discovery aims to discover the causal relationships between any number of variables or component time series, respectively. Bivariate causal discovery often (but not necessarily) assumes \emph{causal sufficiency} (see axis on causal sufficiency below). In the time series case, bivariate causal discovery often (but not necessarily) targets to infer the \emph{summary graph} rather than the \emph{time series graph} or \emph{extended summary graph} (see axis on time series graph discovery below). If time lags are at least partially resolved in the bivariate time series case, that is, if the target of inference is the time series graph or the extended summary graph, then one effectively deals with a multivariate causal discovery problem.

\paragraph{Time series graph discovery vs summary graph discovery vs extended summary graph discovery.}\label{sec:target-graph-axes}

This axis is specific to the temporal setting and concerns the target of inference. Some methods are designed to learn the \emph{time series graph} \citep{runge2012escaping}, also known as \emph{full-time graph} \citep{peters2013causal} and \emph{time series chain graph} \citep{dahlhaus2003causality}, that is, the collection of all causal links $X^i_{t-\tau} \rightarrow X^j_t$ including the respective lags $\tau$ of these links. Part $a)i)$ of Figure~\ref{fig:taxonomy} shows an example of a time series graph with four component time series. As indicated by the grey edges, the pattern of edges in this graph is implicitly assumed to repeat both to the left (past) and right (future). Due to this repetitive structure of the edges, a time series graph is uniquely specified by the collection of edges that point into a vertex at an arbitrary reference time step $t$. Other methods disregard the information about the time lags and instead learn the \emph{summary graph} \citep{Peters2017book}. In the summary graph, there is exactly one vertex per component time series $X^i$ and an edge $X^i \rightarrow X^j$ if and only if there is an edge in the time series graph $X^i_{t-\tau} \rightarrow X^j_t$ at any lag $\tau$. Part $a)iii)$  of Figure~\ref{fig:taxonomy} shows the summary graph associated with the time series graph in part $a)i)$ of the same figure. Another option is to learn \emph{extended summary graphs} \citep{assaad2022discovery}. These graphs go midway between learning time series and summary graphs by distinguishing between contemporaneous and lagged links but disregarding the information about the specific time lags of lagged links. Specifically, the extended summary graph contains exactly two vertices per component time series $X^i$, namely the vertex $X^i_t$ for the present time steps and the vertex $X^{i,-}$ for all past time steps. There is an edge $X^i_t \rightarrow X^j_t$ if and only if, this same edge is also in the time series graph, there is an edge $X^{i,-} \rightarrow X^j_t$ if only if there is at least one $\tau \geq 1$ such that $X^i_{t-\tau} \rightarrow X^j_t$ is in the time series graph, and there is no edge between $X^{i,-}$ and $X^{j,-}$.
By this construction, extended summary graphs do distinguish between $X^i_{t-\tau} \rightarrow X^j_t$ with $\tau > 0$ and $X^i_{t} \rightarrow X^j_t$ but do not distinguish between, for example, $X^i_{t-2} \rightarrow X^j_t$ and $X^i_{t-1} \rightarrow X^j_t$.
Part $a)ii)$  of Figure~\ref{fig:taxonomy} shows the extended summary graph associated with the time series graph in part $a)i)$ of the same figure. Resolving the lag structure does yield more information but also implies a more complex target of inference. Learning more complex graphs (\eg, a time series graph vs a summary graph) is conceptually and statistically more challenging.

Methods for time series causal discovery typically require the user to specify a maximal lag $\tau_{max}$ up to which the method is supposed to be sensitive. If the target of inference is the time series graph, then this choice is apparent as the learned graph has exactly $\tau_{max} + 1$ steps. For example, if the user specifies the maximal lag to be $\tau_{max} = 3$, then the learned graph has exactly $4 = 3+1$ time steps, and the method cannot find causal links of lag $\tau \geq 4$ even if such links are present. If time series graph discovery is combined with the assumption of causal sufficiency, then this combination implicitly implies no causal links with a lag larger than $\tau_{max}$. If causal sufficiency is not assumed, then this additional implicit assumption is not made. 
Also, methods for summary and extended summary graph discovery typically require the specification of a maximal lag up to which the method is sensitive.
As opposed to that, when learning summary graphs or extended summary graphs the choice of $\tau_{max}$ is not apparent from the learned graph.
However, this choice is not apparent from the learned graph in the summary graph and extended summary graph cases.

For the case of causal models for time series, the causal arrow between any two variables can be at any time lag ranging from zero, i.e. a contemporaneous link, to infinity. A causal discovery algorithm can thus aim to either infer the full graph with edges being classified by time lags or to infer the summary graph where the time lag of the causal effect is irrelevant.

\paragraph{Methods based on independence, asymmetry, scores and context.}
\begin{wrapfigure}{r}{6.3cm}
\vspace{-0.5cm}
\begin{mdframed}[backgroundcolor=gray!20] 
\small
{\em ``The wide variety of causal discovery methods can be structured into independence-based, asymmetry-based, score-based, and context-based approaches.''
}
\end{mdframed}
\vspace{-0.5cm}
\end{wrapfigure}
This axis distinguishes causal discovery methods by the type of information/signal that they use to learn the causal graphs from data. In this review, as is common in the literature (for example, see \citep{glymour2019review}), we distinguish the independence-based, asymmetry-based, and score-based approaches. Further, we consider the context-based approach to distinguish those methods that employ the invariance of causal mechanisms across different environments. An exact delineation between these four approaches is not always possible as there are \emph{hybrid} methods that combine more than one approach.

First, \emph{independence-based causal discovery}, sometimes called \emph{constraint-based causal discovery}, utilises marginal and conditional independencies between variables to learn the causal graph or a set of causal graphs consistent with those independencies. 
Recall that an SCM is defined by a collection of structural assignments for each variable, where each assignment is a function of the variable's parents and a noise term. The collection of noise variables is assumed to be jointly independent.
Independence-based causal discovery relies on the fact that, for data generated by an SCM, the structure of the SCM's causal graph imprints some independencies onto the data \citep{verma1990causal, geiger1990identifying, pearl2009causality}. This property is known as \emph{causal Markov condition} \citep{Spirtes2000}. Alternatively, if one does not assume that an SCM generates the data, then the causal Markov condition is not automatically implied but needs to be assumed separately, leading to the so-called \emph{causal Markov assumption}. The $d$-separation criterion \citep{pearl1988probabilistic} allows to graphically determine all independencies that are necessarily implied in a given causal graph \citep{verma1990equivalence, geiger1990identifying, pearl2009causality}. The basic idea then is to run statistical tests of marginal and conditional independencies on the data and, second, use the results of these tests to constrain the causal graph's structure.

For the second of these two steps to hold, one further needs to make the \emph{causal faithfulness assumption} \citep{Spirtes2000}. This assumption says there are no independencies beyond those necessarily implied by the causal Markov condition in the observed data. For example, causal faithfulness excludes that a variable $X$ causally influences another variable $Y$ along multiple pathways which in total cancel out exactly---because otherwise there would be statistical independence, namely the marginal independence of $X$ and $ Y$, that would not be implied by the causal Markov condition.

For linear models and infinite samples, causal faithfulness is guaranteed except for Lebesgue measure zero sets in parameter space \citep{Spirtes2000}. However, \citep{robins2003uniform} shows that causal faithfulness ensures point-wise consistency but not uniform consistency of, for example, the famous PC algorithm \citep{Spirtes2000} (see below for the details on the PC algorithm). Uniform consistency of the PC algorithm is proven under the stronger \emph{strong-faithfulness assumption} \citep{zhang2002strong, kalisch2007estimating}. However, \citep{uhler2013geometry} shows that the set of not strongly-faithful distributions has a non-zero Lebesgue measure for various graph structures. This analysis indicates that causal faithfulness should be regarded as a strong assumption. There are also weaker types of faithfulness assumptions, for example, the \emph{adjacency faithfulness} and \emph{orientation faithfulness} assumptions \citep{ramsey2006adjacency, zhang2008detection}.

Independence-based causal discovery is non-parametric in that no assumption on the SCM's functional relationships and/or noise distributions needs to be made. However, choosing a particular method for (conditional) independence testing may implicitly impose a parametric assumption. For example, testing for (conditional) independence by (partial) correlation implicitly makes the assumption that the data-generating process is linear Gaussian. Conversely, if a parametric assumption can be made, this assumption might favour specific methods for (conditional) independence testing. For example, suppose one can assume linear Gaussian data. In that case, it is reasonable to use a (partial) correlation instead of more general (conditional) independence tests like, for example, a test based on (conditional) mutual information  as given in \citep{runge2018conditional}.

Many independence-based causal discovery are proven to be \emph{sound and complete}, meaning they provably learn the respective target of inference if they are given ground-truth knowledge of (conditional) independencies. However, the statistical tests of (conditional) independence are expected to make errors in finite samples even if all assumptions are met. For the famous PC algorithm \citep{Spirtes2000} (see below for more details), there are also probabilistic finite-sample guarantees \citep{zhang2002strong, kalisch2007estimating}.

Typically, there are multiple graphs that by means of the causal Markov condition, imply the exact same set of (conditional) independencies. For example, the three graphs $X \rightarrow Y \rightarrow Z$ and $X \leftarrow Y \leftarrow Z$ and $X \leftarrow Y \rightarrow Z$ by means of the causal Markov condition all imply exactly the same independence, namely that $X$ and $Z$ are conditionally independent given $Y$ (and no further independencies). Such graphs are said to be \emph{Markov equivalent} to each other and constitute a \emph{Markov equivalence class}. Consequently, independence-based causal discovery algorithms cannot distinguish between Markov equivalent graphs. Instead, these algorithms target learning a graphical representation of the entire Markov equivalence class to which the true causal graph belongs.
Since the graphs $X \rightarrow Y$ and $X \leftarrow Y$ are Markov equivalent, independence-based causal discovery can not infer the causal direction in the fundamental non-temporal bivariate case (see axis I). Independence-based causal discovery becomes meaningful for at least three variables; however, in the time series case, even two variables suffice to make non-trivial causal inferences (because then, if the variables are resolved in time, there are effectively more than just two variables).

Second, \emph{asymmetry-based causal discovery} makes and relies on parametric assumptions on the form of the functional relationships and/or noise distributions of the data-generating SCM \citep{Peters2017book}.

This approach is motivated by the elementary bivariate case, that is, by finding the causal relationship between two variables $X$ and $Y$. As explained above, with independence-based causal discovery, it is not possible to distinguish the Markov equivalent graphs $X \rightarrow Y$ and $X \leftarrow Y$. This impossibility is not a shortcoming of independence-based causal discovery but rather is fundamental unless stronger assumptions are made \citep{peters2012restricted, Peters2017book}. The proof of the impossibility of distinguishing between $X \rightarrow Y$ and $X \leftarrow Y$ works by showing that if the true data-generating SCM goes in the direction $X \rightarrow Y$, then one can always construct an alternative SCM in the direction $X \leftarrow Y$ that gives rise to the same data distribution as the true SCM.

The basic idea for removing this fundamental ambiguity is as follows: For certain choices of \emph{restricted} SCMs, defined by certain restricted parametric assumptions, it is impossible to have a restricted SCM in the first direction $X \rightarrow Y$ and simultaneously a restricted SCM in the second direction $X \leftarrow Y$.
Hence, given the assumption that the true SCM lies in the restricted class of models, it becomes possible to distinguish the causal and anti-causal direction. A restricted class of SCMs with this property is said to be \emph{identifiable}.
This approach to causal discovery relies on the expectation that the SCM in the causal direction generically has lower complexity than any alternative SCM in the anti-causal direction. As explained in Section 4.1.2 of \citep{Peters2017book}, this expectation can be motivated by the principle of \emph{independence of cause and mechanism} \citep{daniusis2010inferring, peters2012restricted}.
There are also asymmetry-based causal discovery methods for learning the causal graph between two or more variables for multivariate causal discovery \citep{Peters2017book}.

Third, \emph{score-based causal discovery} chooses one or multiple best-scoring graphs with respect to a predefined scoring function. This scoring function is typically built on the likelihood of the observed data given a particular graph and an assumed parametric statistical model \citep{Peters2017book}.
This approach requires searching over the space of causal graphs. Even if causal sufficiency (see axis on causal sufficiency) and acyclicity (see axis on cycles below) are assumed, in which case the causal graph is a \emph{directed causal graph (DAG)}, the search space of graphs already grows super-exponentially, \eg~\citep{chickering2002optimal}. An exact search is thus infeasible even for a moderate number of variables. Instead, greedy search techniques are often used, \eg~in the famous GES algorithm \citep{chickering2002optimal, chickering2002learning} (see below for details on this algorithm).
If the assumed statistical model does not yield identifiability beyond the Markov equivalence class, then the scoring function must be chosen such that Markov equivalent graphs have the same score.
Hence, one can search over the space of Markov equivalence classes rather than over the space of graphs.

Fourth, \emph{context-based causal discovery} requires access to data of the same system in different contexts. The term \emph{different context} is understood rather broadly: Its meaning ranges from, for example, observing the same physical system at other locations to, for example, observing a system both before and after an intervention. The basic assumption and idea of this approach to causal discovery are that the causal mechanisms, that is, the functional mappings from causes to effects and hence also the conditional distributions of the effects given their causes, remain unchanged across all contexts (unless the effect variable is the target of an intervention in one of the contexts). In contrast, marginal distributions and hence also the conditional distributions of causes given their effects can change \citep{Peters2017book}. A prime example of a context-based causal discovery method is Invariant Causal Prediction \citep{peters2016causal} (see below for more details). The \emph{joint causal inference (JCI)} framework \citep{mooij2020joint} proposes to model all contexts with one graph by including one or multiple so-called \emph{context variables} whose values determine the context and subsequently pooling the data from the different contexts into one joint dataset. This enlargement of the system by context variables effectively reduces the case of multiple contexts to the standard case of a single context. Consequently, it is possible to apply standard causal discovery algorithms to the pooled data (if the respective assumptions are met). One might thus argue that context-based causal discovery should not be considered a distinct approach to causal discovery but should instead be characterised by the requirement of data from multiple contexts.

\paragraph{Linear or nonlinear dependencies.}
This axis concerns the form of the functional relationships in the data-generating structural causal model.
Broadly, see part $b)$ of Figure~\ref{fig:taxonomy}, one can distinguish between linear and nonlinear functional relationships. In independence-based and score-based causal discovery, an assumption of linearity can enter implicitly by using partial correlation for testing conditional independence (independence-based approach) or by choice of the statistical model (score-based approach). In asymmetry-based causal discovery, an assumption of linearity is, if made, typically explicit by choice of the functional model.
Various asymmetry-based causal discovery methods do not assume linearity but still use restricted functional model classes that do not allow for entirely generic dependencies, for example, the functional model class of nonlinear additive noise models \citep{hoyer2008nonlinear}. However, for simplicity, we here only distinguish the methods by whether or not they assume linearity.

\paragraph{Deterministic vs stochastic systems.}

This axis concerns an assumption on the type of data-generating process.
Some methods assume the data are generated by a deterministic process, for example, a deterministic dynamical system. In contrast, other methods make explicit use of the assumption that the data-generating process is inherently stochastic, see part $c)$ of Figure~\ref{fig:taxonomy}. In the case of stochastic data-generating processes, the stochasticity is interpreted as dynamical noise that arises due to factors outside of the model.
Dynamical noise needs to be distinguished from measurement noise: The former is an inherent property of the data-generating process, and the latter arises from uncertainty in the data-collection process.
The causal inference and discovery frameworks have also been extended to dynamical systems, both deterministic and stochastic, without stable equilibrium distribution \citep{mooij2013deterministic,bongers2018random,rubenstain2018}.

\paragraph{Contemporaneous links.}

This axis is specific to the temporal setting and concerns a connectivity assumption on the causal time series graph.
Some methods make the assumption that all causal links in the time series graph are \emph{lagged}, meaning that all causal links are of the form $X^i_{t-\tau} \rightarrow X^j_t$ with $\tau > 0$, whereas \emph{contemporaneous} links, that is, links of the form $X^i_{t} \rightarrow X^j_t$ are assumed to be absent. Other methods do not make this assumption. For example, in the time series graph in part $a)i)$ of Figure~\ref{fig:taxonomy}, there are the contemporaneous edges $Z_t \rightarrow W_t$ and $Y_t \rightarrow X_t$. Consequently, methods that assume the absence of contemporaneous links would, by assumption, disallow this particular time series graph.
Contemporaneous causal links correspond to causal influences that act on a time scale shorter than the measurement interval; for example, a causal influence on a time scale of six hours in daily measured data.

\paragraph{Causal cycles.}
This axis concerns a connectivity assumption on the causal graph.
Many methods assume the absence of cyclic causal relationships. This assumption means that a variable $X^j_t$ cannot be a causal ancestor of another variable $X^i_{t-\tau}$ if that second variable $X^i_{t-\tau}$ is a causal ancestor of the first variable $X^j_t$. For example, the lower graph in part $d)$ of Figure~\ref{fig:taxonomy} has the causal cycle $X_t \rightarrow Y_t \rightarrow X_t$.
Because causation cannot go backwards in time, the assumption of acyclicity only restricts the contemporaneous section of the causal time series graph. The assumption is thus only relevant for $\tau = 0$. In particular, even with the assumption of acyclicity, it is possible to model temporal feedbacks. For example, although the upper graph in part $d)$ of Figure~\ref{fig:taxonomy} is acyclic, it displays a causal influence of time series $X$ on $Y$ (by the edge $X_t \rightarrow Y_t$) and a causal influence of time series $Y$ on $X$ (by the edge $Y_{t-1} \rightarrow X_t$). For example, if the component time series $X^i$ causally influences another component time series $X^j$ at, say, lag $\tau = 1$ (i.e., in the time series graph, there is the link $X^i_{t-1} \rightarrow X^j_t$ or an indirect directed path from $X^i_{t-1}$ to $X^j_t$), then one can still allow a causal influence of $X^j$ on $X^i$ (\eg, $X^j_{t-2} \rightarrow X^i_t$ or $X^j_t \rightarrow X^i_t$ or indirect directed paths from $X^j_{t-2}$ or $X^j_t$ to $X^i_t$). It is not allowed, however, that $X^i$ and $X^j$ causally influence each other both at lag $\tau = 0$. For example, if both $X^i_t \rightarrow X^j_t$ and $X^j_t \rightarrow X^k_t \rightarrow X^i_t$ are present, then the time series graph is cyclic.
There are also causal discovery methods that allow cyclic causal relationships, \eg~\citep{forre2018constraint, strobl2019constraint, mooij2020constraint, bongers2021foundations}. These methods typically infer less informative graphs than those inferred by methods that do not allow causal cycles. The early work \citep{richardson1996discovery} considers the special case of causal discovery in \emph{linear} cyclic systems.

\paragraph{Causal sufficiency.}
This axis concerns an assumption that can be viewed as an assumption on the data-generating process or the data-collection process.
The assumption of \emph{causal sufficiency} \citep{Spirtes2000} says that there are no \emph{latent confounders}, also called \emph{unobserved confounders} or \emph{hidden common causes}. A latent confounder is an unobserved variable that (potentially indirectly through other unobserved variables) causally influences two observed variables $X^i_{t-\tau}$ and $X^j_t$. %
For example, in the lower graph in part $e)$ of Figure~\ref{fig:taxonomy} the unobserved variable $Z$ acts as a latent confounder of the variables $X$ and $Y$. Consequently, this graph violates causal sufficiency whereas the upper graph in part $e)$ of Figure~\ref{fig:taxonomy} satisfies causal sufficiency.
Methods that do not assume causal sufficiency typically infer graphs that are less informative than the graphs inferred by methods that do assume causal sufficiency. For example, consider the elementary non-temporal bivariate case with two variables $X$ and $Y$. If these variables are dependent and causal sufficiency is assumed, then either $X$ causes $Y$ ($X \rightarrow Y$), or $Y$ causes $X$ ($X \leftarrow Y$). If causal sufficiency is not assumed, then there is the third possibility that neither $X$ causes $Y$ nor vice versa but that there rather is an unobserved variable $L$ which causes both $X$ and $Y$ ($X \leftarrow L \rightarrow Y$ or, for example, $X \leftarrow L^{\prime} \leftarrow L \rightarrow L^{\prime\prime} \rightarrow Y$ with other unobserved variables $L^{\prime}$ and $L^{\prime\prime}$).

\subsection{Description and categorisation of causal discovery methods}\label{sec:list-of-causal-discovery-methods}

\begin{figure}[h!]
    \centering
    \includegraphics[scale=0.9]{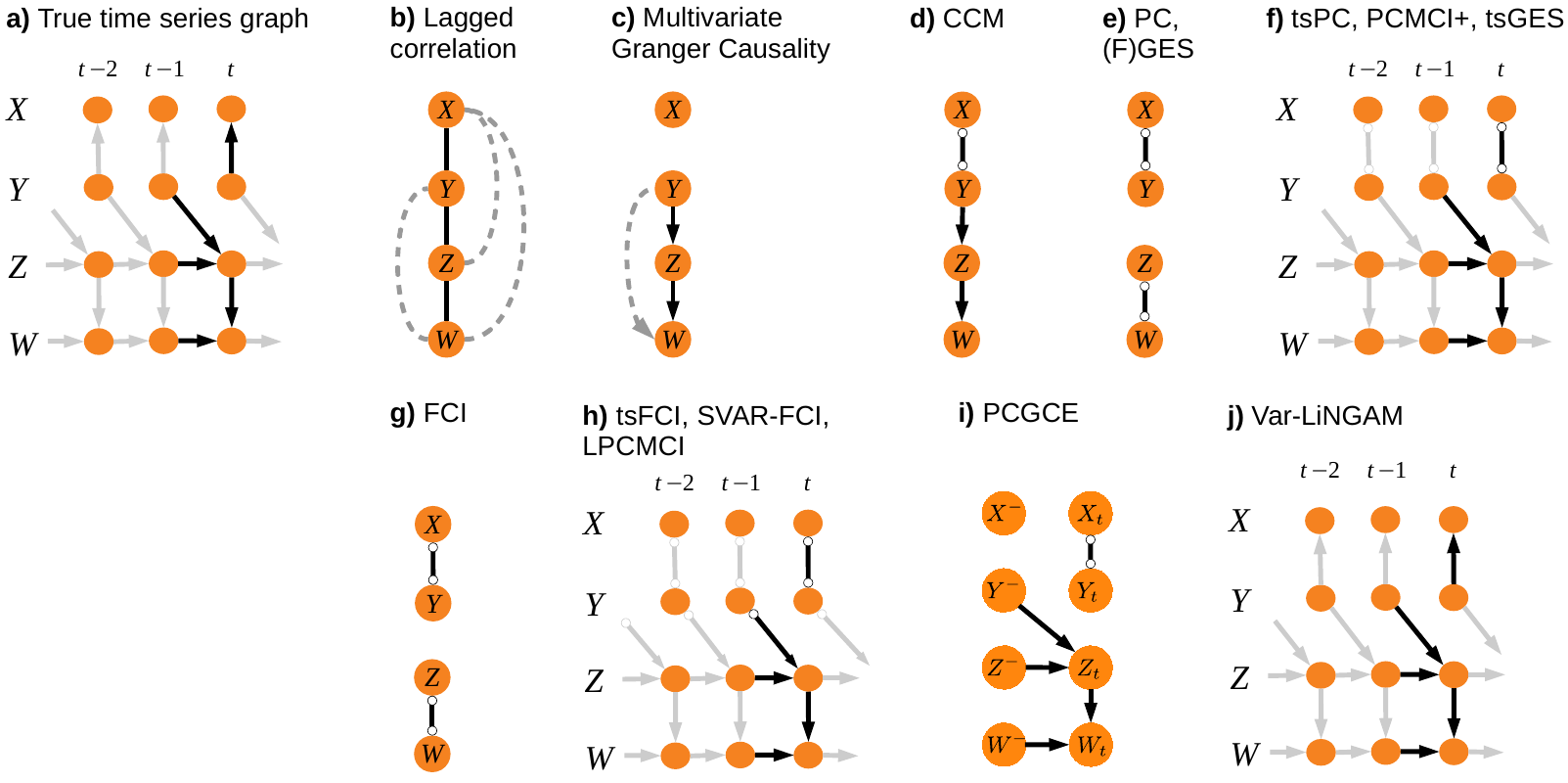}
\caption{Figure illustrating a time series graph (TSG) and the respective graphs discovered by applying various causal discovery methods to data generated from an SCM with that time series graph. (a) Time series graph. (b) The discovered undirected graph by considering (lagged) correlations, where spurious correlations are highlighted as dashed grey lines. (c) The directed graph discovered by multivariate Granger causality does not consider contemporaneous links and retains a spurious link from $Y$ to $W$. (d) Graph discovered by CCM. (e) The graph discovered by applying the plain PC and (F)GES algorithms fail to show lagged links and, in addition, fail to orient a link that the time series adapted algorithms can orient. (f) The time series version of PC (tsPC), PCMCI+, and the time series version of GES (tsGES) discover both lagged and contemporaneous links and orient edges up to the Markov equivalence class. (g) Plain FCI has the same drawbacks as plain PC or (F)GES. (h) FCI-based time series causal discovery algorithms account for latent confounders and thus discover causal arrows up to latent confounding. In particular, the algorithm cannot exclude that the association between $Y_{t-1}$ and $Z_t$ is due to latent confounding rather than a causal relationship. (i) The PCGCE algorithm discovers the extended summary graph up to its Markov equivalence class. (j) Var-LiNGAM discovers all causal relationships correctly if the assumptions of linear relationships and additive non-Gaussian noise are satisfied.}
    \label{fig:toymodel}
\end{figure}

Here, we list and briefly explain several existing causal discovery methods for time series data. We summarise this list and the placement of each method with respect to the axes presented above in Table~\ref{tab:causal_methods}. In Figure~\ref{fig:toymodel}, we illustrate and compare an example time series graph and the respective graphical objects obtained by some of the discussed causal discovery algorithms when applied to data generated from an SCM with that time series graph.

\paragraph{Granger Causality.} Granger Causality (GC)~\citep{granger1969investigating, granger1980testing} is originally a statistical test to decide whether a time series $X_t$ is a \emph{cause} of another time series $Y_t$, in the sense that 
past values of $X_t$ have significant  
predictive power in forecasting $Y_t$. 
GC is thus, in principle, 
a simple test of 
temporal (or lagged) relationship and predictability. Nevertheless, under causal sufficiency and no contemporaneous effects assumptions, it can be formally
shown that GC testing detects actual 
causal links (see \eg \citet{Peters2017book} for a formal derivation of these results in the SEM setting). 

In a multivariate setting, which is seldom the case, testing if $X_t$ causes $Y_t$ requires controlling for all possible confounders. Therefore the conditional GC~\citep{granger1980testing,
geweke1982measurement,chen2004analyzing, barrett2010multivariate},
includes in the restricted and full models the past of all other \emph{relevant} time series in the system that are not $X_t$ and $Y_t$.
Classically, GC considers linear models for which standard $t$-tests or $F$-tests can be employed, but non-linear extensions have been considered both in econometrics
\citep{bell1996non, hiemstra1994testing, abhyankar1998linear, warne2000causality,
diks2006new} and in physical and biological applications~\citep{ancona2004radial,marinazzo2008kernelunivariate,bueso2020explicit}. 

\begin{mdframed}[backgroundcolor=gray!20] 
\small
\vspace{-0.2cm}
\paragraph{Linear and nonlinear Granger causality in dynamic systems.} A unified view of (nonlinear) GC with kernel methods for the physical sciences was introduced in \citep{bueso2020explicit}. Two examples are given here:
\begin{enumerate}
    \item Bivariate system with coupled non-linear and autoregressive relations given by $x_{t+1} = 3.4x_{t}(1-x_{t}^2)$ $\exp(-x_{t}^2)+\varepsilon_{t}^x$ and $y_{t+1} = 3.4y_{t}(1-y_{t}^2)$ $\exp(-y_{t}^2)+{0.5x_{t}y_{t}}{2}+\varepsilon_{t}^y$, where $\varepsilon$ is white Gaussian noise with zero mean and variance $0.4$. The causal direction is $X \to Y$. %
The histogram of the estimated causal index $\delta$ on the left figure reveals GC's insensitivity to the causal direction, the high false positive rate of KGC \citep{marinazzo2008kernel}, and a higher detection power by the XKGC \citep{bueso2020explicit}.
\item Two logistic maps system defined as $x_{t+1} = 1 - 1.8x_{t}^2$ and $y_{t+1} =  (1-\alpha)(1 - 1.8y_{t}^2) +\alpha(1 - 1.8x_{t}^2)$, where $\alpha \in [0,1]$ controls the coupling strength. The causal relationship implemented is $X\to Y$, and the challenge is to assess the detection power of methods without introducing any external variable,  just using $X$ and $Y$. We analyse segments of length $n=2000$ and fixed $p=2$.  
The left figure shows the prediction skills for varying $\alpha$. Note that the system is completely synchronised at $\alpha=0.37$. T
XKGC \citep{bueso2020explicit} improves detection power over GC/KGC for any $\alpha$.
\end{enumerate}
\centerline{\includegraphics[height=3cm]{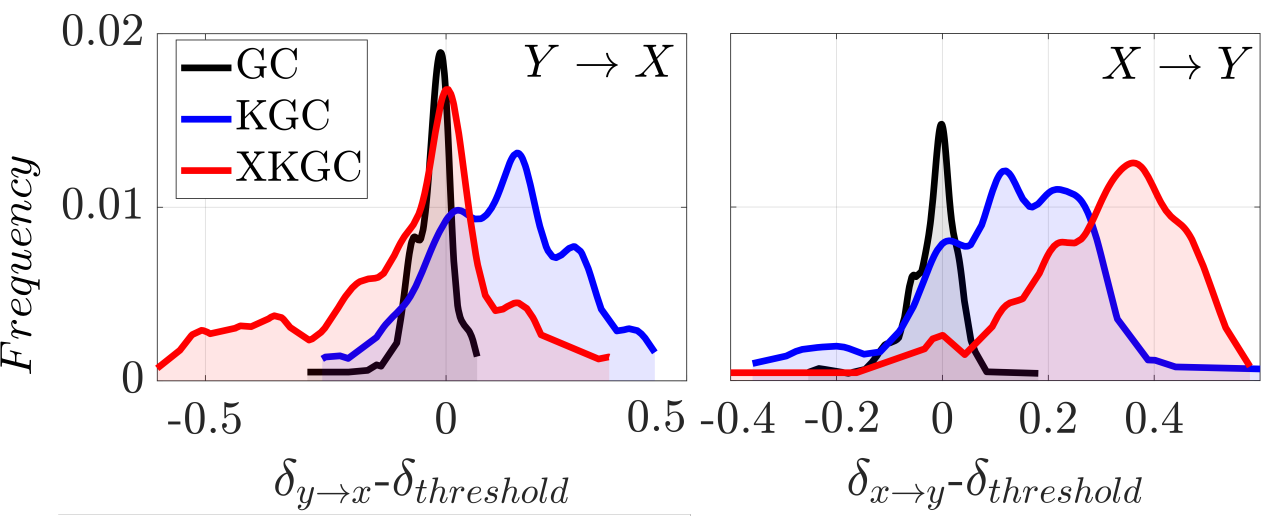}~~\includegraphics[height=3cm]{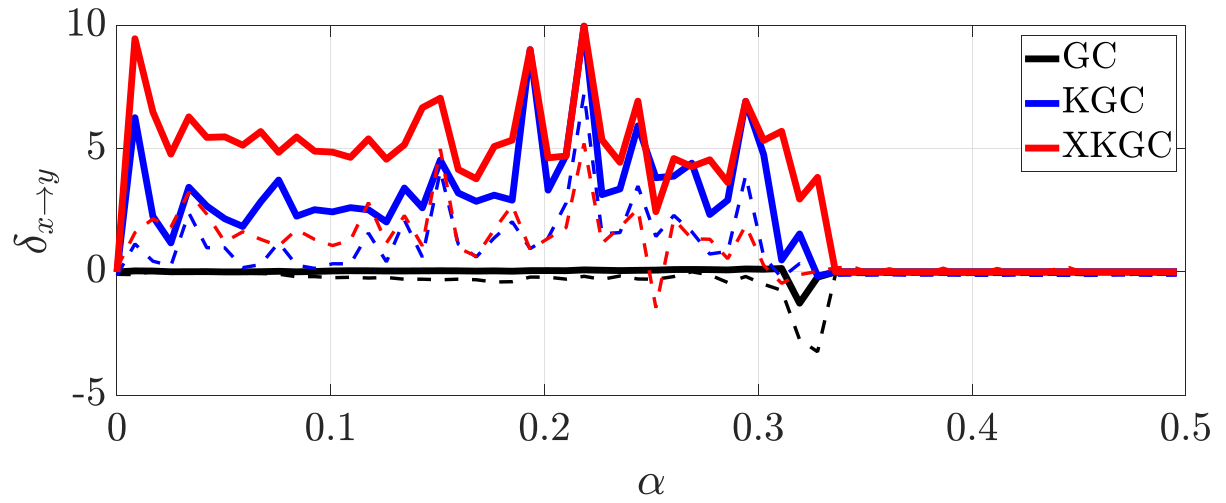}}
\end{mdframed}

Transfer entropy~\citep{schreiber2000measuring} between $Y_t$ and $X_t$ measures the amount of unique information contained in the past of $X_t$, denoted $X^-=\{X_{t-1},X_{t-2},\ldots\}$, about the state of $Y_t$ and is defined as $\mathcal{T}_{X \rightarrow Y|Z}=H(Y_t|Y^-,Z^-)-H(Y_t|Y^-,X^-,Z^-)$. Transfer entropy can be considered as the generalisation of GC by extending the implicit conditional independence test to arbitrary orders of dependence. Indeed, \citet{barnett2009granger} proved that (linear) GC and transfer entropy causality are equivalent under the assumptions of VAR model class and Gaussian error distributions.

\paragraph{CCM.}

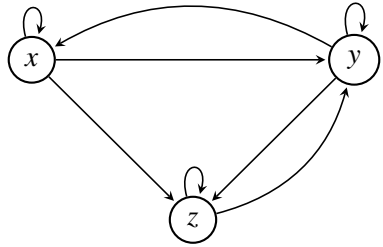
\begin{figure}[h!]%
\centering
  \begin{tikzpicture}[
            > = stealth, %
           shorten > = 1pt, %
            auto,
            node distance = 3cm, %
            semithick %
        ]

        \tikzstyle{every state}=[
            draw = black,
            thick,
            fill = white,
            minimum size = 4mm
        ]

        \node[state] (x) {$x$};
        \node[state] (z) [below right of=x]{$z$};
        \node[state] (y) [above right of=z] {$y$};
        
        \path[->] (x) edge [loop above] node {} (x);
        \path[->] (y) edge [loop above] node {} (y);
        \path[->] (z) edge [loop above] node {} (z);
        \path[->] (y) edge [bend right] node {} (x);
        \path[->] (x) edge node {} (y);
        \path[->] (z) edge [bend right] node {} (y);
        \path[->] (y) edge node {} (z);
        \path[->] (x) edge node {} (z);
    \end{tikzpicture}
    \caption{Summary graph for Lorenz attractor system.} \label{fig:lorenz_DG}
\end{figure}

Convergent cross-mapping (CCM) \citep{sugihara2012detecting} is based on the simple observation that if data from a deterministic dynamical system is generated by a system of ordinary differential equations (ODEs), then the explicit form of these equations directly defines the causes of each variable in the system: $x$ is a cause of $y$ if the dynamics (any of the derivatives of $y$) is expressed in terms of the state of $x$. For example, in the Lorenz attractor system:
\begin{align}
    x' &= \sigma y - \sigma x \nonumber \\ 
    y' &= -xz + \rho x - y \nonumber \\ 
    z' &= xy - \beta z 
\end{align}
$x_t$ is caused by $y_t$, $y_t$ is caused by $x_t$ and $z_t$ and $z_t$ by $x_t$ and $y_t$  as summarised by the summary directed graph of Fig. \ref{fig:lorenz_DG}. 

Learning the generating ODE from time series data would allow us to recover the causal relations and summary-directed graph. Nevertheless, relying on Takens' theorem \citep{Takens81}, \citet{sugihara2012detecting} concluded that it is not necessary to recover the exact ODE to recover its causal properties: if one has a cause and effect variable within an ODE system, then a qualitative description of the dynamics of the cause based on the dynamics of the effect can be recovered. Surprisingly, while one would need a large number of lags to estimate an ODE where no parametric assumptions have been made, Takens' theorem states that a good enough estimate, i.e. one that retains the causal properties of the ODE, can be made with at most $2d + 1$ lags where $d$ is the number of variables in the ODE. 

\begin{mdframed}[backgroundcolor=gray!20]
\small
\vspace{-0.2cm}
\paragraph{The CCM pseudo-algorithm for checking if two variables $X$ and $Y$ are causally related:}
\begin{enumerate}
\item Choose embedding dimension $E$: number of lags to use with $1 \leq E \leq 2d+1$. 
\item Estimate cross-map skill $\rho(l)$ for a sequence of several observations $l_1,\ldots,l_N$ with $l_N\leq L$, $L$ is the maximum number of available observations in the time series. For each $l_i$:
\begin{enumerate}
\item Construct shadow manifold $M_x$: in practice represented by matrix $Y \in R^{l \times E}$ with time series $y_t,y_{t-1},y_{t-2},y_{t-E+1}$
\item Assume the shadow manifold satisfies Takens' theorem condition and retains the metric properties of manifold $M$. Thus estimate euclidean distance $d_i$ of $E$+1 nearest points on manifold $M_x$ to point $(x_t, x_{t-1},\ldots,x_{t-E+1})$. Denote the time indices corresponding to these points as $t_1, t_2,\ldots,t_{E+1}$.
\item Construct estimate of $y_t$ using simplex projection in shadow manifold: weighted average of $E+1$ nearest points (on $M_x$) with weights determined according to the  exponentially weighted distance on $M_x$ of each point (calculated in the previous step): 
\begin{equation}
\hat{y}_t = \sum_{i=1}^{E+1} w_i y_{t_i}~~~~~\text{where}~~~~~w_i = \frac{\exp(-\frac{d_i}{d_1})}{\sum_i \exp(-\frac{d_i}{d_1})}
\end{equation}
\item Construct cross-map skill $\rho(l) = Corr (y_t, \hat{y_t}|M_x$) 
\end{enumerate}
\item Check if cross-map skill $\rho(l)$ converges as $l$ tends to $L$. As the number of observations used increases, the manifold estimation should be denser, so cross-map skill should improve and converge, provided our assumption that $M_x$ retains the metric properties of $M$ is true. 
\end{enumerate}
\end{mdframed}

The algorithm should also be applied symmetrically to establish the convergence of the cross-map skill $\hat{x}_t|M_y$. If both cross-map skills converge, we can establish that both variables belong in the same ODE system, and the causal relations are bi-directional. Note that in step 2c, we only use the shadow manifold $M_x$ to determine which points and with which weights should be used to estimate $y_t$. If the convergence of the cross-map skill happens in only one direction, the proper conclusion is that a uni-directional causal relationship exists between the two variables.  If the cross-map skill of $\hat{y}_t|M_x$ converges, the proper conclusion is that $y$ causes $x$. This is somewhat counterintuitive, at least from the point of view of more classical causal discovery methods, because to establish that $x$ is an effect of the cause $y$, we must be able to predict the cause $y$ using the effect $x$, where for all other methods discussed in this work it is the other way around.

\paragraph{PC-based methods.}
In the following, we start with an exposition of the PC algorithm \citep{Spirtes2000}. %
We then explain both a naive (tsPC) and more sophisticated (PCMCI) time series adaption.%
\begin{enumerate}[leftmargin=0.45cm]
\item {\it PC.} %
The original PC algorithm (named after Peter and Clark's authorship \citep{Spirtes2000}) was constructed for i.i.d random variables and thus, in particular, for non-time series data. Below, we will also describe an extension to the time series case.
The PC algorithm assumes that the underlying causal graph is a (\emph{directed acyclic graph DAG}). A DAG has only directed edges ($\rightarrow$ and $\leftarrow$) and no cycles. As for independence-based algorithms in general, the PC algorithm assumes the causal Markov condition and causal faithfulness to infer $d$-separations on the causal graph from conditional independencies alone. Moreover, the algorithm assumes causal sufficiency. Consequently, the algorithm cannot distinguish between two graphs with the same set of $d$-separations.
The graphical representation of an equivalence class of DAGs with the same $d$-separations
is known as a (\emph{completed partially-directed acyclic graph CPDAG}), which is the object of discovery of PC. As compared to DAGS, CPDAGs can contain undirected edges (${\circ\!{\--}\!\circ}$). These undirected edges signify that both orientations ($\rightarrow$ or $\leftarrow$) are compatible with the set of conditional independencies. 

\begin{mdframed}[backgroundcolor=gray!20]
\small
\vspace{-0.2cm}
\paragraph{The PC algorithm starts from a fully connected undirected graph and consists of three phases:}
\begin{enumerate}
\item The \emph{skeleton phase} uses statistical (conditional) independence tests to infer the adjacencies of the underlying causal graph. %
If two variables $X$ and $Y$ are found to be independent conditional on a (possibly empty) set of variables $\mathbf{Z}$, then the edge between $X$ and $Y$ is removed.
\item The \emph{collider orientation phase} then orients all \emph{collider motifs}, that is, motifs of the form $X \rightarrow Y \leftarrow Z$ where $X$ and $Z$ are non-adjacent. These orientations can be inferred because collider motifs impose a particular pattern of (conditional) (in-)dependencies.
\item The \emph{orientation phase} finally uses graphical rules \citep{meek1995causal} to infer the orientation of as many remaining unoriented edges as possible using the acyclicity assumption and the fact that all colliders have been found in the previous step.
\end{enumerate}
\end{mdframed}

As compared to a brute-force execution of the skeleton phase as done in the so-called SGS algorithm \citep{Spirtes2000}, the PC algorithm improves this phase as follows: Instead of searching for conditioning sets $\mathbf{Z}$ that renders a pair of variables $X$ and $Y$ independent (and thereby non-adjacent) within the set of all variables in the system, it is sufficient to restrict the search to those variables that are adjacent to $X$ and $Y$. The sufficiency of this specified search is a direct consequence of assuming the causal Markov condition. 

The PC algorithm is fully non-parametric by construction and lends itself to various application cases. It performs optimally when the underlying causal graph is sparse and the number of variables is much less than the sample size. However, for sufficiently sparse graphs, PC was shown to be computationally feasible for very high-dimensional graphs as well \citep{kalisch2007estimating}. In the worst-case scenario of very dense graphs, the number of conditional independence tests to be performed grows exponentially with the number of variables.

A variant of PC called \emph{PC-stable} \citep{colombo2014order} was shown to be \emph{order-independent}, that is, invariant under permutations of the variables (this invariance is a desired property because it should not matter which variable is considered to the be ``first'' and which is considered to be the ``last'' variable). Another variant called for \emph{Conservative PC} \citep{ramsey2006adjacency} is sound even if instead of the (full) causal faithfulness assumption, only the weaker \emph{adjacency faithfulness} assumption is made. It is also possible to infer whether or not \emph{orientation faithfulness} is violated \citep{ramsey2006adjacency}.

Although the PC algorithm was originally developed for the acyclic case, the work \citep{mooij2020constraint} shows that PC is also consistent in the presence of cycles if the learned graph is interpreted in a slightly different way using the so-called $\sigma$-separation \citep{bongers2021foundations}. 
The proof of this consistency uses the $\sigma$-separation \citep{bongers2021foundations}, which is a generalisation of $d$-separation to the cyclic case. Subsequently, it imposes a modified version of the Markov and faithfulness assumptions, namely the $\sigma$-Markov and the $\sigma$-faithfulness conditions. Under these conditions, the PC algorithm is proven to correctly learn the $\sigma$-separation equivalence class of the true (potentially cyclic) causal graph.

\item {\it tsPC.} The naive extension of the PC algorithm to the time series case is called tsPC, an example implementation is given in \citep{runge2020discovering}. The general idea is to fix an integer $\tau_{\text{max}}$ that is supposed to be equal to or larger than the maximum time-lag of any edge in the causal time series graph and to learn the finite segment of the time series graph on a time window $[t -\tau_{\text{max}}, t]$. Here, $t$ is an arbitrary reference time step, and samples are created by sliding the time window over all recorded time steps. This approach implicitly assumes that the causal relationships do not change throughout the recorded time steps. Suppose the assumption of no contemporaneous causal influences is made. In that case, it is sufficient only to run the skeleton phase of PC because then the orientation of all edges is determined by time order (an effect cannot precede its cause).
As discussed in \citep{runge2020discovering}, tsPC suffers from a sub-optimal finite-sample performance due to autocorrelation. This issue is remedied by the \emph{PCMCI} algorithm, explained below.
See Figure~\ref{fig:toymodel} for an illustration of the graphs learned by PC and tsPC for the time series case example.

\item {\it PCMCI.} %
The PCMCI algorithm \citep{runge2019detecting} is a time series causal discovery algorithm that addresses some of the shortcomings of the naive time series adaption of PC, in particular the issue of low detection power. PCMCI assumes the time series graph to have no contemporaneous causal influences. This assumption implies the absence of contemporaneous cycles, but feedback cycles involving time lags are possible. Additionally, PCMCI assumes the time series data are generated by a causally stationary process (that is, the causal relationships are assumed to not change over time).%

As discussed in \citep{runge2019detecting,runge2020discovering}, two main challenges with time series data hamper the performance of independence-based discovery algorithms in time series. These challenges are related to autocorrelation, a common feature in time series. First, using non-i.i.d. samples (created in a sliding window fashion as explained above) typically leads to ill-calibrated conditional independence tests, that is, uncontrolled type I errors, because the degrees of freedom are reduced and cannot be easily measured. This ill-calibratedness leads to inflated false positives, that is, the discovery of dependence when, in fact, independence is true. Secondly, high autocorrelation implies that there is little new information in the next time step compared to the previous step. Depending on how the conditioning sets in conditional independence tests are selected, this results in low effect sizes leading to low detection power of true links. The effect size of a (conditional) independence test is defined as the absolute value of the population value of the test statistic; for example, in a (partial) correlation test, the effect size is the absolute value of the population value of the (partial) correlation. While there is a trade-off in addressing both of these challenges, the PCMCI algorithm and its generalisation PCMCI$^+$ (see below) algorithms remedy these challenges to an extent by using a particular choice of conditioning sets in the independent tests that decides about the presence versus the absence of an edge between a given pair of variables. We spell out the details below.

\begin{mdframed}[backgroundcolor=gray!20]
\small
\vspace{-0.2cm}
\paragraph{The PCMCI algorithm unfolds in two phases:}
\begin{enumerate}
    \item The first phase, referred to as $\text{PC}_1$, is a \emph{condition-selection phase} that aims to infer a superset $\hat{\mathcal{P}}(X^j_t)$  of the parents of each variable $X^j_t$ at time step $t$. The PC$_1$ algorithm is a variant of the PC and works as follows: Each sub-step of the skeleton phase is indexed by the integer $p$, starting at $p = 0$ and successively increasing $p$ in increments of one. Within each sub-step, the algorithm tests for independence of $X^i_{t-\tau}$ and $X^j_t$ given a conditioning set that consists of those $p$ potential parents of $X^j_t$ (less $X^i_{t-\tau}$) that have the highest association with $X^j_t$ according to the previous (conditional) independence tests. This particular choice of conditioning sets increases the effect sizes of the (conditional) independence tests, as can be understood information-theoretically~\citep{runge2015quantifying}. A higher effect size leads to a higher statistical power (equivalently, to a lower probability of a type II error), that is, makes it more likely to detect dependence if dependence is true. However, since the effects of autocorrelation have not been dealt with yet, this phase of PCMCI is affected by the same issues as the naive time series extension of PC in terms of false positives.

\item The second phase conducts for each pair of variables $X^i_{t-\tau}$ and $X^j_t$ the so-called \emph{momentary conditional independence (MCI) test} that tests the null hypothesis $$X^i_{t-\tau} \ind X^j_{t} \ | \ \hat{\mathcal{P}}(X^j_t) \backslash  \{X^i_{t-\tau}\}, \hat{\mathcal{P}}(X^i_{t-\tau}) \ .$$ If this hypothesis is not rejected, the edge between $X^i_{t-\tau}$ and $X^j_t$ is removed. While the condition on $\hat{\mathcal{P}}(X^j_t)$ only would suffice to condition out confounded and indirect connections, the additional conditioning on $\hat{\mathcal{P}}(X^i_{t-\tau})$ removes auto-dependencies from $X^i_{t-\tau}$ such that the conditional independence tests are well-calibrated and false positives are controlled at the desired level~\citep{runge2019detecting}. Note that no orientation phase is required because, by the assumption of no contemporaneous causal influences, all edges are time-lagged and oriented by time order.
\end{enumerate}
\end{mdframed}
The numerical studies in \citep{runge2019detecting} show that in combination with the MCI tests in its second phase, PCMCI improves detection power and false positive control compared to the naive time series adaption.

\item \textbf{\it PCMCI$^+$.} %
The PCMCI$^+$ algorithm \citep{runge2020discovering} extends PCMCI to the case where contemporaneous edges are allowed by suitably modifying the MCI phase (but still disallowing contemporaneous causal cycles and latent confounders). 
It consists of three phases: First, a $\text{PC}_1$ lagged phase. Second, an MCI contemporaneous phase. Third, an orientation phase.
The first phase applies the $\text{PC}_1$ algorithm (see above) to the lagged edges. Its goal is to infer a superset of the \emph{lagged} parents of each variable $X^j_t$. However, due to the potential presence of contemporaneous links, the first phase converges to a superset $\hat{\mathcal{P}}(X^j_t)$ of the lagged parents of $X^j_t$ plus the parents of the contemporaneous ancestors of $X^j_t$. The MCI contemporaneous phase is initialised with all the links found in the previous phase and all possible contemporaneous links. It then conducts the conditional independence tests $X^i_{t-\tau} \ind X^j_{t} \ | \ \mathbf{S}, \hat{\mathcal{P}}(X^j_t) \backslash  \{X^i_{t-\tau}\}, \hat{\mathcal{P}}(X^i_{t-\tau})$, where, in addition to the lagged conditions of the MCI tests in PCMCI, $\mathbf{S}$ are sets of contemporaneous adjacencies of $X_t^j$ (and $X_{t-\tau}^i$ for $\tau=0$). This removes all remaining spurious links due to contemporaneous confounding or indirect paths. Finally, the orientation phase orients all lagged links according to time order and then applies PC orientation rules \citep{meek1995causal} to orient as many contemporaneous edges as possible. See Figure~\ref{fig:toymodel} for an illustration. 
\end{enumerate}

\paragraph{FCI-based methods.} 
A major success of causal discovery is the development of methods for learning causal relationships without assuming causal sufficiency. One famous example is the FCI algorithm \citep{spirtes1995causal, Spirtes2000, zhang2008completeness}. %
We review the standard FCI algorithm and several of its time series adapations.%
\begin{enumerate}[leftmargin=0.45cm]
\item {\it FCI.}
The FCI algorithm %
generalises the PC algorithm (see above) to the causally insufficient case.
In addition to latent confounders, the algorithm can also deal with \emph{selection variables}. These variables influence whether a given sample point belongs to the observed population. For example, a certain satellite observation might be more likely to be made if the cloud cover is not too dense. Consequently, the statistical dependence relations will be biased (giving it the name \emph{selection bias}) as only one segment of the entire population of possible satellite observations is being considered. Below, we assume the absence of selection variables and explain the specialisation of FCI in this case. 

\begin{mdframed}[backgroundcolor=gray!20]
\small
\vspace{-0.2cm}
\paragraph{Quick Introduction to Maximal Ancestral Graphs.}
To deal with latent confounders (and selection variables), FCI works with a larger class of graphical models than PC does:
Instead of DAGs, FCI works with \emph{maximal ancestral graphs} (\emph{MAGs}) \citep{richardson2002ancestral}. Maximal ancestral graphs can be interpreted as projections of the underlying DAG (which consists of observed variables, latent confounders, and selection variables) to a graph over the observed variables only. %
When assuming the absence of selection variables (as we do here), it is sufficient to work with a subset of MAGs that are called \emph{directed maximal ancestral graphs} (\emph{DMAGs}) \citep{mooij2020constraint}. Parts of the literature for notational simplicity do not explicitly distinguish between MAGs and DMAGs, that is, speak of MAGs although referring to DMAGs.

Directed maximal ancestral graphs can have two types of edges: directed edges ($\rightarrow$) and bidirected edges ($\leftrightarrow$). A directed edge $X \rightarrow Y$ says that variable $X$ causally influences variable $Y$. This causal influence can be direct or indirect through one or multiple unobserved variables. A bidirected edge $X \leftrightarrow Y$ says that $X$ and $Y$ are subject to latent confounding and that, at the same time, neither $X$ causally influences $Y$ nor the other way around. Being subject to latent confounding means that there is an unobserved variable $L$ that (potentially indirectly through other unobserved variables) causally influences both $X$ and $Y$. In addition, an edge between $X$ and $Y$ (i.e., $X \rightarrow Y$ or $X \leftarrow Y$ or $X \leftrightarrow Y$) means that $X$ and $Y$ are not (conditionally) independent given any set of observed variables.

A subtle part of the interpretation of DMAGs is that directed edges $X \rightarrow Y$ can ``hide'' latent confounding. That is to say, while $X \rightarrow Y$ does say that $X$ causally influences $Y$, it is possible that $X$ and $Y$ are also subject to latent confounding. Even if there is such additional latent confounding, then the DMAG does not, in addition, also contain the edge $X \leftrightarrow Y$ because there is at most one edge between any pair of variables. However, for a certain type of directed edges, called \emph{visible} and which can be determined graphically from the DMAG, one can assert with certainty that there cannot be additional latent confounding \citep{zhang2008causal}.

\end{mdframed}
The FCI algorithm works in a way that is similar to the PC algorithm: First, a sequence of (conditional) independence tests is performed to find the skeleton (that is, the adjacencies) of the graph. Second, several orientation rules are applied to determine the direction of as many links as possible.
For these details, we refer to the original works \citep{spirtes1995causal, Spirtes2000, zhang2008completeness} or to more technical reviews of FCI, for example, see Section S2 in the supplementary material of \citep{gerhardus2020high}.

As in the case of the PC algorithm, FCI does not learn a unique DMAG but rather a Markov equivalence class of DMAGs. These equivalence classes are graphically represented by \emph{directed partial ancestral graphs} (\emph{DPAGs}) \citep{zhang2008causal, zhang2008completeness, mooij2020constraint}. In addition to directed ($\rightarrow$) and bidirected edges ($\leftrightarrow$), DPAGs can also contain edges of the types $X {\circ\!{\rightarrow}} Y$ and $X {\circ\!{\--}\!\circ} Y$. An edge $X {\circ\!{\rightarrow}} Y$ says that $Y$ does not have a causal influence on $X$ while $X$ might or might not have a causal influence on $Y$, whereas an edge $X {\circ\!{\--}\!\circ} Y$ does not make any claim about whether or not $X$ or $Y$ have a causal influence on each other.
See Figure~\ref{fig:toymodel} for an illustration of the graph that the FCI algorithm discovers when applied to time series data. 
The work \citep{mooij2020constraint} has shown that FCI, originally developed with the assumption of acyclicity, can also be consistently applied to data that is generated by a cyclic SCM with certain regularity conditions.

\item {\it tsFCI.} 
The tsFCI algorithm \citep{enter2010causal} adapts FCI to causally stationary time series.
Here, the term \emph{causal stationarity} means that the causal relationship between the variables $X^i_{t-\tau}$ and $X^j_t$ is the same as the causal relationships between the variables $X^i_{s-\tau}$ and $X^j_{s}$ for any time steps $t$ and $s$. In other words, the causal relationships are assumed invariant in time. %
As compared to the FCI algorithm, tsFCI applies the following two conceptual modifications: First, lagged links ($\tau \geq 1)$ are by default oriented as $X^i_{t-\tau} {\circ\!{\rightarrow}} X^j_t$. These default orientations are valid because an effect cannot precede its cause. Note that it would not be valid to orient all lagged links as $X^i_{t-\tau} \rightarrow X^j_t$ because $X^i_{t-\tau} \leftrightarrow X^j_t$ (i.e., latent confounding) is a possibility. Second, so-called \emph{homologous} edges are by default oriented in the same way. That is if the edge between $X^i_{t-\tau}$ and $X^j_t$ has been found to have a certain orientation (for example, $X^i_{t-\tau} \rightarrow X^j_t$ or $X^i_{t-\tau} \leftarrow X^j_t$) and if in addition there is an edge between $X^i_{s-\tau}$ and $X^j_s$ for $s \neq t$, then this latter edge is immediately oriented in the same way as the former edge (for example, oriented as $X^i_{s-\tau} \rightarrow X^j_s$ or $X^i_{s-\tau} \leftarrow X^j_s$). This copying of edge orientations is valid because of causal stationarity.
In addition to these modifications, tsFCI uses the knowledge of time order and causal stationarity to apply further modifications that are useful from a computational and/or statistical point of view. %

There are two versions of tsFCI, both of which have been introduced in the original work \citep{enter2010causal}: One version does not allow for contemporaneous causal influences in the data-generating process and another version in which such influences are allowed. Figure~\ref{fig:toymodel} shows an illustration of the output of tsFCI (its version that allows contemporaneous causal influences) and other FCI-based time series causal discovery algorithms (see below) in the case of an example time series graph.

In the graph learned by tsFCI (both variants), there can be an edge between the pair of variables $X^i_{s-\tau}$ and $X^j_s$ (for example, $X^i_{s-\tau} \leftrightarrow X^j_s$ or, only if $\tau = 0$, $X^i_{s} \leftarrow X^j_s$) and at the same time no edge between the pair of variables $X^i_{t-\tau}$ and $X^j_t$ for $t > s$. At first sight, this non-repetition of edges seems to contradict the assumption of a causally stationary time series. However, there is, in fact, no contradiction: The ``additional'' edge between $X^i_{s-\tau}$ and $X^j_s$ can result from \emph{temporal confounding}, that is, result from confounding by variables that are before the observed time window $[t-\tau_{\max},t]$.

\item{\it SVAR-FCI.} %
The SVAR-FCI algorithm \citep{malinsky2018causal} is another adaption of FCI to causally stationary time series.
As compared to tsFCI (variant with contemporaneous causation), SVAR-FCI removes the ``additional'' edges that have been discussed in the last paragraph in the section on tsFCI. More specifically, whenever SVAR-FCI detects marginal or conditional independence of $X^i_{t-\tau}$ and $X^j_t$ and hence removes the edge between these two variables, then the algorithm automatically and immediately also removes the edges between the variables $X^i_{s-\tau}$ and $X^j_s$ for all time steps $s$. Consequently, in the graph learned by SVAR-FCI, the variables $X^i_{t-\tau}$ and $X^j_t$ are adjacent if and only if for all $s$ the variables $X^i_{s-\tau}$ and $X^j_s$ are adjacent.

The removal of the additional edges is justified by the fact that these edges are known to result from confounding variables that are before the observed time window. Hence, these edges would also disappear in tsFCI (both variants) for an appropriately increased length of the observed time window. Moreover, due to this modification, SVAR-FCI requires a smaller number of (conditional) independence tests than tsFCI. However, as has been realised and explained in \citep{gerhardus2021characterisation}, SVAR-FCI, in theory, discovers fewer edge orientations than tsFCI (variant with contemporaneous causation). To the authors' knowledge, an empirical comparison of tsFCI (variant with contemporaneous causation) and SVAR-FCI on finite samples that would shed light on this trade-off has not yet been performed.

\item{\it SVAR-GFCI.} %
The SVAR-GFCI algorithm \citep{malinsky2018causal} is another adaption of FCI to causally stationary time series.
This algorithm is a hybrid method that combines independence-based and score-based causal discovery. In the first step, the algorithm employs a time series variant of the score-based GES algorithm (see below for an explanation of the GES algorithm). The adjacencies found in this first step are then passed as starting conditions to the independence-based SVAR-FCI algorithm.

\item{\it LPCMCI.} %
The LPCMCI algorithm \citep{gerhardus2020high}, also known as Latent-PCMCI, is another adaption of FCI to causally stationary time series.
As compared to SVAR-FCI, Latent-PCMCI applies the ideas of PCMCI and PCMCI$^+$ (see above) generalised to the causally insufficient case. Simulation studies in \citep{gerhardus2020high} show that Latent-PCMCI strongly outperforms SVAR-FCI on finite samples, especially for strongly autocorrelated and continuously valued variables.

\item{\it Learning extended summary graphs.}
Extended summary graphs \citep{assaad2022discovery} are  time-compressed representations of time series graphs that distinguish between contemporaneous and lagged links (so they are more informative than summary graphs) but that do not distinguish between two different non-zero lags (so they are less informative than time series graph), see \secref{sec:target-graph-axes} above for more details. To learn extended summary graphs, \citep{assaad2022discovery} propose the \textbf{PCGCE} and \textbf{FCIGCE} algorithms. The basic idea of both these algorithms is to group the past time steps of each component time series $X^i$ into a vector variable $X^{i,-}$ and then, using conditional independence tests that can handle vectors of variables, respectively apply the PC and FCI algorithms directly to the extended summary graph. Specifically, \citep{assaad2022discovery} use a conditional independence test based on conditional mutual information. Moreover, for practical reasons, the vector variable $X^{i,-}$ does not contain \emph{all} past time steps of $X^i$ but all those time steps within the time window $[t-\gamma, t-1]$ where $\infty > \gamma \geq 1$ is a user-specified positive integer that corresponds to the maximal lag up to which the method is supposed to be sensitive (this parameter $\gamma$ corresponds to the parameter $\tau_{max}$ in PCMCI, for example).

\end{enumerate}

\paragraph{Greedy Equivalence Search (GES).}
GES~\citep{Meek1997GraphicalMS, chickering2002optimal, chickering2002learning} is probably the most 
famous score-based causal discovery method for 
i.i.d data assuming that the true causal graph is a DAG. 
It performs greedy steps directly on the CPDAG, thus searching in the Markov equivalent class space.
Chickering~\citep{chickering2002optimal} proved that 
an efficient two-phase greedy search, combined with 
the BIC score is sufficient 
to find the true CPDAG in the large sample limit (the Meek conjecture~\citep{Meek1997GraphicalMS}) assuming 
causal sufficiency. Ramsey \textit{et al.}~\citep{ramsey2017million}
developed {Fast GES (FGES)} an optimised and parallelised version of the GES algorithm, which they were able to scale up to a million variables. 
Additionally, GES has been improved by bounding polynomially the score evaluations~\citep{chickering2015selective}; to obtain statistical efficiency~\citep{chickering2020statistically}; to obtain finite-sample correction of confidence intervals~\citep{gradu2022valid} and to deal with latent variables~\citep{claassen22a}.
Similarly to other methods developed for i.i.d.~data,
(F)GES can be applied to uniformly-sampled
causally stationary time series by simply considering 
the transformed lagged variables and imposing 
that the effects cannot precede the causes in time.

\paragraph{Continuous optimisation methods.}
A recent advance in structure learning has been 
the development of so-called continuous optimisation
methods, particularly score-based methods which 
avoid the explosion of the discrete space of DAGs by
employing continuous optimisation. 
The first such method is the 
NOTEARS algorithm proposed by 
Zheng \textit{et al.}~\citep{zheng2018notears} which proposed $h({\bf W}) = \operatorname{tr}(\exp({\bf W} \circ {\bf W}) - d$ as differentiable characterisation of acyclicity for a weighted adjacency matrix ${\bf W}$, that is $h({\bf W})=0$ if and only if the associated graph is a DAG. Such differentiable characterisations allow the plug-in use of different continuous optimisation methods and even complex function parameterisations~\citep{zheng2020learning,ng2022convergence, Lachapelle2020Gradient-Based, ng2020golem, Yu2021curl, bello2022dagma}.
Various implementations of the continuous optimisation framework for structure learning from time series are available:
(1) DYNOTEARS~\citep{pamfil2020dynotears} considers structural linear VAR (SVAR) models, allowing for 
contemporaneous links and enforcing acyclicity 
among the instantaneous edges.  
(2) IDYNO~\citep{gao22IDYNO} is 
designed to perform structural discovery from both observational and interventional data. Moreover, both linear and non-linear relationships are considered. 
(3) NTS-NOTEARS ~\citep{sun2023nts}  models cause-effect relationships through one-dimensional convolution neural networks (CNN) and allows 
prior knowledge to be encoded and exploited directly by the optimisation procedure.

\paragraph{(VAR)LiNGAM.} 

Linear non-Gaussian acyclic model (LiNGAM)~\citep{shimizu2006linear} 
is a classical method for causal discovery which 
assume acyclicity, causal sufficiency, linear relationships and 
non-Gaussian additive independent noises. 
Under those assumptions, the model is shown to
be identifiable thanks to classical results
from independent component analysis (ICA)~\citep{ICA2001}.
Specifically, a linear SEM can be represented by, 
\begin{equation}
    \bm{X} = \mathbf{B}\bm{X} + \bm{\varepsilon},
    \label{eq:SEMlin}
\end{equation} where $\bm{X}$ is the vector of 
system variables, $\bm{\varepsilon}$ the noise vector 
and $\mathbf{B}$ is the matrix of coefficients. 
Thus solving Equation~\ref{eq:SEMlin} for $\bm{X}$ we obtain,
\begin{equation*}
    \bm{X} = (\mathbf{I}-\mathbf{B})^{-1}\bm{\varepsilon}.
\end{equation*}
The above equation is the 
an ICA problem, and
its theory states that when $\bm{\varepsilon}$ are 
non-Gaussian noises variables, the mixing matrix 
$\mathbf{A}=(\mathbf{I}-\mathbf{B})^{-1}$ is identifiable in the large sample 
limit~\citep{ICA2001}.
LiNGAM-ICA works by first 
exploiting the ICA results to
obtain the mixing matrix $\mathbf{A}$, and secondly by permuting and 
normalising $\mathbf{A}$ to obtain the 
appropriate matrix $\mathbf{B}$ and the corresponding causal DAG.
The LiNGAM can also be solved directly 
without using ICA~\citep{shimizu2011directlingam}.
The direct-LiNGAM iteratively selects the variable 
which is the most independent from the residual (in a linear regression onto the remaining variables) and replaces the original data with the residuals matrix.
In the end, the procedure gives a causal order; and ultimately,
simple linear regressions (\eg estimated with least squares) are
used to obtain the final estimation of the lower triangular $\mathbf{B}$ matrix.  

VAR-LiNGAM~\citep{hyvarinen2008causal,zhang2008causal,hyvarinen2010estimation} is the extension of the classical LiNGAM model to time series data. It considers a linear structural vector auto-regressive (VAR) model, with possibly acyclic instantaneous relationships, and assumes 
non-Gaussian disturbances. In particular, the process $\bm{X}_t$ is assumed to evolve following
\begin{equation*}
    \bm{X}_t = \sum_{\tau=0}^k \mathbf{B}_{\tau}\bm{X}_{t-\tau}  + \bm{\varepsilon}_t.
\end{equation*}
Where $\mathbf{B}_0$ is the matrix of instantaneous 
relationships and its sparsity pattern corresponds to a DAG while $\mathbf{B}_\tau$ for $\tau>0$ are the 
matrices of lagged relationships; moreover
the non-Gaussian noise vector $\bm{\varepsilon}_t$ has independent components and is assumed independent over time. 
\citet{hyvarinen2010estimation}
propose two methods to estimate the 
coefficients of the model above: (1) a two-stage method
which combines least-square estimation of the autoregressive model and classical LiNGAM estimation;
(2) a method based on multichannel blind deconvolution. 
In Figure~\ref{fig:toymodel}, we illustrate that if the assumptions are satisfied, VAR-LiNGAM is able to discover and orient all causal links.

\paragraph{Structural equation models for time series, TiMiNo.} 

\citet{peters2013causal} extended the classical structural equation modelling framework to the time series setting by introducing 
time series models with independent noise (TiMINo). 
In particular, a multi-variate time series 
$X_t = (X_t^i)_{i \in V}$ satisfies a TiMINo if 
there exists a $p>0$ and, for every $i \in V$, 
there are subsets $\mathbf{PA}_0^{i} \subseteq X^{V\setminus \{i\} }$ and $\mathbf{PA}_k^{i} \subseteq X^{V}$ for $k=1,\ldots,p$ such that,
\begin{equation}
\label{eq:timino}
    X_t^i = f_i\left((\mathbf{PA}_p^{i})_{t-p}, \ldots, 
    (\mathbf{PA}_1^{i})_{t-1}, (\mathbf{PA}_0^{i})_{t}, N_t^i\right),
\end{equation}
where $N_t^i$ are assumed to be jointly independent over $i$ and time and for each $i$, $N^i_t$ are identically distributed in time.
Thus, the assumed model is extremely general, 
instantaneous relationships are allowed and 
the noise contribution is not restricted by allowing 
arbitrarily mixing through the $f_i$ functions. 
Nevertheless, to prove the identifiability of the 
full-time graph and the causal summary time graph,
additional assumptions have to be made. 
Specifically, in \citet{peters2013causal} it is assumed that either (i) Equations~\ref{eq:timino}  follows an identifiable functional model class (\eg nonlinear functions with additive Gaussian
noise or linear functions with additive non-Gaussian noise~\citep{peters2011identifiability}) or (ii) 
each function $f_i$ exhibits a time structure, that is 
the union of the causal parents of $X^i_t$ contains at least one $X_{t-k}^i$ and moreover the joint
distribution is faithful with respect to the full-time graph with the summary time graph being acyclic.
Under either assumption (i) or (ii), the complete causal graph is proved to be identifiable in the large sample limit. 
Practically, to estimate a TiMINo, methods for causal discovery for additive noise models with i.i.d. data~\citep{mooij2009regression} can be adapted to the
time series setting. As in the direct-LiNGAM, 
it iteratively selects the causal order
between the variables by fitting regression models and
evaluating the independence between the variables and the residuals. To fit the regression models for $f_i$, various methods can be used, such as vector autoregressive models (linear), generalised additive models or 
Gaussian processes.  Moreover, to test independence from the residuals, the HSIC~\citep{gretton2007kernel} can be applied to all possible shifted time series up to the maximum lag order.

\paragraph{Invariant Causal Prediction.}

Invariant causal prediction deals with the setting of independent and identically distributed samples of the random vectors $X=(X_1, X_2,\ldots,X_p)^\top \in \Real^p$, $E=(E_1,E_2,\ldots,E_q)^\top \in \Real^q$, and $Y \in \Real$. For a variable or vector of interest $Y$, $E$ is a set of environment variables that may be causes of $X$ but are not direct causes or effects (direct or indirect) of $Y$.  ICP assumes that $Y$ is generated \emph{causally} from a subset $S^* \subseteq \{1,\ldots,p\}$ of the $p$ variables considered, so that there is causal sufficiency, and $Y$ is generated from a Structural Causal Model obeying:
\begin{align}
    Y = g(X_{S^*}, \epsilon),\phantom{bl} \epsilon \sim F,\phantom{bl} \epsilon \ind X_{S^*},   
\end{align}
where $g$ and $F$ are arbitrary functions and distributions, respectively.  A set $S \subseteq \{1,\ldots,p\}$ is a generic subset of the full set of candidate causes. We refer to $S$ and $X_S$ interchangeably for brevity. The task of ICP is to infer the set $S^*$ of direct causes of the variable of interest $Y$. 

The Invariance Causal Prediction (ICP) framework \citep{peters2016causal} is based on the observation that $Y$ is independent of $E$ given $X_{S^*}$, denoted $Y \ind E | X_{S^*}$. Assuming we have a set of candidate causes $X$ that includes $S^*$, the causal subset, and an environment variable $E$ that we know does not directly cause $Y$ or is an effect of $Y$. We can search for $S^*$ by applying a conditional independence test $Y \ind E|X_S$ on $Y$, $E$ and subsets $S \subseteq \{1,\ldots,p\}$ of $X$. ICP then selects as causal variables the intersection of all those subsets $S$ where the corresponding conditional independence test is not rejected:
\begin{align}
    \hat{S}^* = \bigcap\limits_{S:p_S > \alpha} S.   
\end{align}
Here $p_S$ is the $p$-value associated with the conditional independence test of $Y \ind E|X_S$, with the null hypothesis corresponding to conditional independence. We do not reject conditional independence at significance level $\alpha$ if $p_S>\alpha$.

One way to interpret the problem setting is that causal associations are more robust than other associations. So ICP finds the causes of a variable of interest by investigating which associations are invariant across environments.  Another interpretation is that the environment variables define data generated under different interventions to the system (SCM). 

In physical sciences, regional and temporal variables are good candidates since these often describe changes in environments that alter the conditions  under which physical processes occur. 

\citet{peters2016causal} introduce an algorithm to implement ICP  that assumes linear relationships between causes $X$ and effects $Y$ and a categorical, univariate variable $E$. In  \citep{heinze2018causal}, more general  algorithms are presented that allow for nonlinear relationships between cause and effects and for continuous and multivariate environment variables $E$. \citet{pfister2019invariant} provide an ICP variant for time series data.
The proposed time-series ICP relies on the 
causal invariance assumption across time points, thus 
removing the requirement of environment knowledge. 
This setting is also partially robust to hidden confounders, similar to the original ICP~\citep{peters2016causal} framework, and in general, the ICP is expected to be conservative with respect to violations of its assumptions~\citep{pfister2019invariant}.

\paragraph{Causal frameworks for continuous-time systems.}  

As we already reviewed, discrete-time 
causal systems fit directly as an extension
of i.i.d. and DAG framework. 
When considering discrete-time systems,
we can express the value of a variable at a time 
$t$ as a function of other variables (and itself)
observed at past instants, thus the 
complete causal graph can be seen as a DAG
extended (possibly infinitely) in time. 
Classical methods for causal 
discovery in the i.i.d. setting can 
then be adapted to discrete-time 
systems quite straightforwardly. 

Conversely, 
considering continuous-time systems, 
raise the issue that 
a time-extended DAG is not feasible, 
since the included variables would be 
uncountable~\citep{hansen2014causal}. 
Nevertheless, modelling 
continuous dynamical system
helps in dealing with non-uniform sampling and 
extrapolating among different sampling frequencies, two
major drawbacks of causal discovery methods for discrete-time
systems.
The following are the major causal frameworks available for
continuous-time systems:
\begin{enumerate}
\item {\it Causal interpretation of ODE and SDE.}
Various efforts have been made to describe causal systems with ODEs and SDEs. 
First, causal discourses around ODE were used to obtain different justifications for the cyclic SEM~\citep{hyttinen12a, moij2013ordinary, bongers2018causal, 
peters2022causal}. 
\citet{rubenstain2018} described Dynamical Structural 
Causal Models (DSCM) as extensions of SEM where each 
equation or assignment is a relationship between a 
set of causal parent trajectories and an effect trajectory. 
Under some stability conditions, such DSCM can be obtained from 
ODE systems. 
SDEs have also been studied from a causal perspective~\citep{hansen2014causal, peters2022causal}. 
The advantage of SDEs in modelling physical systems is that they allow incorporating 
an inherent source of stochasticity, a common assumption in 
numerous real-world systems~\citep{abbati2019ares}. 
Graphical parameterisations of SDE equilibrium distributions leading to
models allowing for cycles have been investigated and different structure learning algorithms have been proposed~\citep{varando2020graphical}.

\item {\it Dynamic Causal Models.} 
Dynamic Causal Models (DCM)~\citep{ friston2003dynamic, marreiros2010dynamic, stephan2010ten,abbati2019ares, friston2013analysing}, in short, is a Bayesian framework 
for fitting and comparing 
causal models for coupled
dynamical systems. 
DCM was introduced and applied mostly
in Neuroscience, and  
particularly in the problem of estimating the  
connectivity between brain regions from neuroimaging data, as discussed in detail in Section \ref{subsubsec:causal_methods_neurosci}. Recently, 
and it has been even employed
 to model the COVID-19 pandemic dynamics ~\citep{friston2020dynamic, friston2022dynamic}.

\item {\it Local independence graphs.}
Local independence is a notion of conditional independence for stochastic processes (both discrete and continuous time ones) which can be (in)dependent on each other pasts. 
In detail, for real-valued stochastic processes $X_t=(X^1_t, \ldots, X^p_t)$ and
 $A,B,C \subseteq \{1, \ldots, p)$,
we say that $X^B$ is locally independent of 
$X^A$ given $X^C$ at time $t$ if 
the past of $X^C$ until time $t$ provides the same 
information, to predict $E[X_t^\beta | \mathcal{F}_t^{A \cup C}]$~\footnote{where $\mathcal{F}_t^{A \cup C}$ 
is a right-continuous and
complete filtration which represents the history of the processes $X^{A \cup C}$ see \eg \citep{mogensen2020markov} for a detailed description},
as the past of $X^{A \cup C}$ until time $t$, for each $\beta \in B$.

Didelez~\citep{didelez2008graphical, didelez2006asymmetric} 
studied graphical representations of local independence
with directed graphs together with $\delta$-separation and proved the equivalence of the pairwise and global Markov 
properties for multivariate counting processes.
Directed graphs and $\delta$-separation has been then 
extended to 
mixed graphs and $\mu$-separation~\citep{mogensen2020markov} to 
model partially unobserved systems. 
A constrained-based algorithm has been proposed~\citep{mogensencausal}, which is proven to be sound and complete under faithfulness assumption.
Local independence graphs can be applied to multivariate processes which are solutions of SDEs (such as the multivariate Ornstein-Uhlenbeck process) or event and counting processes such as Hawkes processes~\citep{mogensen2022equality}.
\end{enumerate}

\section{Challenges}\label{sec:causal-discovery-challenges}

In this section, we discuss several challenges for causal discovery that are frequent in real-world applications. Following \citet{runge2019inferring}, we distinguish between challenges related to the data-generating process itself (see \secref{sec:causal-discovery-process-challenges}), challenges associated with the available data (see \secref{sec:causal-discovery-data-challenges}), and challenges of statistical or computational nature (see \secref{sec:causal-discovery-stat-and-comp-challenges}). Users should carefully consider the challenges they face in their application and choose a suitable causal discovery method. More generally, how to reason and deal with typical challenges in causal inference is further discussed in a time series context in \citet{runge2023causal}.

\subsection{Process challenges}\label{sec:causal-discovery-process-challenges}

\begin{wrapfigure}{r}{8cm}
\vspace{-0.5cm}
\begin{mdframed}[backgroundcolor=gray!20] 
\small
{\em ``Discovering causal relations from observational data is impossible without assumptions about the mechanisms and faces important challenges related to data and statistical characteristics. The field will need to incorporate domain knowledge and post-selection inference.''}
\end{mdframed}
\vspace{-0.5cm}
\end{wrapfigure}
Non-linearities pose challenges for causal discovery in both independence-based and score-based methods. Non-linear conditional independence tests, such as the GPDC test \citep{rasmussen2006gaussian} and tests based on conditional mutual information \citep{runge2018conditional}, are computationally more expensive and tend to have lower statistical power than linear tests. Non-linear functional relationships in score-based methods require more complex score functions, which can decrease finite-sample performance. However, non-linear functional relationships can enhance identifiability in some asymmetry-based causal discovery methods \citep{Peters2017book}. Overall, nonlinearities increase model complexity and require careful consideration when selecting appropriate causal discovery methods.

Most time series causal discovery methods assume data generated from a causally stationary process with a unique equilibrium distribution. However, many real-world processes are {non-stationary}. Recently there have been works on devising causal discovery methods that first detect the variables afflicted with non-stationary driving mechanisms and subsequently infer the entire causal graph, including possibly proxy variables corresponding to the driving force of non-stationarity \citep{mooij2020joint, huang2020causal}. Furthermore, techniques exist to detect regime or context changes in non-stationary data \citep{huang2015identification, saggioro2020reconstructing, huang2020causal}, but it remains an %
an active area of research. Domain experts may be able to identify the source of non-stationarity in time series data and preprocess the data to remove it.

Time series data is common in physical sciences, and it has a distinctive feature of {auto-correlation}. Many causal discovery algorithms are not designed for time series data and show decreased performance when applied without modification \citep{runge2019detecting, runge2020discovering}. However, some methods are specifically designed for time series data, reducing the detrimental effect of auto-correlation. See the PCMCI algorithm above for a discussion on the challenges of auto-correlated data. %
In many domains, space adds to time as well. %
In principle, one could feed different spatial locations of the same variable as distinct variables into causal discovery methods. However, this naive approach ignores spatial correlations and quickly results in a high-dimensional problem. Another workaround, employed for example in \citep{tibau2022spatiotemporal,bueso2020explicit}, is to perform dimension-reduction as a preprocessing step and then perform causal discovery on the dimensionally-reduced space. The development of causal discovery inherently designed for {spatio-temporal data} is an active area of research; for example, see \citep{christiansen2022toward}.
Dealing with variables whose dynamics operate on {different time scales}, such as \eg fast atmospheric and slow oceanic processes, pose important challenges. Granger causality in the frequency domain, \eg~\citep{bressler2011wiener, chicharro2011spectral, faes2012measuring}, and a combination of wavelet analysis with transfer entropy \citep{lungarella2007information} are examples of approaches to deal with the time scales of causal influences.

Finally, it is worth mentioning that many causal discovery methods assume the data-generating SCM to be acyclic (see axis VII in \secref{sec:axes-of-distinction}). However, in real-world applications, one can often not exclude the existence of feedback that acts on time scales below the measurement interval. Such feedbacks make the time series graph \textbf{cyclic}. In the non-temporal setting, one might often not be able to exclude the existence of causal cycles. As discussed and referenced above, there are causal discovery methods that can handle cyclic causal graphs.

\subsection{Data challenges}\label{sec:causal-discovery-data-challenges}

Various data challenges arise when tackling the problem of causal discovery in practice. This is mainly due to the discrepancy between the assumed hypothesis needed from each method or framework and the real-world data. One of the most common assumptions of most causal discovery methods is causal sufficiency, which is the hypothesis that all relevant variables are observed. {Unobserved variables} are especially problematic when they are possible confounders between system variables since omitting confounders from the causal discovery could lead to learning spurious or wrong relationships. There are some available methods which do not assume causal sufficiency, such as LPCMCI~\citep{gerhardus2020high}, SVAR-FCI~\citep{malinsky2018causal} and GPS~\citep{claassen22a} (see \secref{sec:list-of-causal-discovery-methods}).

{Missing data} and {selection bias} are other common issues in real-world applications, and there have been some efforts in developing causal discovery methods which are resilient to these challenges~\citep{strobl2018fast,gain2018structure,tu2019causal,versteeg22a}. Not always, especially in the physical realm, the data follow predictable and well-behaved distributions. One such example is {zero-inflated data}, which is common, for instance, in gene expression data, where single-cell expressions lack detectable values of transcripts that appear abundant on bulk (thousands of cells) gene expression experiments. Recent advances have developed graphical models and causal discovery methods in such scenarios~\citep{mcdavid2019graphical,yu2020directed}.

\subsection{Statistical and computational challenges}\label{sec:causal-discovery-stat-and-comp-challenges}

The high dimensionality of data in physical systems, such as spatiotemporal data, and small sample sizes are central statistical challenges for causal discovery. On the other hand, large sample sizes raise issues of unaffordable computational time, which can scale up to cubically for kernel methods typically used for independence testing \citep{kernels2008, gretton2007kernel}. High-dimensional data leads to large conditioning sets in particular algorithms, effectively reducing the sample size available to test the hypothesis. 

As noted in \secref{sec:causal-discovery-process-challenges}, non-linearity is a common characteristic of processes in the physical sciences. For the case of independence-based or hybrid casual discovery techniques, this calls for devising {non-parametric tests of independence}, for instance, tests based on measures of conditional mutual information \citep{runge2018conditional}, or on Gaussian process regression or other kernel-based measures on independence \citep{gretton2007kernel}, or using quantile regression~\citep{Petersen2021testing} and copula-based methods~\citep{bouezmarni2012nonparametric} (also applied to Granger causality), etc.
The no-free-lunch theorem of~\citet{shah2020hardness} states that no single conditional independence test can have power against all alternatives. Here, the challenge lies in devising and applying (a combination of) conditional independence tests that are the most suited for a particular physical system.

The concept of {post-selection inference} \citep{berk2013} involves performing statistical inference on a model that was selected based on data-driven methods rather than being pre-selected. While there are some advances in solving the post-selection inference problem for regression and causal effect estimation, few solutions have been proposed for the inference after causal discovery setting \citep{berk2013,belloni2014inference,rinaldo2019bootstrapping}. One possible solution is sample-splitting, but this is often statistically inefficient. A recent development is the randomised version of the greedy equivalence search (GES) algorithm, which allows for finite-sample correction of classical confidence intervals \citep{gradu2022valid}.

\section{Opportunities for the physical sciences}\label{sec:causal-discovery-opportunities}

The field of causal discovery from observational data is still in its infancy but growing in methodologies, theoretical guarantees of performance, and empirical evidence. Causal inference, in general, is a vast field that offers alternatives and scientific opportunities that we review in what follows. Only revisiting the whole body of empirical science based on association would take a village, but the advances would pay off.

\subsection{Causal hypothesis testing and targeted interventions}

\begin{wrapfigure}{r}{6.3cm}
\vspace{-0.5cm}
\begin{mdframed}[backgroundcolor=gray!20] 
\small
{\em ``Observational causal discovery offers revolutionary opportunities %
to test hypotheses, evaluate the impact of interventions, attribute extreme events with counterfactuals, and characterise complex systems by deriving causal pathways and robust forecasting models.''}
\end{mdframed}
\vspace{-0.5cm}
\end{wrapfigure}
Scientists need a principled way to test different hypotheses against each other. A causal hypothesis is a supposition or theory about how things interact, specifically on whether one thing causes another. Causal studies aim to confirm or reject any given causal hypothesis. The problem is that hypotheses in the physical sciences are often presented as narratives giving a chain of causal factors that lead to the studied phenomenon. Without a causal vocabulary and analytical tools, it is often impossible to precisely state the hypothesis, which leads to several competing hypotheses or, even worse, a false hypothesis, which is accepted as true due to its compelling narrative quality. Testing hypotheses have been conducted in myriad ways \citep{pearl2009causality,Peters2017book,robins2020causal}. 

Causal graphs, as graphical representations of assumed or learned causal relations, provide a more principled way to talk about causal hypotheses. Learned graphs imply causal links and pathways and provide evidence for deciding between rivalling causal hypotheses, in \citet{kretschmer2016using}, for instance, regarding competing hypotheses of Arctic climate teleconnections. 

The conclusiveness and interpretability of discovered causal graphs from purely observational data sets on the often untestable validity of the methods' assumptions and the statistical complexity of the task. But observational causal discovery can help inform more targeted subsequent interventions, which are often too expensive to employ on a large scale~\citep{pearl2009causality,robins2020causal}. Incorporating interventions, if performed meaningfully, could thus make the causal discovery process much more efficient and robust (discovered DAGs not being confined to the Markov equivalence class). Interestingly, interventions could be differentiable, i.e., `learnable' from data \citep{brouillard2020differentiable}. 

\subsection{Cause-effect estimation} \label{sec:causaleffects}

Causal discovery results in qualitative causal graphs, or often Markov equivalence classes of graphs. But often, the target question is a quantitative estimate of a causal effect of one variable $X$ on another variable $Y$, as pioneered by \citet{pearl2009causality}. This topic is discussed in a time series context in \citet{runge2023causal}. The quantity of interest then is the (interventional) distribution of $Y$ given an intervention in $X$, $p(Y=y~\vert~ do(X=x))$. The fundamental problem is that typically $p(Y=y~\vert~ do(X=x))\neq p(Y=y~\vert~ X=x)$. Confounders, for example, can introduce a non-causal association between the treatments and the outcome. Randomised experiments would be the gold standard by eliminating the unwanted non-causal associations \citep{fisher1935design, imbens2015causal, pearl2009causality}. 
The goal of causal effect estimation is to do so without access to interventions by expressing $p(Y=y~\vert~ do(X=x))$ as a function of the observational distribution $p(\mathbf{x})$:
\begin{align}
p(Y=y~\vert~ do(X=x)) &= \text{function of}~p(\mathbf{x}) \,.
\end{align}
If such a re-expression is possible, one calls the causal effect identifiable and obtains a causal estimand, which involves only the observational distribution.
The most well-known method for causal effect estimation from data without parametric assumptions is \emph{covariate adjustment}~\citep{pearl2009causality}, which refers to de-confounding the causal relationship by adjusting for a set of variables $\mathbf{Z}$. In the general case, the adjustment formula is
\begin{align}
p(y~\vert~ do(X=x))=\int p(y~\vert~x,\mathbf{z}) p(\mathbf{z}) d\mathbf{z}\,.
\end{align} 
Recent work has focused on finding statistically optimal adjustment sets~\citep{runge2021necessary}, i.e., for which the estimators have minimal variance.
Using the \textit{do}-calculus \citep{pearl1995causal, huang2006pearls, shpitser2006identification, shpitser2008complete}, it is possible to determine whether a causal effect is, in principle, identifiable from observational data or not. To this end, causal effect estimation requires fully specified causal graphs from causal discovery (with its inherent reliance on further assumptions) or domain expertise that can qualitatively specify a causal graph.
For example, it is known that temperature influences ecosystem respiration, but one may want to quantify how much when given a graph of other observed and unobserved confounding variables. The graph then encodes assumptions about the absence and presence of causal relations.

Different variants of causal effects can be defined based on the interventional distribution $p(y~\vert~ do(X=x))$, and an estimate then involves further parametric assumptions. For example, in a linear model, the total causal effect on the expected value of $Y$ when setting $X$ by intervention to $x^\prime$  as opposed to $x$ is given by 
\begin{equation}
\Delta_{X \rightarrow Y}(x^\prime, x) = \, \Delta x \cdot \beta_{X{\to}Y} \, ,
\end{equation}
where $\Delta x=x^\prime-x$ and $\beta_{X{\to}Y}$ can be estimated as the regression parameter of $X$ in the linear regression of $Y$ on $X \cup \mathbf{Z}$.

\subsection{Causal pathway analysis and mediation}

Next to quantifying the overall causal effect of $X$ on $Y$, a relevant follow-up question is often about the causal pathways: the mechanisms by which this effect propagates. In complex systems, it is often interesting to analyse how perturbations spread throughout the systems and through which subprocesses perturbations are mediated~\citep{runge2015identifying,runge2015quantifying}.
Within the structural causal model framework, mediation formally leads to counterfactual quantities, see, for example, \citet{ vanderweele2015explanation} and briefly below in \secref{sec:counterfactuals}.
But for linear models, the mediated causal effect (MCE) of $X$ on $Y$ that passes through a mediator $M$ (here $X,Y,M\in \mathbf{V}$) can be computed by summing up the contributions along all paths passing through it:
\begin{align}
 {\rm MCE}(X,Y|M) &= \sum_{\pi^M_k} \prod_{\lambda_{i\to j}\in \pi^M_k} \beta_{i{\to}j}  \,,
\end{align}
where the summand iterates over causal paths  $\pi^M_k$ from $X$ to $Y$ through $M$ and the product is over all links $\lambda_{i\to j}$ on each path. The link coefficient $\beta_{i{\to}j}$ can be estimated as the regression coefficient of $V^j_t$ in the  linear regression of $V^j_t$ on the parents of $V^j_t$.
Mediation analysis can also answer the complementary question: how strong is the direct effect of $X$ on $Y$.

\subsection{Identifying causes and pathways leading to anomalies} \label{sec:anomalies}

Anomaly detection~\citep{chandola2009anomaly} is an important problem in engineering and the physical sciences. In Earth sciences, extreme events form a subclass of anomalies and can be structured across different dimensions, such as compound extremes~\citep{zscheischler2020typology}. While detecting anomalies is an important problem, it does not answer the often relevant question of what causes a particular anomaly or, more generally, what causes the anomalous process. In engineering, business science, and healthcare, a related problem is \emph{root cause analysis}~\citep{andersen2006root}.
Causal discovery can address the problem of identifying causal drivers (parents) or indirect mediating pathways and facilitate quantitative analyses to analyse the contribution of different physical drivers in causing an extreme.

\subsection{Causal complex network analysis}

Complex systems are often viewed as networks of interacting subprocesses, for example, the human brain~\citep{bullmore2009complex}, or the Earth system~\citep{Donges2009backbone,ludescher2021network}. Tools of network theory~\citep{boccaletti2006complex} have been used to analyse quantities such as the information flow as it propagates through the system or the stability of subprocesses~\citep{Gozolchiani2011}. A common network measure is the node degree, which quantifies the number of processes linked to a node. A more involved measure is betweenness centrality, which quantifies the number of shortest paths through a particular node.
A crucial question is then to define what these paths mean. In works where the networks are based on pairwise correlation or mutual information~\citep{Donges2009backbone,ludescher2021network}, one may associate paths with a transfer of information.

However, there is a difference between information being transferred versus perturbations propagating through the network. Here a question can be to identify how critical individual subprocesses are in spreading and mediating perturbations in such dynamic complex systems. The propagation of perturbations, aka interventions, relates to a causal question requiring a causal definition of network links able to distinguish direct from indirect interactions. 

In addition, the toolbox of classical network measures is not rich enough for quantifying gateways and mediators of perturbations. Essentially, these measures—with many originating from the social sciences~\citep{freeman1977set}—are based on a different definition of links, for example, two persons knowing each other, as opposed to dynamical interactions in a complex system. Hence, here measures based on causal pathways on which perturbations propagate in a complex system’s interaction network can be utilised, such as those studied in~\citet{runge2015identifying,runge2015quantifying}: Identifying the nodes of the causal graph with the components of the complex system, the average causal effect can be defined as a causal version of the out-degree or closeness centrality, which quantifies by how much an individual component causes any of the remaining components. This serves as a quantitative measure of how much a component is a gateway of perturbations. On the other hand, the average causal susceptibility measures how much a component is changed on average by a perturbation in any of the remaining components as a causal version of the in-degree or in-closeness. Finally, the average mediating causal effect measures how much of the pairwise causal effects between any pair are mediated through a particular variable, which can be seen as a causal version of betweenness centrality. In \citet{runge2015quantifying}, these measures are generalised in an information-theoretic framework.

\subsection{Causally robust forecasting models}

Forecasting a time series from multivariate predictors constitutes another problem where causal knowledge helps. Even considering the case of forecasting inside the same distribution, that is, assuming a stationary distribution, it can be proven information-theoretically that causal predictors maximise the mutual information with the target variable and, by the Markov property, any further predictors do not add further information. More formally~\citep{brown2012conditional}, the negative log-likelihood can be decomposed as follows
\begin{align}
\lim_{n\to\infty}-{l}  &= \lim_{n\to\infty}- \frac{1}{n} \sum_{t=1}^{n} \log \widehat{p} \left(X_{t+1}~|~\mathcal{P} ; \theta \right) \\
&= \mathbb{E} \left[ \log \frac{p(X_{t+1}~|~\mathcal{P} ) }{\widehat{p}(X_{t+1}~|~\mathcal{P} ; \theta  )} \right] \quad +  \quad \underbrace{I\left( X_{t+1}; \mathbf{X}^-_{t+1}\setminus \mathcal{P} ~|~\mathcal{P}\right)}_{=0} \quad + \quad H\left( X_{t+1} ~|~ \mathbf{X}^-_{t+1} \right)\,,
\end{align}
where $n$ is the sample size, $\widehat{p} \left(X_{t+1}~|~\mathcal{P} ; \theta \right)$ the prediction model for $X_{t+1}\in \mathbf{X}_{t+1}$ of the true underlying $p \left(X_{t+1}~|~\mathcal{P} \right)$ given its causal parents $\mathcal{P}$ and model parameters $\theta$. As shown, the log-likelihood decomposes into the model approximation error given $\mathcal{P}$ (first term), the conditional mutual information between the target and unselected variables $\mathbf{X}^-_{t+1}\setminus X_{t+1}$ given $\mathcal{P}$ (second term, zero by the Markov condition), and the irreducible entropy or uncertainty (last term).
This is especially relevant for finding optimal sets of predictors in the case where greedy selection strategies do not work because the predictors cause the target variable synergistically, for example, $X_{t+1}=Z^1_{t}\cdot Z^2_{t}+\eta_{t+1}$. As shown in \citet{runge2015optimal}, an optimal subset selection can be better performed on the smaller subset of causal predictors. In \citet{kretschmer2017early,di2019long}, causal pre-selection was used in a climate context.
Beyond stationary distributions, \citet{pmlr-v97-huang19g} address the task of causally-informed forecasting under nonstationary environments through state-space models.

\subsection{Physical simulation model evaluation} \label{sec:opportunities-eval}

Causal graphs and causal effects can be utilised to intercompare the output of physical models and evaluate and validate them against observations at the level of causal dependencies ~\citep{eyring2020earth,eyring2019taking,nowack2020causal,PerezSuay19shsic}.
One approach in this direction is to compare the causal graph obtained from the observational data to those obtained from simulated data. This procedure has been proposed in the climate sciences to compare climate model simulations and observational data through their corresponding causal graphs derived from PCMCI \citep{nowack2020causal}, cf. Section \ref{sec:casestudies}. The methodology could be adapted and applied to other physical science problems where one typically has complex datasets and simulations to confirm hypotheses.

\subsection{Counterfactual causal attribution of extreme events} \label{sec:counterfactuals}

Counterfactual questions are not about the distributions of a target variable due to possible (future) interventions, but about the distribution of a target variable for an alternative past intervention, given that a particular outcome was observed. 
Formally, just like interventional causal queries are represented by interventional SCMs, counterfactual queries are represented in counterfactual SCMs~\citep{pearl2009causality,Peters2017book}. Given an SCM over variables $\mathbf{V}$ and observations $\mathbf{v}$, in the \emph{counterfactual SCM}, the noise  distribution is updated such that the $\mathbf{V}=\mathbf{v}$ holds. Then the noise terms may not be independent anymore. Counterfactual queries are then \emph{do}-statements in the counterfactual SCM.
One example of a counterfactual distribution query is $p(y'_{x'}\vert y_x)$, which specifies the probability of observing $Y=y'\neq y$ under the hypothetical past intervention $do(X=x')$ when, in fact, $Y=y$ was observed under the intervention $do(X = x)$.
Such queries can be computed in different ways~\citep{pearl2009causality,correa2021nested} and  generally require more assumptions about the underlying structural causal model than causal effect questions or causal discovery.
Next to counterfactual distributional queries, \citet{halpern2016actual} discusses causation and counterfactuals regarding single events.

An example of a counterfactual question in climate is the causal attribution of extreme events~\citep{Hannart2016}.
The above query $p(y'_{x'}\vert y_x)$ is one specific type of a counterfactual question and is sometimes called ``probability of necessity'' (PN), which is typically the quantity of interest in lawsuits. Extreme event attribution requires to study of anthropogenic forcings compared to their absence, that is, solely natural forcings or internal variability of the climate system. If the probability of necessity is high enough, then a human-caused extreme event is established.

\subsection{Signal tracking for the discovery of proximal causes}

Many phenomena, such as extreme events in complex systems, such as El Niño events in the climate system and extreme volatility in the financial system, are caused by an initial anomaly that triggers a travelling cascade of events \citep{press1967compound}. This phenomenon is often called the ``butterfly effect''\footnote{The term is attributed to Lorenz when he noted that a weather model failed to reproduce the results of runs with the unrounded initial conditions. However, the idea was earlier recognised by Poincar\'e and further formalised by Wiener. The analogy became popular and originated the quantitative science of characterising {\em instability} in complex systems undergoing nonlinear dynamics and deterministic chaos.}, characterised by an anomaly in one part of a system having extreme consequences in another space and time. Such cascaded events are challenging to detect, predict, understand and characterise \citep{zscheischler2018future,menzly2004understanding}, and has led to the development of the field of the science surrounding the concept of predictability in complex systems \citep{grunberg1954predictability,bialek2001predictability,boffetta2002predictability}. A potential strategy for uncovering the cause of notable events is causal discovery, for instance, by conducting a simulation which begins from the start of the event in question and tracing the initiating signal back to its source. However, this presents difficulty in determining the initiating trigger. It is challenging to have a dialogue about recognising the drivers of intense impacts because the amount of correlated drivers is usually much greater than the number of causally pertinent drivers, which may only have a substantial effect when combined (synergy) \citep{runge2019inferring}. These kinds of relations are hard to portray with a pairwise network.

\subsection{Causal benchmarks, software and platforms} 

Method development and comparison require benchmark datasets with known causal ground truth for validation. Ideally, such ground truth comes from expert knowledge of real data or actual experiments that can also be used to falsify causal relationships predicted from observational causal inference methods. Unfortunately, in many fields, such as Earth system sciences, such datasets exist only for expert-labelled causal relations among a few variables (\eg, some bivariate examples \citep{mooij2016distinguishing}). A tractable approach is to generate synthetic data with simple model systems that mimic properties and challenges of data from the system under study but where the underlying ground truth is known. These can then be used to study the performance of causal discovery (and causal inference methods more generally) for different challenges in realistic finite sample situations. From a practitioner’s perspective, it is essential to determine which method is best suited for a particular task with particular challenges and for a specific set of assumptions. Synthetic data, adapted to the problem at hand, can be used to choose the suitable method, including method parameters.
A list of key methods for causal discovery and the available software and platforms is given in Table \ref{tab:software}. An example is the SAVAR model~\citep{tibau2022spatiotemporal} that mimics spatio-temporal features of climate data. The website \href{http://www.causeme.net}{causeme.net}~\citep{runge2019inferring,runge2020causality} aims to provide an open platform with synthetic models mimicking real data challenges on which causal discovery methods can be compared. Next to method comparison, the platform also calls for submissions of actual and modelled data sets where the causal structure is known with high confidence and was used on the \href{https://neurips.cc/Conferences/2019/CallForCompetitions}{{Causality 4 Climate NeurIPS competition}}~\citep{runge2020causality}. That competition sparked the investigation of a particular property of synthetic data and models called \emph{var-sortability}, which led to new insights in causal discovery methods~\citep{reisach2021beware}.

\begin{table}[!t]
\small
\begin{center}
\caption{Methods and open-source software for causal discovery.}
\begin{tabular}{p{0.6\textwidth}p{0.35\textwidth}}
\toprule
\textbf{Method}   &  \textbf{Software}  \\
\midrule
Granger causality (GC)~\citep{granger1969investigating}, kernel GC \citep{marinazzo2008kernel}, explicit KGC \citep{bueso2020explicit} & \href{https://github.com/cmu-phil/causal-learn}{causal-learn}, \href{https://www.statsmodels.org/stable/index.html}{statsmodels}, \href{https://github.com/danielemarinazzo/KernelGrangerCausality}{KGC}, \href{https://github.com/DiegoBueso/XKGC}{XKGC}
 \\ 
\midrule
CCM~\citep{sugihara2012detecting,ye2015distinguishing} & \href{https://cran.r-project.org/web/packages/rEDM/index.html}{rEDM} \\  
\midrule
PC~\citep{spirtes1991algorithm, Spirtes2000}, FCI~\cite{enter2010causal} & \href{https://github.com/cmu-phil/tetrad}{Tetrad}, \href{https://github.com/cmu-phil/causal-learn}{causal-learn}, \href{https://CRAN.R-project.org/package=pcalg}{pcalg}, \href{https://github.com/jakobrunge/tigramite}{Tigramite}, \href{https://CRAN.R-project.org/package=MXM}{MXM}, \href{https://CRAN.R-project.org/package=bnlearn}{bnlearn}, \href{https://CRAN.R-project.org/package=dbnlearn}{dbnlearn}, \href{https://www.pywhy.org/}{PyWhy}\\ 
\midrule
PCMCI~\citep{runge2019detecting}, PCMCI$^+$~\citep{runge2020discovering}, LPCMCI~\citep{gerhardus2020high} & \href{https://github.com/jakobrunge/tigramite}{Tigramite}  \\ 
\midrule
DYNOTEARS~\cite{pamfil2020dynotears} & \href{https://github.com/quantumblacklabs/causalnex}{Causalnex} \\
\midrule
TiMINo~\cite{peters2013causal} & \href{http://people.tuebingen.mpg.de/jpeters/onlineCodeTimino.zip}{R script}\\ 
\midrule
VARLiNGAM~\citep{hyvarinen2010estimation} & \href{https://github.com/cmu-phil/causal-learn}{causal-learn}, \href{https://github.com/cdt15/lingam}{lingam},
\href{https://sites.google.com/site/dorisentner/publications/VARLiNGAM}{original R code}
\\ 
\midrule
ICP~\cite{peters2016causal, heinze2018invariant,pfister2019invariant} & \href{https://cran.r-project.org/package=seqICP}{seqICP} \\ 
\bottomrule
\end{tabular}
\label{tab:software}
\end{center}
\end{table}

\section{Perspectives} 

This section reviewed causal discovery in the physical sciences, describing the main methods, challenges, and opportunities for future research. We laid out the fundamental elements of the causal discovery framework --SCMs, graphs, and associated distributions--, and gave an overview of the methodological concepts of learning qualitative causal graphs \citep{runge2019inferring,runge2023causal}. We deliberated on commonplace difficulties encountered in the field, such as determining and preprocessing causal variables, addressing non-stationarity, contemporaneous causation and hidden confounding, and selecting parametric models for nonlinear dependencies and non-Gaussian distributions.
Section \ref{sec:casestudies} will illustrate causal discovery methods in neurosciences and Earth sciences case studies. 

The body of causal inference has traditionally been embedded in several communities, mainly statistics, social sciences, econometrics and health sciences. %
Irreconcilable positions and long-standing discussions exist \citep{dowd2011separated}. Pearl argues that it is essential to distinguish between causal and statistical information, as they refer to two separate concepts\footnote{Statistical information deals with the probability of certain variables being observed. In contrast, causal information deals with hypothetical relationships in new situations.}, and suggests that clear distinctions should be established in the notation used, and each should be subject to different means of calculation \citep{pearl2011statistics}. Arguably, nonparametric SCMs (as a natural generalisation of those used by econometricians and social scientists in the 1950-1960s) have developed the field of causality in new mathematical underpinnings: explicate and enumerate causal assumptions, test implications, decide measurements and experiments, recognise and generate equivalent models, recognise instrumental variables, generalise structural equation models and solve the mediation and external validity problems. These tools, methods and solutions help to determine the accuracy and validity of causal claims in the analysis. 
The machine learning community is approaching the field of causal discovery in innovative ways by leveraging data, assumptions and models collectively. In recent decades, mathematical foundations have been established to address questions of causality in various scientific fields, mainly emerging for statistics and machine learning~\citep{pearl2009causality, Spirtes2000,Peters2017book}. The causal machine learning (CausalML) field has recently introduced \citep{kaddour2022causal} as an umbrella for machine learning methods based on SCMs. It aims to advance the field in several directions: causal supervised learning, causal generative modelling, causal explanations, causal fairness, and causal reinforcement learning. Applications of the new methods are vast and promise advances in computer vision, natural language processing, and graph representation learning. 
Therefore, the field of causal discovery is growing in methods, approaches and impactful applications. A unified agenda for Causal Inference is built and deployed in the wild.

However, despite the significant advances in the last decades, many unresolved philosophical and methodological issues remain for causal discovery from observational data. All such challenges also create avenues of research. 
On the one hand, we identified and discussed algorithmic and data challenges and summarised possible ways to address them in \S\ref{sec:causal-discovery-challenges}. Indeed, we must develop more effective methods for incorporating (uncertain) expert knowledge, determining the spatio-temporal complexity of the underlying dynamic phenomena, and creating more reliable and statistically efficient algorithms. 

On the other hand, perhaps the most critical challenge is the theoretical impossibility of causal discovery from purely observational data \citep{Pearl2000}. There are, however, ways to tackle the challenge. For example, specifying a causal DAG using domain knowledge can help mitigate the potential inaccuracy of their assumptions of sufficiency and faithfulness. Another possibility to learn about a certain equivalence class may consider incorporating domain knowledge into structure learning algorithms by using ``allow lists'' and ``deny lists'' to determine which edges should or should not be included in a DAG or creating a Bayesian prior to assigning varying levels of probability to certain causal relationships \citep{castelo2000priors, ness2017bayesian}. This is very much related to using {\em inductive biases} (such as Occam's razor) \citep{janzing2007causally} and causal invariances (such as parameter modularity and independence of mechanism) \citep{hoyer2008nonlinear} to learn structure beyond likelihood-based scores and conditional independence constraints. Finally, if data from natural experiments, such as $do(A=a)$, is available, this intervention information can be incorporated into the algorithm \citep{cooper2013causal} to automate this reasoning process. Incorporating the abundant domain knowledge within the causal discovery routine can address the identifiability and faithfulness assumptions (very much in line with the  basis or sparsity priors used in equation discovery, cf. \S\ref{sec:equationdiscovery}). By joining forces, both can contribute to resolving pressing scientific issues, ranging from process comprehension to evaluating and upgrading the physics included in physics models. 

Many problems in the physical sciences can be framed as causal questions. Yet many researchers in economics and health services, and even many computer scientists in machine learning, have been trained to be reluctant to use the language of causality \citep{dowd2011separated,pearl2011statistics,hernan2018c,scholkopf2021toward}. This is a cognitive barrier to resolve in the future. 
Besides, the language barrier between the methodological and domain science communities is a significant challenge in the causal discovery endeavour. Bridging this divide by translating domain questions into actionable and precisely stated causal inference tasks seems reasonable. Additionally, hesitance to adopt causal inference can be attributed to the lack of suitable benchmarks to help choose an appropriate method. A benchmarking platform (\href{https://causeme.net}{https://causeme.net}) was introduced that covers the causal discovery problem setting. 
It is necessary to have more of these benchmark platforms and easily accessible databases to facilitate better collaboration between the two communities. To successfully address a causal inference problem, it is essential for the domain scientist and computer scientist to work together; assumptions must be discussed and formalised, data characteristics must be jointly analysed, and conclusions must be assessed from both perspectives. 

It is commonly believed that most research questions in science can be interpreted as causal inference problems. Sentences such as ``we find that $X$ [increases/ decreases/ lags/ leads/ affects/ drives/ impacts] $Y$'' are often found in papers, creating an impression of causality. Nevertheless, scientists should be more explicit and transparent when making assumptions which lead to causal conclusions. This is not only sensible but is also necessary when analysing complex systems like the Earth, the Brain, or the Economy, since the outcomes of such research may have significant economic, environmental, and social implications.
The field of Causal Inference provides a comprehensive arsenal of tools grounded in rigorous mathematical principles and a vibrant interdisciplinary milieu to confront the challenge at hand successfully.

%% file: 03_LearningPhysics_book.tex
\chapter{Learning physical laws from data}\label{sec:equationdiscovery}

\begin{wrapfigure}{r}{6.3cm}
\vspace{-0.5cm}
\begin{mdframed}[backgroundcolor=gray!20]
\small
{\em ``For centuries, scientists observed Nature to extract simple laws and equations to explain the world mechanisms, to anticipate and predict behaviours and gain faith in their interventions/actions.''}
\end{mdframed}
\vspace{-0.5cm}
\end{wrapfigure}
Distilling mechanistic models of the world is how physicists have successfully understood and explained the natural world. The prototypical process starts with an experiment or observation. A mathematical model is hypothesised, which can predict a new experiment's outcome. The observations will either support or falsify the hypothesis, leading to more experiments and refined models.

\begin{figure}[h!]
    \centering
    \includegraphics[width=14cm]{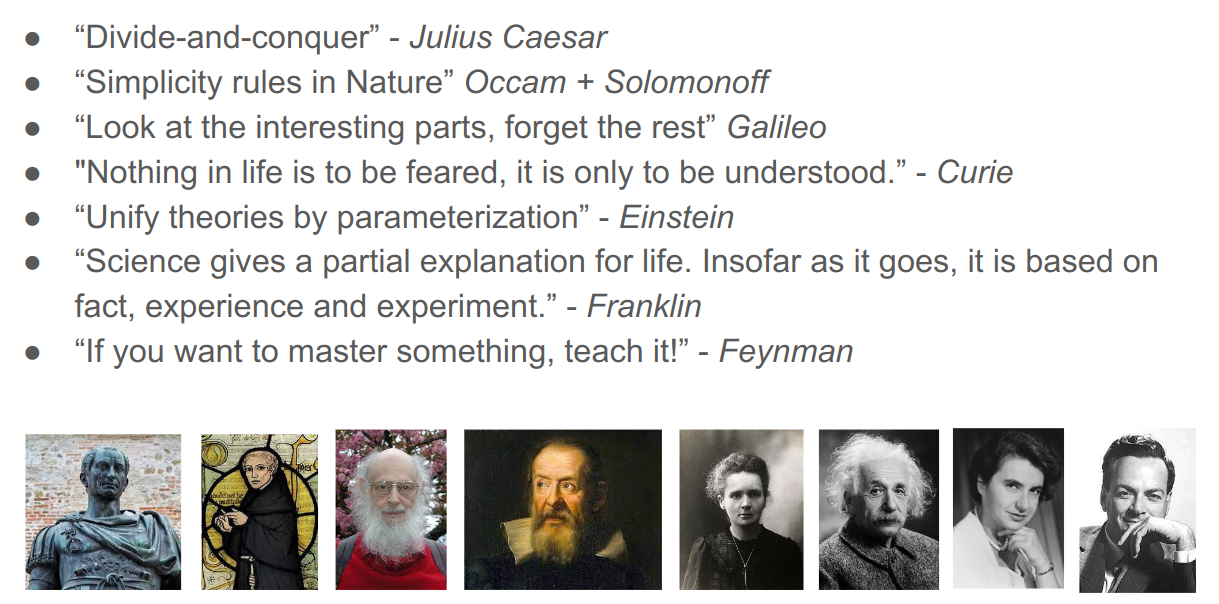}
    \caption{Historical Inspirations/Motivations in Law/Equation Discovery.}
    \label{fig:fathers-mothers}
\end{figure}

For many complex systems, we have poor models because certain interaction terms are unknown. Another case is when we know some microscopic interaction laws. Still, the emergent properties at a larger scale do not directly follow, such that predictions at the larger scale need new coarse-scale interactions.
To cope with these situations and make sense of the sheer amount of data produced by modern instruments, researchers have been looking into automating the processes involved in model building and creating new insights. Many motivations and inspirations have been adopted to guide the scientific method development (see Fig.~\ref{fig:fathers-mothers}): \eg\ a physical law or equation describing the system should be compositional, thus following the ``divide and conquer'' rule, should be as simple as possible (but not simpler) thus following Occam's razor, further developed and formalised by Solomonoff, eventually focus on the interesting parts of the system and disregard the rest, create understanding, generalise and unify different working theories and models, and the learned representation (and its parameters) should be self-explainable, amenable and intuitive.

This section reviews the state-of-the-art in equation discovery from data. Unlike in traditional law discovery {\em \`a la Kepler} where trial-and-error was dominant, modern statistics and machine learning techniques exploit the regularities found in the data to discover plausible, simple and explainable equations, to learn feature representations that describe (typically dynamic) systems. We will first consider the {\em explicit} discovery of equations that describe observed data, also called \emph{Symbolic Regression}. Second, we look into {\em implicit} discovery through dimensionality reduction techniques and transfer operators. We finish the section by discussing the main challenges and research opportunities.

\section{Explicit equation discovery with symbolic regression}

Symbolic Regression (SR) refers to a class of machine learning techniques that aim to discover mathematical relationships and patterns in data. SR aims to find a compact, human-readable mathematical expression that accurately reflects the underlying relationships in the data.
This approach is particularly useful in cases where the relationships between variables are complex or unknown, and traditional statistical methods may not provide adequate explanations.
SR operates under the assumption that the underlying data-generating mechanism can be described by a sparse and algebraic input-output relationship.

There is a major divide in the method to achieve that.
One class of methods use genetic algorithms or other discrete search methods to find mathematical terms, typically represented by a graph of mathematical operations.
Another class of methods uses continuous space search methods and solves a relaxation of the discrete problem of finding compact equations.
A third and most recent addition to the family of symbolic regression methods uses massive amounts of synthetic data for pretraining a system that can quickly guess a suitable expression at test time.

More formally, we are trying to solve the following optimisation problem:
\begin{align}
    \argmin_f \mathbb E_{(x,y)\sim \mathcal D} \|f(\boldsymbol{x}) - \boldsymbol{y}\|^2 + \lambda C(f) \label{eqn:SR}
\end{align}
where we attempt to find a low-complexity function (equation) $f$ that best\footnote{We use the squared error here for simplicity, but other notions of distance are possible.} maps the inputs $\boldsymbol{x}\in\Real^n$ to their corresponding outputs $\boldsymbol{y}\in\Real^m$ in the data distribution $\mathcal D$,
$C(f)$ refers to a measure of complexity of $f$, and $\lambda$ is the weighting factor.
For instance, the complexity measure could be the number of terms in the equation.

\begin{figure}[h!]%
    \centering
    \includegraphics[width=5cm]{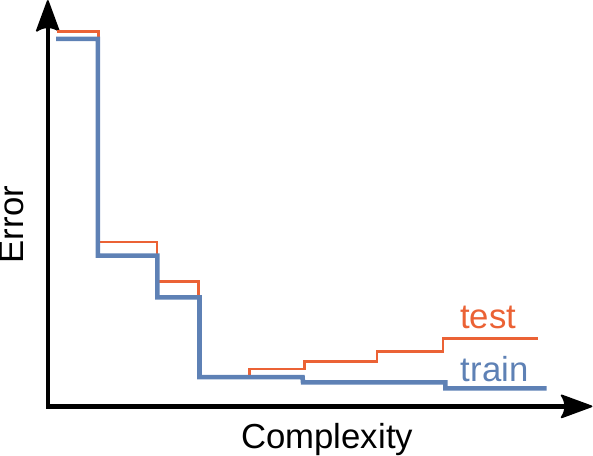}
    \caption{Illustration of a Pareto curve of solutions to an SR problem.
    }\label{fig:pareto-illustration}
\end{figure}

A central problem when performing symbolic regression is selecting an appropriate weighting factor. More generically, the question is which level of complexity is right. 
There is probably no definite answer to this question. Instead, we consider the solution to a symbolic regression problem as a family of Pareto-optimal solutions:
\begin{align}
    f^{(c)} = \argmin_{f} \mathbb E_{(\boldsymbol{x},\boldsymbol{y})\sim \mathcal D} \|f(\mathbf{x}) - \mathbf{y}\|^2 \qquad \text{s.t.}\quad C(f)=c \label{eqn:SR_complexity}
\end{align}
where $f^{(c)}$ refers to the best fitting expression with complexity $c$.
A more complex expression will be able to fit the given data at least as well as a less complex expression.
Fig.~\ref{fig:pareto-illustration} illustrates a typical  Pareto curve along the optimisation objectives: goodness of fit (error reduction) and function complexity. In addition to the training error, which monotonically decreases with function complexity, the illustration also shows a hypothetical test error that shows the overfitting of too complex functions.

We will now look more closely into different methods that have been proposed to solve the optimisation problem \eqnpref{eqn:SR_complexity} in practice.

\subsection{Symbolic regression using discrete search methods}
The problem of symbolic regression is, at its core, a search for suitable functions $f$ in \eqref{eqn:SR_complexity}.
Since those functions should have low complexity, it is natural to attempt to perform a search for functions.
The first attempt to do that was proposed by \citet{icga85:cramer} by inventing Genetic Programming, which got popularised and applied through Koza \citep{Koza1990:GP,koza1994genetic}.
The idea is simple: search for computer programs to solve a particular problem by iteratively creating many random programs and selecting the best fit, and create a new pool of candidates by recombination and random modification. This mimics the biological evolution process of nature to create the genetic material of living organisms.
Applied to symbolic regression:
The functions are represented as a graph of input variables, operators and basic algebraic functions.

\begin{figure}[t!]
    \centering
\scriptsize
\renewcommand{\tabcolsep}{0.1cm}
\begin{tabular}{c|c|c|c|c|c}
\makecell{{\bf 1} Observational\\time series\\$\{x(t),y(t),z(t)\}$} &
\makecell{{\bf 2} Partial derivatives\\ $\Big\{\dfrac{\Delta x}{\Delta y},\dfrac{\Delta z}{\Delta x},\dfrac{\Delta y}{\Delta z}\Big\}$} &
\makecell{{\bf 3} Candidate\\symbolic\\functions} &
\makecell{{\bf 4} Symbolic\\partial\\derivatives} &
\makecell{{\bf 5} Compare\\predictive\\par.der.}&
\makecell{{\bf 6} Accurate\\and sparsest\\solutions} \\
\hline
\includegraphics[width=.17\linewidth]{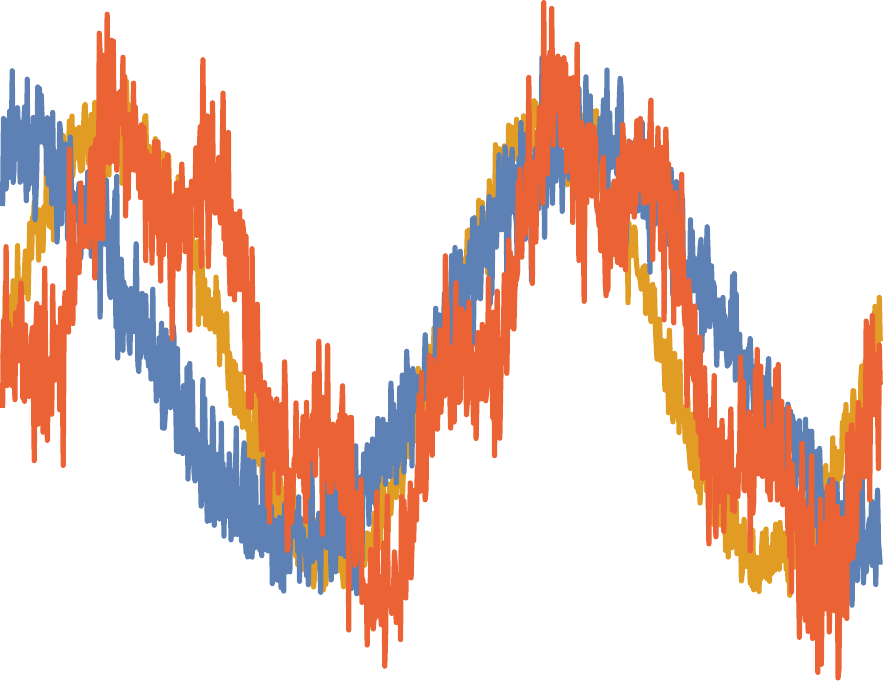}  &
\includegraphics[width=.12\linewidth]{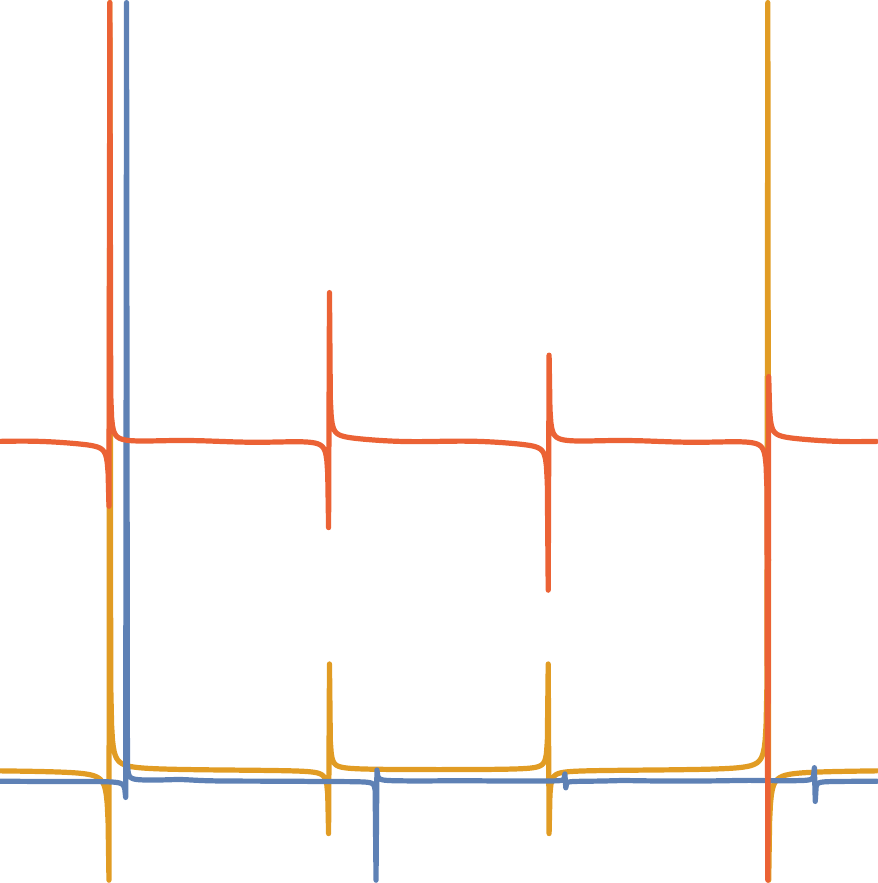}  &
\raisebox{1cm}{\makecell{$f = x\cos(y)$\\
$f = 0.7\exp(y/z)$\\
$f = y^2 - 9.8\cos(x)$\\ \ldots}} &
\raisebox{1cm}{\makecell{$\dfrac{\partial f}{\partial y}= y + \sin(x)\dfrac{\Delta x}{\Delta y}$\\
$\dfrac{\partial y}{\partial x}\bigg|_{f(x,y)} = \dfrac{\partial f}{\partial x}/\dfrac{\partial f}{\partial y}$}} &
\raisebox{1cm}{$\dfrac{\Delta y}{\Delta y}\bigg|_{D_i} = \dfrac{\partial y}{\partial x}\bigg|_{f_i}$} &
\raisebox{1cm}{\makecell{$f = z+9.8\sin(x)$\\
$f=y^2-9.8\cos(x)$}}\\
\end{tabular}
    \caption{Schematic view of the symbolic regression method for discovering physical laws in \citep{schmidt2009}. Starting from observational data (1), partial derivatives are computed numerically for all pairs of variables (2). A set of candidate symbolic functions $f$ is derived (3), whose symbolic partial derivatives are computed (4) and compared to the predictive ones (5). The process 3-5 is iterated until, finally, a small set of the most accurate and simple equations is returned (6).}
    \label{fig:schmidt_lipson}
\end{figure}

In the paper by \citet{schmidt2009}, this approach was refined and applied to the discovery of physical laws.
As the method was indeed able to discover Lagrangian and Hamiltonian formulations from data, it stimulated a growing interest in symbolic regression and sparked the development of many methods. These general search methods are also referred to as \emph{evolutionary algorithms}.
The approach of \citet{schmidt2009} is illustrated in \figref{fig:schmidt_lipson}.
The general method for symbolic regression was implemented in a tool called \emph{Eureqa} \citep{eureqa} that is now only available as an online service \citep{datarobot}.

There are several publicly available and open-source implementations, such as the PySR \citep{pysr}, gplearn \citep{gplearn}, Glyph \citep{Glyph} and Operon \citep{operon}. A more detailed overview of genetic algorithm-based methods and their combination with gradient descent can be found in \citet{kommenda2020:SRLS}.

\paragraph{Feynman AI.}
An approach that exploits physical knowledge such as units and makes reasonable assumptions for equations in physics
 is Feynman AI \citep{udrescu2020ai,Udrescu2020:AI-feynman2_0}.
The method augments the genetic algorithm searching for expressions
by enforcing fitting physical units, decomposing the problem using symmetries and checking separability.
To check for symmetries, a neural network is trained on the data to allow accessing whether the underlying function is symmetric. It is worth noting that a large amount of data is used here.
From a set of 100 equations taken from the Feynman Lectures, the method was able to recover all of them whereas
Eureqa only solves 71.

\paragraph{Search with Deep Reinforcement Learning.}
A method that uses Deep Reinforcement Learning to search for a suitable solution to the symbolic regression problem is Deep Symbolic Regression (DSR) \citep{petersen2021deep}.
The key idea is to treat the search for expressions as an exploration problem in reinforcement learning (RL).
The functional expressions are represented as a sequence of tokens corresponding to a depth-first graph traversal and are generated by a recurrent neural network. Numerical constants are fitted using the BFGS optimiser. %
This generative network is trained on the given dataset using RL to find a highly fitting solution. An interesting contribution is a formulation of a risk-seeking policy gradient that tries to optimise for the best-case scenario (a good solution can be found) rather than the typical average case.
The method was able to solve 83\% of the standard Nguyen-1 dataset.

\subsection{Sparse linear regression and neural network approach} \label{sec:sindy}

The symbolic regression problem can also be tackled via traditional regression methods. In contrast to the search in a discrete set of functions, the search is performed in a dense set, typically represented by a real-valued parameterised function.
So \eqref{eqn:SR} is solved by choosing a large enough function class described by $f_{w}$ with $\boldsymbol{w} \in \Real^p$. The optimisation is then performed over the space of parameter-values $w$.
Linear regression is a special case, where the function $f_w(\boldsymbol{x}) = \boldsymbol{w} \cdot \boldsymbol{x}^\T$.

\begin{figure}[h!] %
\centering
    \includegraphics[width=0.4\linewidth]{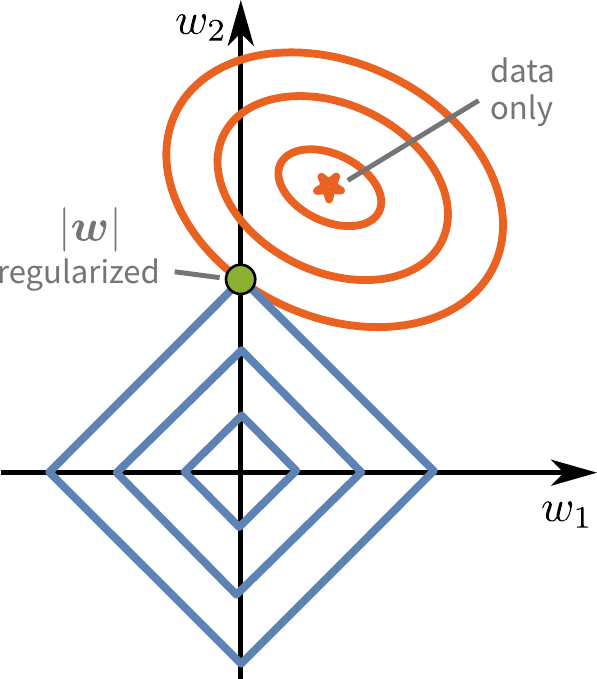}
    \caption{$L_1$ regularisation typically leads to sparse solutions.
      The lines show the isolines of quadratic loss (red) and $|w|$ (blue).
      Instead of the data-only solution (red star), a sparse solution (green) is found. }\label{fig:l1reg}
\end{figure}

What about the complexity regularisation term $C(f_w)$ in \eqref{eqn:SR}?
Ideally, the term should count the number of non-zero parameters in $\boldsymbol{w}$, expressed as $|\boldsymbol{w}|_0$ and referred to as $L_0$ norm. However, this term jeopardises the efficient solution of the regression problem because it is non-linear and non-differentiable. 
One practical alternative is to use the $L_1$ norm instead, \ie the sum of absolute values, which also leads to sparse solutions (see Fig.~\ref{fig:l1reg}).
In the case of linear regression, this is termed LASSO regression \citep{tibshirani96regression}.

The methods differ in the function class $f_w$, the regularisation term and the optimisation method used.

\paragraph{SINDy: Sparse identification of dynamical systems.}
In some cases, the class of building blocks that might occur as summands in the analytical description of the data are known.
Then a rather simple but effective method can be employed
that is called \emph{sparse identification of dynamical systems}, SINDy for short.
It was proposed in \citet{brunton2016discovering} to find differential equations of dynamical systems from observations. For the general symbolic regression problem, the FFX method by \citet{McConaghy:2011:GPTP} was already earlier proposing the same idea.
The input data is passed through a predefined library of base functions and interaction terms. Then the resulting high-dimensional representation is fit to the data using sparse linear regression.
All relevant terms keep a non-zero weight and constitute the final expression.

Let us unpack this in more detail for a dynamical system of $n$ variables described by the system of ordinary differential equations $\frac{d}{dt} \boldsymbol{x} = \boldsymbol{g}(\boldsymbol{x})$, where $\boldsymbol{x}(t) \in \mathbb{R}^n$.
Each component of $\boldsymbol g$ can now be substituted by a linear combination of library functions:
\begin{equation*}
\begin{array}{rclcl}
\dfrac{d}{dt} x_1 & =& g_1(x_1,x_2,\ldots,x_n) & = & w_{11} l_1 (x_1,...,x_n) +  \ldots + w_{1m} l_m (x_1,...,x_n) \\[.5em]
\dfrac{d}{dt} x_2 & =& g_2(x_1,x_2,\ldots,x_n) & = & w_{21} l_1 (x_1,...,x_n) +  \ldots + w_{2m} l_m (x_1,...,x_n) \\
 & \vdots & \\
\dfrac{d}{dt} x_n & =& g_n(x_1,x_2,\ldots,x_n) & = & w_{n1} l_1 (x_1,...,x_n) +  \ldots + w_{nm} l_m (x_1,...,x_n),
\end{array}
\end{equation*}
where $l_1,\ldots,l_m$ is the \emph{predefined finite library of candidate functions} and $w_{ij}$ are the scalar coefficients to be learned following our objective \eqref{eqn:SR}.
To obtain a sparse solution, the $L_1$ regularisation (LASSO) can be used, \ie $C(f) = |\boldsymbol{w}|_1$, as described above.
Alternatively, the (squared) $L_2$ norm of the weights $\|\boldsymbol{w}\|^2_2$ can be used, corresponding to classical ridge regression that permits a closed-form solution. However, an iterative pruning of small weights must be used to obtain a sparse solution.

\begin{mdframed}[backgroundcolor=gray!20]
\small
{\bf Illustration of sparse identification of dynamical systems (SINDy)} \citep{brunton2016discovering}
using a synthetic dataset of the well-known Lotka-Volterra system, as shown in \citet{Adsuara2020:diffeqnfromEOD}.
The Lotka-Volterra system models the interaction between prey and its predator in ecology and is given by the following equations:
\begin{equation}
\begin{array}{rcl}
\dfrac{d}{dt}x_1 &=& \alpha x_1 - \beta x_1x_2 \\[.5em]
\dfrac{d}{dt}x_2 &=& -\gamma x_2 + \delta x_1x_2 \nonumber
\end{array}
\end{equation}
with the coefficients, $\alpha$ and $\gamma$, being the intrinsic growth/decrease rates of $x_1$ and $x_2$, and $\beta$ and $\delta$ are cross terms taking into account the interaction between species.
In our particular case, we will set $\alpha = {3}/{2}$, $\beta = 1$, $\gamma = 3$, and $\delta = {1}/{2}$.
We show the results of the identification of these parameters using SINDy in the table below for two levels of additive white Gaussian noise of the signal-to-noise ratio of $5$ ({high noise level}) and $40$ dB {(low noise level)}.
As usual, the created data was split into train/test data ($75\%,25\%$), respectively.
The ODE coefficients are recovered sufficiently well to achieve a high correlation coefficient $R$ but are generally underestimated due to the sparsity regularisation.
\small
\begin{center}
\renewcommand{\tabcolsep}{0.1cm}
\begin{tabular}{ccc@{\qquad}cc@{\quad}@{\quad}cc}
\toprule
 & \multicolumn{4}{c}{\bf Learned Coefficients} & \multicolumn{2}{c}{\bf True} \\
{\bf Library functions} & \multicolumn{2}{c}{40 dB\ \ } & \multicolumn{2}{c}{5 dB\ \ } &  \multicolumn{2}{c}{\bf Coefficients} \\
 & $\frac{d}{dt}x_1$ & $\frac{d}{dt}x_2$ & $\frac{d}{dt}x_1$ & $\frac{d}{dt}x_2$ & $\frac{d}{dt}x_1$ & $\frac{d}{dt}x_2$\\
\midrule
$x_1$   & 1.3822  & 0       & 1.1404  & 0       & 1.5 & 0  \\
$x_2$   & 0       & -2.9123 & 0       & -2.7946 & 0   & -3 \\
$x_1x_2$&-0.9797  & 0.4849  & -0.9520 & 0.4710  & -1  & 0.5\\
$x_2^3$ & 0       & 0       & 0       & -0.0001 & 0  & 0 \\
\midrule
R & \multicolumn{2}{c}{$0.9999$} & \multicolumn{2}{c}{$0.8674$} \\
\bottomrule
\end{tabular}
\end{center}
\end{mdframed}

For fitting dynamical systems,
the temporal derivatives need to be computed.
Finite differences are often too sensitive to noise
 such that kernel regression (aka Gaussian processes), which allow for explicit derivative computation, are preferred \citep{CampsValls16grsm, Emman2018}.
 A more recent approach is to solve noise estimation and model identification in one joint optimisation
 procedure \cite{kaheman2022:SindyNoise}.
Intuitively, for every data point, the corruption by noise is estimated.
Since this optimisation problem is highly underdetermined, an additional constraint is used, namely
that when integrating the estimated dynamical system model a small error should occur.
This trick separates noise from the signal and leads to an improved estimation quality.

\begin{figure}[h!]\centering
\begin{minipage}{0.61\linewidth}
\includegraphics[width=\linewidth]{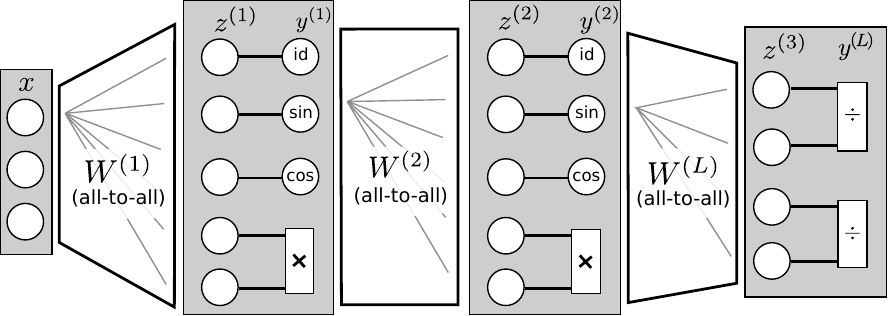}
\end{minipage}\hfill
\begin{minipage}{0.35\linewidth}
\centering\footnotesize
Example:\\
$y = \nicefrac{1}{3} \left( (1+x_2)  \sin(\pi x_1) + x_2  x_3  x_4\right)$\\[.5em]
\includegraphics[width=\linewidth]{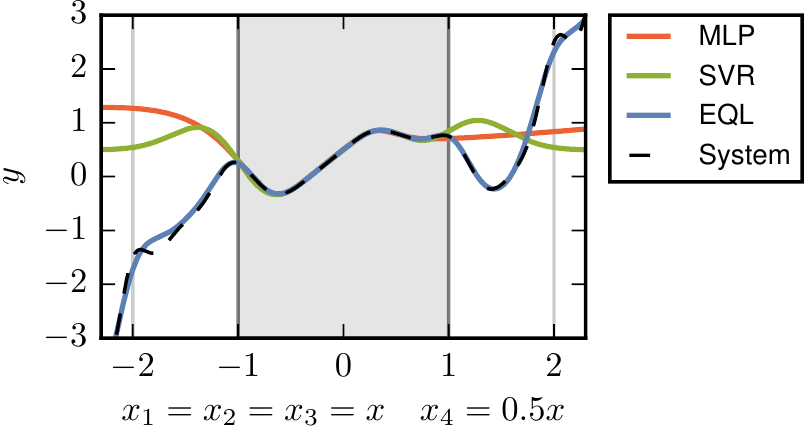}
\end{minipage}
\caption{Equation Learning Architecture and example equation. Left: Illustration of the EQL Architecture, reproduced from \citep{SahooLampertMartius2018:EQLDiv}, a feed-forward neural network with special \emph{activation} functions (sin, multiplication etc.). Note that each unit type will occur many times.
Right: Example system with four inputs $x_{1,2,3,4}$ and one output $y$.
Training is only in the $[-1,1]^4$. EQL recovers the equation and extrapolates \citep{SahooLampertMartius2018:EQLDiv}.
}\label{fig:eql-architecture}
\end{figure}

\paragraph{Neural network approach: Equation Learner.}
Enlarging the function class $f$ is possible using neural networks.
Probably the first work in this direction is the Equation Learner (EQL) introduced in \citet{MartiusLampert2017:EQL} that uses a neural network with algebraic base functions and a particular regularisation scheme to solve \eqref{eqn:SR} and \ref{eqn:SR_complexity}.
The function $f$ is represented by a neural network, modified only to contain elementary operations that should appear in a potential solution.
\Figref{fig:eql-architecture}(left) shows the architecture of the Equation Learner (EQL) in a simplified form.

The input variables are mapped with a dense layer to multiple instances of trigonometric functions, identity, multiplication and division, but more base functions, such as squares or exponentials, are possible.
The resulting values are again mapped with a matrix to another layer of elementary functions, and so forth, until the last layer corresponds to the output (containing only division operators in the picture).

The network is trained using stochastic gradient descent (\eg Adam) on the mean squared error loss and $L_1$ regularisation on the weights to induce sparsity:
\begin{align}
    \mathbb E_{(\boldsymbol{x},\boldsymbol{y})\sim \mathcal D} \|f_W(\boldsymbol{x}) - \boldsymbol{y}\|^2 + \lambda |\boldsymbol{W}| \label{eqn:EQL:loss}
\end{align}
where $f_W$ denotes the neural network with parameters $\boldsymbol{W}$.

\begin{figure}[h!]
    \centering
    Toy data\\
    \includegraphics[width =0.32\textwidth]{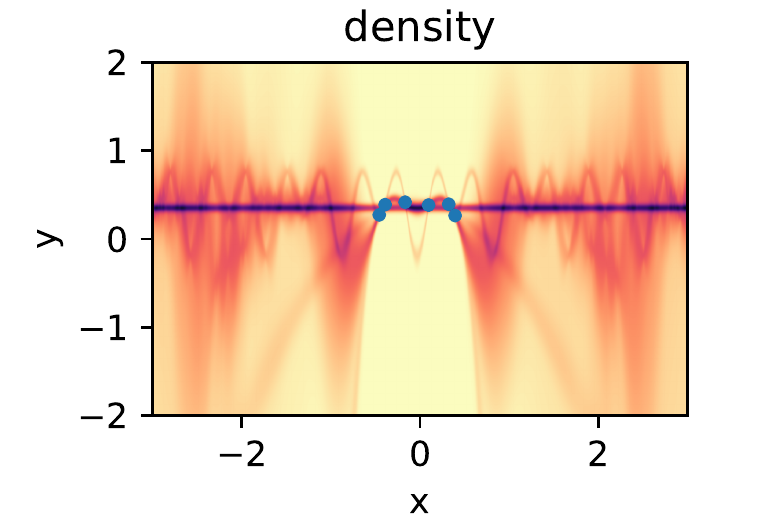}
    \hfill
    \includegraphics[width =0.32\textwidth]{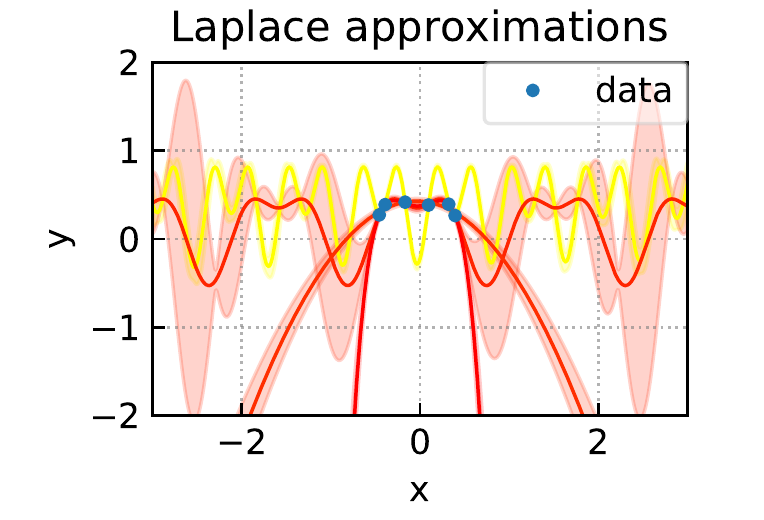}
    \hfill
    \includegraphics[width =0.32\textwidth]{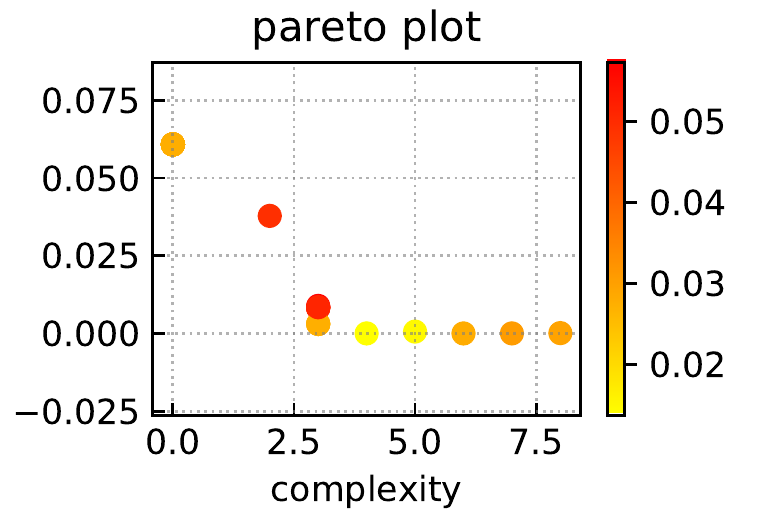}\\
    Mauna Atmospheric CO$_2$ concentration dataset\\
    \includegraphics[width =0.32\textwidth]{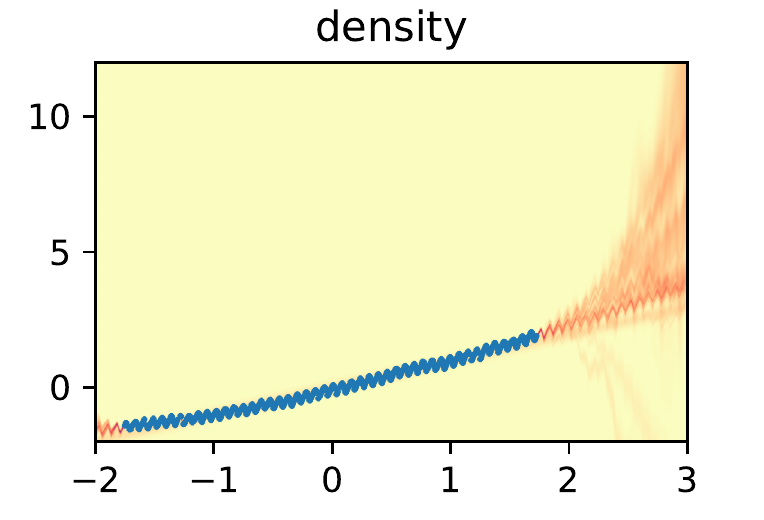}
    \hfill
    \includegraphics[width =0.32\textwidth]{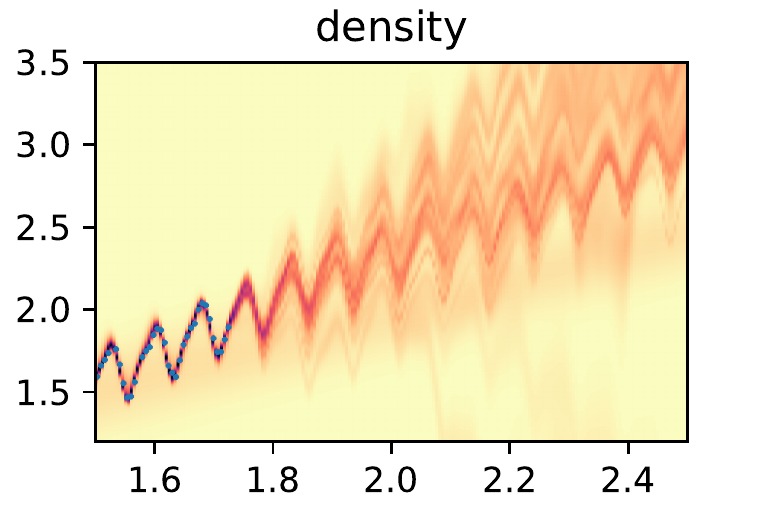}
    \hfill
    \includegraphics[width =0.32\textwidth]{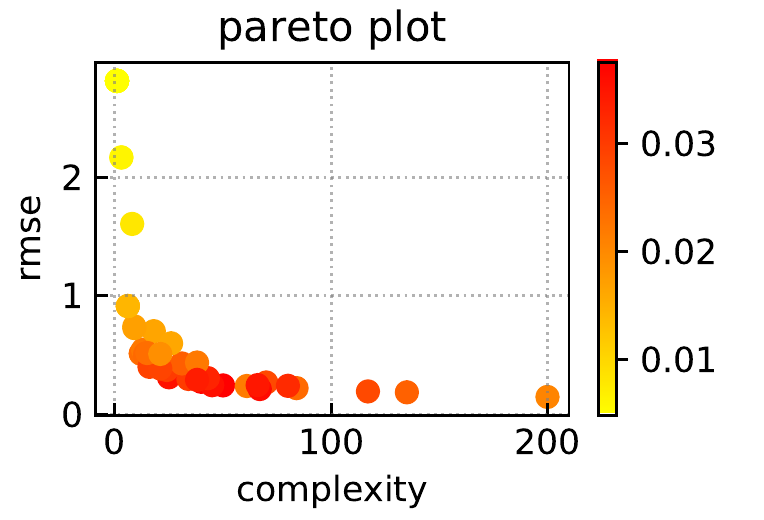}\\[-0.8em]

    \caption{Illustration of uncertainty estimates using a mixture of Laplace approximations of learned equations.
    Top row: toy example $y =0.8 \cos x -0.4+ \epsilon$
    where $\epsilon \sim\mathcal N(0,0.03^2)$ with just 6 datapoints.
    Bottom row: Atmospheric CO$_2$ concentration at Mauna Loa Observatory \citep{mauna:CO2data} (concentration vs. time, both in arbitrary units).
    The left panels show the predictive distributions. The panel in the
    middle shows individual local Laplace approximations with $2\sigma$ (shaded area) for the toy data and a zoomed density for the Mauna dataset.
    The colour represents the weight and aligns with the Pareto plots on the right side, showing RMSE over the complexity of each equation. Reproduced from \citep{Werner2022:UncertaintyEQL}.
    }
    \label{fig:SR:uncertainty}
\end{figure}
Note that the system needs to be differentiable for training, such that a pure complexity term, such as $L_0$ regularisation that would count the number of non-zero weights, does not work out of the box. Although methods have been developed since then \cite{louizos2018learning}, the $L_1$ regularisation in \eqref{eqn:EQL:loss} is effective but creates an undesired trade-off between error and sparsity. Something that we also encountered in the illustrating example when applying SINDy.
The EQL method introduces an additional regularisation phase after converging with $L_1$ that clamps all $|W_i|<\epsilon \ll 1$ to zero and optimises without regularisation. This yields a practical approach to optimising for sparsity without trade-offs. In \citet{Kim2021:EQL-L05} an alternative to $L_1$ with $L_{0.5}$ was used.

The reader may wonder how the system is successfully trained with elementary functions such as division or square root.
Indeed, a naive application would fail due to exploding values or gradients. In \citet{SahooLampertMartius2018:EQLDiv} and
\citet{WernerEtal2021:InformedEQL}, suitable parameterisations and training steps are proposed.
Choosing different $\lambda$ \eqnpref{eqn:EQL:loss} will create differently sparse resulting networks. Each represents a particular symbolic expression resembling the Pareto curve illustrated in \figref{fig:pareto-illustration}.
In \figref{fig:eql-architecture} (right), a synthetic example system is shown. The training data is only generated in the $[-1,1]^4$ hypercube. The correct equation was discovered, and perfect extrapolation is possible in this case, see \citet{SahooLampertMartius2018:EQLDiv} for details.

Instead of manually selecting a particular solution, which might be a good procedure when structural insights are to be obtained when investigating some unknown phenomenon, one can also use several or all solutions along the Pareto curve to estimate uncertainty about the predictions for extrapolation, as proposed in \citet{Werner2022:UncertaintyEQL}. \Figref{fig:SR:uncertainty} illustrates this approach. For each found equation, a Laplace approximation allows to approximate the uncertainty due to parameter estimation errors and yields a Gaussian posterior.
Combining these using a weight based on the validation error and the complexity yields the estimated density. Note how the uncertainty in extrapolation shows clearly the structure of the discrete set of automatically generated hypotheses.

\subsection{Learning to solve symbolic regression}\label{sec:sr:pretrained}
All methods so far treat every symbolic regression problem in isolation --
the search or optimisation algorithm was applied to a new dataset from scratch.
We are now looking into the idea of learning to solve a particular problem quickly by using
data from a whole class of symbolic regression instances.
Generally, the idea is to approximate the inverse mapping from data to a suitable equation.
The Dreamcoder paper by \citet{ellis2021:Dreamcoder} showed the first instantiation of this idea.
Provided with the language of algebraic expressions (arithmetic operations, variables, base functions)
and a \emph{simulator} to generate data for a particular equation instance,
the method learns a probabilistic mapping from data to equation terms and a library of common equation building blocks.
Given a particular instance of data, the system can relatively quickly guess and verify suitable explaining equations.
Developing the idea of pretraining further and specialising it for symbolic regression was done by the following method.

\paragraph{NeSymReS.}
The approach in \citet{Biggio2021:NeSymReS} is to use a high-capacity transformer model pretrained to solve the symbolic regression problem. The method is called \emph{Neural Symbolic Regression that Scales} (NeSymReS).
The method uses a large set of symbolic regression problems to approximate the inverse mapping from data to equations.
After this pretraining phase, given new data, the inverse mapping can generate likely candidate equations.
Intuitively, an experienced data scientist might solve the problem similarly: looking at the data and postulating a particular functional form that might explain it, testing it, and potentially trying a different plausible hypothesis.

\begin{figure}[h!]
    \centering
    \includegraphics[width=0.9\linewidth]{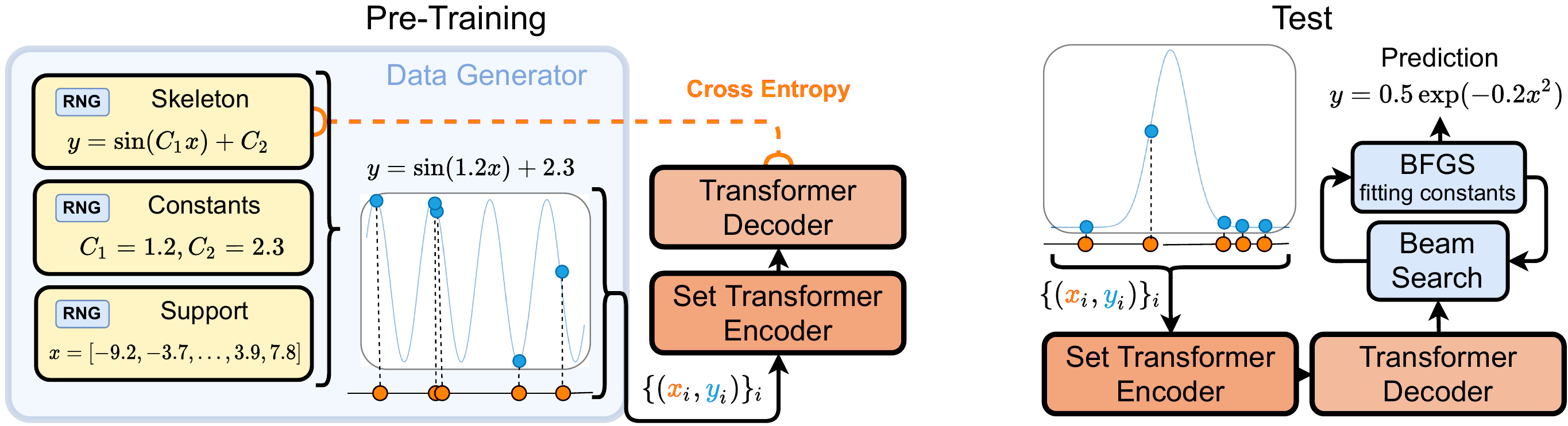}
    \caption{Overview of the NeSymReS method.
    Left: step with randomly generated training data.
    Right: inference of candidate equations for unseen data.
    Reproduced from \citep{Biggio2021:NeSymReS}.}
    \label{fig:nesymres:overview}
\end{figure}

Let us look closer at the method. As visualised in \figref{fig:nesymres:overview}, the core is a transformer\footnote{The transformer architecture is the basic building block of many large-scale machine learning systems, such as \mbox{GPT-3}~\citep{brown2020:GPT-3}.} 
architecture \citep{vaswani2017:transformers} that can generate algebraic expressions symbol-by-symbol given a set of data points.

The input is represented as a set of $(\boldsymbol{x},\boldsymbol{y})$ pairs (1024 in \citet{Biggio2021:NeSymReS}), which are processed through a set-encoder. The latter is invariant to permutations of the data points.
The output of the transformer are tokens that correspond to the typical symbols of input variables, base functions, operators and constant placeholders, which resemble the skeleton of the predicted function.
Importantly, the transformer does not have to guess the right constants, just their location in the expression, as these constant placeholders are fit to the data using non-linear optimisation (here BFGS).

Trained on millions of synthetically generated pairs of random expressions with corresponding data, the transformer does a remarkable job guessing likely equations. Importantly, the prediction is not deterministic but allows sampling of possible functions.
Thus, new potential solutions can be generated and validated when new data is presented at test time until a sufficiently good fit is found or the Pareto.

\begin{figure}[h!] %
\centering
    \includegraphics[width=0.35\linewidth]{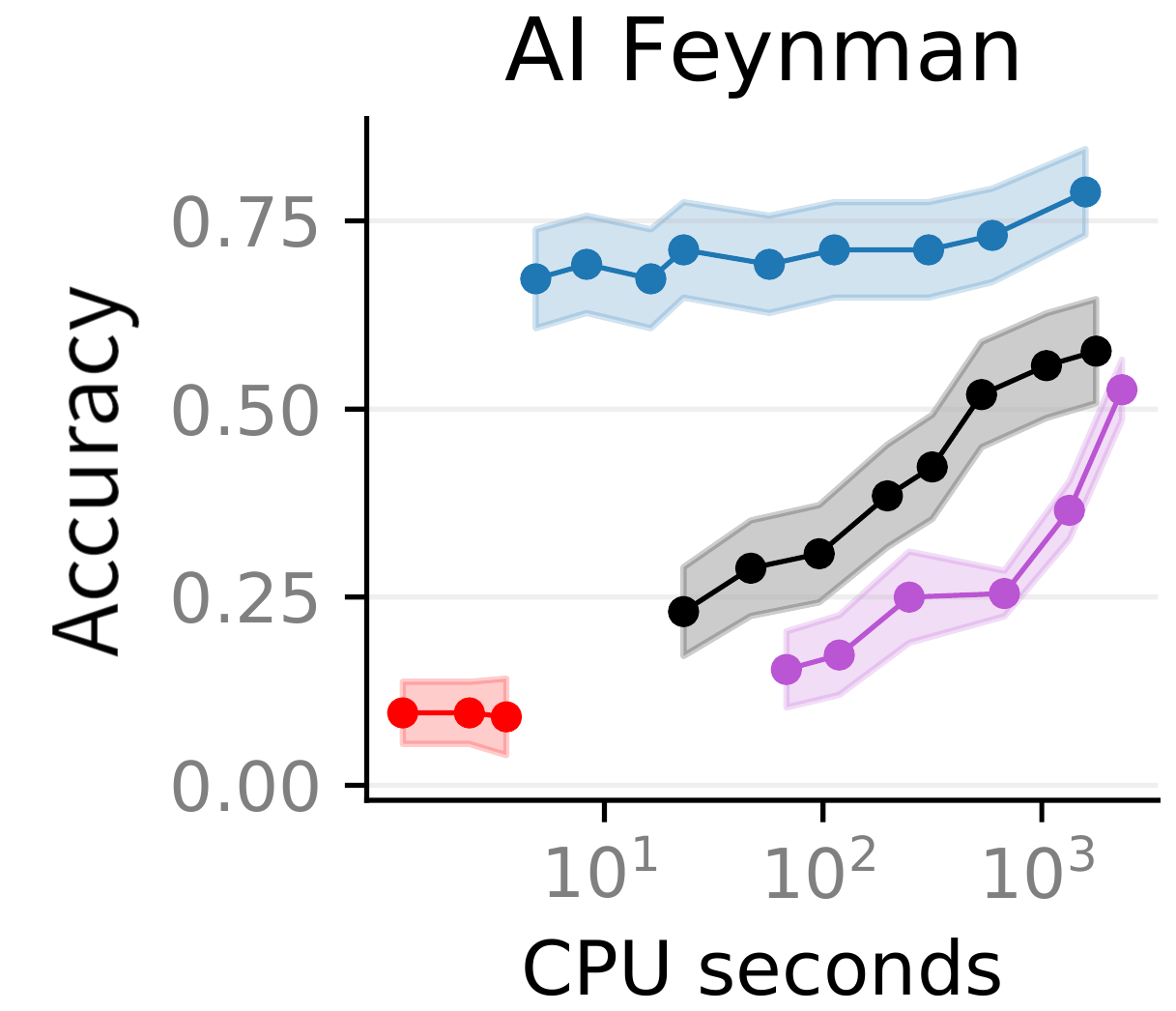}
    \includegraphics[width=0.35\linewidth]{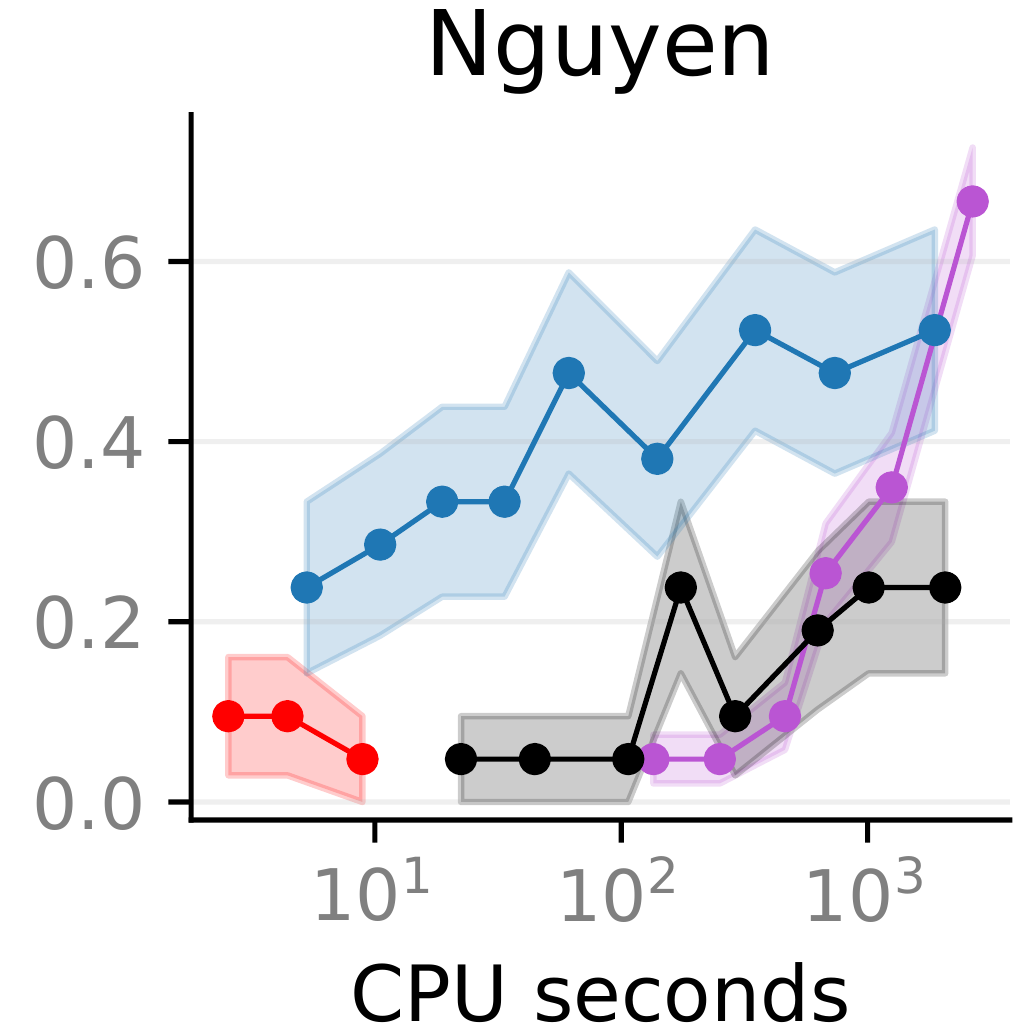}\\
    \includegraphics[width=0.8\linewidth]{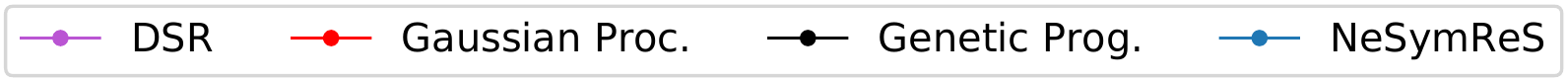}
    \caption{Performance of NeSymReS, DSR, classical SR (using gplearn \citep{gplearn}), and Gaussian processes on the AI Feynman and Nguyen datasets (equations unseen during training). Reproduced from  \citep{Biggio2021:NeSymReS}.}\label{fig:nesymres:performance}
\end{figure}

\Figref{fig:nesymres:performance} shows the accuracy of different SR methods for unseen equations from the Feynman and Nguyen benchmarks.
The performance is presented in dependence on wall-clock time.
NeSymReS is remarkably fast at finding a well-fitting expression for the data.

As a downside, the method was only shown for three input variables, and it remains to be seen how much it can be scaled in this respect.
Also, the dataset used to guess an equation at test time cannot be big (currently in the order of 1000 data points) because the set transformer encoder cannot yet handle larger sets well.

\subsection{Comparison}\label{sec:SR:comparison}
As there are quite a number of methods, we aim to discuss their differences, strengths and weaknesses by comparing them along a set of axes.
We start with using domain knowledge, as we seldom face a completely uninformed setting in physics. We continue with aspects of the embedding, scaling, speed and usability, summarised in \tabref{tab:SR:comparison}.

\paragraph{Using domain knowledge.}
A common form of domain knowledge is the base functions and their approximate frequency of occurrence in describing the system under consideration.
In standard symbolic regression with genetic algorithms, the number and kind of base functions are very flexible, and each term can have its individual penalty in terms of complexity. More specific domain knowledge, such as monotonicity, function image constraints and derivative constraints, can also be considered, as presented in \citet{Kronberger2022:shape-constraintSR}. Another recent work is presented in \citet{Cornelio2023:AI-Descartes} that allows to incorporate axiomatic contraints.

In FFX/SINDy, the library of functions is the prime way to specify domain knowledge. Relative preferences could be implemented by different regularisation strengths.

For the EQL framework, the choice of base functions is a bit more complicated, as the systems need to remain optimisable with gradient descent. In \citet{WernerEtal2021:InformedEQL}, a suitable relaxation for functions with divergences (in function value or derivative) is proposed, and a way to specify preferences among base functions is analysed. The control of the relative frequency of used terms is possible but less direct than in genetic algorithm-based methods.
In NeSymReS, domain knowledge can be embedded by selecting/generating the training set with appropriate synthetic problems, although this was not explicitly demonstrated.

\paragraph{Scaling.}
Most SR methods are for small-scale problems with a few hundred to a few thousand data points and low-dimensional problems, \ie 1-10 input and output variables.
Classical search methods scale unfavourably with dimensionality as the search space grows exponentially.
That is why most SR methods are good at finding relatively small and compact equations for low-dimensional systems but fail for both high-dimensional systems or those where larger equations are the most compact solution.

FFX/SINDy can handle large output dimensions easily, using some form of ridge regression. However, it also suffers from large input dimensions as the library bank becomes exponentially large unless a factorisation or other simplifying structure is known, for instance, a strong locality assumption in PDEs.

NeSymReS is also limited to several variables and small data sets. Although, the limitation of the data set size can be lifted by sampling a smaller subset of data points for guessing the skeletons and using all data for the parameter tuning with BFGS.

EQL is the only scalable method, as gradient descent works on all dimensions simultaneously, and machine learning methods are developed to scale. A larger initial neural network should be used for larger systems, but no specific adaptations are required.

\begin{table}[t!]
    \caption{Comparison of symbolic regression methods. See \secref{sec:SR:comparison} for more details.}
    \label{tab:SR:comparison}
    \centering
    \small
    \begin{tabular}{p{0.2\linewidth}|cclp{.2\linewidth}p{.15\linewidth}}
    \toprule
        &  embeddable & scaling & speed   & restriction & domain knowledge \\

        \midrule
     Genetic Programming \citep{schmidt2009}
        &  \xmark  & \xmark     & slow    & for small systems  &  base-functions, complexity of terms\\
        \midrule
     AI Feynman \citep{udrescu2020ai}
        &  \xmark  & \xmark     & slow    & for physical systems in canonical form & physic: units, symmetries\\
        \midrule
     DSR \cite{petersen2021deep}
        &  \xmark  & \xmark     & medium  & small input dim    &  training domain\\
        \midrule
     FFX \citep{McConaghy:2011:GPTP}, SINDy \citep{brunton2016discovering}
        &  \cmark  & \phantom{$^1$}\cmark\tablefootnote{It scales well  to large output sizes; for high-dimensional input  strong structural assumptions are required.}
                                & blazing & needs known library &  training domain\\
        \midrule
     EQL \citep{SahooLampertMartius2018:EQLDiv} &  \cmark  & \cmark & slow & base functions limited, sometimes less concise &  base-functions, complexities\\
        \midrule
     NeSymReS \citep{Biggio2021:NeSymReS}
        &  \xmark  & \xmark     & fast   & small input dim    &  training set\\
         \bottomrule
    \end{tabular}
\end{table}

\paragraph{Differentiability and Embeddability.}
An interesting feature is whether the SR system can be used as a module in a larger computational pipeline.
For instance, if observations are available as images and the causal variables have to be first extracted from the images before they can be used for a concise SR prediction module.
An example using SINDy inside an autoencoder architecture \citep{champion2019data} learns the coordinate frame and dynamics equations at the same time, as detailed in \secref{sec:eq:discover:latent} below.
EQL is conceptually easy to embed into larger architectures, such as deep networks, as it is end-to-end differentiable.
An example is discovering PDEs~\citep{long2019:PDE-Net}, or learning an energy function given observations of the  dynamics in the context of density functional theory \citep{lin2020:DFTEQL}, discussed in more detail in \secref{sec:case:DFT-EQL}.
The other methods are more difficult to embed.

\paragraph{Speed.}
As shown in \figref{fig:nesymres:performance}, symbolic regression methods are best compared in performance per compute-time,
because search methods can, in principle, find the global optimum given enough time (although this time might be longer than the age of the universe), so just the ``final'' performance is difficult to measure.
SINDy is not in this comparison because it requires knowledge of the base functions. However, it would be the fastest method, followed by NeSymReS, DSR and classic GPs. EQL is likely the slowest method on small systems because it requires a long training time. However, the time does not significantly increase with system size and amount of data.

\paragraph{When to use which method?}
For systems where we have a good idea about the occurring modules of the functional form, FFX and SINDy are probably the method of choice. They are simple and effective.
For dynamical systems SINDy is the most specialized method.
The other SR methods are good if the functional building blocks are unknown or nested structures are expected.
Genetic Programming based methods generally shine on small problem settings. Modern implementations can also fit constants, but they result is sometimes complex nested structures.
DSR and NeSymReS are faster than standard SR methods and can yield potentially less complex equations. However, less software is built for them, and they are less easy to use.
For high-dimensional data or when high fitting accuracy is required, EQL might be the right choice.
Also, when SR should be embedded, only FFX, SINDy and EQL are practically usable.

\section{Implicit equation discovery:  dimensionality reduction and transfer operators} \label{sec:dimensionality_reduction}

This section reviews state-of-the-art methods for recovering implicit feature representations of systems from data. We will review connections among methods and emphasise the role and examples in the broad discipline of physics. Unlike in the previous section, an explicit equation is not discovered but rather an operator that encapsulates the system's characteristics (typically spatio-temporal dynamics). The field is tightly related to dimensionality reduction and feature extraction in machine learning and signal processing \citep{Arenas13}, but also to transfer operators in functional analysis \citep{koopman}.

\subsection{Reduced-order models}

In many domains, the goal is to study system dynamics from model simulations. In this case, equations are encapsulated in the model itself, but large-scale, high-fidelity nonlinear models can be challenging to simulate and require significant computational power. In such cases, reduced order models (ROMs) can simplify analysis and control design by trading off model accuracy for computational complexity reduction. ROM can combine complex component-level simulation models into system-level simulations used for control analysis and design.

Two main classes of techniques for building ROMs: model-based and data-driven. Model-based methods rely on a mathematical or physical understanding of the underlying model and are designed for specific PDE-based models. In contrast, data-driven methods use input-output data from the original high-fidelity first-principles model to construct a ROM that accurately represents the underlying system. Data-driven ROMs can be either static or dynamic models. Static ROMs can be developed using techniques such as curve fitting and lookup tables (LUT), while dynamic ROMs can be developed using deep learning techniques such as LSTM, feedforward neural nets, and neural ODEs.
The obtained ROM ideally contains the essential physical mechanisms of the original system while exhibiting simpler dynamics that can enhance interpretability. In addition to the simpler physics, the ROM would be much cheaper from the computational point of view than the numerical integration of the governing equations, which can help obtain more efficient control and optimisation techniques.

Developing a ROM typically requires two steps. The first step is to find a set of coordinates where the original dynamics can be expressed in a compact form. This is typically done in terms of {\it modes}, associated with coherent features in the original system~\citep{taira,rowley}. In the second step, it is necessary to find a set of differential equations governing the temporal evolution of the amplitudes of the aforementioned modes. These equations, which constitute a dynamical system, enable shedding light on the physics of the (reduced) phenomenon under study. A widely-used approach to perform the modal decomposition is the so-called proper-orthogonal decomposition (POD)~\citep{lumley}, which is also known as principal component analysis (PCA) in statistics and empirical orthogonal function (EOF) analysis in meteorology \citep{Arenas13}, and is closely connected with the singular-value decomposition (SVD)~\citep{brunton2022data}.

\begin{mdframed}[backgroundcolor=gray!20]
\small
{\bf Proper-orthogonal decomposition (POD), aka PCA or EOF } \citep{lumley,Arenas13} decomposes a dataset or a high-dimensional (spatio-temporal) field into a set of orthogonal basis functions called modes or eigenfunctions, which capture the dominant features of the data. The first few modes explain most of the variability in the data, while the later modes explain smaller and smaller amounts of variability. By truncating the number of modes, one can obtain a low-dimensional representation of the data that preserves the essential features of the original system. %

Given a set of data snapshots $\mathbf{X}=[\mathbf{x}_1, \mathbf{x}_2, \ldots, \mathbf{x}_n] \in \mathbb{R}^{m\times n}$, where $m$ is the dimension of the data and $n$ is the number of snapshots, we seek to decompose the data into a set of $r$ orthogonal modes $\{{\mathbf{u}_i}\}_{i=1}^{r}$, such that $\mathbf{X}\approx\mathbf{U}\bm{\Sigma}\mathbf{V}^\top$, where $\mathbf{U}=[\mathbf{u}_1, \mathbf{u}_2, \ldots, \mathbf{u}_r]\in \mathbb{R}^{m\times r}$, $\bm{\Sigma}\in \mathbb{R}^{r\times r}$ is a diagonal matrix containing the singular values, and $\mathbf{V}^\top\in \mathbb{R}^{r\times n}$ is the matrix of temporal coefficients.
The modes $\{\mathbf{u}_i\}_{i=1}^{r}$ can be computed by performing a singular value decomposition (SVD) of the data matrix  $\mathbf{X}=\mathbf{U}\bm{\Sigma}\mathbf{V}^\top=\sum_{i=1}^{r}\sigma_i\mathbf{u}_i\mathbf{v}_i^\top$ where $\sigma_i$ is the $i$-th singular value, and $\mathbf{v}_i$ is the $i$-th right singular vector. The modes are then given by $\mathbf{u}_i=\frac{1}{\sqrt{\sigma_i}}\mathbf{X}\mathbf{v}_i$.
\end{mdframed}

The POD/PCA framework enables decomposing spatio-temporal data (\eg flow velocities, weather or climate variables which depend on the spatial coordinates and time) into a set of spatial modes (which only depend on the spatial coordinates), multiplied by their temporal coefficients (which define their change of amplitude with time). POD/PCA ensures that components are orthogonal and optimality with respect to the variance explained by a reduced number of modes, cf.\ Fig.~\ref{fig:pca_dmd} for an example of PCA on a toy spatio-temporal data flow. Other alternative multivariate methods, like partial least squares (PLS) or canonical correlation analysis (CCA) seek projections that maximise covariance or correlation, respectively \citep{Arenas13}. Still, all these projection methods are linear and thus cannot cope with nonlinear spatio-temporal feature relations and complex dynamics. This can be addressed with kernel machines \citep{Arenas13}. Oblique and nonlinear transformations can also be learned by embedding Varimax in Reproducing Kernel Hilbert Spaces (RKHS) explicitly \citep{Bueso20rock,bueso2020explicit}.
Other ways to obtain non-linear transformations from observation space to ROM space, \eg~using neural networks, will be discussed below.

\subsection{Transfer operators for learning nonlinear dynamics} %

Transfer operators are related to the abovementioned methods and allow the characterisation and modelling of complex dynamic systems. These operators' eigenfunctions can decompose a system given by an ergodic Markov process into fast and slow dynamics and identify modes of the stationary measure called metastable sets.

The Koopman operator is a linear operator that describes the dynamics of a system by lifting the state variables into an infinite-dimensional Hilbert space. Thus, it enables us to effectively linearise complex temporal trajectories and hence is a compelling approach in dynamical systems research \citep{koopman}. Its application is expanding in both theoretical \citep{koopman,brunton_koopman}, and practical domains from molecular dynamics and fluid dynamics, atmospheric sciences, and control theory~\citep{Schmid10, BBPK16, KNKWKSN18}.

The advantages of the Koopman operator are numerous. First, it is a powerful tool for analysing and predicting the behaviour of a system over time. By lifting the state variables into a higher dimensional space, the Koopman operator can identify patterns in a system's behaviour that may otherwise be difficult to detect. This can be especially useful for uncovering hidden dynamical structures in chaotic systems.

Second, the Koopman operator allows us to develop data-driven models of dynamical systems. Using the operator's eigenfunctions as basis functions, it is possible to develop models of dynamical systems operating on these summarised coordinates, without solving or even understanding the underlying equations of motion. This makes the Koopman operator an attractive tool for model-based control and optimisation.
Third, the Koopman operator is useful to discern key properties of highly nonlinear dynamical systems. In short, with the expansion of original state variables to infinite dimensions, the operator can uncover subtle nonlinear behaviour in a system that would otherwise be difficult or impossible to detect \citep{koopman}. %

\begin{mdframed}[backgroundcolor=gray!20]
\small
{\bf Koopman operator} \citep{koopman}
The Koopman operator is a linear operator that describes the evolution of an observable function of a dynamical system. Let $\mathcal{M}$ be a manifold of dimension $n$ and $f:\mathcal{M}\rightarrow \mathbb{R}$ be a real-valued function. The Koopman operator, denoted by $\mathcal{K}$, is defined as an infinite-dimensional linear operator that acts on the space of observable functions $f$ such that for any $f\in L^2(\mathcal{M})$,
$${\mathcal K}f({\bf x})=f(T({\bf x})),$$
where $T:\mathcal{M}\rightarrow \mathcal{M}$ is the evolution operator that maps each point ${\bf x}\in\mathcal{M}$ to its next iterate in time. Now, let $f:\mathcal{M}\rightarrow \mathbb{R}$ be a bounded, measurable observable and $\mathcal{K}$ be the Koopman operator associated with a dynamical system. Then, there exists a sequence of eigenfunctions $\psi_j:\mathcal{M}\rightarrow \mathbb{C}$ and a corresponding sequence of eigenvalues $\lambda_j\in\mathbb{C}$ such that
$${\mathcal K}\psi_j({\bf x})=\lambda_j\psi_j({\bf x}).$$
The Koopman operator preserves the linear structure of the space of observables, provides a linear representation of nonlinear dynamics, and its eigenfunctions provide a useful basis for approximating dynamical systems. %
\end{mdframed}

There are, however, some drawbacks associated with the Koopman operator. First, it is intrinsically an infinite-dimensional operator, and although there are efficient finite-dimensional approximations available, the accurate computation of their eigenvalues and eigenfunctions can be computationally expensive \citep{kaiser2020data}. This can be a problem for real-time applications, such as model-based control \citep{kais21}. Second, related to this drawback, the operator's eigenfunctions are often difficult to interpret, hampering the capacity of the learned representations to explain the underlying system's dynamics. Finally, the Koopman operator assumes that a locally-linear behaviour can represent nonlinearities sufficiently accurately.
Thus, the Koopman operator is an important theoretical and applied research tool for understanding and predicting the behaviour of complex dynamical systems. Its data-driven approach to model-based control and optimisation has opened up new possibilities for real-time applications \citep{kais21}. Moreover, its ability to uncover subtle nonlinear behaviour in chaotic systems has made it invaluable for studying chaotic dynamical systems \citep{takeishi2017learning}.

\begin{figure}[h!]
    \centering
\begin{tabular}{ccccc}
(a) Space & (b) Time & (c) Data & (d) PCA & (e) DMD \\
\hline
$x_1(s,\cdot)$ & $x_1(\cdot,t)$ & $x(s,t=10)$ & Mode 1 & Mode 1\\
     \includegraphics[width=2.5cm]{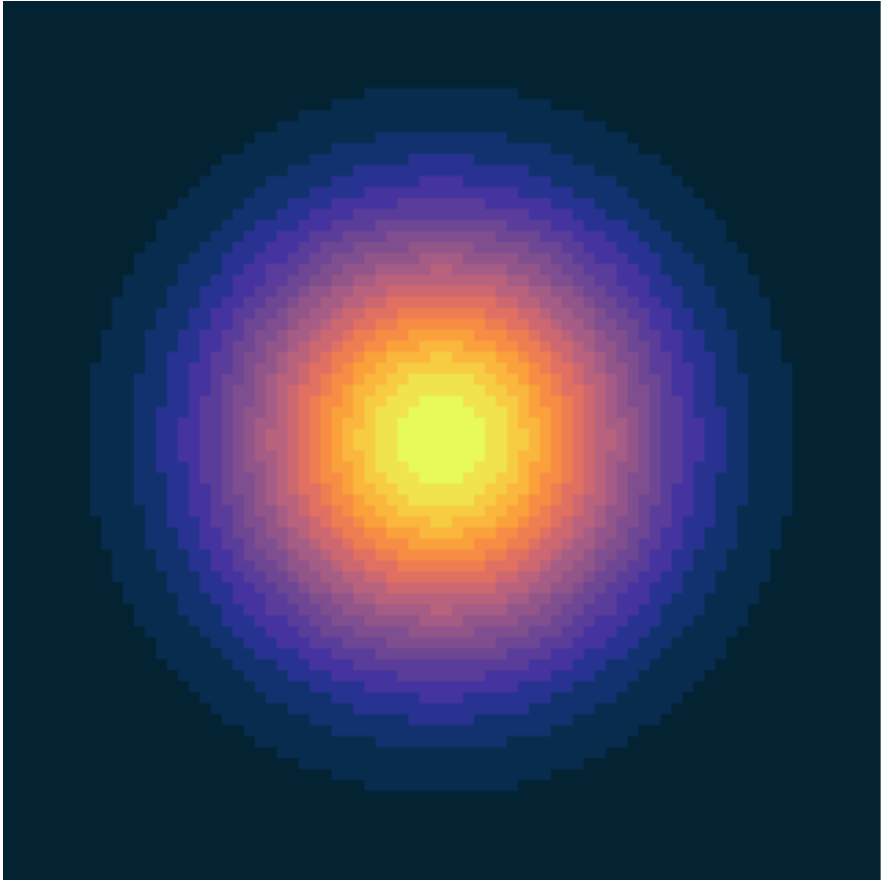} &
     \includegraphics[width=2.5cm]{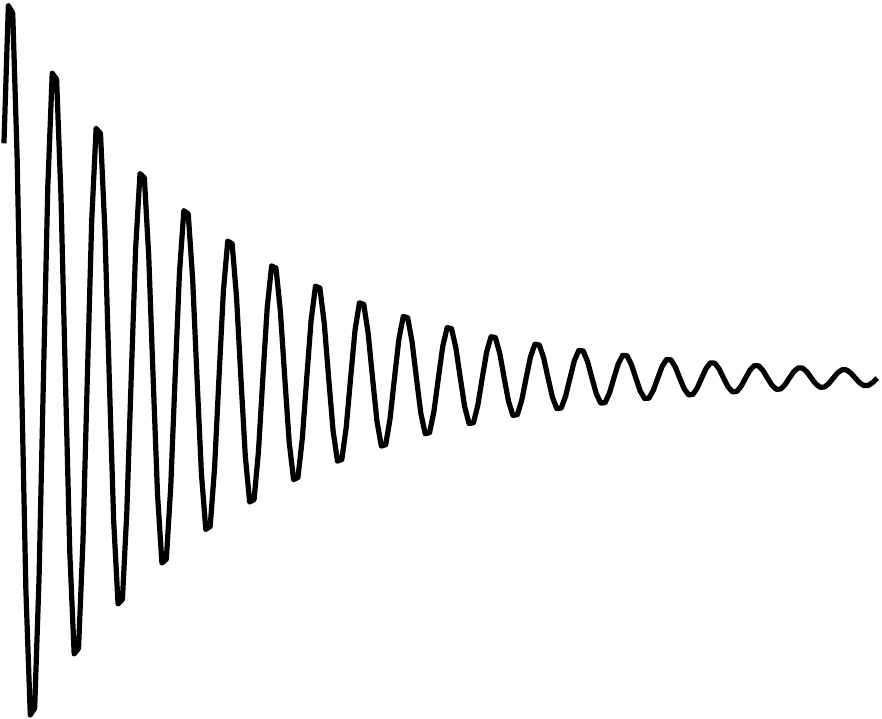} &
     \includegraphics[width=2.5cm]{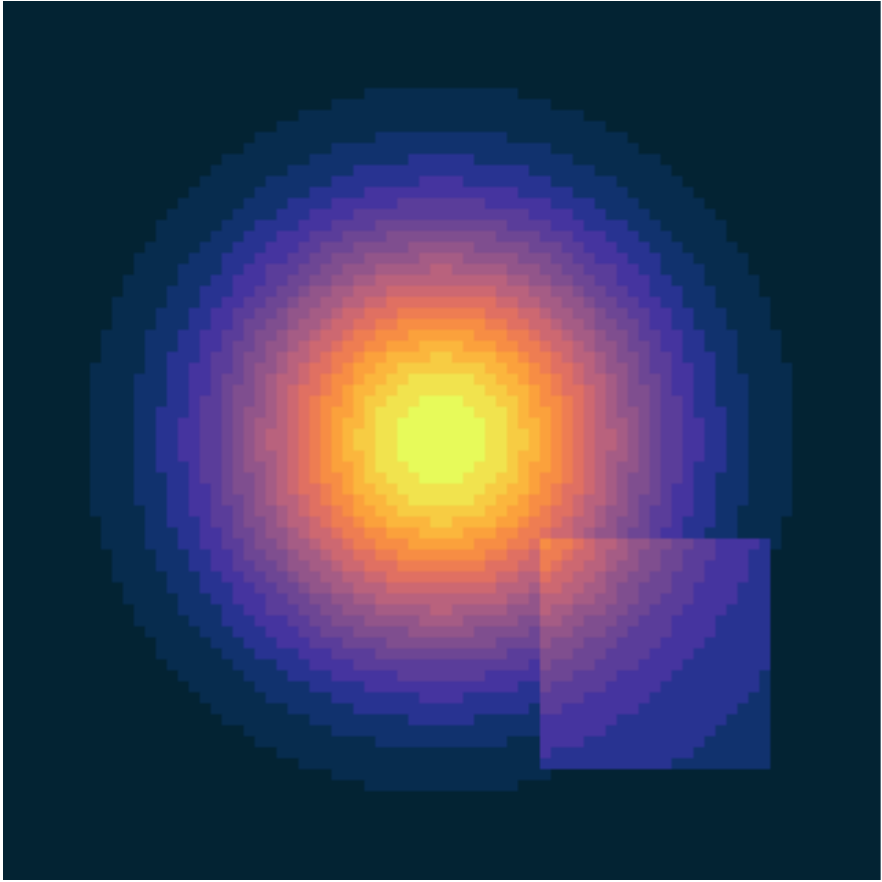} &
     \includegraphics[width=2.5cm]{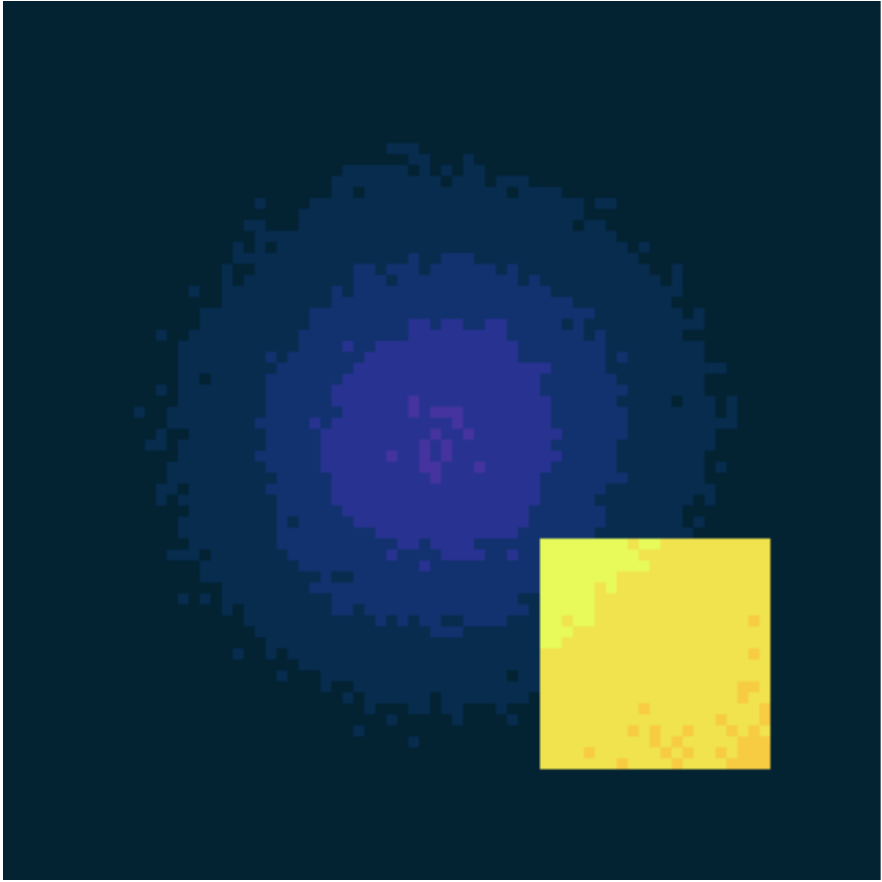} &
     \includegraphics[width=2.5cm]{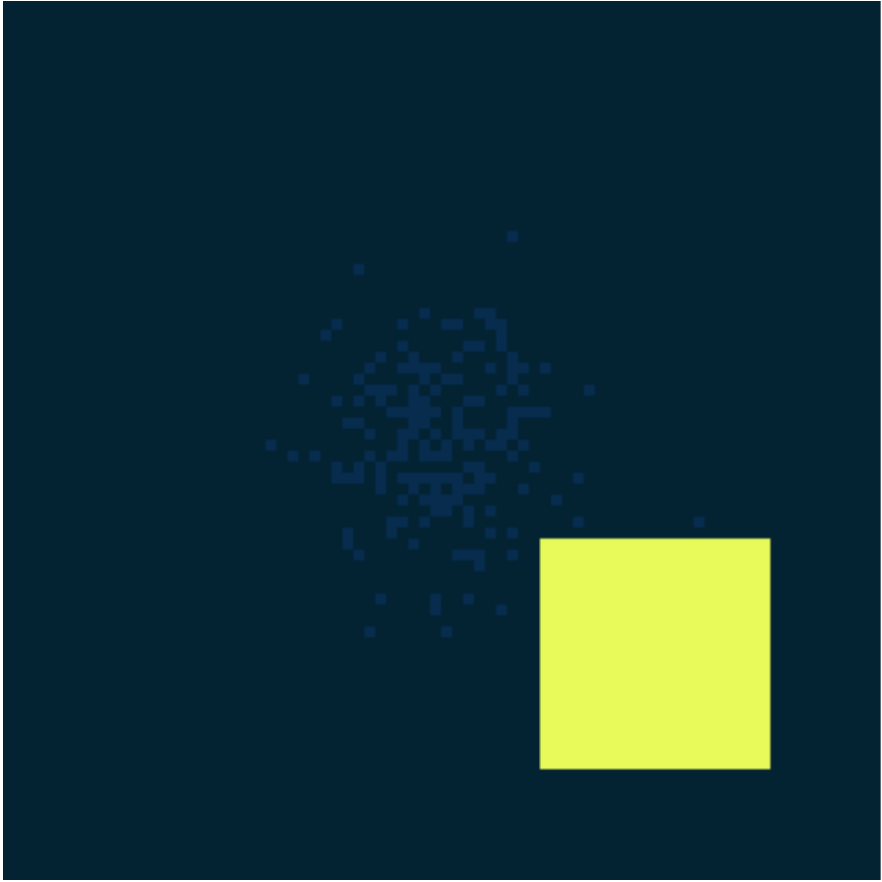} \\
$x_2(s,\cdot)$ & $x_2(\cdot,t)$ & & Mode 2 & Mode 2\\
     \includegraphics[width=2.5cm]{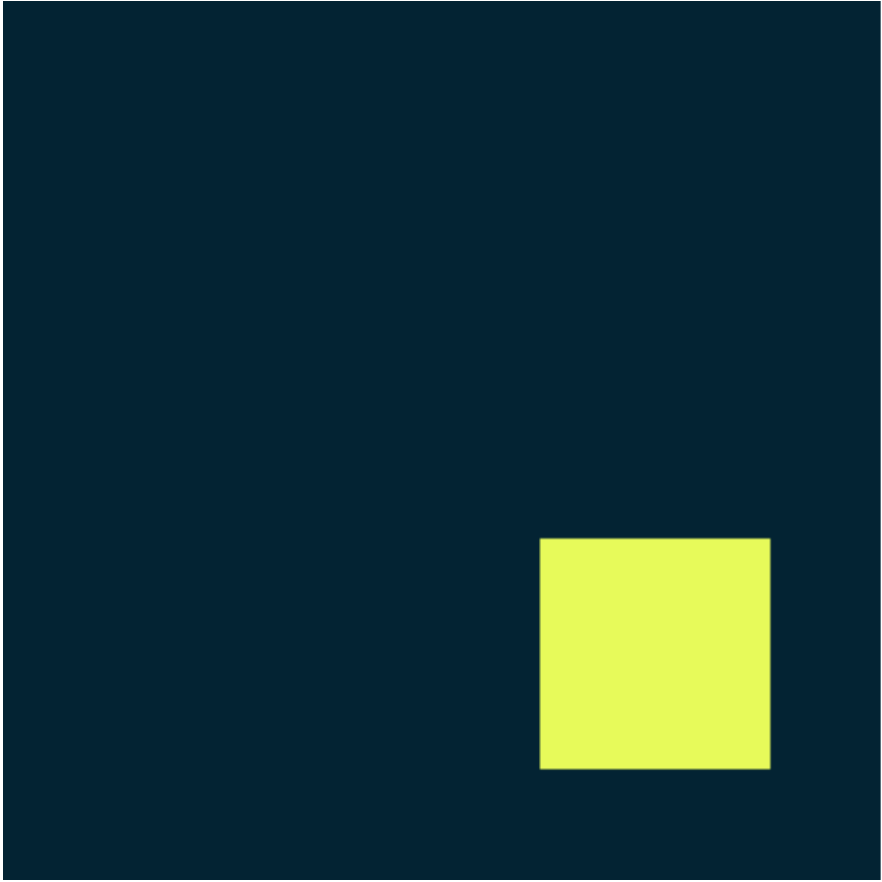} &
     \includegraphics[width=2.5cm]{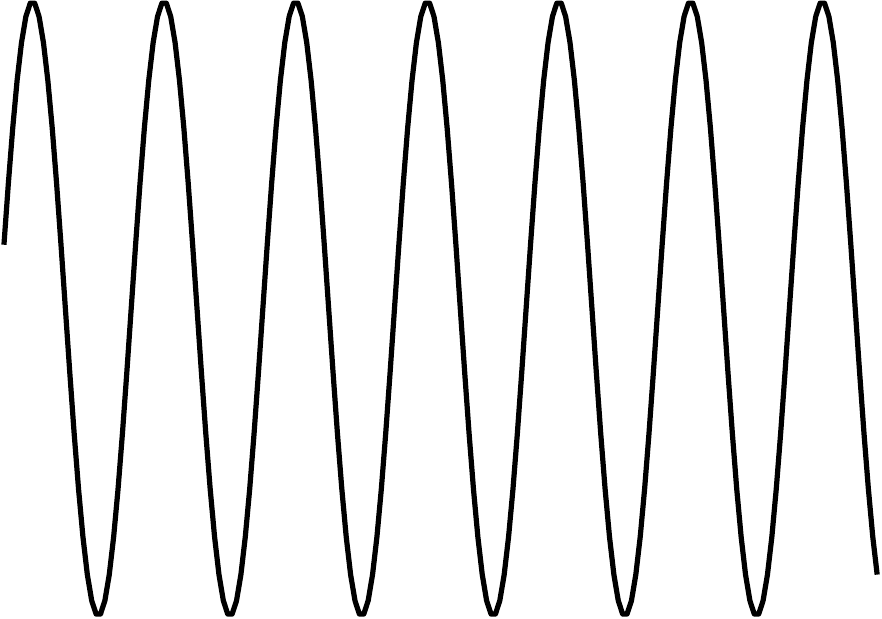} & &
     \includegraphics[width=2.5cm]{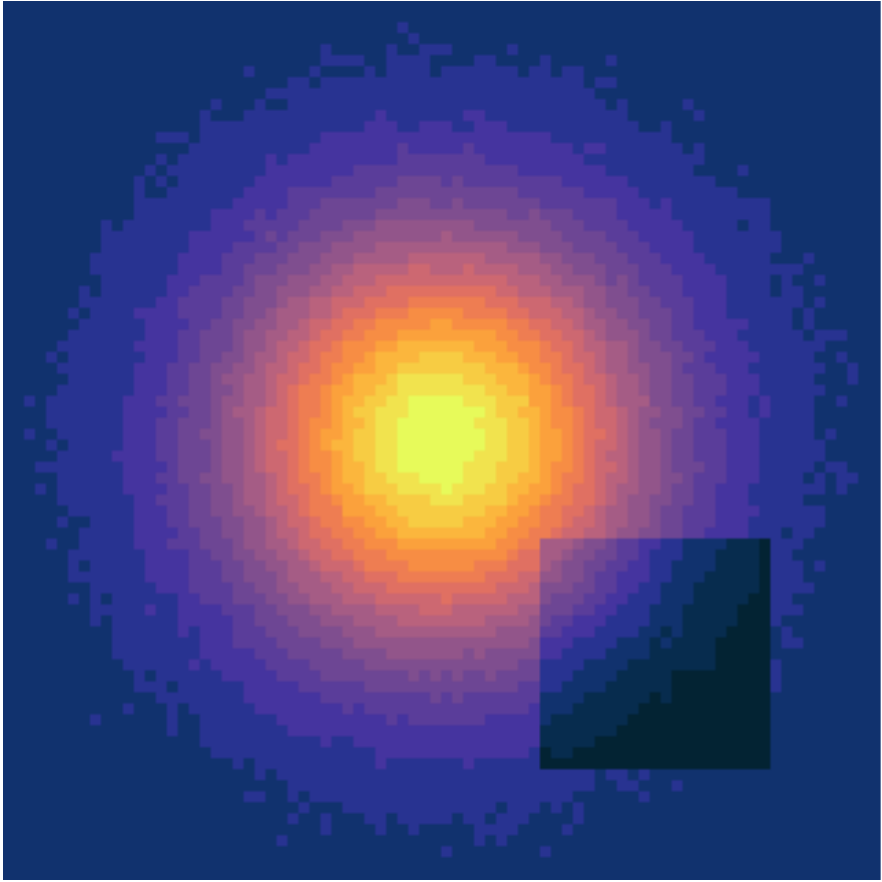} &
     \includegraphics[width=2.5cm]{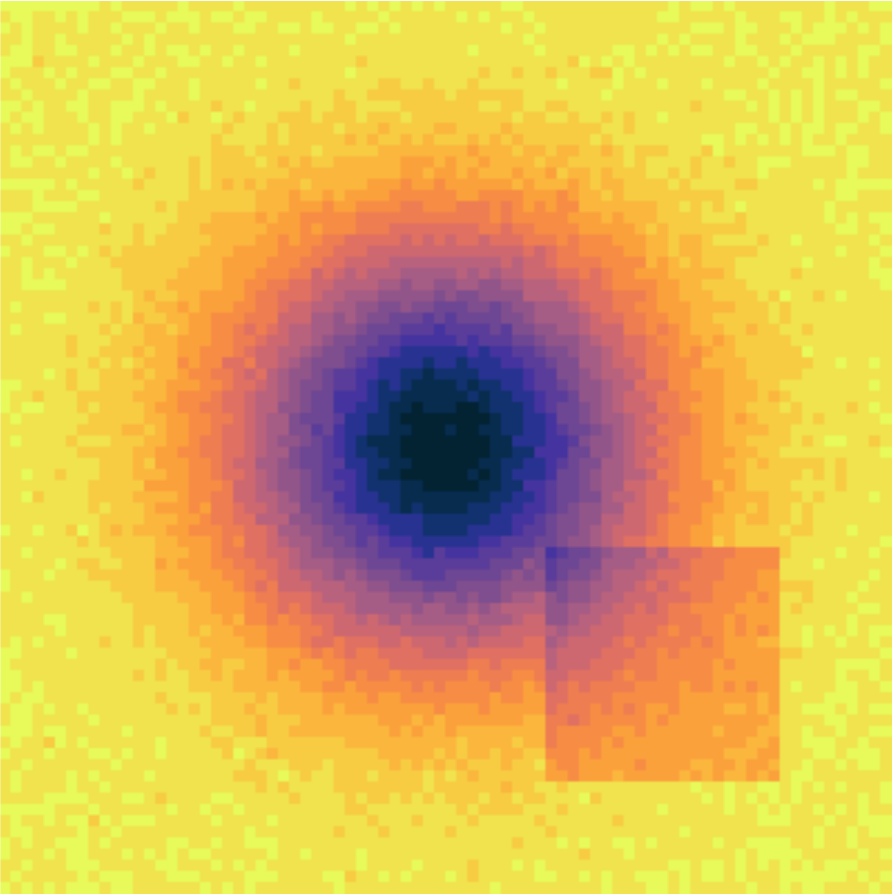} \\
\end{tabular}
    \caption{Comparison between DMD and PCA with synthetic spatio-temporal data. The signal under analysis $x(s,t)$ (c) is the sum of two generative signals (a,b): $x_1(s)$ is a Gaussian that decays exponentially, and $x_2(s,t)$ is a square that oscillates at a lower frequency. The projections onto the two top components of PCA (d) and DMD (e) show that DMD extracts cleaner spatial coherence patterns from the data.}
    \label{fig:pca_dmd}
\end{figure}

More specifically, given a space in which the dynamics is linear, a successful approach to approximate transfer operators (such as the Koopman operator) from the data is called dynamic-mode decomposition (DMD) \citep{Schmid10}. It is also possible to obtain ROMs using DMD~\citep{Schmid10}, which is also based on concepts from linear algebra and assumes that the system's state can be advanced in time via a linear operator $\mathbf{A}$. While the POD modes are orthogonal in space, the DMD ones are orthogonal in time, and each mode is associated with a particular frequency and a growth rate. Therefore, DMD may help to identify temporal patterns in the data more clearly than POD. In contrast, POD may lead to a more compact low-order representation of the original system due to its optimality property. See an illustrative example in Fig.~\ref{fig:pca_dmd}.

\begin{mdframed}[backgroundcolor=gray!20]
\small
{\bf Dynamic-mode decomposition (DMD)} \citep{Schmid10} is a technique used to approximate the normal modes and eigenvalues of a linear system. Additionally, these modes can be associated with a damped or driven sinusoidal behaviour in time. DMD is useful for identifying a system's frequency and decay/growth rate. Let us define a dynamical process formulated as $\frac{d{\mathbf{x}}}{dt} = f({\mathbf{x}}, t, \mu),$ where ${\mathbf{x}}$ defines a measurement, $t$ is a time, $\mu$ is a parametric dependence, and $f$ indicates an unspecified system but from which we obtain many data.
Therefore, the complex dynamical system $f$ can be approximated as follows $\frac{d{\mathbf{x}}}{dt} \approx {\bf A}{\mathbf{x}}$, where $x \in \mathbb{R}^n$, $n\gg 1$ and ${\bf A}$ defines a linear dynamical system. %
Then its general solution is the `exponential solution' defined as ${\mathbf{x}}={\mathbf{v}}e^{\lambda t}$, where ${\mathbf{v}}$ and $\lambda$ are eigenvectors and eigenvalues of the linear system ${\bf A}$. The problem of finding the eigenvectors ${\mathbf{v}}$ and the eigenvalues $\lambda$ is a eigenvalue problem defined as $\lambda {\mathbf{v}} = {\bf A}{\mathbf{v}}.$ %

Yet, we are interested in obtaining ${\bf A}$, not its eigendecomposition. This is what the so-called `exact DMD' does. DMD uses observations/measurements $x_j = {\mathbf{x}}(t_j)$, defined at a time point $j$ to construct two matrices: the first  concatenating the data from the first snapshot to $(m-1)$-th snapshot, and the second with the shifted-by-1-time-step samples, ${\bf X}$ and ${\bf Y}$, respectively.
The goal is thus building a linear dynamical system $A$ fitted with $\frac{d{\mathbf{x}}}{dt} = {\bf A} {\mathbf{x}}$, and thus {\em learn} the linear dynamical system ${\bf A}$ that takes the data ${\mathbf{x}}$ from current state $(j-1)$ to future state $(j)$, that is ${\bf Y} = {\bf A} {\bf X}$. The linear dynamical system $A$ can be extracted using a pseudo-inverse ${\bf X}^{\dagger}$ of ${\bf X}$, that is ${\bf A} = {\bf Y} {\bf X}^{\dagger}.$ Intuitively, the linear dynamical system ${\bf A}$ performs a least-square fitting from the current state ${\bf X}$ to the future state ${\bf Y}$.

\end{mdframed}
Over the last decades, different numerical methods have been introduced: Ulam's method \citep{Ulam60}, extended dynamic-mode decomposition (EDMD) \citep{WKR15, WRK15, KKS16}, and the variational approach of conformation dynamics (VAC) \citep{NoNu13, NKPMN14}. The advantage of purely data-driven methods is that they can be applied to simulation and observational data. Hence, information about the underlying system itself is not required. An overview and comparison of such methods can be found in~\citep{KNKWKSN18}. Applications and variants of these methods are also described in \citep{RMBSH09, TRLBK14, MP15}, while kernel-based reformulations of the methods above have been proposed before in~\citep{WRK15,SP15}. Note that the framework of higher-order dynamic-mode decomposition (HODMD)~\citep{Sole} enables relaxing the linear assumption by including several temporal snapshots to build the operator by exploiting Takens' delay-embedding theorem~\citep{Takens81}. The HODMD approach requires additional hyper-parameter tuning, but it has led to very insightful results, for instance, in the context of complex turbulent flows, where this method has enabled identifying the coherent structures responsible for the concentration of pollutants in cities~\citep{lazpita_pof}. Another relevant application of HODMD includes cardiovascular flows~\citep{Blood}.

\subsection{Dynamic modes in neural-network latent spaces}

\begin{figure}[h!]
    \centering
\includegraphics[width=15cm]{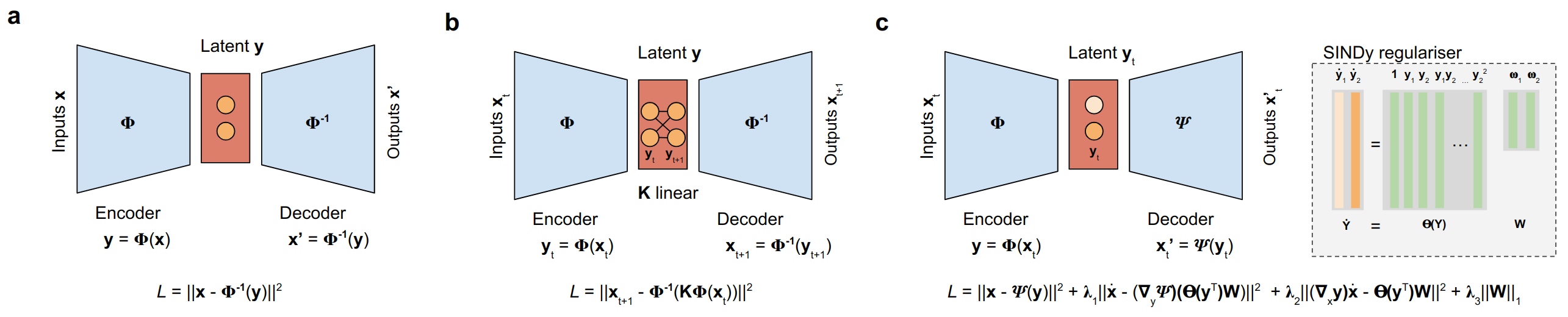}~~~\\
    \caption{Approximating Koopman operator with explicit nonlinear mappings using autoencoders. (a) An autoencoder neural network learns a mapping $\boldsymbol{\Phi}$ compressed latent representation ${\bf y}$ from input data ${\bf x}$ by minimising the reconstruction error $L$. (b) One can incorporate the Koopman linear operator ${\bf K}$ operating in the latent representation ${\bf y}_t$, which can then be used for prediction ${\bf x}_{t+1}$ from the transformed ${\bf y}_{t+1}$ %
    \citep{brunton_koopman}. (c) Those representations are not necessarily physically consistent. This can be addressed by enforcing equation dictionaries using SINDy in the loss for the simultaneous discovery of coordinates and parsimonious dynamics \citep{champion2019data}. The loss now accounts for the reconstruction of the input data, as in a regular autoencoder, and the temporal dynamics (gradients $\nabla_x, \nabla_y$) of ${\bf x}$ and ${\bf y}$ projected onto SINDy bases.}%
    \label{fig:neuralkoopman}
\end{figure}

Despite the interesting properties of POD and DMD, their inherent linearity typically leads to the requirement of very large numbers of modes to reconstruct most of the variance of the original signal, for example, in three-dimensional turbulent flows~\citep{baars}. Neural networks, especially autoencoders (AEs), have been proposed to obtain a reduced-order nonlinear representation of the original data. AEs exploit non-linear activation functions to produce significantly more compact representations in the latent space than those with, \eg POD~\citep{r1}. Figure \ref{fig:neuralkoopman}[a-b] shows the use of AEs for learning (dynamic) feature representations.

AEs have been used in fluid mechanics to obtain compact modal decompositions of the flow around a two-dimensional cylinder~\citep{r2} and in more complex turbulent flows, \eg the flow in a simplified urban environment~\citep{r3}. Interestingly, when restricting neural networks to linear activation functions, one recovers the POD modes, as shown by \citet{r4} with a multilayer perceptron (MLP) in turbulent channel flow. Shallow NNs have been used for flow reconstruction, in this case from sparse measurements, as illustrated by \citet{r5} for several flow cases. A more general illustration of the potential of AEs based on convolutional neural networks (CNNs) was presented by \citet{r6}, and an application to spectral submanifolds was developed by \citet{r7}. The reader is referred to Refs.~\citep{r8,r9} for a survey of classical methods applicable to linear subspaces.

Despite the superior compression performance of AEs compared with POD, the former does not have two very interesting properties of the latter, namely the optimality and orthogonality of the resulting modes. These are important properties due to their connection with interpretable and parsimonious ROMs.
Regarding optimality, \citet{r10} proposed an interesting approach based on hierarchical autoencoders (HAEs). They first trained a CNN-based AE fixing the dimension of the latent space to just one, obtaining one latent vector. Then, they trained another CNN-AE with a latent dimension of two and fixed the first latent variable to the one obtained in the previous NN, thus obtaining a second latent vector. Through this recursive strategy, they obtained a sequence of latent vectors exhibiting progressively less contribution to the reconstruction of the original signal, allowing them to establish a ranking in the resulting modes. This approach was tested in the flow around a two-dimensional cylinder, although it is important to note that the resulting modes were not orthogonal. This was addressed by \citet{r3}, who used $\beta$-variational autoencoders ($\beta$-VAEs), which enable introducing stochasticity in the latent space to impose orthogonality in the resulting AE modes, a phenomenon that was explained in \citet{RolinekZietlowMartius:VAERecPCA} among a connection of $\beta$-VAEs to PCA. Also in the case of the $\beta$-VAEs, the modes can also be ranked in terms of their contribution to the reconstruction.

\subsection{Equation discovery in latent representations}\label{sec:eq:discover:latent}

Perhaps the biggest challenge in data-driven model discovery is balancing model efficiency with descriptive capabilities. Parsimonious models with the fewest terms required to capture essential interactions promote interpretability and generalizability. However, obtaining parsimonious models is linked to the coordinate system in which the dynamics are measured. The previous methods based on dimensionality reduction, \eg ROM, DMD or AE, extract expressive components without simultaneously discovering coordinates. In \citep{champion2019data}, AEs were trained for data reconstruction and to recover a parsimonious dynamical system model through sparse regression using SINDy (see \secref{sec:sindy}). See Fig.~\ref{fig:neuralkoopman}[c]. The joint goal of discovering models and coordinates is critical for understanding many modern systems. Using SINDy as an explicit equation discovery regulariser in the latent space balances simplifying coordinate transformations and nonlinear dynamics to identify coordinate transformations where only a few nonlinear terms are present.

\subsection{Discovering fundamental variables} \label{sec:eq:discover:fundamental_vars}

Despite advances in equation discovery (either through implicit or explicit representations), the main core problem is  identifying state variables. The discovery typically refers to the identification of the governing equations, not the identification of the physical forces or variables. The vast majority of data-driven models of discovery rely, however, on pre-existing knowledge of the state variables, \eg, the position and velocity of a rigid body object. This relies on deep domain knowledge and strong assumptions. In addition, such assumptions cannot work properly for new physical systems or when those state variables cannot be measured. The work \citep{chen2022automated} proposed a principle for determining the number and identity of state variables in a system from high-dimensional data and demonstrated high effectiveness using video recordings of physical systems. The algorithm discovered the intrinsic dimension of the observed dynamics and could identify candidate sets of state variables without prior knowledge of the underlying physics. Alternatively, other studies sought to identify the fundamental state variables via manifold learning in ambient RKHS (termed {\em Diffusion} maps, \eg, \citep{keme22,thie20}, see also Section \ref{subsubsec:empirical_neuronal_trajec}).

In short, the field of {\em variable discovery} is filled with many opportunities in the physical, biological and chemical sciences \citep{pukrittayakamee2009simultaneous,keme22,schutt2017quantum}, as well as many challenges \citep{chen2022automated}. Finally, and interestingly, we want to emphasise that variable discovery is intimately related to revealing latent confounders in the field of causal inference \citep{probaLatZ,autoReg,Diaz22noiseimputation}.

\section{Perspectives} %

Let us indulge ourselves with a brief overview of the main challenges (both conceptual and technical) and the opportunities for future research in the field of equation discovery for the physical sciences.

\subsection{Challenges}

\begin{figure}[h!]%
    \centering
    \includegraphics[width=9cm]{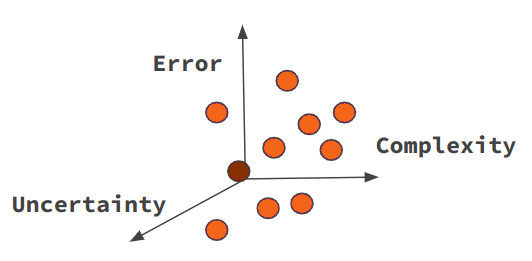}
    \caption{On the quest for the optimal model in the sparsity-extrapolation-generalisation space.} %
    \label{fig:thequest}
\end{figure}

The field of equation discovery from data is very prolific and is situated at the intersection of many communities: statistics, machine learning, computational fluid dynamics, Bayesian inference, dynamical systems and control theory, functional analysis and causal inference. The field has occupied scientists for centuries at all levels. The quest for optimal and automated solutions has traditionally considered moving in one or several subspaces in the sparsity-extrapolation-generalisation space, i.e., models should be simple, generalisable/robust, and capable of extrapolating outside the sample space (Fig.~\ref{fig:thequest}). These are very ambitious goals, implying both theoretical and practical challenges. 

\paragraph{Theoretical challenges.} 
From a more theoretical perspective, both the {\em identifiability} of the system's equations \citep{fajardo2023fundamental,antonelli2022data} and the role (or preference) for {\em sparsity} (simplicity) have been questioned \citep{fuentes2021equation,guimera2020bayesian}. In addition, there is a long-standing debate on the {\em evaluation} of the obtained solution, where many criteria can be adopted. %
Is it only about invariances and robustness in space and time? Is Nature always simple and compositional, such that compactness and sparsity rule in natural systems? %
The issue of model's (\ie hypotheses) intercomparison and evaluation also speaks to the more elusive question of how to reconcile  solutions offered by different equation discovery methods. 

Another important theoretical challenge is related to the fact that, very often, one (1) assumes that all involved state variables are given/observed, which resembles the sufficiency assumption in causal inference, and (2) selects a subset of representative states, assumes a particular basis to express the solution, or can operate on a manifold subspace \citep{champion2019data,brunton2022data,daniels2015automated,waltz2009automating}. 
Both are strong assumptions that challenge the process of discovering equations and raise many questions. 
How do we choose the right variables to include in the equation discovery method? How much does the solution change when a variable is omitted or added? Here, {\em identifiability} issues and {\em latent confounders} play a substantial role. 
And foremost, what if we cannot measure the underlying state variables? Can we identify them automatically? Several methods have arguably proposed to discover the latent variables \citep{chen2022automated}, and other efforts exploit the link between RKHS techniques and a plausible master equation underlying the spatiotemporal evolution of the data probability density function \citep{coif06} to automatically learn a set of fundamental coordinates of the system, cf. \secref{sec:eq:discover:fundamental_vars}. 
The inferred subspace has shown effectiveness in identifying the dynamics of (partial) differential equations \citep{keme22,thie20}, see examples in \secref{subsubsec:empirical_neuronal_trajec}. However, like other explicit methods discussed in this section, it requires prior guidelines on the differential equation to model and incorporates a range of heuristics during its processing pipeline \citep{keme22}.

\paragraph{Practical challenges.} 
Important practical challenges are related to the {\em model development} and {\em data characteristics}: (1) {\em high dimensionality}, (2) {\em nonlinear relationships} and (3) {\em risk of overfitting}. In many cases, the number of variables and parameters can be very large, making it difficult to find the most relevant features and relationships. This is the scenario where non-identifiability arises. Another challenge is that real-world systems often exhibit highly nonlinear relationships between input and output variables. This makes finding a good representation of the underlying dynamics difficult, and the search space for equations can become very large. While several nonlinear methods that capture complex dynamics exist (kernels, neural networks), performance evaluation and hyperparameter tuning are still important challenges. Finally, symbolic regression approaches can also suffer from overfitting, where the model fits the training data well but performs poorly on new data, which speaks to the trade-off between the model's accuracy and complexity. 

When working with spatiotemporal data, it is relatively unclear how to incorporate information about time-lagged relations and interventional data. %
Both the explicit and the implicit approximations show particular challenges, though:
On the one hand, symbolic regression techniques (such as SINDy) face significant problems in defining the basis functions, working in high-dimensional problems, and the impact of (even a limited) amount of noise. Other methods, like those based on AI Feynman, even if they incorporate sensible criteria to guide the equation discovery (like compositionally, reversibility or physical unit consistency), almost predict completely different equations when changing constants in the true equation \citep{suseela2022comparative}.
On the other hand, implicit methods (like DMD or Koopman operators) do not provide an explicit equation but a latent feature representation to explain system dynamics. These methods struggle with nonstationarities, nonlinearities, and gaps noise, which remain unresolved problems in the literature \citep{wu2021challenges}. Note that similar challenges to those in causal discovery remain here, cf. \secref{sec:causal-discovery-challenges}. Besides, DMD methods fail to generalise outside the training data and violate basic physical laws. To alleviate this, integrating domain knowledge (such as symmetries, invariances and conservation laws) in DMD has been recently introduced as an effective, robust approach \citep{baddoo2023physics}. 

\subsection{Opportunities}

While symbolic regression presents challenges, it offers exciting research opportunities for the physical sciences. %
Three main opportunities can be identified: (1) {\em model interpretability}, (2) {\em model compression and evaluation}, and (3) {\em model selection}. Equation discovery is a step forward in the system's understanding. The field leverages fully interpretable models, and unlike causal models, equation discovery (symbolic regression) models are directly applicable predictive models. The discovered equations from data can provide insights into the underlying dynamics of a system, thus helping researchers better understand how different factors interact and contribute to a particular outcome. Even with implicit latent representations, interpretability can be accomplished with interventional analysis. Another interesting opportunity is model compression, as symbolic regression models can also be used to compress large datasets into simple equations that capture the system's essential features; this can make it easier to analyse and visualise the data and make it more computationally efficient to work with. Another practical opportunity of symbolic regression models is that they offer a reduced set of possible solutions typically ranked in amenable Pareto fronts, which of course, trigger difficulties in choosing the right model but also fruitful scientific discussions about the plausibility of identified relations. This can save researchers time and effort and help identify unexpected patterns and relationships in the data.

\paragraph{Model interpretability and intervention analysis.} 
The discovered explicit models are interpretable in nature. However, when complexity cannot be traded for accuracy or whenever an implicit feature representation is learned, intervention (or sensitivity) analysis offers opportunities for interpretability. %
For example, one can (1) {\em ignore or simplify the problem} by performing small perturbations away from real-world dynamics, which might help identify the proper relationship between variables; %
(2) {\em intervene on exogenous variables} (\eg wind or solar irradiation, mixing coefficients, initial conditions in climate sciences, or targeted, direct brain stimulation in neurosciences) which is equivalent to collecting more data; %
(3) {\em create a library of trajectories} under different conditions and select those which match the desired intervention; %
and 
(4) {\em intervene in the learned latent space} and decode the intervention back to input space, thus allowing us to generate interventions that follow the system's natural trajectories. 
Further analysis methods, \eg for studying interventions of the learned ODE, might be fruitful, as they can access long-term dynamical properties, such as how distortion in the eigenvalues affects the system's stability as the phase space changes.

\paragraph{Model compression and evaluation.} 
Scientists frequently use metrics to evaluate new ideas or distinguish between competing hypotheses. As we have seen before, a governing equation should be simple but not simpler, accurate for prediction, robust under distortions and changes, and invariant in space and time. Equation discovery offers a {\em direct} way to learn plausible models and an {\em indirect} way to contrast and evaluate derived models. For this, one typically assesses 
(1) the {\em predictive accuracy} when answering how well the (simplest) hypothesis explains the data; 
(2) the model's {\em invariance} under distributional shift to account for the causal mechanisms; and 
(3) the {\em robustness} under interventions to study how the proposed process (or descriptive equation --representation) is consistent with interventions on the model dynamics, such as deactivation of components or targeted modification of exogenous variables. The latter is the rarest form of validation due to its high computational cost and difficulty in experimental design. 

\paragraph{Model selection.} 
Enforcing sparsity in model selection can lead to unrealistically too simple models. That is why methods that can provide solutions along the Pareto line (\figref{fig:pareto-illustration}) are needed to capture complex relationships and offer subsets of plausible model solutions. Alternative regularisation schemes will likely be important alongside profound estimates of uncertainty and extrapolation indicators.

%% file: 04_CaseStudies_book.tex
\chapter{Case studies in the physical sciences\label{sec:casestudies}} 
This section gives concrete examples of applying different data-driven causal and equation discovery in important fields of the physical sciences: neuroscience, Earth and climate sciences, and fluid and mechanical dynamics, cf. Table~\ref{tab:casestudies}.

\begin{table}[h!]
    \centering
    \small
   \caption{Case studies presented and the main methods used in this section.}
    \label{tab:casestudies}
    \begin{tabular}{|c|c|c|c|}
      \hline
      & Neuroscience  & Earth \& climate  & Fluid dynamics\\
      \hline
      \makecell{Causal\\discovery} & \makecell{Causal connectivity\\ \scriptsize{\sf(DCM, GC, TE, SCM)}}  & 
      \makecell{Carbon-water interactions\\Climate model comparison\\
      \scriptsize{\sf(CCM, PCMCI)}} & -- \\
      \hline
      \makecell{Equation\\discovery}  & \makecell{Learning trajectories\\ \scriptsize{\sf(kFDA, GP, Variational Bayes} \\
      \scriptsize{\sf RNN, Diffusion Maps)}} &  \makecell{Ocean Mesoscale closures\\ \scriptsize{\sf(RVM, DMD, SINDy)}} & \makecell{Turbulence understanding\\Vortex shedding\\ \scriptsize{\sf(SINDy, Genetic Programming)}}\\
      \hline
    \end{tabular}
\end{table}

\section{Neuroscientific applications of physics-based machine learning \label{subsec:neurosci}}

\subsection{Overview of parsimonious models for neural population dynamics \label{subsubsec:parsimonious_models_neural}}
Neuroscientific modelling falls within the remit of the field known as computational or theoretical neuroscience, which studies the transmission of information in the nervous system at multiple spatiotemporal scales (ranging from neuronal to whole-brain levels) in relation to perception, cognition, and behaviour (\eg, \citep{gern14,koch04}). Thus, an ongoing challenge in computational neuroscience is to link biophysically detailed models operating at microscopic levels with meso/macroscopic theories of cortical processing \citep{rabi18}. This enterprise is often addressed by deducting low-dimensional systems of partial differential equations or maps describing \emph{coarse-grained} variables derived from collective neural responses. Such different neurobiologically plausible simplifications are commonly termed ensemble, population, neural-mass, or simply \emph{firing-rate} models (see, for instance, \citep{byrn21,brun00}).

These synthesis efforts are intimately connected with empirically discovering a reduced dynamical system generating the observed neuronal activity. However, neurocomputational modelling traditionally focused on analytical, deductive approaches mapping realistic cortical networks to \emph{tissue-level} descriptions, as opposed to data-driven model discovery, reviewed in \secref{sec:equationdiscovery}. Thus, ensemble models are typically principles-based, often hinged on assumptions about dynamical interactions arising within homogeneous pools of neurons (\eg, \citep{amar77,byrn20,taba19,wils21}). 
 
Early neural ensemble models stemmed from applying statistical mechanical principles to the interaction of homogeneous pools of (excitatory and inhibitory) populations \citep{wils72,pott13}. Later, physics formalisms like the Fokker-Planck approach for describing the spatiotemporal evolution of the probability distribution of neuronal activity enabled theoretical neuroscientists to take a more holistic approach to identify mean-field approximations of networks of spiking neurons (\eg, \citep{gern14,matt02,brun00}). These and other nonlinear dynamical systems tools \citep{rabi08} provided closed-from, exact solutions for the collective behaviour of neural populations \citep{mont15,matt19,byrn20} capable of an extensive dynamical repertoire, although strongly dependent on universal theoretical assumptions, given their deductive nature (see \secref{sec:introduction}). Alternatively, a Laplacian assumption on this probability distribution resulted in neural mass descriptions \citep{marreiros2010dynamic}, recently proposed as building blocks for whole-brain models with translational applications \citep{schi22}.       

Overall, these chiefly deductive approaches rendered compact models of differential equations based on \emph{a priori} assumptions about neural and synaptic variables, fostering the interpretability of high-complex neuronal networks. By contrast, inferential approaches in neuroscience have been typically utilised to empirically identify neural dynamics underlying cognition and behaviour, as discussed next.     

\subsection{Empirical reconstruction of neuronal trajectories \label{subsubsec:empirical_neuronal_trajec}} 

There is an increasing focus on applying classic and deep machine learning approaches to reconstruct attracting and transient dynamics of cerebral cortex responses \citep{dunc21}. A neural trajectory $T \; (n \times d)$ is often defined as the sequence of $n$ neural response vectors $\boldsymbol{x}(t)$ embedded in a $d$-dimensional state-space (the \emph{ambient} space), spanned by neural ensemble activity or proxies thereof (\eg, firing-rates, electromagnetic potentials), their lags and nonlinear transformations \citep{galg23,dunc21,balaguer2011attracting}. 

Traditionally, standard dimensionality reduction techniques (\eg, PCA, Multi-dimensional Scaling, Discriminant analysis etc., see \secref{sec:dimensionality_reduction}) were directly applied for the visualisation of high-dimensional neural trajectories, showcasing coarse-grained aspects of firing-rate dynamics concerning, for example, cognitive decisions or motor functions \citep{cunn14,hym12}. More recently, Gaussian processes provided a flexible approach to derive a low-dimensional manifold representing the dynamical systems generating the observed activity. They just require a reasonable hypothesis on temporal correlations between observations (the prior covariance function \citep{rasmussen2006gaussian}). For instance, Gaussian process-based factor analysis (GPFA) \citep{yu09}, and other latent-variable methods \citep{aoi20}, provide such low-dimensional subspace while simultaneously approximating the probability of spiking -without the compelling need for probability density estimation \citep{yu09,gokc22}. These and related approaches can identify latent neural trajectory manifolds in prefrontal and motor cortices underlying decision-making \citep{rutt20,aoi20,dunc18}. Specifically, recent GPFA variants were able to discern between competing models for the contribution of upstream areas to recurrent dynamics supporting decision-making in the monkey prefrontal cortex \citep{galg23}. 

In general, new developments in deep auto-encoders such as Latent Factor Analysis via Dynamical Systems (LFADS) enabled the successful reconstruction of single-trial spiking activity \citep{pand18}. Moreover, related approaches such as preferential subspace identification (PSI) incorporate behavioural labels to inform the dimensionality reduction algorithm; effectively discerning \emph{behaviourally relevant} dynamics from the general neuro-dynamical landscape \cite{sani21}. Remarkably, recent self-supervised methods, also guided by behavioural observations, leverage contrastive learning and nonlinear ICA to produce consistent sub-spaces across experimental sessions and subjects \cite{schn23}. In addition, like previous approaches, they can operate in a supervised fashion, fostering decoding with respect to competing methods in a range of calcium and electrophysiological recordings in different species \cite{schn23}.

\begin{figure}[t!]
\centering 
\includegraphics[width=0.85\textwidth]{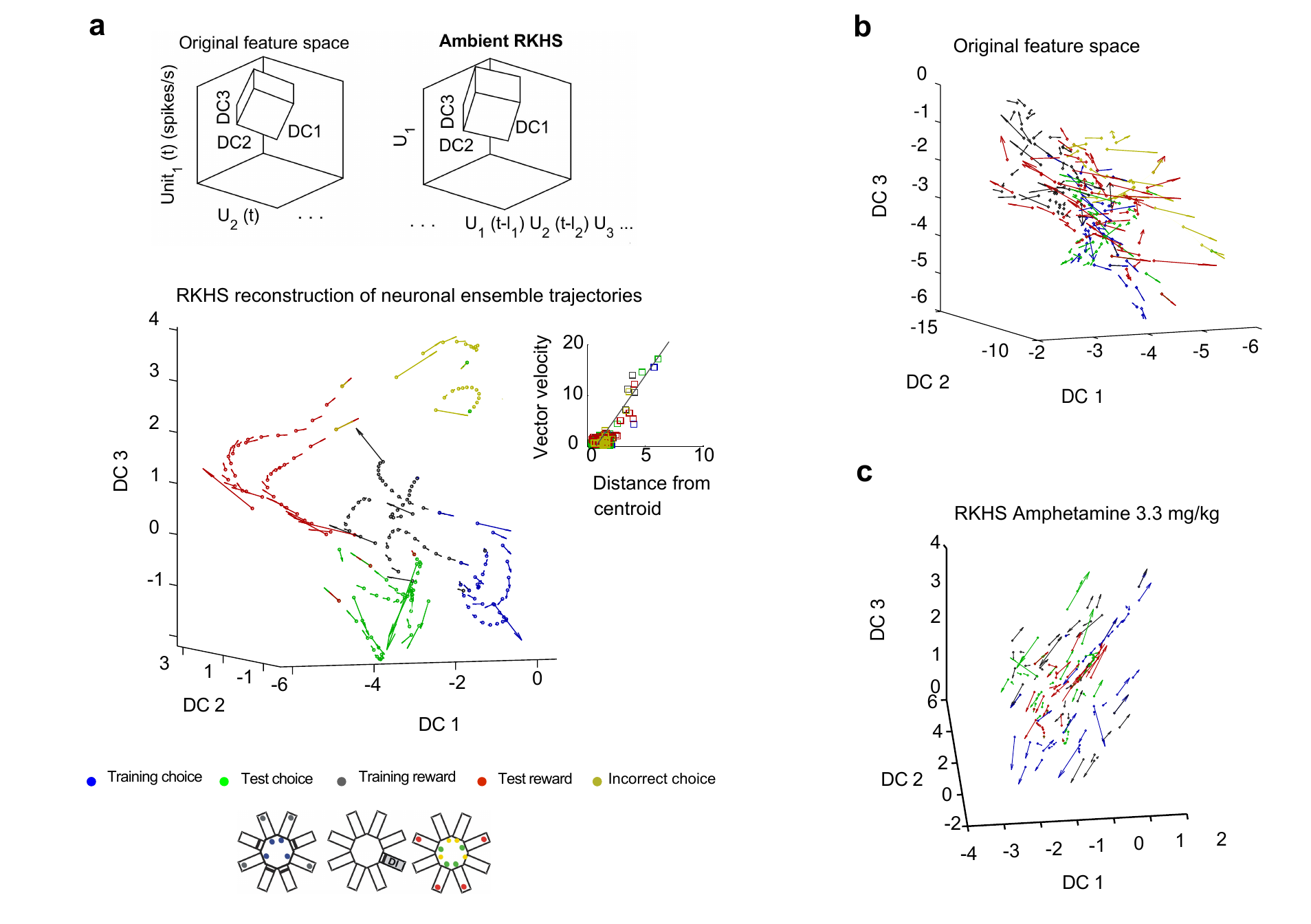}

   \caption{Example of the reconstruction of neural trajectories in an RKHS derived from rodent anterior cingulate cortex (ACC) multi-array recordings, taken from \cite{lapbal15}. \textbf{a}. The flow field stems from projecting an ACC ensemble firing rates onto the three main coordinates of a discriminant subspace (DC1-DC3, computed here by kernel-fisher discriminant analysis orthogonalised for a faithful representation, details in, \eg, \citep{balaguer2011attracting,bala20,bala14,lapbal15}). The DC1-DC3 is embedded into an optimal ambient high-dimensional RKHS (top schematics), spanned by neuronal ensemble firing rate and its higher-order correlations (up to 3rd order in this example) operating on a delay-coordinate map. The colour code (bottom) corresponds to rat choices in this experiment (the schematic of the radial-arm maze used in this experiment is taken from \citep{balaguer2011attracting,bala14}), which occupy distinct regions of the subspace. The flow field indicates faster shifts (large vector lengths) during the transition points, while it slows downs nearer the centroids of the clusters, suggesting an attracting-like dynamic landscape. The inset quantifies this uneven distribution of flow field speeds as a function of the distance to the centroid, further supporting this observation on flow convergence. \textbf{b}. This ordered phase-space structure cannot be achieved in the original feature space or with delay-coordinate maps. \textbf{c}. It is also destroyed when the animal receives a high dose of amphetamine, even in an optimal ambient space. Figure adapted from \citep{lapbal15} and from \citep{balaguer2011attracting} with publishers' permission (the Society for Neuroscience and the Public Library of Science).}
     \label{fig:recons1e}
\end{figure}

Covariance (kernel) function methods can also recreate salient facets of cortical dynamics like attracting sets. To this end, they leverage delay-embedding techniques in RKHS spanned by neuronal correlations and their temporal structure \citep{balaguer2011attracting,bala14,lapbal15,bala20} for identifying compact manifolds mapping animal's choice with attracting sets of ensemble trajectories \citep{lapbal15,bala20}. Figure \ref{fig:recons1e} presents an illustrative example of these RKHS techniques for recreating neuronal trajectories underlying the effect of dopamine at the circuit level. This approach facilitated the evaluation of mechanistic theories of dopamine modulation during decision-making, which was challenging given the limitations of direct experimental manipulations \citep{lapbal15}. The figure shows the flow field of trajectories derived from the activity of neuronal constellations in the rodent anterior cingulate cortex. Interestingly, the dynamic landscape depicted during working memory tasks in an optimal RHKS can be approximately described as transients connecting multiple attracting sets mapping spatial choices (Fig. \ref{fig:recons1e}a). This robust multi-stable scenario is completely disrupted by high doses of amphetamine (a well-known trigger of dopamine release, Figure \ref{fig:recons1e}b), while it is enhanced by low doses (see \citep{lapbal15}), in line with long-standing theoretical predictions of biophysical models \citep{durs08}.

Further facets of brain dynamics, such as chaotic attractors, have been recently addressed with recurrent, piecewise-linear architectures, amenable to optimisation via back-propagation variants or Bayesian variational inference \citep{durs17,bren22}. In these approaches, tractability is promoted by leveraging units' linearisation for approximating trajectory inference in a parsimonious, transparent fashion. Empowered by these characteristics, such linearised recurrent networks could infer pathological whole-brain dynamics from functional magnetic resonance imaging (fMRI) recordings \citep{kopp19}. Moreover, recent developments of these methods embody biophysically-inspired computations such as dendritic processing, fostering their reconstruction capabilities of nonlinear dynamical systems \citep{bren22}.

More broadly, classic and deep architectures can facilitate the Bayesian inference of the optimal range of biophysically realistic model parameters. This is a challenging task given the potentially high sensitivity of realistic networks to different parametrisations \citep{barr19}. Along these lines, approximate Bayesian computation (ABC) has been combined with connectionist approaches to identify parameters in models operating at multiple spatial scales, ranging from microscopic-level Hodking-Huxley-type single neurons \citep{luec17} to macroscopic, cognitive-level decision-making models \citep{boel22}. 

For instance, Sequential Neural Posterior Estimation (an ABC method) alternates between deep learners and variational Bayes approaches for parameter approximation. First, a standard (non-biophysical plausible) classifier is used to constrain the range of initial parameters $\theta$ generated from the prior $p(\theta)$ by predicting their suitability. Subsequently, a deep learner operating on multivariate data $\boldsymbol{x}$ and parameters sampled from such constrained prior $\hat{p}(\theta)$ estimates the likelihood $\hat{p}(\boldsymbol{x}|\theta)$, enabling a progressively more refined Bayesian computation of the posterior over neuronal and synaptic parameters $p(\theta|\boldsymbol{x})$ \citep{gonc20}. These approaches were able, \eg, to discern between neuronal model configurations, essentially indistinguishable in observed activity, the ones metabolically optimal in the pyloric network in crustacean \citep{dein22}; or to infer reaction times and choices in classic (drift-diffusion-type), descriptive models of decision-making \citep{boel22}.

Alternatively, neural trajectory reconstruction of Hodking-Huxley ensembles has been recently tackled with nonlinear manifold learning in RKHS \citep{thie20,keme22}. These methods, like common dimensionality reduction-based techniques (see \secref{sec:dimensionality_reduction}), identify first an optimally reduced set of coordinates from an original higher-dimensional ambient space embedding the time series $\boldsymbol{y}$. However, by contrast with other approaches, components spanning the low-dimensional representation are typically lead eigenvectors of a discretised Laplace operator governing the spatiotemporal evolution of the underlying $p(\boldsymbol{y}(\boldsymbol{x},t))$, where $\boldsymbol{x}$ is homologous to a spatial coordinate. Thus, the reduced subspace spanned by the main non-redundant eigenvectors is often termed a Diffusion Map \citep{coif06} (see also \secref{sec:eq:discover:fundamental_vars}) given by the set of $i^{th}$ emergent coordinates $\{ \phi(\boldsymbol{x})_{i} \}$.

In a subsequent stage, a fully connected (deep/shallow) architecture learns the dynamical system on the diffusion map, in other words, estimates the function $f$ that maps the temporal flow of the time series $\boldsymbol{y}$ to its derivatives w.r.t. emerging coordinates, for instance, in a one-dimensional diffusion map $\phi_1$, 
\begin{equation} \label{eq:DifMaps}
    \frac{ \partial y (\phi_1,t) }{\partial t} \approx f \bigg( y, \frac{\partial y }{ \partial \phi_{1} },\frac{\partial^2 y }{ \partial \phi_{1}^{2}  }, \ldots ,  \frac{\partial^n y }{ \phi_{1}^{n}   }, \gamma  \bigg),
\end{equation}
\noindent where $\gamma$ is a set of parameters which enables the learned map to reproduce bifurcations. 

Recently, diffusion maps and related approaches were also utilised for the one-dimensional reduction of neural recordings at multiple spatial scales, ranging from single-unit to macroscopic levels imaging data \citep{bren23,nieh21}. Such drastic simplification facilitates, for instance, the interpretability of decision-making models, while preserving a competitive prediction accuracy \citep{bren23}.  

Interestingly, when the argument of $f$ contains no derivatives and incorporates additive Gaussian noise, the system's dynamics reduces to the well-known Langevin stochastic differential equation, stemming from a Fokker-Planck process governing the temporal evolution of a probability distribution \citep{genk20}. Thus, by approximating a discretised Fokker-Plank operator, it is possible to empirically infer parameters of the stochastic process leveraging conventional likelihood estimation techniques \citep{genk20,genk21}. Langevin dynamics fit many natural phenomena and is of special interest in high-level decision-making models in neuroscience. This formalism was recently used in \citep{genk21} for model discovery based on a one-dimensional Langevin equation and an additional stochastic spike generator,
\begin{equation} \label{eq:Langevin}
\begin{split}
  \frac{ d x(t) }{d t} \approx D \cdot F (x) + \sqrt{2 \, D} \cdot \psi(t;0,1)), \\
   y \sim Poisson(x;\lambda(t)),
\end{split}
\end{equation}
\noindent where $x(t)$ is a latent trajectory, $D$ is the diffusion constant, $\psi$ is white normal noise, $F$ is the deterministic map to be inferred, and $\lambda(t)$ is the parameter of an inhomogeneous Poisson process generating the observed spike train time series $y(t)$. This approach is capable of identifying a parsimonious model (that is, a set of $\{D, F(x), \lambda(t)\}$) from the spike train time series. Thus, it was used to discern between competing models of perceptual decision-making by comparing probability distributions underpinning such alternative parameter sets via standard Kullback-Lieber divergence \citep{genk21}.

In short, neuroscientific studies typically conceive the discovery of biophysical laws as the inference of deterministic dynamics embedded in essentially stochastic neural processes. This goal has been interpreted either as empirically reconstructing attracting and transient components of neural activity (disentangled from coupled noise \eg, \citep{balaguer2011attracting,rutt20,keme22}); or as identifying parameters of parsimonious, \emph{a priori} model shapes \citep{genk21,luec17}. These approaches provide valuable insights on the latent neural dynamical landscape. 

However, a liking theme in such inferential methods is that, despite their advances in tractability (\eg, \citep{bren22,genk21}) and interpretability (\eg, \citep{bala20,keme22}), they are often not designed to empirically discern a unique set of differential equations, as it is popular in other areas of physics (\eg, \citep{brunton2016discovering,sindy2,sindy1}, see examples in \secref{sec:casestudies}). This is at odds with the chief goal for deductive approaches, outlined in Section \ref{subsubsec:parsimonious_models_neural}. Key reasons for this shortcoming might be found in the high-noise levels arising in intrinsically stochastically-dominated neural processes, in which most hidden variables are not experimentally accessible (especially in \emph{in-vivo}) \citep{genk21,balaguer2011attracting}. This challenging scenario hinders the direct application of approaches common in other fields and poses an intriguing question for future research endeavours.

\section{Learning causally interacting brain regions}%
\label{subsec:causal_brain_regions}

\subsection{Causality in the connected brain \label{subsubsec:connected_brain}}

The debate on the causal role of brain connectivity has a long-standing tradition (see \eg \citep{razi2016connected}). The classic view of functional segregation (mapping functions to physical brain regions) veered to connectionism, that is, brain functions result from interactions between neurocomputational units \citep{razi2016connected}. Consistently, the focus gradually shifted from functional segregation (the study of regionally-specific brain activation) to functional integration (the study of the connectivity between cortical areas \citep{razi2016connected}). 

Historically, connectivity studies establish the distinction between structural (the anatomical location of white matter, axonal tracts), functional and {\em effective} connectivity (and sometimes with normative connectivity, in contrast to individual-specific connectomes) \citep{siddiqi2022causal}. This classification is relevant to determine the type of causality questions that can -or {\em cannot}- be addressed \citep{reid19,ross15}. Typically, functional connectivity methods estimate statistical dependencies such as spatiotemporal correlations or coherence measures between brain ensembles. In contrast, a subset of these methods, commonly termed effective connectivity approaches, refer to the quantification of directed interactions between brain circuits (\eg, \citep{siddiqi2022causal,reid19,tognoli2014metastable,bala18}). In this arena, the quest for demonstrating causality relationships in neuroscience has attracted much attention over the recent decades \citep{siddiqi2022causal,ross15}, and its plausibility has been widely debated \citep{weichwald2021causality,barack2022call,barnett2018misunderstandings,ross15,Haufe2014}. For instance, in neuroimaging, multiple issues such as confounding factors \citep{Woolgar2014,Todd2013} and varying temporal delays (intrinsic to, for instance, fMRI) challenge estimates of network information flow directionality (\eg, \citep{siddiqi2022causal,weichwald2021causality,Haufe2014,lohm12,Davis2014} among many others). 

Methodologically, a key characteristic of causal approaches -in difference with conventional probabilistic modelling- is the need for predicting how the system reacts under interventions \citep{weichwald2021causality}; in other words, for defining counterfactual models (see \secref{sec:counterfactuals}). Problematically, a large amount of interventional data is necessary to falsify the wide range of causal hypotheses in a high-dimensional system like the cerebral cortex \citep{weichwald2021causality}. For instance, \emph{targeted} brain interventions via intracranial electrical stimulation (iES) in conscious patients is typically a robust approach for testing causality \citep{siddiqi2022causal}, but large-scale datasets using this experimental protocol are scarce, given ethical and experimental limitations of invasive techniques \citep{weichwald2021causality}. However, comprehensive, high-quality interventional data would be fundamental to falsify as many competing causal scenarios as possible. This is especially important in cognitive neuroscience given the lack of experimental access to some fundamental variables, which increases the number of plausible causal models underlying observable behaviour \citep{weichwald2021causality}.

This shortage of comprehensive targeted lesion/stimulation datasets, and the improvement of whole-brain registration techniques, led to the development of analytical methods (or adaptations of existing ones) to better understand causality in cortical circuits. Most notably, Granger Causality and related approaches (GC, originated in the field of Economics \citep{granger1969investigating}), Structural Causal Modelling (SCM \citep{pearl2009causality,bongers2018causal}), and Dynamic Causal Modelling (DCM \citep{friston2003dynamic}) have been extensively used, as will be discussed next.

\subsection{Causal methods in neuroscience \label{subsubsec:causal_methods_neurosci}}

GC and an extension of this concept, Transfer Entropy (TE), are perhaps the most common {\em model-free} methods for assessing causal relations in neuroscience. These two generalist approaches estimate the direction of causality between interacting neural populations by analysing the time series derived from brain responses \citep{ding2006granger,friston2013analysing,barnett2018misunderstandings}. They are regular statistical tools for studying orchestrated interactions between brain regions via magneto/electroencephalography (M/EEG) and fMRI recordings (\eg, \citep{friston2013analysing,stokes2017study,bassett2017network}). At microscopic levels, they have also been applied to detect synaptic connections between neurons \citep{shei18}. Specifically, GC is based on the assumption that time series prediction leveraging its past values significantly improves by inputting historical values from another, causally connected time series (see details in \secref{sec:list-of-causal-discovery-methods}). Thus, the presence of causal relationships is detected by testing the hypothesis that one time series autocorrelations have predictive power for the other time series \citep{ding2006granger}. 
 
 TE expands this idea to accommodate broader types of nonlinear temporal interactions by computing the amount of information that one time series \emph{transfers} to another. Similarly to GC, it conjectures that the current value of one time series can be better estimated by conditioning the predictive probability to past values of both itself and another time series, inferring causality direction \citep{barnett2009granger}. Alternatively, SCM and its recent variants are Bayesian approaches for assessing plausible causal graphs in brain networks. They have been applied, for instance, to foster interpretability in behavioural decoding approaches \citep{weic15}. However, their use in cognitive neuroscience is still challenging, given the key difficulties discussed in Section \ref{subsubsec:connected_brain}, and the indirect nature of most neuroimaging measurements (reviewed in \citep{weichwald2021causality}).

Accompanying these model-free approaches, perhaps the most standard model-based technique for connectivity inference between brain regions is Dynamic Causal Modelling (DCM). DCM is a Bayesian method incorporating different degrees of {\em a priori} biological plausibility for understanding mechanisms underlying neuroimaging data (see \eg,\citep{penny2004comparing,stephan2010ten,marreiros2010dynamic,Cooray2015}). It has been employed to study neural pathways of effective connectivity in \eg, motor control, attention, learning, decision-making, emotion, and other higher cognitive functions \citep{fris19}; and even to model EEG seizure activity dynamics in epilepsy \citep{Cooray2015}. 

In general, whole-brain modelling methods like DCM or other more recent models \citep{cabra22} can provide a more nuanced understanding of the underlying mechanisms of brain function than model-free approaches \citep{fris05}. However, the need for large datasets w.r.t. the complexity of the range of alternative models hampers the interpretation of the estimated connectivity \citep{siddiqi2022causal,penny2004comparing,stephan2010ten,marreiros2010dynamic}. Indeed, classic DCMs \citep{penny2010comparing} have been criticised for the difficulty in falsifying their model selection approach \citep{lohm12} and perhaps for this reason, they were not extensively tested in clinical settings \citep{penny2004comparing}. Specifically \citep{lohm12} suggested the ambiguity of DCM inference in generating a unique optimal connectivity map due to, \eg, known  challenges in model fitting and selection in such a large space of possible architectures \citep{lohm12,chic14}. 

These caveats of DCM as a robust approach for causality assessment led to the development of variants such as spectral DCMs, the canonical microcircuit DCM -introducing higher degrees of laminar-specific, biophysical detail towards more informative priors for E/MEG modelling-, or the stochastic dynamic causal model, sDCM (see a review in \citep{fris19}). sDCM incorporates random processes to the basic DCM equations, enhancing its fitting capability to hemodynamic responses and hence alleviating excessive dominance of priors in Bayes model selection \citep{bernal2013multi}. In a classic DCM for fMRI data, the neural state $\boldsymbol{x}(t) \in \{1,n\}$ (for $n$ interacting brain regions) corresponding to a single task-based input $u(t)$, is determined using the simple first-order differential equation
$\frac{ d \boldsymbol{x}(t)}{d t} = (\boldsymbol{A}+u(t) \cdot \boldsymbol{B} ) \: \boldsymbol{x}(t) +u(t) \cdot \boldsymbol{c} $; where the matrix $\boldsymbol{A}$ encodes (endogenous) connections between brain regions, $\boldsymbol{B}$ the strength in which inputs modulate each connection (\emph{modulatory} inputs) and $\boldsymbol{c}$ the gain of the \emph{driving} inputs to each region. sDCM expands this approach by adding intrinsic $\boldsymbol{\beta}(t)$ and extrinsic $\gamma(t)$ stochastic fluctuations to account for the incomplete observability of both states and inputs to brain areas relevant to the cognitive task:
\begin{equation}  \label{eq:sDCM}
\begin{split} 
  \frac{ d \boldsymbol{x}(t)}{d t} = (\boldsymbol{A}+\nu(t) \cdot \boldsymbol{B} ) \: \boldsymbol{x}(t) + \nu(t) \cdot \boldsymbol{c} + \boldsymbol{\beta}(t),\\
   \nu(t)=u(t) + \gamma(t),\\
   \boldsymbol{y}(t)=g(\theta) * \boldsymbol{x}(t) + \boldsymbol{\epsilon}(t),
\end{split}
\end{equation}  
where $\nu(t)$ is a \emph{hidden} input cause masked by fluctuations (univariate here for simplicity), and the last equation represents a hemodynamic model (present in all DCMs variants) of non-neural parameters $\{\theta, \boldsymbol{\epsilon}(t)\}$. Finally, their convolution with the neural state $\boldsymbol{x}(t)$, yields the observed fMRI blood-oxygenation level-dependent (BOLD) response $\boldsymbol{y}(t)$ in relevant brain areas (termed regions of interest, ROI) \citep{fris19}.  

\begin{figure}[t!]
\centering 
\includegraphics[width=0.9\textwidth]{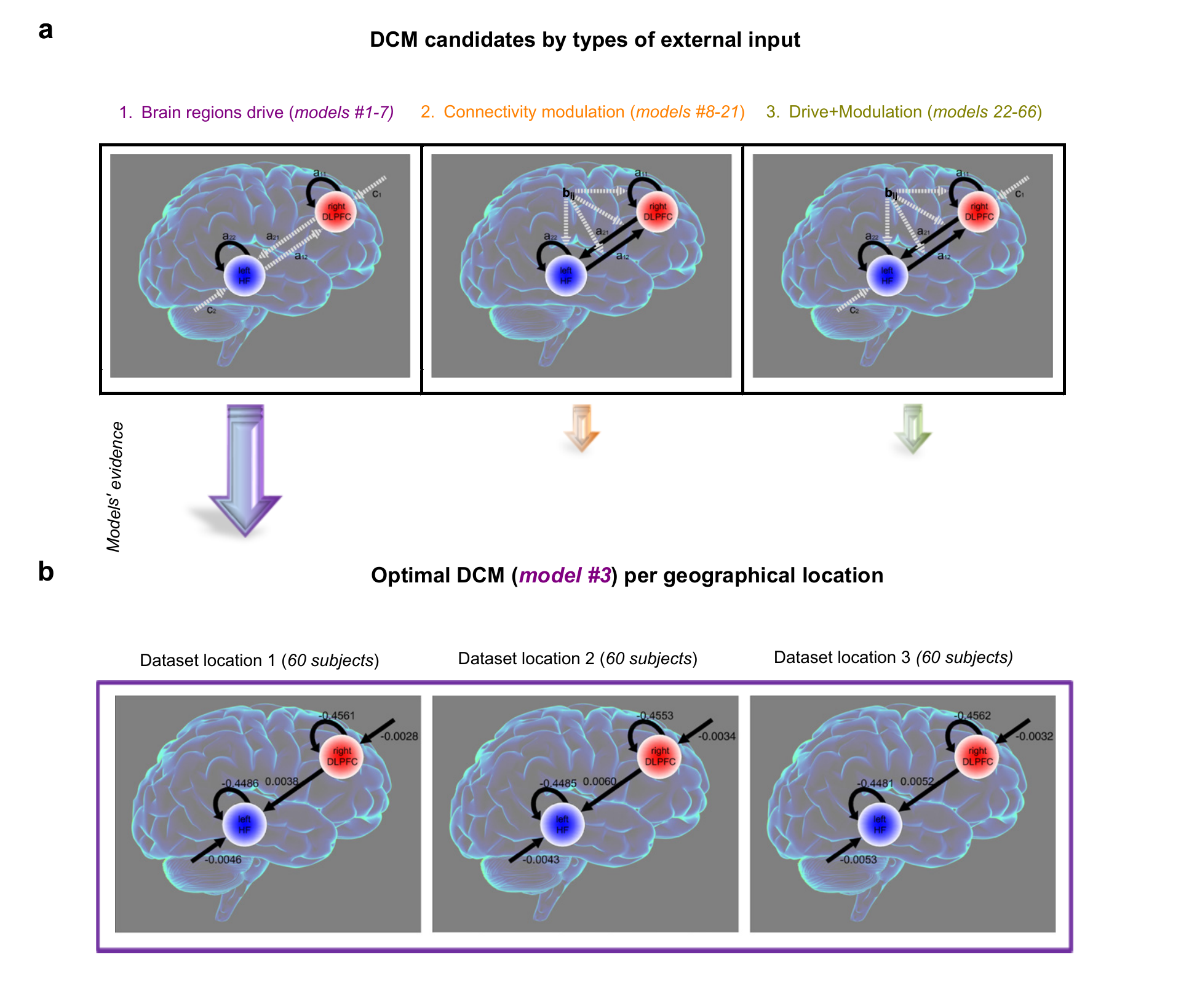}
   \caption  {Example of consistency assessment for stochastic DCM, taken from \citep{bernal2013multi}. \textbf{a}. Connectivity hypotheses associated with the performance of a 2-Back working memory task by healthy human participants \citep{bernal2013multi}. Left: input fluctuations to the two ROIs (right DLPFC and left HF; 7 distinct model combinations). Centre: input variance modulates the connections themselves (14 models). Right: combinations of both mechanisms (44 models). Connectivity hypotheses are tested on independent datasets collected from three locations (Bonn, Berlin and Mannheim, 60 subjects each). Model evidence (log-likelihood marginalised over model's free parameters \citep{penny2004comparing}) is much stronger for a model of the first family (\#3, random effects Bayes factor 98\%+ in favour of this model). \textbf{b}. Remarkably, connectivity parameters do not differ across sites (Friedman non-parametric test, $p>0.2$), and interaction between sites and model parameters were not found ($p>0.8$), supporting the model's robustness. Figure adapted from \citep{bernal2013multi} with the publisher's permission (Elsevier).}
     \label{fig:dcm}
\end{figure}

Figure \ref{fig:dcm} summarises an illustrative reliability study from \citep{bernal2013multi}, specially designed for assessing sDCM robustness. In this example, a large sample of participants ($n$=180) was recruited from three different geographical locations. fMRI recordings were obtained from healthy subjects from the same age range while performing a classic 2-Back working memory task (to recall numbers shown two trials before). 
 This N-Back task activates the dorsolateral prefrontal cortex-hippocampal formation (DLFC-HF) network connectivity, which is abnormal in schizophrenia patients \citep{meye05}. 
 \citet{bernal2013multi} used sDCM to identify the DLPFC-HF effective connectivity and compared the consistency of the models in this multi-centre setting (Figure \ref{fig:dcm}). Three {\em a priori} likely mechanisms to explain the BOLD responses to this task were implemented in three different families of models: with only driving inputs to the two regions ($B \equiv 0$, Figure \ref{fig:dcm}a, left), only connectivity modulation ($C \equiv 0$, Figure \ref{fig:dcm}a, centre) and both mechanisms combined (Figure \ref{fig:dcm}a, right). Noticeably, the random effects Bayes Model selection process strongly favours a specific connectivity model belonging to the driving inputs family (Figure \ref{fig:dcm}b) over all the rest, consistently for the three independent locations. Specifically, the DLPFC-HF connectivity parameters were statistically indistinguishable across datasets (Figure \ref{fig:dcm}b), supporting the reliability of sDCM results.          
 
In line with these whole-brain analyses, other studies showcased the consistency of  causality methods at microscopic levels. As a representative example, \citep{chen23} recently proposed the effectiveness of GC in inferring information directionality in zebrafish motor circuits from single-cell calcium imaging signals. Causally strong, interventional data was inaccessible in this setting. Despite this, results were in full agreement with the known physiology of this species. In addition, and besides these standard methods (GC, TE and DCM), recent approaches have also addressed the causality robustness question from different angles, for instance, by focusing on changes in information {\em reversibility} as a sign of aberrant resting-state brain dynamics -which could subserve as a biomarker of Alzheimer's disease \citep{cruz23}.

These and other even more indirect causality measures (like standard statistical approaches \citep{reid19}) have provided useful insights when operating on neurophysiological recordings with high temporal precision. The ideal recording modalities are thus those capable of directly recording local (electrical) field potentials (LFP), such as intracranial electroencephalography (iEEG) or neuronal-level techniques (like in \citet{chen23}). However, when indirect causality measures are estimated from other modalities -especially from functional imaging- the multiple confounders discussed earlier rank them in a weak position in a causality scale when compared with approaches based on interventions \citep{siddiqi2022causal}. Therefore, their capability for providing a reliable indication of causality interactions is highly disputed \citep{Mehler2018,siddiqi2022causal,lohm12}.

Nevertheless, limitations for assessing causal relationships in neuroimaging do not preclude indirect analytical approaches to constrain the universe of plausible causal graphs for a specific scenario \citep{reid19,siddiqi2022causal}. Thus, there is a reasonable consensus in considering them as valuable contributors to strengthen causality claims, provided they are combined with more direct measures of causality based on interventional data \citep{reid19,siddiqi2022causal}. Indeed, the ideal scenario from a causality perspective occurs when its inference is consistent throughout different approaches; in other words, when different methods having complementary views provide synergistic evidence \citep{siddiqi2022causal}. For instance, converging evidence between a causal fMRI/EEG model, a targeted lesion, and the stimulation of a specific cortical circuit would score high on a causality scale than either of these methods alone; since the counterfactual could be established by focal stimulation \citep{neum23} combined with real-time neuroimaging recordings \citep{siddiqi2022causal}, enriching the conclusions of the lesion study. 

This optimal coalescence of multiple causal approaches for effective connectivity inference was termed Convergent Causal Mapping and is the recommended approach for designing new experiments \citep{siddiqi2022causal}. Thus, from this concerted perspective, studies considering a single approach in isolation -especially if it is not based on interventions- should not make strong translational claims, like suggesting direct therapeutic applications \citep{siddiqi2022causal,neum23}. In addition, future works should consider testing robustness to different environments \citep{weichwald2021causality} (like in the example shown in Figure \ref{fig:dcm} \citep{bernal2013multi}) for further reinforcing the credibility of the inferred causal flow.

\section{Learning causal graphs of carbon and water fluxes}

\subsection{Introduction}

The Earth is a highly complex, dynamic, and networked system where very different physical, chemical and biological processes interact in and across several spheres. Land and atmosphere are tightly coupled systems interacting at different spatial and temporal scales \citep{Diaz21rccm}. The main challenge to quantifying such relations globally comes from the lack of sufficient in-situ measurements and the fact that some of these variables are latent and not directly observable with remote sensing systems. One can, for example, measure SM but not GPP directly. As an alternative, many studies have relied on model simulations to investigate SM-precipitation \citep{koster2006glace}, GPP-SM \citep{green2019large} and ET-SM relations \citep{milly1992potential,jung2010recent}, to name just a few. However, assuming a model implies assuming the knowledge of the causal mechanisms and relations governing the system. This is not necessarily a correct assumption, especially in model misspecification, non-linearities and non-stationarities. Discovering such relations from data is of paramount relevance in these cases. In the following, we review the performance of two standard methods of causal discovery from time series data to learn the relationships between environmental factors and carbon and heat and energy fluxes at the local (site, flux tower) level and the global (planetary, product-derived) level. At the local level, we exploit data acquire by eddy-covariance instruments estimating fluxes exchange. At the global level, we exploit Earth observation data from satellite observations.

\subsection{Clustering of biosphere-atmosphere causal graphs at the site level}

\begin{figure}[t!]
\centering 
\includegraphics[width=1\textwidth]{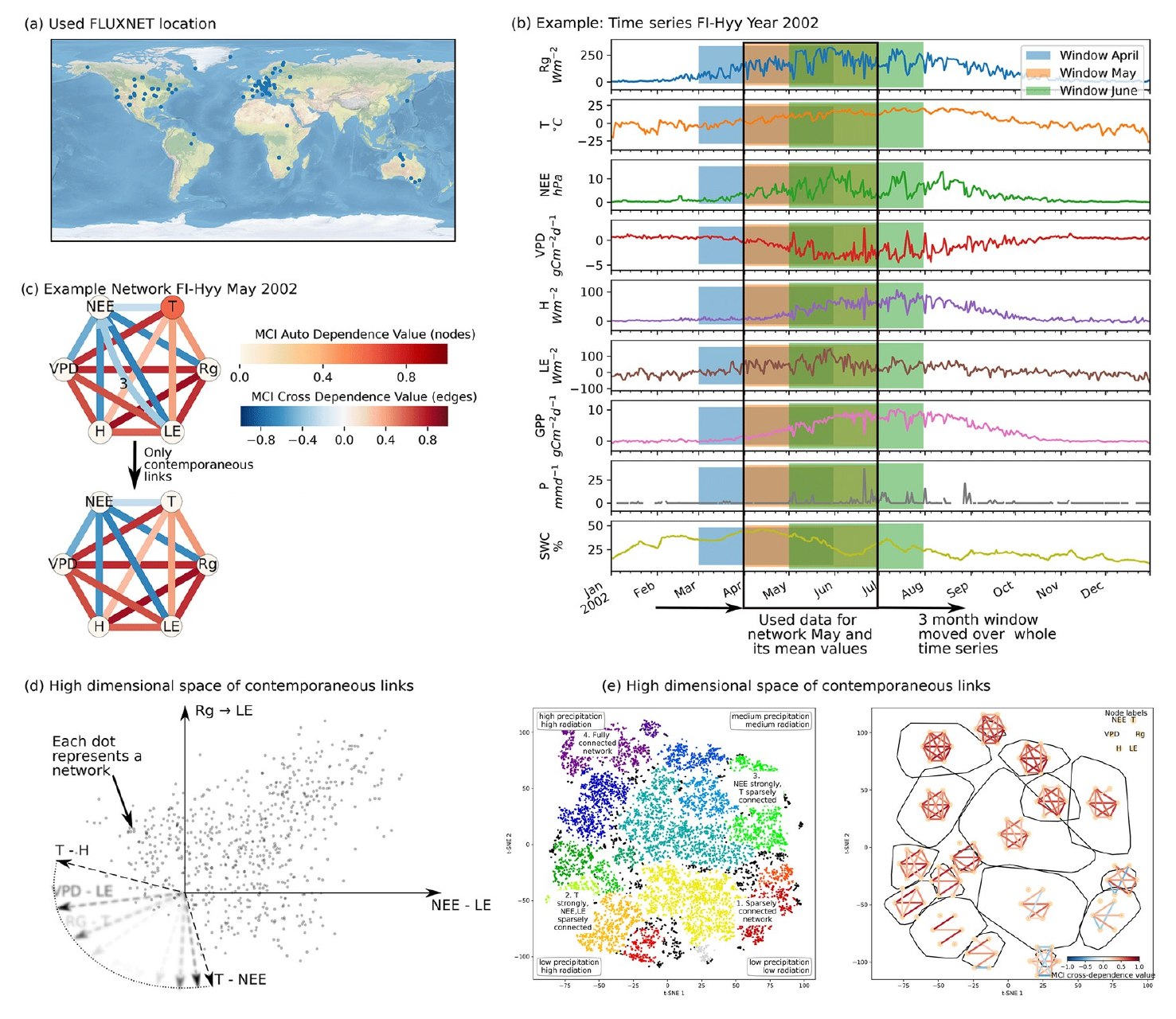}
  \caption{Clustering causal graphs of local measurement stations of biosphere-atmosphere interactions (adapted from \citet{krich2021functional}). See the main text for explanations.
  }
  \label{fig:krich2021}
\end{figure}

The atmosphere and terrestrial ecosystems constitute another closely interconnected complex system where processes interact across a range of temporal and spatial scales. Further, causal relations also depend on vegetation types, climatic regions, and the season. Fortunately, measurement campaigns of the past decades have resulted in good coverage of measurement sites, available in the FLUXNET database~\citep{BALDOCCHI2014}, a collection of long-term global observations of biosphere-atmosphere fluxes measured via the eddy covariance method. \citet{runge2023causal} discuss a similar case study in-depth.

Here we review the study of \citet{krich2021functional} that analysed causal networks for different seasons at eddy covariance flux tower observations in the FLUXNET network and how they depend on  meteorological conditions. 
Figure~\ref{fig:krich2021} explains the methodological setup. From a selection of 119 FLUXNET sites (Fig.~\ref{fig:krich2021}(a)) daily time series data of the following variables were considered (see Fig.~\ref{fig:krich2021}(b) for one site):  short-wave downward radiation (or global radiation, Rg), air temperature (T), net ecosystem exchange (NEE) (inverted), vapour pressure deficit (VPD), sensible heat (H), latent heat flux (LE), gross primary productivity (GPP), precipitation (P), and soil water content (SWC). For details on data processing, we refer to \citet{krich2021functional}.

Causal networks were then estimated with PCMCI~\citep{runge2019detecting} (time lags from 0 to 5 days) in sliding windows of 3 months to capture the temporal evolution of biosphere-atmosphere interactions. Based on findings in \citet{krich2020estimating}, a smoothed seasonal mean was subtracted to remove the common driver influence of the seasonal cycle. This results in 10.038 networks for the different months and sites (an example network is shown in Fig.~\ref{fig:krich2021}(c)). Node colours indicate the level of autocorrelation (auto-MCI-partial correlation~\citep{runge2019detecting}), and link colours the cross-link strength (cross-MCI); time lags are indicated by small labels, and straight edges are contemporaneous. Since the strongest and most consistent links are contemporaneous, further analysis focused on these $15$ links.

A previous study~\citep{krich2020estimating} discussed individual networks in more detail; the scope of \citet{krich2021functional} was to apply a dimension reduction, here t-distributed stochastic neighbour embedding (t-SNE \citep{van2008visualizing}) which considers each of the causal graphs as an observation in a high-dimensional space of the contemporaneous MCI partial correlation values (Fig.~\ref{fig:krich2021}(d)). t-SNE allows projecting this high-dimensional space onto two dimensions (Fig.~\ref{fig:krich2021}(e, left)) that are the dominant features of transitions between different states of biosphere-atmosphere interactions. The coloured clusters in Fig.~\ref{fig:krich2021}(e, left) are based on the OPTICS approach~\citep{ankerst1999optics}, and the four corners indicate the four archetypes of network connectivity and the networks' underlying meteorological conditions (averages taken over the sliding windows in Fig.~\ref{fig:krich2021}(b)). Finally, Fig.~\ref{fig:krich2021}(e, right) shows the convex hulls of clusters and their average network.

Each point of the low-dimensional embedding represents a specific ecosystem's biosphere-atmosphere interactions at a specific time and allows us to investigate their behaviour. A main finding of \citet{krich2021functional} was that ecosystems from different climate zones or vegetation types have similar biosphere-atmosphere interactions if their meteorological conditions are similar. For example, temperate and high-latitude ecosystems feature similar networks to tropical forests during peak productivity.  During droughts, both ecosystems behave more like typical Mediterranean ecosystems during their dry season. 
Such meta-analyses of causal networks allow for another perspective on understanding ecosystems, including an analysis of anomalous changes in network structure as indicators of ecosystem shifts (see \secref{sec:anomalies}).

\subsection{Causal relations at global scale}

As an alternative to Granger causality, the work \citep{sugihara2012detecting} presented the convergent cross-mapping (CCM) method, which may deal with the issues of non-stationary and nonlinear processes and deterministic relations in dynamic systems with weak to moderate cause-effect variable coupling. CCM assesses the reconstruction of a variable's state space using time embeddings to determine if $X \to Y$. This method has been extended to account for causal relations operating at different time lags and applied to various research areas. However, it is sensitive to noise levels, hyperparameter selection, and false detections in strong, unidirectional variable coupling cases. To address these issues, the robust CCM (RCCM) \citep{Diaz21rccm} alternatively relies on bootstrap resampling through time and the derivation of more stringent cross-map skill scores. The method also exploited the information-geometric causal inference (IGCI) method in \citep{monster2017causal} to infer weak and strong causal relationships between variables and estimate the embedding dimension to derive global maps of causal relations. 

Let us exemplify the RCCM method to discover interactions of three key variables in the carbon cycle: moisture, photosynthesis and air temperature (Tair). For that, we use data compiled in the \href{http://www.earthsystemdatalab.net}{Earth System Data Lab (ESDL)}, which contains harmonised products with a spatial resolution of $0.25^{\circ}$ and a temporal resolution of $8$ days, spanning over $11$ years from $2001$ to $2011$. The RCCM method is applied in each grid cell, which allows us to infer spatial patterns of causal relations between several key variables of the carbon and water cycles.

\begin{figure}[h!]
    \centering
    \includegraphics[width=12cm]{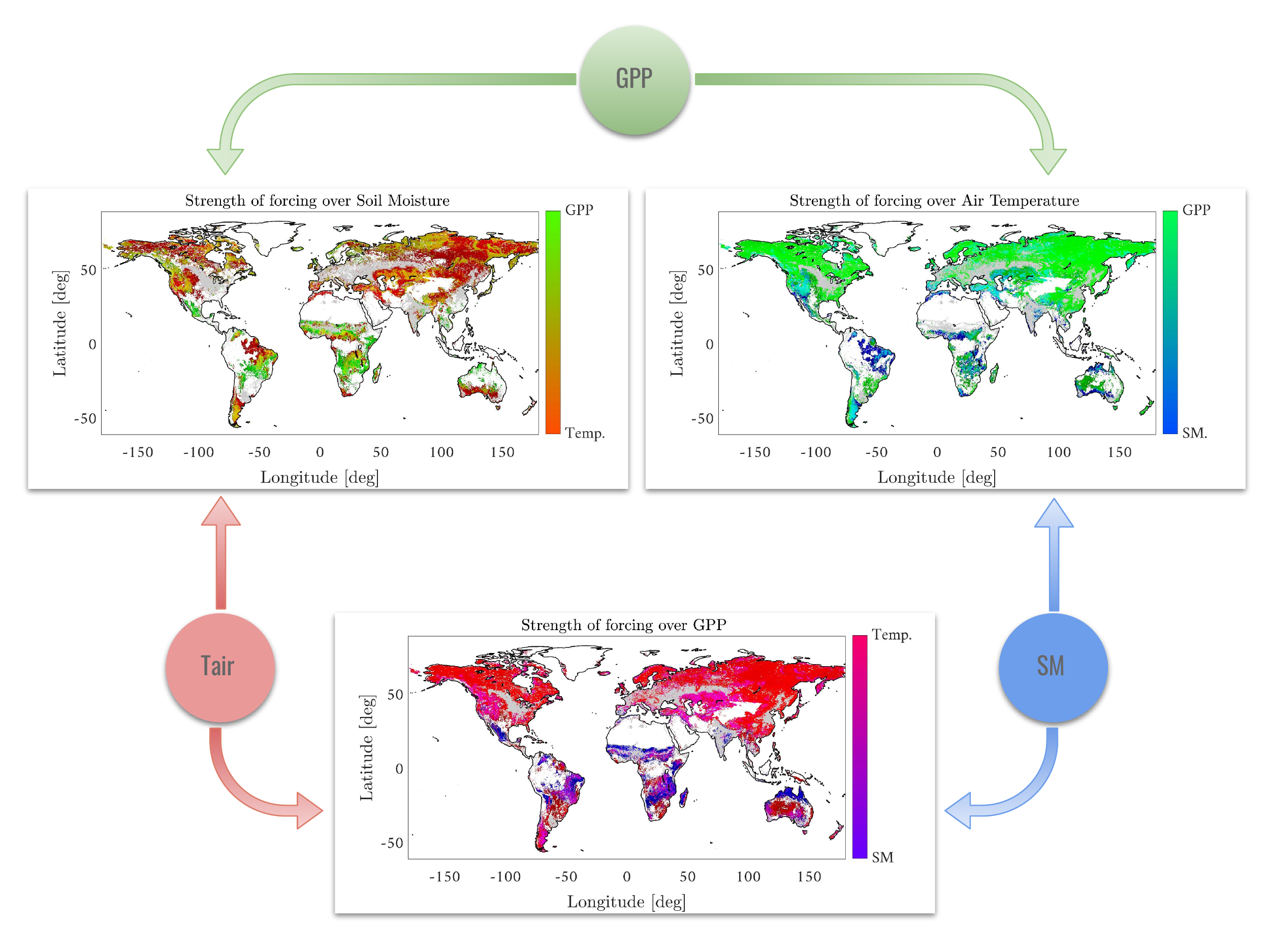}
    \caption{Applying RCCM in \citep{Diaz21rccm} to discover causal relations between GPP, Tair and SM. GPP drives Tair in cold ecosystems; Tair controls SM in water-limited areas; GPP dominates SM. Croplands were masked to avoid interference from human activity. }
    \label{fig:PhotosynthesisTempSM}
\end{figure}

Fig.~\ref{fig:PhotosynthesisTempSM} shows GPP drives Tair mostly in cold ecosystems due to changes in land surface albedo. Results show GPP is an important forcing of local temperature in many areas. Recent studies have found temperature is an important factor of GPP, driven by radiative factors in cold climates and turbulent energy fluxes in warmer, drier ecosystems. SM and Tair are closely linked, limiting evaporation and raising Tair under dry conditions. This could explain the significant impact of Tair in high latitudes. GPP is mainly influenced by Tair in water-limited regions, especially in high northern latitudes where cold temperatures limit photosynthesis and plant growth. GPP and ET are tightly related as carbon assimilation in plants is linked with water losses through transpiration \citep{field1995stomatal}. Low water availability reduces GPP and ET, causing increased air and surface temperatures and a drier atmosphere. SM being stronger than GPP is mostly seen in transitional wet/dry climates \citep{koster2004regions}. No strong forcings in tropical rainforest areas indicate GPP is mostly driven by solar radiation and affected by high VPD values \citep{madani2020recent}.

\section{Causal climate model intercomparison}

\begin{figure}[t!]
\centering 
\includegraphics[width=0.85\textwidth]{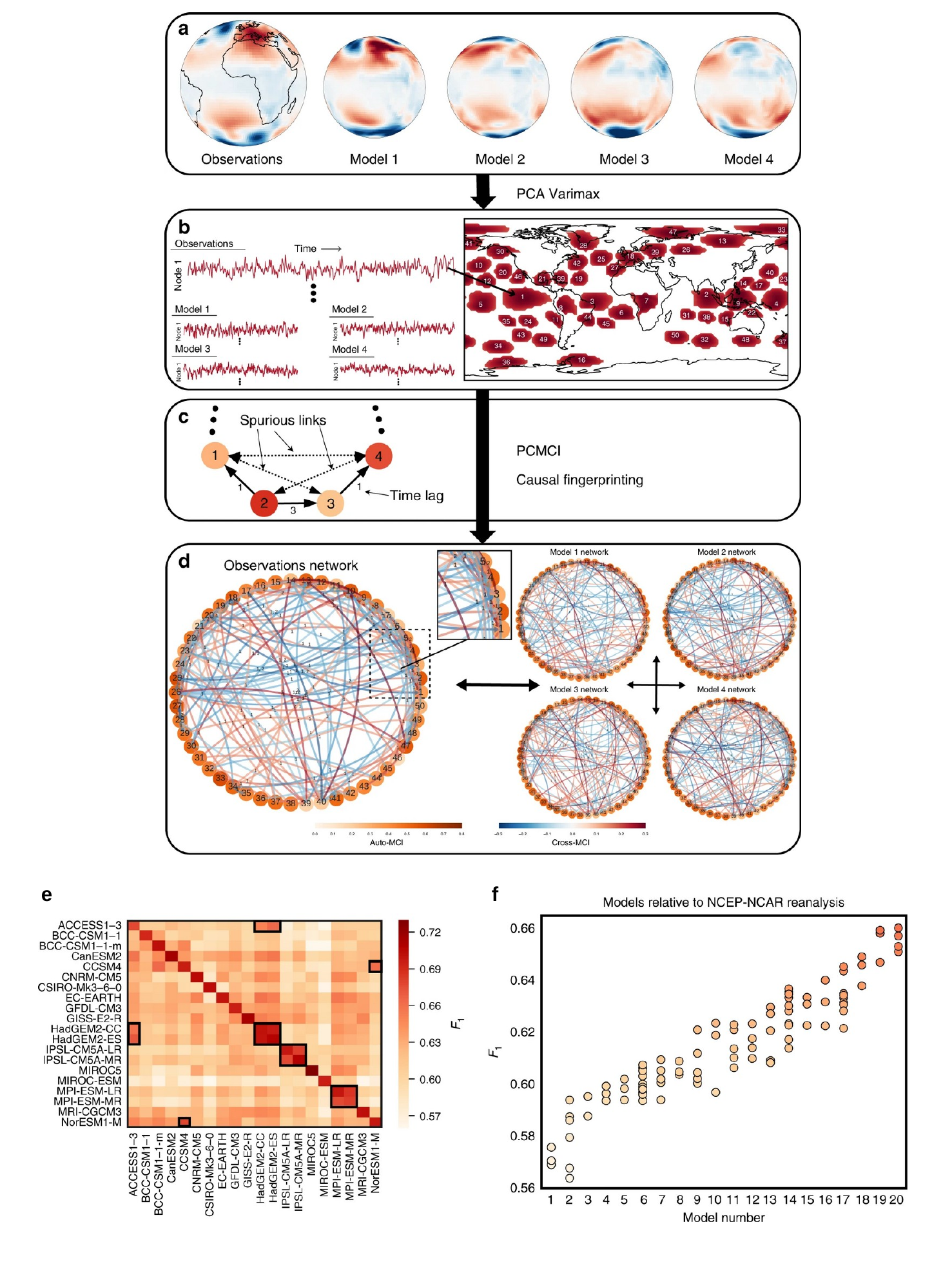}
  \caption{Causal climate model evaluation (adapted from \citet{nowack2020causal}). See the main text for explanations.
  }
  \label{fig:nowack}
\end{figure}

As introduced in \secref{sec:opportunities-eval}, causal inference can help to assess the output of physical models and evaluate and compare them against observations at the level of causal dependencies ~\citep{eyring2020earth,eyring2019taking,nowack2020causal,PerezSuay19shsic}.
Climate models~\citep{eyring2016overview} provide short-term predictions and future climate projections under given anthropogenic emission scenarios and are the basis for climate-related decision-making. As models can only provide an approximation to the real system, it is essential to evaluate them against observations. Such climate model evaluation is largely based on means, climatologies, or spectral properties \citep{stocker2013climate,eyring2020earth}. Here the problem of equifinality may occur: even though a particular model might well fit descriptive statistics of the data, the model might not well simulate the causal physical mechanisms that produce this statistic, given that multiple model formulations and parameterisations, even when wrong, can fit the observations equally well. The issue is that such models would lead to erroneous future projections--a causal problem of out-of-distribution prediction. Causal model evaluation~\citep{runge2019inferring} can evaluate the ability of models to simulate the causal interdependencies of its subprocesses in a process-based model evaluation framework  \citep{maraun2017towards}. 

Here we briefly summarise one approach in this direction~\citep{nowack2020causal}. The author aimed to compare causal networks among regional subprocesses in sea-level pressure between observations and climate models of the CMIP ensemble~\citep{eyring2016overview}. Figure~\ref{fig:nowack}\textbf{a-d} illustrates the method's steps. First, the regional subprocesses were constructed from gridded climate time series (daily-mean sea level pressure from the NCEP-NCAR reanalysis~\citep{kalnay1996ncep}) using Varimax principal component analysis (PCA) to obtain a set of regionally confined climate modes of variability (Fig.~\ref{fig:nowack}\textbf{b}).
The Varimax-PCA weights were then applied to the pressure data from each climate model (the regional weights' cores are indicated in red). 
Each component is associated with a time series (3-day averaged) and is one of the causal network nodes. Then the causal discovery method PCMCI~\citep{runge2019detecting} was applied to these time series to reconstruct the lagged time series graph among these nodes, which constitute characteristic causal fingerprints (Fig.~\ref{fig:nowack}\textbf{c,d}) for the observational data as well as the individual models. Node colours indicate the level of autocorrelation (auto-MCI-partial correlation~\citep{runge2019detecting}), and link colours the cross-link strength (cross-MCI); time lags are indicated by small labels. Only the around 200 most significant links are shown.

These causal fingerprints can then be used for model evaluation and intercomparison. Figure~\ref{fig:nowack}\textbf{e} depicts a comparison of models among each other, that is, the matrix of average F1-scores for pair-wise network comparisons between ensemble members of 20 climate models (labelled following CMIP-nomenclature in capital letters) for simulations spanning approximately the historical period from 1948 to 2017 and two surrogate models (Random, Independent). The rows show the models taken as references in each case, and the columns indicate the models compared to these references. Higher scores imply a better agreement between networks, i.e., that the two models are more similar regarding their causal fingerprint.
One can see that causal fingerprints from different ensemble members of the same model (diagonal in Fig.~\ref{fig:nowack}\textbf{e}) are more consistent than networks estimated from two different models (off-diagonal). The blocks are consistent with different models sharing a common development background.
In Figure~\ref{fig:nowack}\textbf{f}, the models' causal fingerprints are each compared to the fingerprint of the observational data (ordered F1-scores). The result is a continuum of more- and less-similar models (but models have significantly different causal fingerprints). The networks can be further investigated to analyse which regional interactions the models differ more from observations. 

Causal model evaluation can provide important information to model developers on where their models can be improved. Furthermore, \citet{nowack2020causal} show that more realistic fingerprints also affect projected changes in land surface precipitation. Hence, causal model analyses could be used to constrain climate change projections. The assumption is that the underlying physical processes (\eg, large-scale circulation) lead to dynamical coupling mechanisms captured in the causal fingerprints. One may now argue that high modelling skill on historical data is also relevant for modelling future changes if the physical processes remain important under future climate change.

\section{Learning density functionals}\label{sec:case:DFT-EQL} 

Being able to describe many-body systems is exciting and important for many applications. Density functional theory \citet{Evans79:DFT} (DFT) is an approach to creating a description for classical and quantum many-body systems in equilibrium.
The aim is to find a unique (free) energy functional that gives rise to the particle density profile. 
The analytical form of the (free) energy functional is generally unknown, except for a handful of particular model systems. 
One way to treat more complex systems is to perform computer simulations and learn the energy functional via machine learning. 
The first attempts in classical DFT used a convolutional network~\citep{Lin2019}, which does not allow much theoretical insight.

In \citet{lin2020:DFTEQL}, the above-mentioned symbolic regression method, EQL \citep{SahooLampertMartius2018:EQLDiv}, was adapted to represent part of the energy function.
This is an interesting application, as the problem contains known parts of the computational pipeline that we do not want to replace and other parts that should be replaced via the data-driven approach. The fact that EQL can be embedded into any differentiable computational structure is crucial here.

\begin{figure}\centering
\begin{tabular}{@{}lll@{}}
\includegraphics[height=0.22\linewidth]{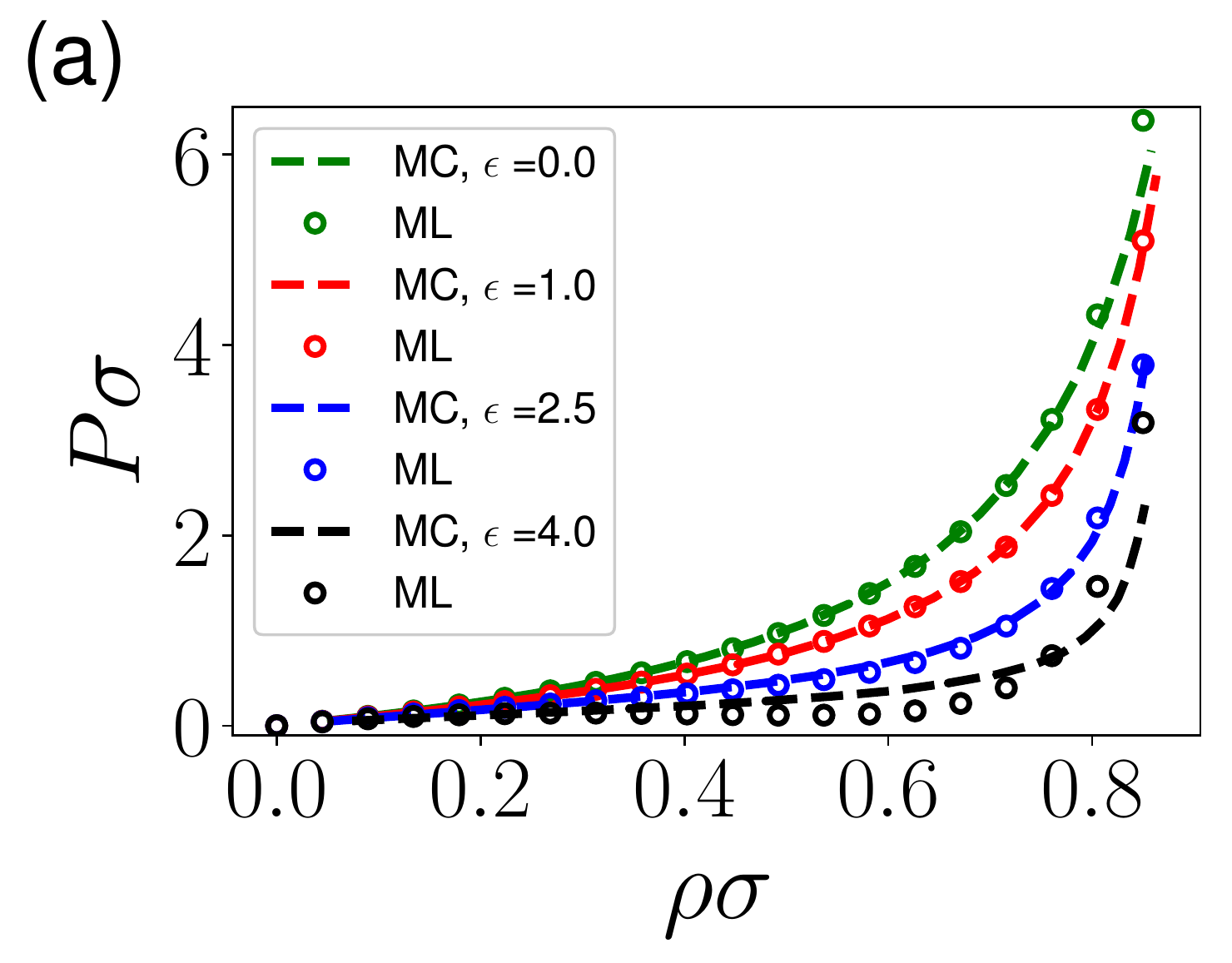} &
\includegraphics[height=0.22\linewidth]{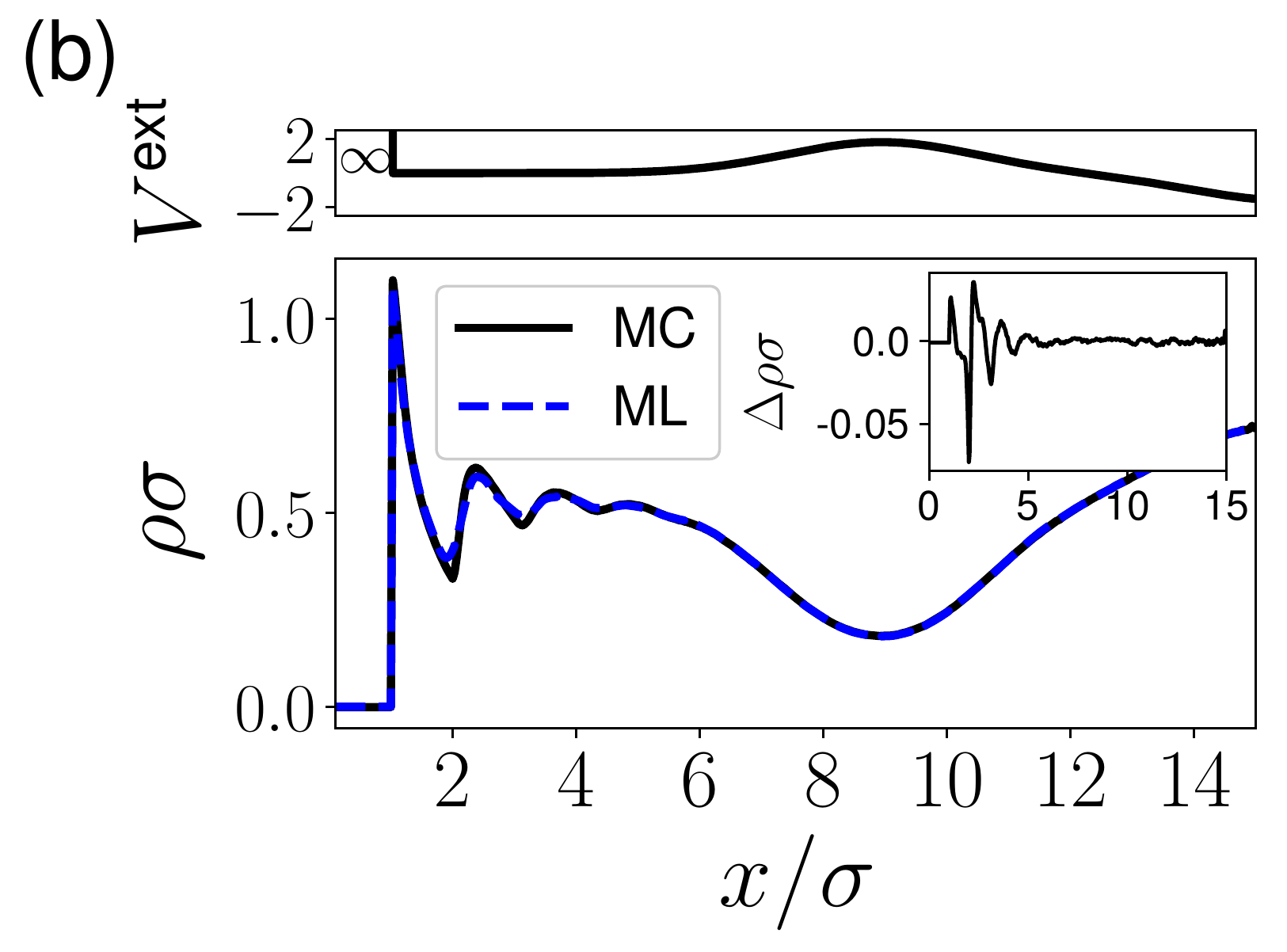}&
\includegraphics[height=0.22\linewidth]{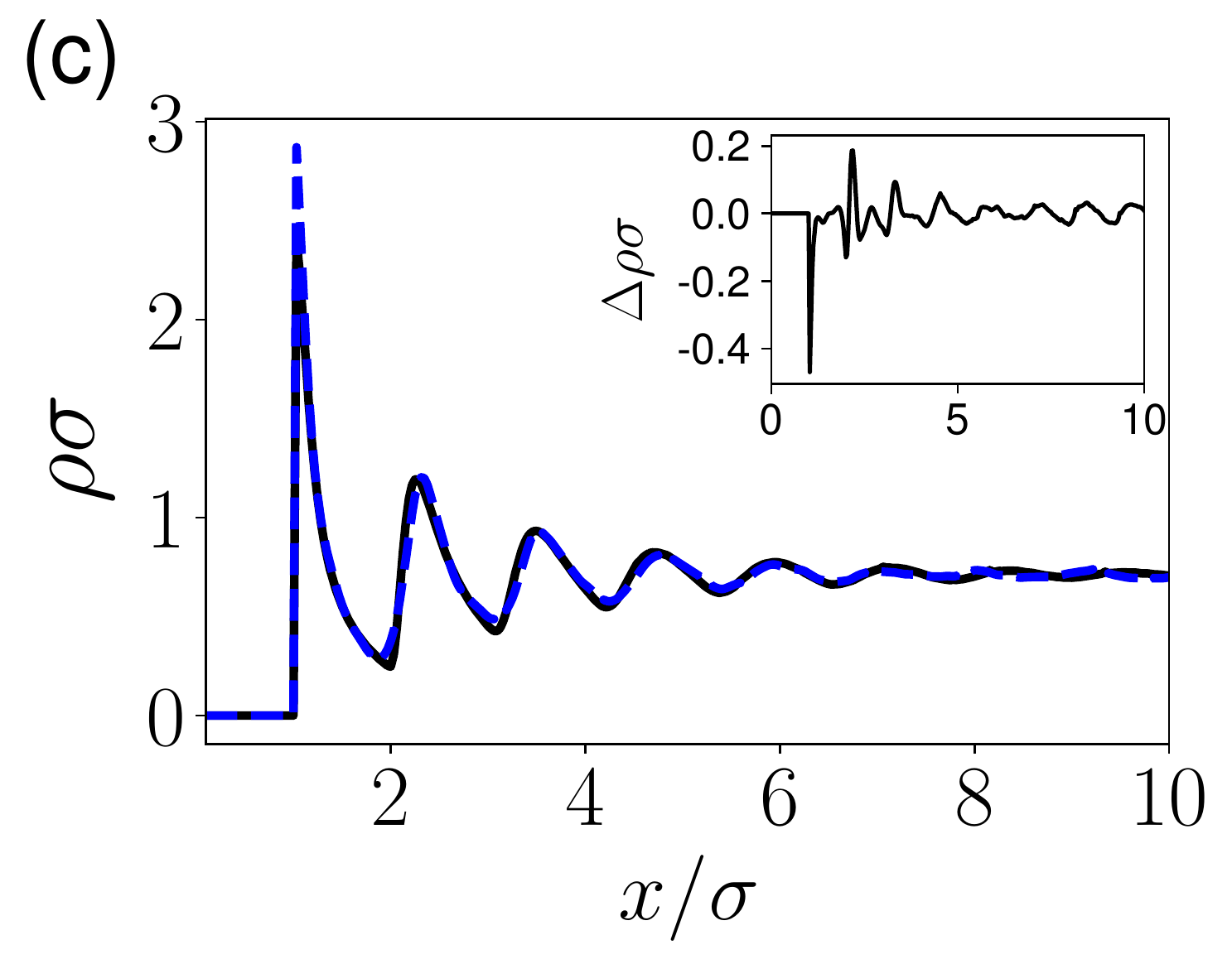}
\end{tabular}\vspace{-1em}
\caption{Results for learning the density functional for Lennard-Jones fluids. 
(a) shows the equation of state $P(\rho)$ (pressure) for different interaction strengths $\epsilon$ comparing Monte-Carlo simulations (MC) with the Machine learning (ML) results. (b) density profile for $\epsilon=1.25$, $\mu=\ln(1.15)$
inside the training region, but $V$ is not in the training data. (c) density profile at a hard wall for $\epsilon=1.9$, $\mu=\ln(1.9)$ (outside the training region $\epsilon\in[0.5,1.5]$).
Dark solid lines are simulation profiles, and blue dashed lines are ML results.
Insets in (b) and (c) show $\Delta\rho=\rho^\mathrm{mc}-\rho^\mathrm{ml}$.}
\label{fig:dft:LJ}
\end{figure}

The problem can be formulated as a self-consistency equation: 
\begin{equation}
\label{eqn:rho_ML}
\rho(x) = \exp\left(\left.\mu-\frac{\delta F(\rho(x))}{\delta \rho}\right|_{\rho=\rho^{\mathrm eq}}-V\right),
\end{equation}
where $\rho$ is the particle density, $F$ is the external free energy functional that needs to be learned, and $\mu$ and $V$ are chemical and external potential, respectively. Notice that the derivative of $F$ (which is represented by an EQL network) is used in the equation.
An analytical description for $F$ can be obtained using symbolic regression on simulation data. 
In \citet{lin2020:DFTEQL}, for the case of hard rod particles and Lennard-Jones fluids, solutions were found that extrapolate well to unseen situations (different external potential or mean density), as shown in Fig.~\ref{fig:dft:LJ}.
It is a promising approach to gain more theoretical insights when applied to less studied systems. 

\section{Discovering %
governing equations in boundary-layer transition to turbulence} \label{sec:casestudy:discovering}

A classical approach to discovering the governing equations of a reduced-order model (ROM) describing a particular phenomenon, for which the governing partial differential equations (PDEs) are known, is to perform Galerkin projection~\citep{wang_galerkin,noack_galerkin}. In Galerkin projection, a set of orthogonal basis modes (obtained, for instance, via POD) are used to develop a ROM of the system from data. Then, the governing PDEs are projected onto these modes, transforming the PDEs into a system of ordinary differential equations (ODEs) governing the dynamics of the temporal coefficients associated with those modes~\citep{brunton2022data}. 

For incompressible fluid flows, the spatiotemporal velocity vector $\pmb{u}(\pmb{x},t)$ (where $\pmb{x}$ are the spatial coordinates and $t$ time) can be expressed as follows after performing POD:
\begin{equation}\label{eq:pod_rv}
\pmb{u}(\pmb{x},t) \simeq \pmb{u}_0 (\pmb{x}) + \sum_{k=1}^r a_k(t) \pmb{u}_k (\pmb{x}),
\end{equation} 
where $\pmb{u}_0$ is the mean flow, $\pmb{u}_k$ are the spatial modes, $a_k(t)$ are the temporal coefficients and $r$ is the number of retained modes in the ROM. The expansion (\ref{eq:pod_rv}) is then substituted into the governing PDEs, {\it i.e.} the incompressible Navier--Stokes equations, taking advantage that the POD modes are linear combinations of the instantaneous flow realisations (thus satisfying the boundary conditions) and are solenoidal ({\it, i.e.} divergence-free, due to the incompressibility condition). It is then possible to take an inner product in space with $\pmb{u}_i (\pmb{x})$. Since the POD modes are orthogonal, a set of ODEs can be obtained for the time derivatives of the temporal coefficients ${\rm d}a_i(t) / {\rm d}t$ as a function of the spatial modes and also $a_i(t)$. Despite being a widely-used method, it has the limitations of requiring knowledge of the underlying PDEs, and it also may exhibit convergence problems in more challenging scenarios. As discussed in \secref{sec:dimensionality_reduction}, another alternative to produce a ROM for physical systems in a purely data-driven way is dynamic-mode decomposition (DMD), in which the obtained modes are orthogonal in time~\citep{brunton_dmd}; note that this approach, in its compressive version~\citep{brunton_cdmd}, shares similarities with the eigensystem-realisation algorithm (ERA)~\citep{era}. In this sense, DMD and its connections with the Koopman operator~\citep{rowley_et_al} were exploited by Eivazi {\it et al.}~\citep{eivazi2020recurrent} to reproduce the dynamics of a near-wall model of turbulence~\citep{moehlis_et_al} by using external forcing to reproduce the nonlinear behaviour of the system~\citep{koopman,brunton_koopman}. 

In addition to the approaches mentioned above, other techniques enable learning the equations of motion just from data, as discussed in the early work by Crutchfield {\it et al.}~\citep{Crutchfield_et_al}. These approaches typically rely on a library of candidate functions to build the resulting governing equation and solve an optimisation problem to obtain the expression that best represents the data. Note that it is essential to use any knowledge on the physical properties of the analysed data to inform the library (\eg whether non-linearities, periodicities, etc. are present in the system that produced the data under study), as well as to define the relevant state variables, sampling rate, the initial set of parameters defining relevant trajectories, etc. Embedding prior physical information into the obtained model is crucial for the success of these approaches, and failing to do so may lead to rate and even incorrect models~\citep{antonelli_et_al}. Furthermore, these approaches typically suffer from the curse of dimensionality~\citep{curse}, making it even more important to make the right choices in the library of candidate functions to ensure convergence; thus, being able to define the best basis functions to reduce the dimensionality of the system while retaining the most relevant physics is also is critical. Generally, only after solving the optimisation problem is it possible to assess which terms in the library are necessary and which ones may be combined, a fact that complicates {\it a-priori} equation discovery.

SINDy has been successfully applied to boundary-value problems~\citep{shea_et_al} using forcing functions and performs well even with noise. Furthermore, SINDy has produced very successful results in a wide variety of fluid-mechanics problems, ranging from thermal convection~\citep{sindy1}, chaotic electroconvection~\citep{sindy2}, the so-called ``fluidic-pinball'' problem~\citep{sindy3}, turbulent wakes~\citep{sindy4} and ROM development~\citep{sindy5}. Interestingly, SINDy has also been successfully combined with autoencoders to discover low-dimensional latent spaces~\citep{champion2019data}, benefiting from the non-linear data-compression capabilities of the latter and the interpretability of the former. This is certainly a promising direction to discover hidden complex relations in fluid-flow data and other high-dimensional physical systems, which requires further investigation, particularly when obtaining deeper insight into the interpretation of the latent variables. 

Besides the methods above based on discovering nonlinear dynamical systems, other strategies exist to obtain equations from data. For instance, gene-expression programming (GEP), a branch of evolutionary computing~\citep{gep}, is based on having a population of candidate functions to build the solution that best approximates the data and progressively improving this population by the survival of the fittest. The main advantage of this approach is that it leads to closed-form equations, even for data where the governing equation is unknown. In principle, it leads to interpretable solutions (although, in some cases, the resulting equations are so convoluted that interpretability is complicated). GEP has been used to model turbulence~\citep{sandberg1}, particularly in the context of the so-called Reynolds-averaged Navier--Stokes (RANS) equations. In short, the RANS equations are obtained after decomposing the instantaneous velocity into a mean and a fluctuation component (Reynolds averaging~\citep{reynolds}), and although this simplifies the flow-simulation process (RANS approaches are widely used in industry), the so-called closure problem emerges~\citep{pope,tennekes}. This problem is associated with the unknown impact of turbulent fluctuations on the mean flow. All the existing models for these stresses are empirical, which precludes RANS simulations from producing accurate results for arbitrary flow cases. In this context, Weatheritt~and~Sandberg~\citep{sandberg2} used GEP to obtain general expressions for these turbulent stresses in various cases, including turbulent ducts~\citep{vinuesa_duct}, which are challenging for RANS models due to the presence of secondary flows. They achieved quite successful RANS models for the secondary flows. GEP effectively obtained more general expressions for the turbulent stresses than those in the classical literature~\citep{boussinesq}, a critical step for RANS models to produce a reasonable performance for complex flows~\citep{spalart}.

\begin{figure}[h!]
\centering 
\includegraphics[width=0.85\textwidth]{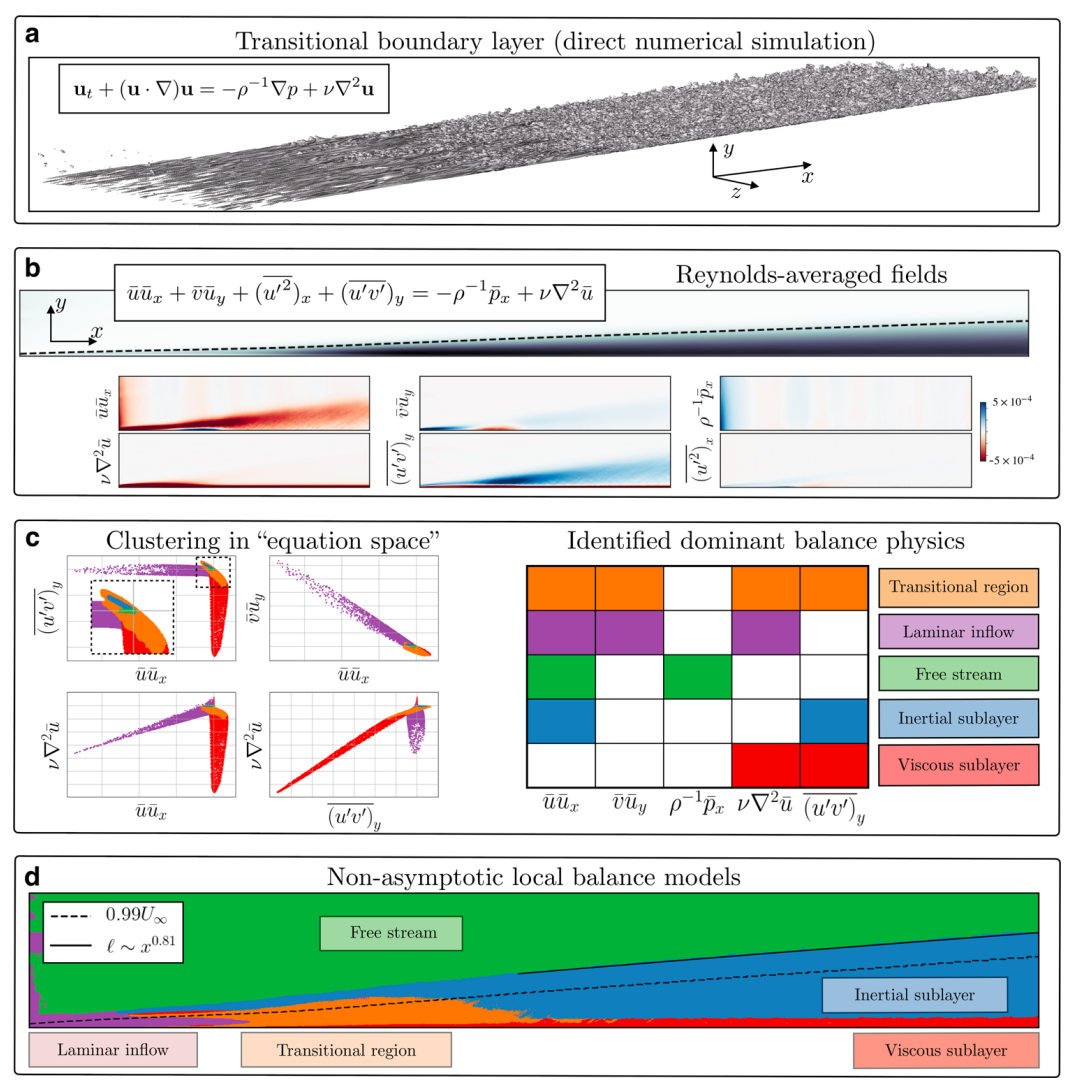}
   \caption{Data-driven balance model by Callaham {\it et al.}~\citep{Callaham_et_al} applied to a boundary layer undergoing transition to turbulence. a) Instantaneous data from high-fidelity simulations~\citep{dns_data} and b) terms in the RANS equations obtained from the turbulence statistics. c) Covariance of the various terms grouped into clusters, labelled based on their physical meaning. d) Representation of the various clusters in the flow field, together with various boundary-layer quantities. Figure reproduced from Ref.~\citep{Callaham_et_al} with permission of the publisher (Springer Nature).}
   \label{fig:balance_models}
\end{figure}

Finally, we conclude with a technique that is not aimed at discovering equations from data but rather focuses on identifying the dominant terms in the equations for various geometrical regions in the domain under study, given the available data. This method is based on data-driven balance models~\citep{Callaham_et_al}. It can help improve our system's physical interpretation by understanding the most relevant terms defining various mechanisms in the data, particularly in non-asymptotic cases where the negligible terms are not obvious. Using unsupervised learning, the authors sought clusters of points in the domain with negligible covariance in directions that represent terms with a negligible contribution to the physics, a condition equivalent to stating that the equation is satisfied by a few dominant terms within the cluster. In particular, they used Gaussian-mixture models (GMMs)~\citep{gmms} to cluster the data. Then they obtained a sparse approximation in the direction of maximum variance using sparse principal-component analysis (SPCA)~\citep{spca}. Callaham {\it et al.}~\citep{Callaham_et_al} show the applicability of this framework to a wide range of problems, including turbulence transition, combustion, nonlinear optics, geophysical fluids, and neuroscience. In all these cases, they obtained relevant insight into the governing equations, which can help uncover novel and unexpected physical relations. In particular, their application to the case of transition to turbulence is very illustrative, as shown in Figure~\ref{fig:balance_models}. This figure shows that starting from high-fidelity turbulence data, the RANS equations~\citep{pope,tennekes} mentioned above are obtained and their terms analysed. Visualisation of the clustering in equation space reveals some interesting relations, such as the high covariance of the so-called viscous and Reynolds shear-stress-gradient terms, $\nu \nabla^2 \overline{u}$ and $\overline{\left ( u^\prime v^\prime \right )}_y$ respectively, which identify the viscous sublayer in the domain. Note that subscripts here denote partial derivatives with respect to the corresponding spatial variable, the overbar indicates time averaging, and the prime is used for fluctuating quantities. The inertial sublayer, which would correspond to the turbulent region in this boundary-layer flow, would be dominated by the convection of the mean flow $\overline{u} \: \overline{u}_x$ and $\overline{\left ( u^\prime v^\prime \right )}_y$, which are again correctly identified through a strong covariance and highlighted in the corresponding region of the domain.

\section{Learning reduced-order models for vortex shedding behind an obstacle}

In this section, we illustrate the possibility of learning ROMs in the case of flow around an obstacle, focusing on the wake. One possibility is to perform a modal decomposition, for instance, based on POD, and then carry out Galerkin projection of the governing Navier--Stokes equations onto the POD modes, as discussed in \secref{sec:casestudy:discovering}. This would lead to differential equations governing the temporal evolution of the POD coefficients associated with the spatial modes. This approach may exhibit two main problems in the case of turbulent flows, namely the possible numerical challenges of performing Galerkin projection and the need for many modes to reconstruct a significant fraction of the flow energy. As stated above, autoencoders can provide a compressed version of the original data by exploiting non-linearities, thus exhibiting the great potential to express high-dimensional turbulence data in a few non-linear modes. As shown by Eivazi {\it et al.}~\citep{r3}, it is possible to learn a reduced representation of the original data where the latent vectors expressed in physical space exhibit orthogonality. This is achieved by promoting the learning of a latent space with disentangled latent vectors, which also enables learning parsimonious latent spaces. This is done by regularising the loss function, where the associated hyperparameter $\beta$ gives the name to the $\beta$-VAE framework discussed in \secref{sec:dimensionality_reduction}. Larger values of $\beta$ give more weight to the term in the loss responsible for learning statistically-independent latent variables, therefore, when $\beta = 0$ one obtains the standard reconstruction loss function.
In contrast, larger values of $\beta$ lead to higher orthogonality of the learned modes. At the same time, larger $\beta$ values will yield a worse reconstruction for the set number of latent vectors in the model. Based on this trade-off, it is possible to obtain a good balance between reconstruction and orthogonality. Eivazi {\it et al.}~\citep{r3} illustrated this on the turbulent flow around two wall-mounted obstacles and showed that with only 5 AE modes, it is possible to reconstruct around $90\%$ of the turbulent kinetic energy (TKE) with over $99\%$ orthogonality of the modes.
In comparison, 5 POD modes only reconstruct around $30\%$ of the TKE. This is very interesting because the $\beta$-VAE, which, unlike other AE-based methods, produces orthogonal modes, yields a reduced representation that can be interpreted from a physical point of view. The first AE and POD modes are very similar, identifying the shear layers around the obstacles and the wake shedding. However, the AE modes exhibit a broader range of scales, incorporating additional higher-frequency turbulent fluctuations into the basic identified features (similar to those in the POD results). Consequently, there is great potential for this type of method to shed light on the physics of complex turbulent flows, in particular when using novel data-driven methods, such as transformers~\citep{vaswani2017:transformers,heechang}, to predict the dynamics of the latent space. 

Another linear approach discussed above to obtain low-dimensional representations of the flow is dynamic-mode decomposition (DMD), which is based on building a linear operator connecting the instantaneous snapshots in time. Unlike POD, the DMD modes are orthogonal in time, {\it i.e.} they are associated with a single frequency, which helps identify temporal features in fluid flows. HODMD enables establishing more complex relationships among snapshots, and although it requires additional hyper-parameter tuning, it can help to identify more detailed patterns in the flow. Mart\'inez-S\'anchez {\it et al.}~\citep{alvaro_urban} used HODMD to study the turbulent flow in a simplified urban environment, emphasising the structures behind a wall-mounted obstacle. In this type of flow, a number of flow features emerge around the obstacle~\citep{monnier_old}, where a very important feature is the so-called arch vortex. This vortex, where the legs exhibit wall-normal and the spanwise roof vorticities, is responsible for the high concentration of pollutants in urban environments; therefore, understanding its formation mechanisms can have important implications for urban sustainability. In this context, HODMD enabled identifying two types of modes, namely vortex-generator and vortex-breaker features. The former is associated with low frequency, whereas the latter exhibits higher frequency, and both play important dynamic roles in flow physics. Another extension of the HODMD method also applied to the flow around a wall-mounted obstacle, was proposed by Amor {\it et al.}~\citep{amor_et_al}. This study featured the so-called on-the-fly version of HODMD. The data is analysed dynamically as the simulation is run, without storing massive amounts of data for post-processing.
Furthermore, more refined criteria for convergence of the modal decomposition were proposed, thus yielding a more effective way to analyse the data. Consequently, this on-the-fly approach reduces up to $80\%$ in memory requirements compared with the traditional offline method. This is a big advantage when applied to large-scale numerical databases.

\begin{figure}
\centering 
\includegraphics[width=\textwidth]{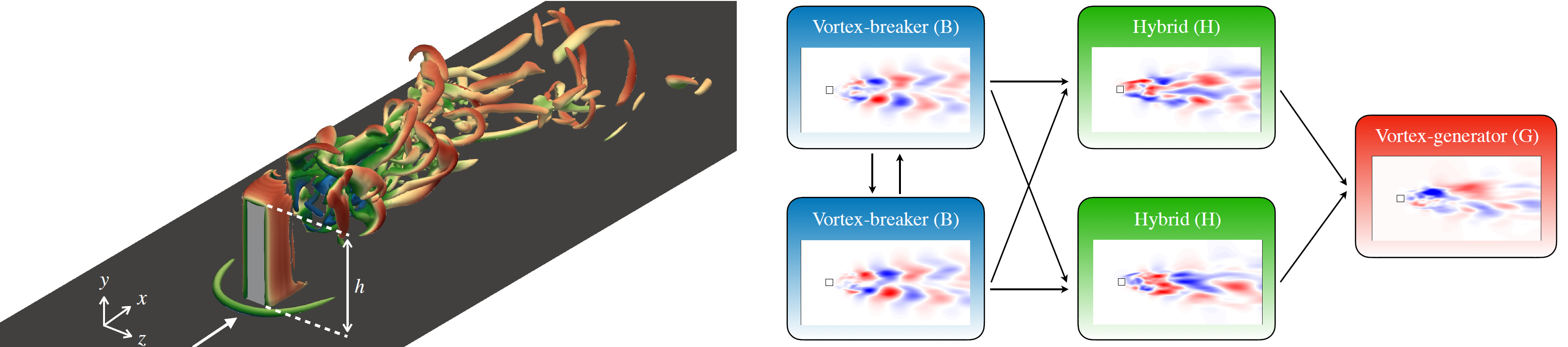}
   \caption{(Left) Instantaneous snapshot of the flow around a wall-mounted square cylinder, where the vortex clusters are identified with the $Q$ criterion~\citep{q_Ref}. The structures are coloured by their streamwise velocity, ranging from (dark blue) negative to (dark red) positive values. (Right) Schematic representation of the causal relations among modes, where two vortex-breaker (B), two hybrids (H), and one vortex-generator (G) modes are shown. Figure adapted from Ref.~\citep{alvaro_causality}.}
   \label{fig:causality}
\end{figure}

 Causality maps, discussed in \secref{sec:causality}, have been used to study the dynamic interactions present in turbulent flows, focusing on the physical roles of various features. In particular, Lozano-Dur\'an {\it et al.}~\citep{lozano_causality} studied the time series of the first Fourier modes in a turbulent channel. They found the following strong causal relations among modes: i) wall-normal modes causing streamwise modes, a phenomenon very closely connected with the well-known lift-up mechanism~\citep{orr,Landahl} in near-wall turbulence; ii) wall-normal modes causing spanwise modes, which is associated with the roll generation, also connected with the lift-up process and the incompressibility of the flow; iii) streamwise modes causing spanwise ones, and spanwise modes causing wall-normal ones; both phenomena are connected with the mean-flow instability, including spanwise meandering and breakdown of the streaks~\citep{b1,b2}. These causal relations were also identified~\citep{alvaro_causality} in other simplified models of near-wall turbulence, such as the nine-equation model by Moehlis {\it et al.}~\citep{moehlis_et_al}, a fact that confirms the robustness of the causality framework utilised to study turbulence phenomena. Regarding the flow around a wall-mounted obstacle, the various modes discussed above and their connection with the arch vortex were assessed by Mart\'inez-S\'anchez {\it et al.}~\citep{alvaro_causality} also using causality analysis. As can be observed in Figure~\ref{fig:causality}~(left), the flow under consideration exhibits large-scale separation at the sharp edges of the obstacle and very prominent vortical structures in the wake. Figure~\ref{fig:causality}~(right) exhibits the vortex-generator and breaker modes discussed above (associated with low and high frequencies, respectively), as well as an additional type of mode of intermediate frequency, denoted as hybrid mode. Clear causal relations are identified between the vortex-breaker and hybrid modes, closely connected with developing vortex-generator modes. This is of great interest because these causal relations define a sequence of events required for the production of the arch vortex (and the subsequent accumulation of pollutants in urban environments); thus, being able to control and inhibit this sequence of events may lead to novel sustainability solutions in cities (as well as to a deeper physical understanding of these complex turbulent flows).

\section{Uncovering new physical understanding in wall-bounded turbulence}

\begin{figure}[t!]
\centering 
\includegraphics[width=0.8\textwidth]{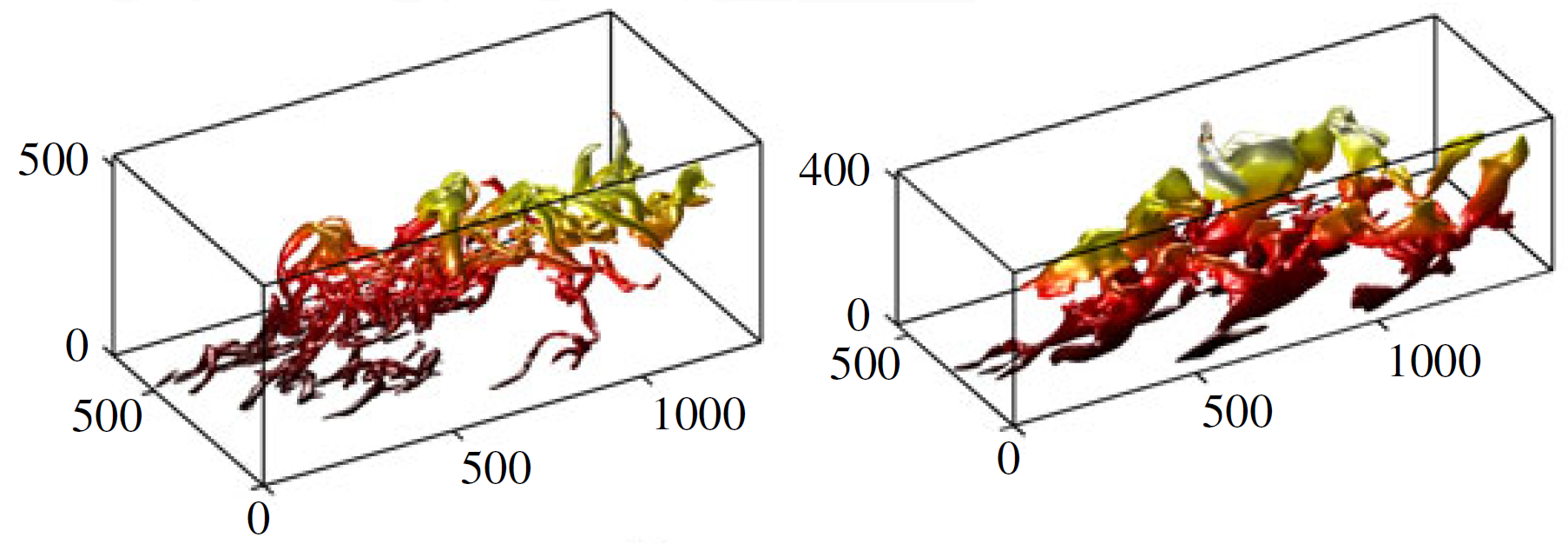}
   \caption{Coherent structures in a turbulent channel flow. We show (left) a vortex cluster and (right) an intense Reynolds-stress event. Figure adapted from Ref.~\citep{adrian} with permission from the publisher (Cambridge University Press).}
   \label{fig:structures}
\end{figure}

Turbulent flow is one of the most elusive areas of study within fluid mechanics. The wide range of spatial and temporal scales present in turbulence, and the highly non-linear behaviour that characterises it, significantly complicate the possibilities of gaining a deep physical understanding of the main mechanisms within turbulence; this becomes even more complicated in the case of wall-bounded turbulence, which is ubiquitous in science and engineering. Turbulence is characterised by coherent structures, three-dimensional regions that instantaneously satisfy certain physical properties. Note that this term sometimes refers to the features extracted by modal analysis. Still, we will consider the above definition in this work's context. A very important coherent structure in wall-bounded turbulence is the near-wall streak, extensively studied in the 1960s by Kline {\it et al.}~\citep{kline_et_al}. As reported by Kim {\it et al.}~\citep{kim_et_al}, the near-wall production of turbulence is very closely connected with the dynamics of these streaks. Another important quantity in wall turbulence is the Reynolds shear stress, briefly introduced in \secref{sec:casestudy:discovering}. This quantity is essentially a correlation between stream-wise ($u^{\prime}$) and wall-normal ($v^{\prime}$) fluctuations and is responsible for the wall-normal momentum transport. Studying the coherent structures most relevant to the development of the Reynolds stresses is a critical goal for reaching a deeper understanding of turbulence. Several studies in the 1970s~\citep{wallace_et_al,lu_willmarth} focused on the quadrant analysis to carry out this task; in this analysis, different near-wall events are classified in terms of the sign of their fluctuations, where the most dominant events are the so-called sweeps ($u^{\prime}>0$, $v^{\prime}<0$) and ejections ($u^{\prime}<0$, $v^{\prime}>0$). More recently, del \'Alamo {\it et al.}~\citep{del_alamo} have studied vortex clusters in turbulent channels, and Lozano-Dur\'an {\it et al.}~\citep{adrian} have analysed extreme Reynolds-stress events in the same flow case. The latter is defined as the three-dimensional connected regions satisfying:
\begin{equation}
|u^{\prime} v^{\prime} | > H u_{{\rm rms}} v_{{\rm rms}},
\end{equation}
where the subscript `rms' denotes root-mean-squared quantities, and $H$ is an empirical threshold denoted as a hyperbolic hole. In Figure~\ref{fig:structures}, we show both types of coherent structure in a turbulent channel flow. Additional insight into the role of both types of structures can be obtained by tracking their evolution in time, such that the various structure interactions (advection, merges, splits, and dissipation) can be assessed~\citep{adrian_tracking}. Convolutional neural networks (CNNs) have been used to predict the temporal evolution of the structures in turbulent channels~\citep{structure_pred}, an approach that enables a deeper understanding of their dynamic behaviour. In turbulence, there is a direct cascade of energy from the larger, energy-containing structures towards the smaller dissipative ones; however, there is also an energy path in the opposite direction~\citep{Cardesa_science}. This picture, observed in homogeneous isotropic turbulence, becomes even more complicated in the case of wall-bounded turbulence~\citep{jimenez_cascades}. Each wall-normal location has a different energy cascade because the wall segregates the flow by introducing wall-normal inhomogeneity. A comprehensive review of coherent structures in turbulence was provided by Jim\'enez~\citep{jimenez_perspective}, who highlighted the potential and challenges of this perspective on turbulence. A very interesting open question raised in this work is the multi-scale organisation and interaction among the various individual structures and how they can dynamically produce the underlying physics of the flow.  

Despite the extensive body of work on coherent structures in turbulence, there are still a number of open questions regarding the objective identification of the structures which play the most important role in the dynamics of turbulent flows. This fundamental question has implications for the theoretical knowledge of the physics of turbulence and the potential of flow-control strategies. If it is possible to identify these structures and they can be suppressed, there may be potential for novel and effective drag-reduction techniques. The vortex clusters and Reynolds-stress structures were defined based on historical reasons and physical intuition. Although they play an important role in the flow, it is unclear whether these are the most relevant motions. A new type of structure that maximises the momentum transfer in a turbulent channel was identified by Jim\'enez~\citep{fluxes}, and he reported significant differences between these and the Reynolds stresses. An extension of this idea was implemented in two-dimensional decaying turbulence by removing subregions of the domain and assessing their relative influence in the future evolution of the flow~\citep{jimenez_2d}. The idea is to quantify the ``significance'' of the various regions, and the result confirmed the initial physical intuition regarding this case: the most significant regions were vortices. The least significant ones exhibited high strain. In this direction, Cremades {\it et al.}~\citep{cremades_et_al} proposed an approach to exploit the explainability of neural networks to assess the relevance of the coherent structures in turbulent flows. In this study, the SHapley Additive exPlanations (SHAP) framework~\citep{shapley,shap} was used on the coherent structures identified in a turbulent channel; more concretely, the intense Reynolds stresses were first identified, and then a CNN was used to predict the location of those structures in the next time step~\citep{structure_pred}. The SHAP technique allows for identifying the impact of each of the features in the input (in this case, the three-dimensional Reynolds-stress events) on the prediction of the next step, thus enabling an assessment of their relevance to the future evolution of the flow. This framework could be used to find new ways of objectively identifying coherent regions in the flow. Another way to gain insight into the detailed mechanisms of turbulence via neural networks is to perform flow estimation, \eg from the quantities measured at the wall to the turbulent fluctuations above~\citep{guastoni2,guemes_gans}. After training a neural network to make this prediction, detailed knowledge of the connection between the scales at the wall and the ones above can be gained through neural-network interpretability~\cite{vinuesa_interp}. This approach allows us to discover a symbolic equation that can reproduce the predictive capabilities of the network. This can be achieved through the methodology developed by Cranmer {\it et al.}~\citep{cranmer_et_al}, which relies on symbolic regression (\eg based on genetic programming) to obtain the equation relating input and output; see \secref{sec:casestudy:discovering} for a related discussion. By analysing such an equation, it is possible to identify the characteristics of the scales relevant to this wall-normal interaction in wall-bounded turbulent flows.

\section{Discovery of ocean mesoscale closures}\label{sec:rvm}

The closure problem, described above in RANS equations, apply to many ocean and atmosphere modelling. In climate modelling, we must resolve (spatial) scales from meters to thousand kilometres. However, due to computational limitations, we need to truncate the spatial spectrum at a given scale - equivalent to the grid spacing of the numerical climate model. Therefore, all processes occurring below the spatial scales need to be approximated - this is the so-called parameterisation or closure problem for subgrid processes. The closure problem, described above in RANS equations, apply to many ocean and atmosphere modelling. 

While RANS separates terms into time-averaged and fluctuating components, the most common approach is based on Large Eddy Simulation (LES), in which the filtering separates into a resolved scale and a sub-grid scale. 
The LES decomposition is based on the self-similarity of small-scale turbulent structures. The resolved scales are defined using a convolution integral with associated physical width, usually the grid cell size. Commonly used filters are box filters, normalised Gaussian, or a combination of both filters. Applying the filtering to the governing equations of the fluid (momentum and buoyancy) gives rise to a set of equations for the resolved scale, with a term - coined subgrid scale forcing - which depends on the fine scale. For the momentum equation, this term subgrid term would be expressed as 
\begin{eqnarray}
\mathbf{S}&=& \begin{pmatrix}
{S}_x \\
{S}_y
\end{pmatrix} =(\overline{\mathbf{u}}\cdot\overline{\nabla})\overline{\mathbf{u}} - \overline{ (\mathbf{u}\cdot\nabla)\mathbf{u}},
\label{eq:subgrid_forcing}
\end{eqnarray}
where $\nabla$ is the horizontal 2D gradient operator, and the horizontal velocity ${\mathbf{u}}=\left({u},{v}\right)$, and the overline denotes the filtered (hence resolved) velocity on the grid. 

Therefore $\mathbf{S}$ must be approximated with only resolved scales $\overline{\mathbf{u}}$ since the total variable $\mathbf{u}$ is not available to the model. Typically turbulence subgrid closures in a fluid are ad-hoc, such as Smagorinsky-type closures, in which the form of the closure is based on some physical argument that is assumed to hold across scale and regimes. This is rarely the case. 

For ocean and atmosphere problems, the closure idea can also be boiled down to finding an expression for multiscale interaction that only depends on the resolved scale of the fluid. 
Similarly to traditional fluid problems, closures or parameterisations in the ocean and atmosphere modelling is often empirical and a source of error in simulations.   
Instead, equation discovery algorithms, as discussed in previous sections, can be used to uncover relationships between variables. 
For the closure problem, these algorithms can be applied to derive equations that describe the behaviour of the subgrid scales using resolved variables based on simulated data. 
The goal is to find the simplest (in some sense) mathematical relationship that accurately captures the behaviour of the subgrid-scale model, which can then be used for prediction. 
The main advantage of equation discovery algorithms is that they can uncover relationships that may not be immediately apparent, reducing the need for expert knowledge and human intuition in model building, as typically done for parameterisation in ocean and atmospheric modelling.

\citet{zanna2020data} used Relevance Vector Machine (RVM), a sparse Bayesian regression algorithm, to find closure models for momentum and buoyancy subgrid forcing. 
The RVM algorithm finds the most relevant input features - functions of the resolved scales - that will describe the subgrid-scale model. The RVM starts with many basis functions and iteratively removes irrelevant basis functions, arriving at a compact set of basis functions that best represent the data. Compared to other methods, the RVM algorithm has the advantage of handling noisy and redundant data and high-dimensional input spaces. Finally, it provides a probabilistic output, which can be a useful measure of uncertainty. 

Below is a closure found by \citet{zanna2020data}, using data from an ocean primitive equation model
\begin{equation}
\mathbf{S} \approx \kappa_{BT}\overline{\nabla}\cdot 
\begin{pmatrix} 
\zeta^2 - \zeta D & \zeta\tilde{D} \\
\zeta\tilde{D} & \zeta^2 + \zeta D 
\end{pmatrix},
\label{eqn:barotropiw_rvm_momentum_b}
\end{equation}
where 
$\zeta = {\overline{v}}_x - {\overline{u}}_{y}, D =  {\overline{u}}_{y} + {\overline{v}}_{x}, \tilde{D} =  {\overline{u}}_{x} - {\overline{v}}_{y}$, the short-hands $()_{x,y} \equiv \frac{\partial}{\partial x,y}$ are used for spatial derivatives, $\zeta$ is the relative vorticity, and $D$ and $\tilde{D}$ are the shearing and stretching deformation of the flow field, respectively.  The authors were able to relate the found expression to energy transfer across scales, which mimics the impact of unresolved scales on large-scale energetics. 

However, sparse linear regression entails trade-offs between the size and expressiveness of the feature library and the complexity and cost of sparse regression, as discussed in \citet{zanna2020data} and above.  If we wish to include a deep library of functions, the number of different expressions needed will grow exponentially and might be limited by accurately taking derivatives of functions. Finally, many expressions might be highly correlated, preventing convergence \citep{hastie2015statistical}.

\begin{figure}[t!]
\centering 
\includegraphics[width=1\textwidth]{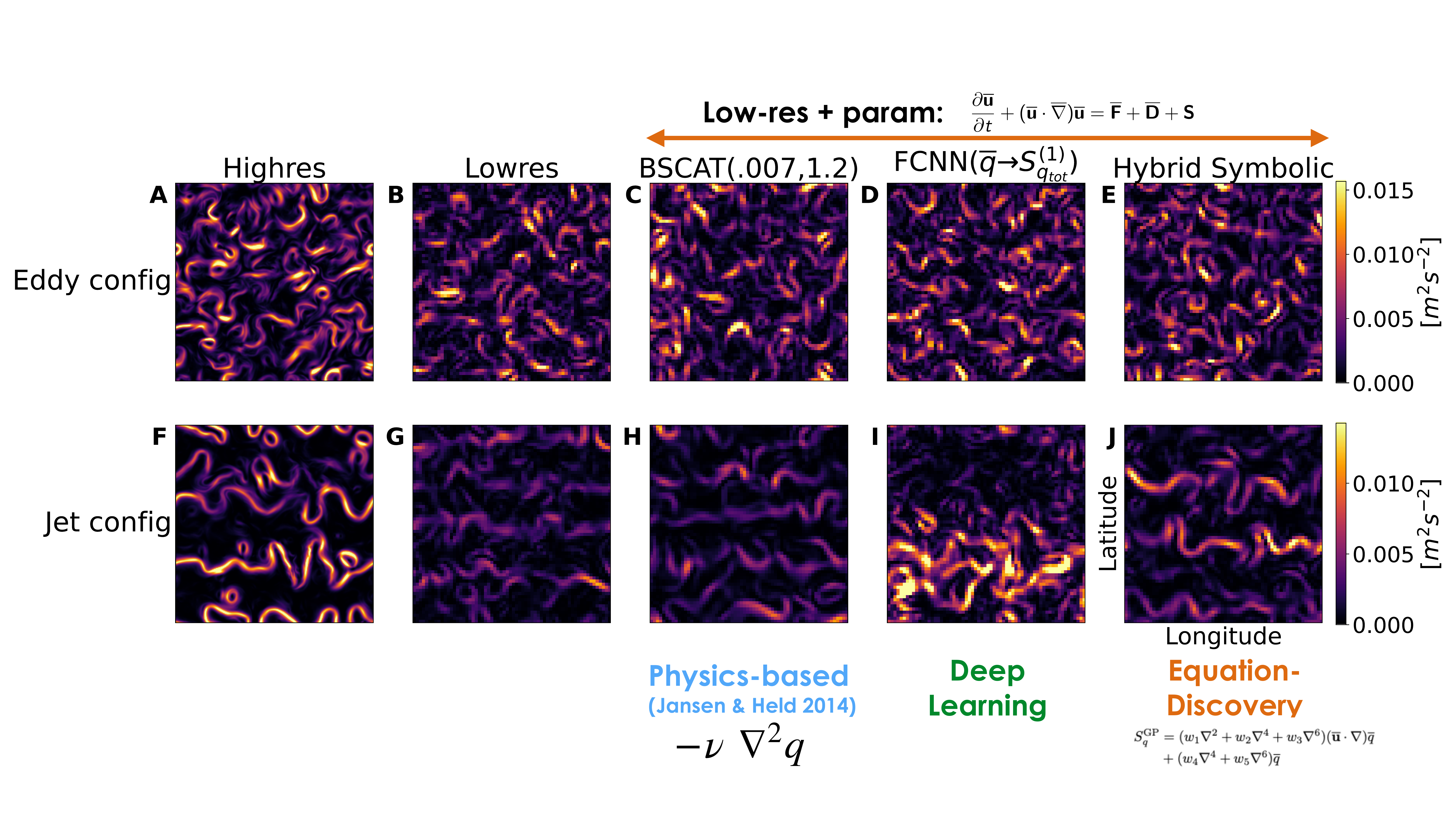}
\vspace{-1cm}
   \caption{Snapshot of potential vorticity in two different simulations: Top = Eddy (mostly isotropic turbulence), Bottom = Jet (some elongated sharp features mixed with isotropic turbulence features). A, F: High-Resolution simulations; B, G: Coarse Resolution; C, H: Coarse Resolution with physics-based parameterisations; D, I: Coarse Resolution with a Neural Network-based Dramatisation; E, J: Coarse Resolution with parameterisation discovered with symbolic regression. The data-driven parameterisations are trained on eddy configuration, and only the equation-discovery lead to robust generalisation in different regimes without retraining. \citep{Ross-et-al2022}}
   \label{fig:mesoscale_closure}
\end{figure}

As discussed above, genetic programming (GP)~\citep{koza1994genetic} is an alternative approach. GP algorithms, unlike sparse regression, do not require a defined library of functions. 
\citep{Ross-et-al2022} used GP with some modifications, including building spatial derivatives in spectral space and combining them with sparse regression to find robust expressions in turbulent datasets generated by idealised simulations. 
Focusing on results from \citep{Ross-et-al2022}, they look for the missing subgrid forcing for potential vorticity, $q$ - a variable that combines momentum and buoyancy effects in geophysical flows, and related to $\nabla \times \mathbf{u}$.
In the first few iterations, the algorithm discovered quadratic expressions proportional to $(\overline{\mathbf{u}} \cdot \nabla)\overline{q}$, similarly to previous theoretical studies \citep{meneveau2000,anstey2017deformation}. Often these expressions cannot be used as standalone parameterisations implemented in coarse-resolution models due to numerical stability constraints. 
The next few iterations of the GP-sparse regression algorithm led to eddy-viscosity models that dissipate energy at small scales, $\nabla^4 \overline{q}$ and redistribute energy to larger scales, i.e. kinetic energy backscatter $\nabla^6 \overline{q}$ \citep{jansen2014parameterizing}. 
Additional terms, which are cubic in model variables and contain a double-advection operator, $(\overline{\mathbf{u}} \cdot \nabla)^2$, can ensure dissipation of enstrophy \citep{marshall2010parameterization}, helping with model stability. 
In addition, there were additional terms that we were not discovered previously.
In summary, our discovered closure contains elements of existing subgrid parameterisations, which have pros and cons when used as standalone ones but, when combined, could capture all necessary properties for stable implementation and accurate representation of momentum, energy and enstrophy fluxes missing at coarse resolution.  

To test our discovered closure,  we implement it in a coarse-resolution simulation (see Fig.~\ref{fig:mesoscale_closure}). 
The goal is to improve the physics of the coarse-resolution model (panel B) relative to the high-resolution model simulation (panel A). 
To this end, we run the coarse-resolution simulation with a physics-based (empirical parameterisation; panel C), with data-driven parameterisation learned using a convolutional neural network (panel D), and with the equation-discovery parameterisation (panel E). All simulations are improving the flow, and some aspects of the statistics are also improved.  
However, generalisation is vastly different without retraining the data-driven driven parameterisations or tuning the physics-based parameterisations. 
We test our parameterisations in the same model in which we changed the rotation rate in order to form jets and less isotropic turbulence (panels F for high resolution and G for low resolution without any parameterisation). The physics-based parameterisation has little impact on the flow  (panel H), but the implementation of the deep learning parameterisation has a very detrimental effect on the flow, most trying to make the flow more isotropic (panel I). On the other hand, the implementation of the equation discovery-based parameterisation substantially improves the flow (panel J)- reinforcing the need to discover relationships from data that encapsulate the necessary laws of physics to mimic the scale interactions which are internal to the fluid  and not dependent on the configuration of the simulations.  

This symbolic parameterisation includes up to the seventh spatial derivative of $\overline{q}$, which may be unrealistic to implement into a climate model. However, it might be more realistic than a fully non-local approach, such as the convolutional neural network parameterisations considered in other studies or extremely local physics-based parameterisations (such as anti-viscosity).
Most importantly, the sparse model can generalise well without retraining, while the neural network-based parameterisations perform poorly. 

%% file: 05_Conclusions_book.tex
\chapter{Concluding remarks}\label{sec:conclusions}

The fields of causal and equation discovery have emerged in recent years as important research areas that apply artificial intelligence and machine learning to analyse complex systems \citep{peters2017elements,schmidt2011automated,runge2023causal}. 
The fields respectively aim to identify causal relationships and discover equations that can be used to predict the behaviour of the system, including the effects of interventions. 

In this paper, we have reviewed the state-of-the-art in both fields and discussed their respective approaches and techniques. 
Causal discovery aims to discover the qualitative cause-and-effect relationships between the variables in a system. In order to achieve this task using non-experimental data, causal discovery employs certain enabling assumptions (see \secref{sec:causality}). Among these assumptions is that the data-generating process can be described by a structural causal model and the corresponding causal graph. Methods for causal discovery are manifold and can be partitioned into constraint-based, score-based, asymmetry-based, and context-based methods. Data from the physical world typically comes in the form of time series with autocorrelation and potentially non-stationary behaviour. Autocorrelation and potential non-stationarity pose statistical challenges for many causal discovery methods as they are typically designed for i.i.d. data. In addition, the true functional relationship between variables can be highly non-linear, and the variables can be high-dimensional, both increasing model complexity and affecting the efficiency of causal discovery methods. All these challenges are compounded by the fact that the data acquired from real-world processes are often far from ideal, with problems such as missing data and inherent selection bias that might lead to the observed data not being representative of the process underlying it. These and many other challenges are avenues for future research in causal discovery and many of its sister fields, such as Bayesian networks and conditional independence testing, to name a few.
   
In equation discovery, the focus is on understanding the structure of a system by discovering equations, state variables and laws that can be used to predict (and, more importantly, to understand) its behaviour (see \secref{sec:equationdiscovery}). The main techniques used in this field are symbolic regression, evolutionary algorithms, and deep learning. These methods offer the potential to discover both linear and nonlinear equations but suffer from the need for large datasets and the difficulty of finding accurate equations in complex systems. More relevant challenges have to do with identifiability issues and the impossibility of evaluating the generality of the equations or even the criteria to select the most general ones. Broader (and perhaps more philosophical) questions need to be addressed, such as compressibility or sparsity, confronted with expressive power, the role of physical units and modularity, to name a few.

Thus, causality studies and equation inference approaches have synergistic goals. Both fields have made significant advances in recent years and offer considerable promise for further research. In particular, techniques from both fields can be combined to create hybrid models capable of uncovering causal relationships and equations. Additionally, developing more efficient algorithms and better methods for dealing with the challenge of overfitting could lead to further progress in both fields. More specifically, the question remains as to which current approaches provide stronger guarantees for the uniqueness/equifinality of the discovered equation or inferred causal graph. This key aspect in the inferential discovery of physical models should receive more attention in future research enterprises. 

A wide range of case studies in many areas of interest in the physical sciences (neurosciences, Earth and climate sciences, fluid mechanics) has illustrated the performance of causal discovery and equation discovery algorithms (see \secref{sec:casestudies}). We noted that specific methods and techniques reside in particular fields and do not permeate to others, mainly because of the needed assumptions and data characteristics. Yet, as has been the case for centuries, there is a lack of transdisciplinary in science. Directly stemming from this review, it is evident that analysing complex systems require an inter/trans-disciplinary approach that combines method and domain expertise. Techniques from artificial intelligence, machine learning, and control theory can be combined to better understand a system's behaviour and make accurate predictions. As such, future research in causal and equation discovery should consider the potential benefits of a more integrative and fused approach to analysing complex systems. This is perhaps especially important for recent developments designed for answering critical questions in specific areas but which, given their fundamental nature, have a wider appeal. For instance, the progress in the empirical inference of transfer operators in fluid mechanics or chemical reaction pathways has unexplored implications in understanding the metastable dynamics of neuronal network responses. 

Overall, the fields of causal discovery and equation discovery are rapidly advancing, and there is a growing synergy between them. Despite the remaining challenges, researchers have made great strides in uncovering the underlying structure of complex systems. With continued research and development, we can look forward to further advances in both fields and unlocking complex systems' mysteries.

%% file: 06_Acks_book.tex
\chapter*{Acknowledgements \&  Contributions}

G.C-V. received funding from the European Research Council (ERC) under the ERC Synergy Grant USMILE (grant agreement 855187), the Fundaci\'on BBVA with the project `Causal inference in the human-biosphere coupled system (SCALE)', the European Union’s Horizon 2020 research and innovation programme under Marie Skłodowska-Curie grant agreement No 860100 (IMIRACLI), the GVA PROMETEO AI4CS project on `AI for complex systems' (2022-2026) with CIPROM/2021/056, and the European Union’s Horizon 2020 research and innovation program within the project `XAIDA: Extreme Events - Artificial Intelligence for Detection and Attribution', under grant agreement No 101003469. G.V. and E.D. were partly supported by the ERC USMILE project (grant agreement 855187) and the GVA PROMETEO AI4CS (CIPROM/2021/056).

J.R. has received funding from the European Research Council (ERC) Starting Grant CausalEarth under the European Union’s Horizon 2020 research and innovation program (Grant Agreement No. 948112), from the European Union’s Horizon 2020 research and innovation programme under grant agreement No 101003469 (XAIDA), from the European Union’s Horizon 2020 research and innovation programme under Marie Skłodowska-Curie grant agreement No 860100 (IMIRACLI), and by the Helmholtz AI project CausalFlood (grant no. ZT-I-PF-5-11). J.R., A.G., and L.Z. were supported in part by the National Science Foundation under Grant No. NSF PHY-1748958. U.N. was supported by grant no. 948112 Causal Earth of the European Research Council (ERC).

G.M. is a member of the Machine Learning Cluster of Excellence, EXC number 2064/1 – Project number 390727645
He acknowledges the support from the German Federal Ministry of Education and Research (BMBF) through the T\"ubingen AI Center (FKZ: 01IS18039B). G.M. thanks Gabriel Kornberger for discussions on symbolic regression.

E.B-B. received funding from the Royal Society International Exchanges programme 2021 Round 1 (Ref. IES$\backslash$R1$\backslash$211062, ``Validating Cortical Network Models at the Edge of Asynchrony''), and from the EU H2020 Research and Innovation Programme under the Grant Agreement No. 945539 (Human Brain Project SGA3), via the Voucher awarded to the Partnering Project ``Async-Neuromorph''. Funders had no role in study design, data collection and analysis, decision to publish, or preparation of the manuscript.

R.V. received funding from the European Research Council (ERC) under the ERC Consolidator Grant DEEPCONTROL (``2021-CoG-101043998''). 

L.Z. received M$^2$LInES research funding by the generosity of Eric and Wendy Schmidt by recommendation of the Schmidt Futures program. LZ acknowledges funding from the NSF Science and Technology Center, Center for Learning the Earth with Artificial Intelligence and Physics (LEAP).

%% file: 07_Contributions_book.tex
GCV and JR conceptualised and defined the structure of the paper. GCV wrote the introduction and conclusions and participated in writing all sections, especially on equation discovery. 
GV, AG, UN, ED, and JR wrote the causal discovery section with the guidance of JR. 
GM, RV and GCV wrote the equation discovery section. 
EBB wrote the neuroscientific case studies, 
JR, ED and GCV contributed to the case studies on the discovery of causal relations among carbon and water fluxes, 
GM contributed to the case study on learning density functionals, 
LZ wrote the ocean mesoscale closures, 
RV wrote the three case studies related to fluid mechanics, 
JR wrote the case study on climate model intercomparison. 
All authors read, revised, commented and approved the manuscript.

%% file: main_book_format_for_arxiv.bbl
\begin{thebibliography}{558}
\providecommand{\natexlab}[1]{#1}
\providecommand{\url}[1]{\texttt{#1}}
\expandafter\ifx\csname urlstyle\endcsname\relax
  \providecommand{\doi}[1]{doi: #1}\else
  \providecommand{\doi}{doi: \begingroup \urlstyle{rm}\Url}\fi

\bibitem[Abbati et~al.(2019)Abbati, Wenk, Osborne, Krause, Sch{\"o}lkopf, and
  Bauer]{abbati2019ares}
G.~Abbati, P.~Wenk, M.~A. Osborne, A.~Krause, B.~Sch{\"o}lkopf, and S.~Bauer.
\newblock {AR}e{S} and {M}a{RS} adversarial and {MMD}-minimizing regression for
  {SDE}s.
\newblock In K.~Chaudhuri and R.~Salakhutdinov, editors, \emph{Proceedings of
  the 36th International Conference on Machine Learning}, volume~97 of
  \emph{Proceedings of Machine Learning Research}, pages 1--10. PMLR, 09--15
  Jun 2019.

\bibitem[Abhyankar(1998)]{abhyankar1998linear}
A.~Abhyankar.
\newblock {Linear and nonlinear Granger causality: Evidence from the UK stock
  index futures market}.
\newblock \emph{The Journal of Futures Markets (1986-1998)}, 18\penalty0
  (5):\penalty0 519, 1998.

\bibitem[Adsuara et~al.(2020)Adsuara, Pérez-Suay, Moreno-Martínez,
  Camps-Valls, Kraemer, Reichstein, and Mahecha]{Adsuara2020:diffeqnfromEOD}
J.~E. Adsuara, A.~Pérez-Suay, {\'A}.~Moreno-Martínez, G.~Camps-Valls,
  G.~Kraemer, M.~Reichstein, and M.~Mahecha.
\newblock Discovering differential equations from {E}arth observation data.
\newblock In \emph{IGARSS 2020 - 2020 IEEE International Geoscience and Remote
  Sensing Symposium}, pages 3999--4002, 2020.

\bibitem[Amari(1977)]{amar77}
S.~Amari.
\newblock Dynamics of pattern formation in lateral-inhibition type neural
  fields.
\newblock \emph{Biol. Cybern.}, 27\penalty0 (2):\penalty0 77--87, 8 1977.

\bibitem[Amor et~al.(2023)Amor, Schlatter, Vinuesa, and
  Le~Clainche]{amor_et_al}
C.~Amor, P.~Schlatter, R.~Vinuesa, and S.~Le~Clainche.
\newblock Higher-order dynamic mode decomposition on-the-fly: {A} low-order
  algorithm for complex fluid flows.
\newblock \emph{J. Comput. Phys.}, 475:\penalty0 111849, 2023.

\bibitem[Ancona et~al.(2004)Ancona, Marinazzo, and
  Stramaglia]{ancona2004radial}
N.~Ancona, D.~Marinazzo, and S.~Stramaglia.
\newblock Radial basis function approach to nonlinear {G}ranger causality of
  time series.
\newblock \emph{Physical Review E}, 70\penalty0 (5):\penalty0 056221, 2004.

\bibitem[Andersen and Fagerhaug(2006)]{andersen2006root}
B.~Andersen and T.~Fagerhaug.
\newblock \emph{Root cause analysis: simplified tools and techniques}.
\newblock Quality Press, 2006.

\bibitem[Ankerst et~al.(1999)Ankerst, Breunig, Kriegel, and
  Sander]{ankerst1999optics}
M.~Ankerst, M.~M. Breunig, H.-P. Kriegel, and J.~Sander.
\newblock {OPTICS}: {O}rdering points to identify the clustering structure.
\newblock \emph{ACM Sigmod record}, 28\penalty0 (2):\penalty0 49--60, 1999.

\bibitem[Anstey and Zanna(2017)]{anstey2017deformation}
J.~A. Anstey and L.~Zanna.
\newblock A deformation-based parametrization of ocean mesoscale eddy
  {R}eynolds stresses.
\newblock \emph{Ocean Modelling}, 112:\penalty0 99--111, 2017.

\bibitem[Antonelli et~al.(2022{\natexlab{a}})Antonelli, Chiaverini, and
  Di~Lillo]{antonelli2022data}
G.~Antonelli, S.~Chiaverini, and P.~Di~Lillo.
\newblock {On data-driven identification: Is automatically discovering
  equations of motion from data a Chimera?}
\newblock \emph{Nonlinear Dynamics}, pages 1--12, 2022{\natexlab{a}}.

\bibitem[Antonelli et~al.(2022{\natexlab{b}})Antonelli, Chiaverini, and
  Di~Lillo]{antonelli_et_al}
G.~Antonelli, S.~Chiaverini, and P.~Di~Lillo.
\newblock On data-driven identification: {I}s automatically discovering
  equations of motion from data a {C}himera?
\newblock \emph{Nonlinear Dyn}, 2022{\natexlab{b}}.

\bibitem[Aoi et~al.(2020)Aoi, Mante, and Pillow]{aoi20}
M.~C. Aoi, V.~Mante, and J.~W. Pillow.
\newblock {Prefrontal cortex exhibits multidimensional dynamic encoding during
  decision-making}.
\newblock \emph{Nat. Neurosci.}, 23\penalty0 (11):\penalty0 1410--1420, Nov.
  2020.

\bibitem[Arenas-Garc\'{i}a et~al.(2013)Arenas-Garc\'{i}a, Petersen,
  Camps-Valls, and Hansen]{Arenas13}
J.~Arenas-Garc\'{i}a, K.~Petersen, G.~Camps-Valls, and L.~Hansen.
\newblock Kernel multivariate analysis framework for supervised subspace
  learning: {A} tutorial on linear and kernel multivariate methods.
\newblock \emph{IEEE Signal Processing Magazine}, 30\penalty0 (4):\penalty0
  16--29, 2013.

\bibitem[Assaad et~al.(2022{\natexlab{a}})Assaad, Devijver, and
  Gaussier]{assaad2022discovery}
C.~K. Assaad, E.~Devijver, and E.~Gaussier.
\newblock Discovery of extended summary graphs in time series.
\newblock In J.~Cussens and K.~Zhang, editors, \emph{Proceedings of the
  Thirty-Eighth Conference on Uncertainty in Artificial Intelligence}, volume
  180 of \emph{Proceedings of Machine Learning Research}, pages 96--106. Pmlr,
  01--05 Aug 2022{\natexlab{a}}.

\bibitem[Assaad et~al.(2022{\natexlab{b}})Assaad, Devijver, and
  Gaussier]{assaad2022survey}
C.~K. Assaad, E.~Devijver, and E.~Gaussier.
\newblock Survey and evaluation of causal discovery methods for time series.
\newblock \emph{Journal of Artificial Intelligence Research}, 73:\penalty0
  767--819, 2022{\natexlab{b}}.

\bibitem[Baars and Tinney(2014)]{baars}
W.~J. Baars and C.~Tinney.
\newblock Proper orthogonal decomposition-based spectral higher-order
  stochastic estimation.
\newblock \emph{Phys. Fluids}, 26:\penalty0 055112, 2014.

\bibitem[Baddoo et~al.(2023)Baddoo, Herrmann, McKeon, Nathan~Kutz, and
  Brunton]{baddoo2023physics}
P.~J. Baddoo, B.~Herrmann, B.~J. McKeon, J.~Nathan~Kutz, and S.~L. Brunton.
\newblock Physics-informed dynamic mode decomposition.
\newblock \emph{Proceedings of the Royal Society A}, 479\penalty0
  (2271):\penalty0 20220576, 2023.

\bibitem[Bai et~al.(2020)Bai, Kaiser, Proctor, Kutz, and Brunton]{brunton_cdmd}
Z.~Bai, E.~Kaiser, J.~Proctor, J.~Kutz, and S.~Brunton.
\newblock Dynamic mode decomposition for compressive system identification.
\newblock \emph{Aiaa J.}, 58:\penalty0 561, 2020.

\bibitem[Balaguer-Ballester et~al.(2011)Balaguer-Ballester, Lapish, Seamans,
  and Durstewitz]{balaguer2011attracting}
E.~Balaguer-Ballester, C.~C. Lapish, J.~K. Seamans, and D.~Durstewitz.
\newblock Attracting dynamics of frontal cortex ensembles during memory-guided
  decision-making.
\newblock \emph{PLoS Computational Biology}, 7\penalty0 (5):\penalty0 e1002057,
  2011.

\bibitem[Balaguer-Ballester et~al.(2014)Balaguer-Ballester, Tabas-Diaz, and
  Budka]{bala14}
E.~Balaguer-Ballester, A.~Tabas-Diaz, and M.~Budka.
\newblock Can we identify non-stationary dynamics of trial-to-trial
  variability?
\newblock \emph{PLoS ONE}, 9\penalty0 (4):\penalty0 1--13, 04 2014.

\bibitem[Balaguer-Ballester et~al.(2018)Balaguer-Ballester, Moreno-Bote, Deco,
  and Durstewitz]{bala18}
E.~Balaguer-Ballester, R.~Moreno-Bote, G.~Deco, and D.~Durstewitz.
\newblock Editorial: {M}etastable dynamics of neural ensembles.
\newblock \emph{Frontiers in Systems Neuroscience}, 11, 2018.

\bibitem[Balaguer-Ballester et~al.(2020)Balaguer-Ballester, Nogueira, Abofalia,
  Moreno-Bote, and Sanchez-Vives]{bala20}
E.~Balaguer-Ballester, R.~Nogueira, J.~M. Abofalia, R.~Moreno-Bote, and M.~V.
  Sanchez-Vives.
\newblock Representation of foreseeable choice outcomes in orbitofrontal cortex
  triplet-wise interactions.
\newblock \emph{PLOS Computational Biology}, 16\penalty0 (6):\penalty0 1--30,
  06 2020.

\bibitem[Baldocchi(2014)]{BALDOCCHI2014}
D.~Baldocchi.
\newblock Measuring fluxes of trace gases and energy between ecosystems and the
  atmosphere – the state and future of the eddy covariance method.
\newblock \emph{Global Change Biology}, 20\penalty0 (12):\penalty0 3600--3609,
  2014.

\bibitem[Barack et~al.(2022)Barack, Miller, Moore, Packer, Pessoa, Ross, and
  Rust]{barack2022call}
D.~L. Barack, E.~K. Miller, C.~I. Moore, A.~M. Packer, L.~Pessoa, L.~N. Ross,
  and N.~C. Rust.
\newblock A call for more clarity around causality in neuroscience.
\newblock \emph{Trends in neurosciences}, 2022.

\bibitem[Barnett et~al.(2009)Barnett, Barrett, and Seth]{barnett2009granger}
L.~Barnett, A.~B. Barrett, and A.~K. Seth.
\newblock Granger causality and transfer entropy are equivalent for {G}aussian
  variables.
\newblock \emph{Physical review letters}, 103\penalty0 (23):\penalty0 238701,
  2009.

\bibitem[Barnett et~al.(2018)Barnett, Barrett, and
  Seth]{barnett2018misunderstandings}
L.~Barnett, A.~B. Barrett, and A.~K. Seth.
\newblock Misunderstandings regarding the application of {G}ranger causality in
  neuroscience.
\newblock \emph{Proceedings of the National Academy of Sciences}, 115\penalty0
  (29):\penalty0 E6676--e6677, 2018.

\bibitem[Barrett et~al.(2010)Barrett, Barnett, and
  Seth]{barrett2010multivariate}
A.~B. Barrett, L.~Barnett, and A.~K. Seth.
\newblock Multivariate {G}ranger causality and generalized variance.
\newblock \emph{Phys. Rev. E}, 81\penalty0 (4):\penalty0 41907, 2010.

\bibitem[Barrett et~al.(2019)Barrett, Morcos, and Macke]{barr19}
D.~G. Barrett, A.~S. Morcos, and J.~H. Macke.
\newblock Analyzing biological and artificial neural networks: {C}hallenges
  with opportunities for synergy?
\newblock \emph{Current Opinion in Neurobiology}, 55:\penalty0 55--64, 2019.
\newblock Machine Learning, Big Data, and Neuroscience.

\bibitem[Bassett and Sporns(2017)]{bassett2017network}
D.~S. Bassett and O.~Sporns.
\newblock Network neuroscience.
\newblock \emph{Nature neuroscience}, 20\penalty0 (3):\penalty0 353--364, 2017.

\bibitem[Bell et~al.(1996)Bell, Kay, and Malley]{bell1996non}
D.~Bell, J.~Kay, and J.~Malley.
\newblock A non-parametric approach to non-linear causality testing.
\newblock \emph{Economics Letters}, 51\penalty0 (1):\penalty0 7--18, 1996.

\bibitem[Bello et~al.(2022)Bello, Aragam, and Ravikumar]{bello2022dagma}
K.~Bello, B.~Aragam, and P.~Ravikumar.
\newblock {DAGMA}: Learning {DAGs} via {M}-matrices and a log-determinant
  acyclicity characterization.
\newblock In \emph{Advances in Neural Information Processing Systems}, 2022.

\bibitem[Belloni et~al.(2014)Belloni, Chernozhukov, and
  Hansen]{belloni2014inference}
A.~Belloni, V.~Chernozhukov, and C.~Hansen.
\newblock Inference on treatment effects after selection among high-dimensional
  controls.
\newblock \emph{The Review of Economic Studies}, 81\penalty0 (2):\penalty0
  608--650, 2014.

\bibitem[Benner et~al.(2015)Benner, Gugercin, and Willcox]{r8}
P.~Benner, S.~Gugercin, and K.~Willcox.
\newblock {A survey of projection-based model reduction methods for parametric
  dynamical systems}.
\newblock \emph{SIAM Rev.}, 57\penalty0 (4):\penalty0 483--531, 2015.

\bibitem[Berk et~al.(2013)Berk, Brown, Buja, Zhang, and Zhao]{berk2013}
R.~Berk, L.~Brown, A.~Buja, K.~Zhang, and L.~Zhao.
\newblock Valid post-selection inference.
\newblock \emph{The Annals of Statistics}, 41\penalty0 (2):\penalty0 802 --
  837, 2013.

\bibitem[Bernal-Casas et~al.(2013)Bernal-Casas, Balaguer-Ballester, Gerchen,
  Iglesias, Walter, Heinz, Meyer-Lindenberg, Stephan, and
  Kirsch]{bernal2013multi}
D.~Bernal-Casas, E.~Balaguer-Ballester, M.~F. Gerchen, S.~Iglesias, H.~Walter,
  A.~Heinz, A.~Meyer-Lindenberg, K.~E. Stephan, and P.~Kirsch.
\newblock Multi-site reproducibility of prefrontal-hippocampal connectivity
  estimates by stochastic {DCM}.
\newblock \emph{Neuroimage}, 82:\penalty0 555--563, 2013.

\bibitem[Bialek et~al.(2001)Bialek, Nemenman, and
  Tishby]{bialek2001predictability}
W.~Bialek, I.~Nemenman, and N.~Tishby.
\newblock Predictability, complexity, and learning.
\newblock \emph{Neural computation}, 13\penalty0 (11):\penalty0 2409--2463,
  2001.

\bibitem[Biggio et~al.(2021)Biggio, Bendinelli, Neitz, Lucchi, and
  Parascandolo]{Biggio2021:NeSymReS}
L.~Biggio, T.~Bendinelli, A.~Neitz, A.~Lucchi, and G.~Parascandolo.
\newblock Neural symbolic regression that scales.
\newblock In M.~Meila and T.~Zhang, editors, \emph{Proceedings of the 38th
  International Conference on Machine Learning}, volume 139 of
  \emph{Proceedings of Machine Learning Research}, pages 936--945. Pmlr, 18--24
  Jul 2021.

\bibitem[Bishop(2006)]{gmms}
C.~Bishop.
\newblock Pattern recognition and machine learning.
\newblock \emph{Springer New York}, 2006.

\bibitem[Boccaletti et~al.(2006)Boccaletti, Latora, Moreno, Chavez, and
  Hwang]{boccaletti2006complex}
S.~Boccaletti, V.~Latora, Y.~Moreno, M.~Chavez, and D.-U. Hwang.
\newblock Complex networks: {S}tructure and dynamics.
\newblock \emph{Physics reports}, 424\penalty0 (4-5):\penalty0 175--308, 2006.

\bibitem[Boden(2014)]{boden2014creativity}
M.~A. Boden.
\newblock Creativity and artificial intelligence: {A} contradiction in terms.
\newblock \emph{The philosophy of creativity: New essays}, pages 224--46, 2014.

\bibitem[Boelts et~al.(2022)Boelts, Lueckmann, Gao, and Macke]{boel22}
J.~Boelts, J.-M. Lueckmann, R.~Gao, and J.~H. Macke.
\newblock Flexible and efficient simulation-based inference for models of
  decision-making.
\newblock \emph{eLife}, 11:\penalty0 e77220, July 2022.

\bibitem[Boffetta et~al.(2002)Boffetta, Cencini, Falcioni, and
  Vulpiani]{boffetta2002predictability}
G.~Boffetta, M.~Cencini, M.~Falcioni, and A.~Vulpiani.
\newblock Predictability: {A} way to characterize complexity.
\newblock \emph{Physics reports}, 356\penalty0 (6):\penalty0 367--474, 2002.

\bibitem[Bollen(1989)]{bollen1989structural}
K.~A. Bollen.
\newblock \emph{Structural Equations with Latent Variables}.
\newblock John Wiley \& Sons, New York, NY, USA, 1989.

\bibitem[Bongard and Lipson(2007)]{Bongard2007}
J.~Bongard and H.~Lipson.
\newblock Automated reverse engineering of nonlinear dynamical systems.
\newblock \emph{Proceedings of the National Academy of Sciences of the United
  States of America}, 104\penalty0 (24):\penalty0 9943--9948, 2007.

\bibitem[Bongers and Mooij(2018)]{bongers2018random}
S.~Bongers and J.~M. Mooij.
\newblock From random differential equations to structural causal models: {T}he
  stochastic case.
\newblock \emph{arXiv preprint arXiv:1803.08784}, 2018.

\bibitem[Bongers et~al.(2018)Bongers, Blom, and Mooij]{bongers2018causal}
S.~Bongers, T.~Blom, and J.~M. Mooij.
\newblock Causal modeling of dynamical systems, 2018.

\bibitem[Bongers et~al.(2021)Bongers, Forr{\'e}, Peters, and
  Mooij]{bongers2021foundations}
S.~Bongers, P.~Forr{\'e}, J.~Peters, and J.~M. Mooij.
\newblock Foundations of structural causal models with cycles and latent
  variables.
\newblock \emph{The Annals of Statistics}, 49\penalty0 (5):\penalty0 2885 --
  2915, 2021.

\bibitem[Bouezmarni et~al.(2012)Bouezmarni, Rombouts, and
  Taamouti]{bouezmarni2012nonparametric}
T.~Bouezmarni, J.~V. Rombouts, and A.~Taamouti.
\newblock Nonparametric copula-based test for conditional independence with
  applications to {G}ranger causality.
\newblock \emph{Journal of Business \& Economic Statistics}, 30\penalty0
  (2):\penalty0 275--287, 2012.

\bibitem[Boussinesq(1923)]{boussinesq}
J.~V. Boussinesq.
\newblock \emph{Th\'eorie analytique de la chaleur: mise en harmonie avec la
  thermodynamique et avec la th\'eorie m\'ecanique de la lumi\`ere T. 2,
  {R}efroidissement et \'echauffement par rayonnement conductibilit\'e des
  tiges, lames et masses cristallines courants de convection th\'eorie
  m\'ecanique de la lumi\`ere}.
\newblock Gauthier-Villars, 1923.

\bibitem[Bradley et~al.(2001)Bradley, Gold, and Silverman]{bradley2001}
P.~Bradley, S.~Gold, and S.~Silverman.
\newblock Constructive induction from incomplete data: {A} comparative study.
\newblock \emph{Machine Learning}, 42\penalty0 (1):\penalty0 7--48, 2001.

\bibitem[Brennan et~al.(2023)Brennan, Aggarwal, Pei, Sussillo, and
  Proekt]{bren23}
C.~Brennan, A.~Aggarwal, R.~Pei, D.~Sussillo, and A.~Proekt.
\newblock One dimensional approximations of neuronal dynamics reveal
  computational strategy.
\newblock \emph{PLOS Computational Biology}, 19\penalty0 (1):\penalty0 1--27,
  01 2023.
\newblock \doi{10.1371/journal.pcbi.1010784}.
\newblock URL \url{https://doi.org/10.1371/journal.pcbi.1010784}.

\bibitem[Brenner et~al.(2022)Brenner, Hess, Mikhaeil, Bereska, Monfared, Kuo,
  and Durstewitz]{bren22}
M.~Brenner, F.~Hess, J.~M. Mikhaeil, L.~F. Bereska, Z.~Monfared, P.-C. Kuo, and
  D.~Durstewitz.
\newblock Tractable dendritic {RNNs} for reconstructing nonlinear dynamical
  systems.
\newblock In K.~Chaudhuri, S.~Jegelka, L.~Song, C.~Szepesvari, G.~Niu, and
  S.~Sabato, editors, \emph{Proceedings of the 39th International Conference on
  Machine Learning}, volume 162 of \emph{Proceedings of Machine Learning
  Research}, pages 2292--2320. Pmlr, 17--23 Jul 2022.

\bibitem[Bressler and Seth(2011)]{bressler2011wiener}
S.~L. Bressler and A.~K. Seth.
\newblock {Wiener-Granger} causality: {A} well established methodology.
\newblock \emph{Neuroimage}, 58\penalty0 (2):\penalty0 323--329, 2011.

\bibitem[Brouillard et~al.(2020)Brouillard, Lachapelle, Lacoste,
  Lacoste-Julien, and Drouin]{brouillard2020differentiable}
P.~Brouillard, S.~Lachapelle, A.~Lacoste, S.~Lacoste-Julien, and A.~Drouin.
\newblock Differentiable causal discovery from interventional data.
\newblock \emph{Advances in Neural Information Processing Systems},
  33:\penalty0 21865--21877, 2020.

\bibitem[Brown et~al.(2012)Brown, Pocock, Zhao, and
  Luj{\'a}n]{brown2012conditional}
G.~Brown, A.~Pocock, M.-J. Zhao, and M.~Luj{\'a}n.
\newblock Conditional likelihood maximisation: {A} unifying framework for
  information theoretic feature selection.
\newblock \emph{The journal of machine learning research}, 13:\penalty0 27--66,
  2012.

\bibitem[Brown et~al.(2020)Brown, Mann, Ryder, Subbiah, Kaplan, Dhariwal,
  Neelakantan, Shyam, Sastry, Askell, Agarwal, Herbert-Voss, Krueger, Henighan,
  Child, Ramesh, Ziegler, Wu, Winter, Hesse, Chen, Sigler, Litwin, Gray, Chess,
  Clark, Berner, McCandlish, Radford, Sutskever, and Amodei]{brown2020:GPT-3}
T.~Brown, B.~Mann, N.~Ryder, M.~Subbiah, J.~D. Kaplan, P.~Dhariwal,
  A.~Neelakantan, P.~Shyam, G.~Sastry, A.~Askell, S.~Agarwal, A.~Herbert-Voss,
  G.~Krueger, T.~Henighan, R.~Child, A.~Ramesh, D.~Ziegler, J.~Wu, C.~Winter,
  C.~Hesse, M.~Chen, E.~Sigler, M.~Litwin, S.~Gray, B.~Chess, J.~Clark,
  C.~Berner, S.~McCandlish, A.~Radford, I.~Sutskever, and D.~Amodei.
\newblock Language models are few-shot learners.
\newblock In H.~Larochelle, M.~Ranzato, R.~Hadsell, M.~F. Balcan, and H.~Lin,
  editors, \emph{Advances in {Neural} {Information} {Processing} {Systems}},
  volume~33, pages 1877--1901. Curran Associates, Inc., 2020.

\bibitem[Brunel(2000)]{brun00}
N.~Brunel.
\newblock Dynamics of sparsely connected networks of excitatory and inhibitory
  spiking neurons.
\newblock \emph{J. Comput. Neurosci.}, 8\penalty0 (3):\penalty0 183--208, May
  2000.

\bibitem[Brunton et~al.(2017)Brunton, Brunton, Proctor, Kaiser, and
  Kutz]{brunton_koopman}
S.~Brunton, B.~Brunton, J.~Proctor, E.~Kaiser, and J.~Kutz.
\newblock Chaos as an intermittently forced linear system.
\newblock \emph{Nat. Commun.}, 8:\penalty0 19, 2017.

\bibitem[Brunton and Kutz(2022)]{brunton2022data}
S.~L. Brunton and J.~N. Kutz.
\newblock \emph{Data-driven science and engineering: {M}achine learning,
  dynamical systems, and control}.
\newblock Cambridge University Press, 2022.

\bibitem[Brunton et~al.(2016{\natexlab{a}})Brunton, Brunton, Proctor, and
  Kutz]{BBPK16}
S.~L. Brunton, B.~W. Brunton, J.~L. Proctor, and J.~N. Kutz.
\newblock Koopman invariant subspaces and finite linear representations of
  nonlinear dynamical systems for control.
\newblock \emph{PLoS ONE}, 11, 2016{\natexlab{a}}.

\bibitem[Brunton et~al.(2016{\natexlab{b}})Brunton, Proctor, and
  Kutz]{brunton2016discovering}
S.~L. Brunton, J.~L. Proctor, and J.~N. Kutz.
\newblock Discovering governing equations from data by sparse identification of
  nonlinear dynamical systems.
\newblock \emph{Proceedings of the national academy of sciences}, 113\penalty0
  (15):\penalty0 3932--3937, 2016{\natexlab{b}}.

\bibitem[Bueso et~al.(2020)Bueso, Piles, and Camps-Valls]{Bueso20rock}
D.~Bueso, M.~Piles, and G.~Camps-Valls.
\newblock Nonlinear {PCA} for spatio-temporal analysis of {E}arth observation
  data.
\newblock \emph{IEEE Transactions on Geoscience and Remote Sensing},
  58\penalty0 (8):\penalty0 5752--5763, Aug. 2020.

\bibitem[Bullmore and Sporns(2009)]{bullmore2009complex}
E.~Bullmore and O.~Sporns.
\newblock Complex brain networks: {G}raph theoretical analysis of structural
  and functional systems.
\newblock \emph{Nature reviews neuroscience}, 10\penalty0 (3):\penalty0
  186--198, 2009.

\bibitem[Burlacu et~al.(2020)Burlacu, Kronberger, and Kommenda]{operon}
B.~Burlacu, G.~Kronberger, and M.~Kommenda.
\newblock {Operon C++: An Efficient Genetic Programming Framework for Symbolic
  Regression}.
\newblock In \emph{Proceedings of the 2020 Genetic and Evolutionary Computation
  Conference Companion}, Gecco '20, page 1562–1570, New York, NY, USA, 2020.
  Association for Computing Machinery.

\bibitem[Butterfield(1965)]{butterfield1965origins}
H.~Butterfield.
\newblock \emph{The origins of modern science}, volume 90507.
\newblock Simon and Schuster, 1965.

\bibitem[Byrne et~al.(2020)Byrne, O’Dea, Forrester, Ross, and
  Coombes]{byrn20}
A.~Byrne, R.~D. O’Dea, M.~Forrester, J.~Ross, and S.~Coombes.
\newblock {Next-generation neural mass and field modeling}.
\newblock \emph{Journal of Neurophysiology}, 123\penalty0 (2):\penalty0
  726--742, 2020.
\newblock Pmid: 31774370.

\bibitem[Byrne et~al.(2021)Byrne, Ross, Nicks, and Coombes]{byrn21}
A.~Byrne, J.~Ross, R.~Nicks, and S.~Coombes.
\newblock Mean-field models for {EEG/MEG}: {F}rom oscillations to waves.
\newblock \emph{Brain Topography}, 35, 05 2021.

\bibitem[Cabral et~al.(2022)Cabral, Castaldo, Vohryzek, Litvak, Bick,
  Lambiotte, Friston, Kringelbach, and Deco]{cabra22}
J.~Cabral, F.~Castaldo, J.~Vohryzek, V.~Litvak, C.~Bick, R.~Lambiotte,
  K.~Friston, M.~L. Kringelbach, and G.~Deco.
\newblock Metastable oscillatory modes emerge from synchronization in the brain
  spacetime connectome.
\newblock \emph{Communications Physics}, 5\penalty0 (1):\penalty0 184, 2022.

\bibitem[Callaham et~al.(2021{\natexlab{a}})Callaham, Koch, Brunton, Kutz, and
  Brunton]{Callaham_et_al}
J.~L. Callaham, J.~V. Koch, B.~W. Brunton, J.~N. Kutz, and S.~L. Brunton.
\newblock Learning dominant physical processes with data-driven balance models.
\newblock \emph{Nat. Commun.}, 12\penalty0 (1):\penalty0 1--10,
  2021{\natexlab{a}}.

\bibitem[Callaham et~al.(2021{\natexlab{b}})Callaham, Rigas, Loiseau, and
  Brunton]{sindy4}
J.~L. Callaham, G.~Rigas, J.-C. Loiseau, and S.~L. Brunton.
\newblock {An empirical mean-field model of symmetry-breaking in a turbulent
  wake}.
\newblock \emph{Sci. Adv.}, 8:\penalty0 eabm4786, 2021{\natexlab{b}}.

\bibitem[Callaham et~al.(2022)Callaham, Brunton, and Loiseau]{sindy5}
J.~L. Callaham, S.~L. Brunton, and J.-C. Loiseau.
\newblock {On the role of nonlinear correlations in reduced-order modeling}.
\newblock \emph{J. Fluid Mech.}, 938:\penalty0 A1, 2022.

\bibitem[Camps-Valls et~al.(2016)Camps-Valls, Verrelst, Munoz-Mari, Laparra,
  Mateo-Jimenez, and Gomez-Dans]{CampsValls16grsm}
G.~Camps-Valls, J.~Verrelst, J.~Munoz-Mari, V.~Laparra, F.~Mateo-Jimenez, and
  J.~Gomez-Dans.
\newblock A survey on {G}aussian processes for {E}arth-observation data
  analysis: {A} comprehensive investigation.
\newblock \emph{IEEE Geoscience and Remote Sensing Magazine}, 4\penalty0
  (2):\penalty0 58--78, 2016.
\newblock \doi{10.1109/MGRS.2015.2510084}.

\bibitem[Cardesa et~al.(2017)Cardesa, Vela-Mart\'in, and
  Jim\'enez]{Cardesa_science}
J.~I. Cardesa, A.~Vela-Mart\'in, and J.~Jim\'enez.
\newblock The turbulent cascade in five dimensions.
\newblock \emph{Science}, 357:\penalty0 782--784, 2017.

\bibitem[Carlberg et~al.(2017)Carlberg, Barone, and Antil]{r9}
K.~Carlberg, M.~Barone, and H.~Antil.
\newblock {Galerkin v. least-squares Petrov--Galerkin projection in nonlinear
  model reduction}.
\newblock \emph{J. Comput. Phys.}, 330:\penalty0 693--734, 2017.

\bibitem[Castelo and Siebes(2000)]{castelo2000priors}
R.~Castelo and A.~Siebes.
\newblock Priors on network structures. {B}iasing the search for {B}ayesian
  networks.
\newblock \emph{International Journal of Approximate Reasoning}, 24\penalty0
  (1):\penalty0 39--57, 2000.

\bibitem[Cenedese et~al.(2022)Cenedese, Ax\r{a}s, B\"auerlein, Avila, and
  Haller]{r7}
M.~Cenedese, J.~Ax\r{a}s, B.~B\"auerlein, K.~Avila, and G.~Haller.
\newblock {Data-driven modeling and prediction of nonlinearizable dynamics via
  spectral submanifolds}.
\newblock \emph{Nat. Commun.}, 13:\penalty0 872, 2022.

\bibitem[Champion et~al.(2019)Champion, Lusch, Kutz, and
  Brunton]{champion2019data}
K.~Champion, B.~Lusch, J.~N. Kutz, and S.~L. Brunton.
\newblock Data-driven discovery of coordinates and governing equations.
\newblock \emph{Proceedings of the National Academy of Sciences}, 116\penalty0
  (45):\penalty0 22445--22451, 2019.

\bibitem[Chandola et~al.(2009)Chandola, Banerjee, and
  Kumar]{chandola2009anomaly}
V.~Chandola, A.~Banerjee, and V.~Kumar.
\newblock Anomaly detection: {A} survey.
\newblock \emph{ACM computing surveys (CSUR)}, 41\penalty0 (3):\penalty0 1--58,
  2009.

\bibitem[Chatfield(2013)]{chatfield2013analysis}
C.~Chatfield.
\newblock \emph{The analysis of time series: {T}heory and practice}.
\newblock Springer, 2013.

\bibitem[Chen et~al.(2022)Chen, Huang, Raghupathi, Chandratreya, Du, and
  Lipson]{chen2022automated}
B.~Chen, K.~Huang, S.~Raghupathi, I.~Chandratreya, Q.~Du, and H.~Lipson.
\newblock Automated discovery of fundamental variables hidden in experimental
  data.
\newblock \emph{Nature Computational Science}, 2\penalty0 (7):\penalty0
  433--442, 2022.

\bibitem[Chen et~al.(2023)Chen, Ginoux, Carbo-Tano, Mora, Walczak, and
  Wyart]{chen23}
X.~Chen, F.~Ginoux, M.~Carbo-Tano, T.~Mora, A.~M. Walczak, and C.~Wyart.
\newblock Granger causality analysis for calcium transients in neuronal
  networks, challenges and improvements.
\newblock \emph{eLife}, 12:\penalty0 e81279, Feb. 2023.

\bibitem[Chen et~al.(2004)Chen, Rangarajan, Feng, and Ding]{chen2004analyzing}
Y.~Chen, G.~Rangarajan, J.~Feng, and M.~Ding.
\newblock Analyzing multiple nonlinear time series with extended {G}ranger
  causality.
\newblock \emph{Physics letters A}, 324\penalty0 (1):\penalty0 26--35, 2004.

\bibitem[Chicharro(2011)]{chicharro2011spectral}
D.~Chicharro.
\newblock On the spectral formulation of {G}ranger causality.
\newblock \emph{Biological cybernetics}, 105\penalty0 (5):\penalty0 331--347,
  2011.

\bibitem[Chicharro and Panzeri(2014)]{chic14}
D.~Chicharro and S.~Panzeri.
\newblock Algorithms of causal inference for the analysis of effective
  connectivity among brain regions.
\newblock \emph{Frontiers in Neuroinformatics}, 8, 2014.

\bibitem[Chickering(2002{\natexlab{a}})]{chickering2002learning}
D.~M. Chickering.
\newblock Learning equivalence classes of {B}ayesian-network structures.
\newblock \emph{Journal of Machine Learning Research}, 2:\penalty0 445–498,
  Mar. 2002{\natexlab{a}}.

\bibitem[Chickering(2002{\natexlab{b}})]{chickering2002optimal}
D.~M. Chickering.
\newblock Optimal structure identification with greedy search.
\newblock \emph{Journal of machine learning research}, 3\penalty0
  (Nov):\penalty0 507--554, 2002{\natexlab{b}}.

\bibitem[Chickering and Meek(2015)]{chickering2015selective}
D.~M. Chickering and C.~Meek.
\newblock Selective greedy equivalence search: {F}inding optimal {B}ayesian
  networks using a polynomial number of score evaluations.
\newblock In \emph{Proceedings of the Thirty-First Conference on Uncertainty in
  Artificial Intelligence}, pages 211--219, 2015.

\bibitem[Chickering(2020)]{chickering2020statistically}
M.~Chickering.
\newblock Statistically efficient greedy equivalence search.
\newblock In J.~Peters and D.~Sontag, editors, \emph{Proceedings of the 36th
  Conference on Uncertainty in Artificial Intelligence (UAI)}, volume 124 of
  \emph{Proceedings of Machine Learning Research}, pages 241--249. Pmlr, 03--06
  Aug 2020.

\bibitem[Christiansen et~al.(2022)Christiansen, Baumann, Kuemmerle, Mahecha,
  and Peters]{christiansen2022toward}
R.~Christiansen, M.~Baumann, T.~Kuemmerle, M.~D. Mahecha, and J.~Peters.
\newblock Toward causal inference for spatio-temporal data: {C}onflict and
  forest loss in {C}olombia.
\newblock \emph{Journal of the American Statistical Association}, 117\penalty0
  (538):\penalty0 591--601, 2022.

\bibitem[Claassen and Bucur(2022)]{claassen22a}
T.~Claassen and I.~G. Bucur.
\newblock Greedy equivalence search in the presence of latent confounders.
\newblock In J.~Cussens and K.~Zhang, editors, \emph{Proceedings of the
  Thirty-Eighth Conference on Uncertainty in Artificial Intelligence}, volume
  180 of \emph{Proceedings of Machine Learning Research}, pages 443--452. Pmlr,
  01--05 Aug 2022.

\bibitem[Coifman and Lafon(2006)]{coif06}
R.~R. Coifman and S.~Lafon.
\newblock Diffusion maps.
\newblock \emph{Applied and Computational Harmonic Analysis}, 21\penalty0
  (1):\penalty0 5--30, 2006.
\newblock Special Issue: Diffusion Maps and Wavelets.

\bibitem[Colombo et~al.(2014)Colombo, Maathuis, et~al.]{colombo2014order}
D.~Colombo, M.~H. Maathuis, et~al.
\newblock Order-independent constraint-based causal structure learning.
\newblock \emph{J. Mach. Learn. Res.}, 15\penalty0 (1):\penalty0 3741--3782,
  2014.

\bibitem[Cooper and Yoo(1999)]{cooper2013causal}
G.~F. Cooper and C.~Yoo.
\newblock Causal discovery from a mixture of experimental and observational
  data.
\newblock In \emph{Proceedings of the Fifteenth Conference on Uncertainty in
  Artificial Intelligence}, UAI'99, page 116–125, San Francisco, CA, USA,
  1999. Morgan Kaufmann Publishers Inc.

\bibitem[Cooray et~al.(2015)Cooray, Sengupta, Douglas, and Friston]{Cooray2015}
G.~K. Cooray, B.~Sengupta, P.~Douglas, and K.~Friston.
\newblock Dynamic causal modelling of electrographic seizure activity using
  {B}ayesian belief updating.
\newblock \emph{Neuroimage}, July 2015.

\bibitem[Copernicus et~al.(1965)Copernicus, Lerner, Segonds, Verdet, Luna,
  Savoie, and Toulmonde]{Copernicus1543}
N.~Copernicus, M.-P. Lerner, A.~P. Segonds, J.-P. Verdet, C.~Luna, D.~Savoie,
  and M.~Toulmonde.
\newblock \emph{De revolutionibus orbium coelestium}, volume~1.
\newblock Johnson Reprint Corporation, 1965.

\bibitem[Cornelio et~al.(2023)Cornelio, Dash, Austel, Josephson, Goncalves,
  Clarkson, Megiddo, Khadir, and Horesh]{Cornelio2023:AI-Descartes}
C.~Cornelio, S.~Dash, V.~Austel, T.~R. Josephson, J.~Goncalves, K.~L. Clarkson,
  N.~Megiddo, B.~E. Khadir, and L.~Horesh.
\newblock Combining data and theory for derivable scientific discovery with
  {AI-Descartes}.
\newblock \emph{Nature Communications}, 14:\penalty0 1777, 2023.

\bibitem[Correa et~al.(2021)Correa, Lee, and Bareinboim]{correa2021nested}
J.~Correa, S.~Lee, and E.~Bareinboim.
\newblock Nested counterfactual identification from arbitrary surrogate
  experiments.
\newblock In M.~Ranzato, A.~Beygelzimer, Y.~Dauphin, P.~Liang, and J.~W.
  Vaughan, editors, \emph{Advances in Neural Information Processing Systems 34
  (NeurIPS 2021)}, volume~34, pages 6856--6867, 2021.

\bibitem[Cramer(1985)]{icga85:cramer}
N.~L. Cramer.
\newblock A representation for the adaptive generation of simple sequential
  programs.
\newblock In J.~J. Grefenstette, editor, \emph{Proceedings of an International
  Conference on Genetic Algorithms and the Applications}, pages 183--187,
  Carnegie-Mellon University, Pittsburgh, PA, USA, 1985.

\bibitem[Cranmer(2020)]{pysr}
M.~Cranmer.
\newblock {PySR: Fast \& Parallelized Symbolic Regression in Python/Julia},
  Sept. 2020.

\bibitem[Cranmer et~al.(2020)Cranmer, Sanchez-Gonzalez, Battaglia, Xu, Cranmer,
  Spergel, and Ho]{cranmer_et_al}
M.~Cranmer, A.~Sanchez-Gonzalez, P.~Battaglia, R.~Xu, K.~Cranmer, D.~Spergel,
  and S.~Ho.
\newblock Discovering symbolic models from deep learning with inductive biases.
\newblock In \emph{Advances in Neural Information Processing Systems (NeurIPS
  2020)}, 2020.

\bibitem[Cremades et~al.(2023)Cremades, Hoyas, Quintero, Lellep, Linkmann, and
  Vinuesa]{cremades_et_al}
A.~Cremades, S.~Hoyas, P.~Quintero, M.~Lellep, M.~Linkmann, and R.~Vinuesa.
\newblock Explaining wall-bounded turbulence through deep learning.
\newblock \emph{Preprint arXiv:2302.01250}, 2023.

\bibitem[Crutchfield and McNamara(1987)]{Crutchfield_et_al}
J.~Crutchfield and B.~McNamara.
\newblock Equations of motion from a data series.
\newblock \emph{Complex Syst.}, 1:\penalty0 417--452, 1987.

\bibitem[Cruzat et~al.(2023)Cruzat, Herzog, Prado, Sanz-Perl, Gonzalez-Gomez,
  Moguilner, Kringelbach, Deco, Tagliazucchi, and Iba{\~n}ez]{cruz23}
J.~Cruzat, R.~Herzog, P.~Prado, Y.~Sanz-Perl, R.~Gonzalez-Gomez, S.~Moguilner,
  M.~L. Kringelbach, G.~Deco, E.~Tagliazucchi, and A.~Iba{\~n}ez.
\newblock Temporal irreversibility of large-scale brain dynamics in
  {A}lzheimer's disease.
\newblock \emph{Journal of Neuroscience}, 43\penalty0 (9):\penalty0 1643--1656,
  2023.

\bibitem[Cunningham and Yu(2014)]{cunn14}
J.~P. Cunningham and B.~M. Yu.
\newblock Dimensionality reduction for large-scale neural recordings.
\newblock \emph{Nat. Neurosci.}, 17\penalty0 (11):\penalty0 1500--1509, Nov.
  2014.

\bibitem[Dahlhaus and Eichler(2003)]{dahlhaus2003causality}
R.~Dahlhaus and M.~Eichler.
\newblock Causality and graphical models in time series analysis.
\newblock \emph{Oxford Stat. Sci. Ser}, 27, 01 2003.

\bibitem[Daniels and Nemenman(2015)]{daniels2015automated}
B.~C. Daniels and I.~Nemenman.
\newblock Automated adaptive inference of phenomenological dynamical models.
\newblock \emph{Nature communications}, 6:\penalty0 8133, 2015.

\bibitem[Daniu\v{s}is et~al.(2010)Daniu\v{s}is, Janzing, Mooij, Zscheischler,
  Steudel, Zhang, and Sch\"{o}lkopf]{daniusis2010inferring}
P.~Daniu\v{s}is, D.~Janzing, J.~Mooij, J.~Zscheischler, B.~Steudel, K.~Zhang,
  and B.~Sch\"{o}lkopf.
\newblock Inferring deterministic causal relations.
\newblock In \emph{Proceedings of the Twenty-Sixth Conference on Uncertainty in
  Artificial Intelligence}, Uai'10, page 143–150, Arlington, Virginia, USA,
  2010. AUAI Press.
\newblock ISBN 9780974903965.

\bibitem[Darwin(1859)]{Darwin1859}
C.~Darwin.
\newblock \emph{On the Origin of Species}.
\newblock John Murray, London, 1859.

\bibitem[{DataRobot Inc}(2023)]{datarobot}
{DataRobot Inc}.
\newblock Eureqa as part of {DataRobot}'s service.
\newblock \url{https://www.datarobot.com/nutonian/}, 2023.

\bibitem[Davis et~al.(2014)Davis, LaRocque, Mumford, Norman, Wagner, and
  Poldrack]{Davis2014}
T.~Davis, K.~F. LaRocque, J.~A. Mumford, K.~A. Norman, A.~D. Wagner, and R.~A.
  Poldrack.
\newblock What do differences between multi-voxel and univariate analysis mean?
  {H}ow subject-, voxel-, and trial-level variance impact {fMRI} analysis.
\newblock \emph{NeuroImage}, 97:\penalty0 271--283, 2014.

\bibitem[Deistler et~al.(2022)Deistler, Macke, and Gonçalves]{dein22}
M.~Deistler, J.~H. Macke, and P.~J. Gonçalves.
\newblock Energy-efficient network activity from disparate circuit parameters.
\newblock \emph{Proceedings of the National Academy of Sciences}, 119\penalty0
  (44):\penalty0 e2207632119, 2022.

\bibitem[del \'Alamo et~al.(2006)del \'Alamo, Jim\'enez, Zandonade, and
  Moser]{del_alamo}
J.~C. del \'Alamo, J.~Jim\'enez, P.~Zandonade, and R.~D. Moser.
\newblock Self-similar vortex clusters in the turbulent logarithmic region.
\newblock \emph{J. Fluid Mech.}, 561:\penalty0 329--358, 2006.

\bibitem[Deng et~al.(2020)Deng, Noack, Morzynski, and Pastur]{sindy3}
N.~Deng, B.~R. Noack, M.~Morzynski, and L.~R. Pastur.
\newblock {Low-order model for successive bifurcations of the fluidic pinball}.
\newblock \emph{J. Fluid Mech.}, 884:\penalty0 A37, 2020.

\bibitem[Dennett(1991)]{dennett1991real}
D.~C. Dennett.
\newblock Real patterns.
\newblock \emph{The journal of Philosophy}, 88\penalty0 (1):\penalty0 27--51,
  1991.

\bibitem[Di~Capua et~al.(2019)Di~Capua, Kretschmer, Runge, Alessandri, Donner,
  van Den~Hurk, Vellore, Krishnan, and Coumou]{di2019long}
G.~Di~Capua, M.~Kretschmer, J.~Runge, A.~Alessandri, R.~Donner, B.~van
  Den~Hurk, R.~Vellore, R.~Krishnan, and D.~Coumou.
\newblock Long-lead statistical forecasts of the indian summer monsoon rainfall
  based on causal precursors.
\newblock \emph{Weather and Forecasting}, 34\penalty0 (5):\penalty0 1377--1394,
  2019.

\bibitem[Diaz et~al.(2023)Diaz, Johnson, Varando, and
  Camps-Valls]{Diaz22noiseimputation}
A.~Diaz, J.~Johnson, G.~Varando, and G.~Camps-Valls.
\newblock Learning latent functions for causal discovery.
\newblock \emph{Submitted}, 2023.

\bibitem[Diaz et~al.(2022)Diaz, Adsuara, Moreno-Martinez, Piles, and
  Camps-Valls]{Diaz21rccm}
E.~Diaz, J.~Adsuara, A.~Moreno-Martinez, M.~Piles, and G.~Camps-Valls.
\newblock Inferring causal relations from observational long-term carbon and
  water fluxes records.
\newblock \emph{Scientific Reports}, 12:\penalty0 1610, 2022.

\bibitem[Dickerson(2003)]{dickerson2003kant}
A.~B. Dickerson.
\newblock \emph{Kant on representation and objectivity}.
\newblock Cambridge University Press, 2003.

\bibitem[Didelez(2006)]{didelez2006asymmetric}
V.~Didelez.
\newblock Asymmetric separation for local independence graphs.
\newblock In \emph{23rd Annual Conference on Uncertainty in Artifical
  Intelligence}, 2006.

\bibitem[Didelez(2008)]{didelez2008graphical}
V.~Didelez.
\newblock Graphical models for marked point processes based on local
  independence.
\newblock \emph{Journal of the Royal Statistical Society. Series B (Statistical
  Methodology)}, 70\penalty0 (1):\penalty0 245--264, 2008.

\bibitem[Diego~Bueso(2020)]{bueso2020explicit}
G.~C.-V. Diego~Bueso, Maria~Piles.
\newblock Explicit {G}ranger causality in kernel {H}ilbert spaces.
\newblock \emph{Physical Review E}, 102:\penalty0 062201, 2020.

\bibitem[Diks and Panchenko(2006)]{diks2006new}
C.~Diks and V.~Panchenko.
\newblock A new statistic and practical guidelines for nonparametric {G}ranger
  causality testing.
\newblock \emph{Journal of Economic Dynamics and Control}, 30\penalty0
  (9-10):\penalty0 1647--1669, 2006.

\bibitem[Ding et~al.(2006)Ding, Chen, and Bressler]{ding2006granger}
M.~Ding, Y.~Chen, and S.~L. Bressler.
\newblock Granger causality: {B}asic theory and application to neuroscience.
\newblock \emph{Handbook of time series analysis: recent theoretical
  developments and applications}, pages 437--460, 2006.

\bibitem[Donges et~al.(2009{\natexlab{a}})Donges, Zou, Marwan, and
  Kurths]{Donges2009backbone}
J.~Donges, Y.~Zou, N.~Marwan, and J.~Kurths.
\newblock The backbone of the climate network.
\newblock \emph{Epl}, 87:\penalty0 48007, 2009{\natexlab{a}}.

\bibitem[Donges et~al.(2009{\natexlab{b}})Donges, Zou, Marwan, and
  Kurths]{donges2009complex}
J.~F. Donges, Y.~Zou, N.~Marwan, and J.~Kurths.
\newblock Complex networks in climate dynamics.
\newblock \emph{The European Physical Journal Special Topics}, 174\penalty0
  (1):\penalty0 157--179, 2009{\natexlab{b}}.

\bibitem[Dowd(2011)]{dowd2011separated}
B.~E. Dowd.
\newblock Separated at birth: {S}tatisticians, social scientists, and causality
  in health services research.
\newblock \emph{Health Services Research}, 46\penalty0 (2):\penalty0 397--420,
  2011.

\bibitem[Ducasse(1951)]{ducasse1951whewell}
C.~J. Ducasse.
\newblock Whewell's philosophy of scientific discovery. {II}.
\newblock \emph{The Philosophical Review}, 60\penalty0 (2):\penalty0 213--234,
  1951.

\bibitem[Duncker and Sahani(2018)]{dunc18}
L.~Duncker and M.~Sahani.
\newblock Temporal alignment and latent {G}aussian process factor inference in
  population spike trains.
\newblock In S.~Bengio, H.~Wallach, H.~Larochelle, K.~Grauman, N.~Cesa-Bianchi,
  and R.~Garnett, editors, \emph{Advances in Neural Information Processing
  Systems}, volume~31. Curran Associates, Inc., 2018.

\bibitem[Duncker and Sahani(2021)]{dunc21}
L.~Duncker and M.~Sahani.
\newblock Dynamics on the manifold: {I}dentifying computational dynamical
  activity from neural population recordings.
\newblock \emph{Current Opinion in Neurobiology}, 70:\penalty0 163--170, 2021.
\newblock Computational Neuroscience.

\bibitem[Durstewitz(2017)]{durs17}
D.~Durstewitz.
\newblock A state space approach for piecewise-linear recurrent neural networks
  for identifying computational dynamics from neural measurements.
\newblock \emph{PLOS Computational Biology}, 13\penalty0 (6):\penalty0 1--33,
  06 2017.

\bibitem[Durstewitz and Seamans(2008)]{durs08}
D.~Durstewitz and J.~K. Seamans.
\newblock The dual-state theory of prefrontal cortex dopamine function with
  relevance to catechol-o-methyltransferase genotypes and schizophrenia.
\newblock \emph{Biological Psychiatry}, 64\penalty0 (9):\penalty0 739--749,
  2008.
\newblock Neurodevelopment and the Transition from Schizophrenia Prodrome to
  Schizophrenia.

\bibitem[Dzeroski and Todorovski(2007)]{dzeroski2007}
S.~Dzeroski and L.~Todorovski.
\newblock \emph{Inductive Logic Programming: {T}echniques and Applications}.
\newblock Springer Science+Business Media, 2007.

\bibitem[Džeroski and Todorovski(1995)]{dzeroski1995}
S.~Džeroski and L.~Todorovski.
\newblock Reliable induction of recursive production rules.
\newblock \emph{Machine Learning}, 20\penalty0 (3):\penalty0 229--256, 1995.

\bibitem[Ebert-Uphoff and Deng(2017)]{ebert2017causal}
I.~Ebert-Uphoff and Y.~Deng.
\newblock Causal discovery in the geosciences—{U}sing synthetic data to learn
  how to interpret results.
\newblock \emph{Computers \& Geosciences}, 99:\penalty0 50--60, 2017.

\bibitem[Eivazi et~al.(2021)Eivazi, Guastoni, Schlatter, Azizpour, and
  Vinuesa]{eivazi2020recurrent}
H.~Eivazi, L.~Guastoni, P.~Schlatter, H.~Azizpour, and R.~Vinuesa.
\newblock Recurrent neural networks and {K}oopman-based frameworks for temporal
  predictions in a low-order model of turbulence.
\newblock \emph{Int. J. Heat Fluid Flow}, 90:\penalty0 108816, 2021.

\bibitem[Eivazi et~al.(2022)Eivazi, Le~Clainche, Hoyas, and Vinuesa]{r3}
H.~Eivazi, S.~Le~Clainche, S.~Hoyas, and R.~Vinuesa.
\newblock {Towards extraction of orthogonal and parsimonious non-linear modes
  from turbulent flows}.
\newblock \emph{Expert Syst. Appl.}, 202:\penalty0 117038, 2022.

\bibitem[Ellis et~al.(2021)Ellis, Wong, Nye, Sablé-Meyer, Morales, Hewitt,
  Cary, Solar-Lezama, and Tenenbaum]{ellis2021:Dreamcoder}
K.~Ellis, C.~Wong, M.~Nye, M.~Sablé-Meyer, L.~Morales, L.~Hewitt, L.~Cary,
  A.~Solar-Lezama, and J.~B. Tenenbaum.
\newblock {DreamCoder}: {B}ootstrapping inductive program synthesis with
  wake-sleep library learning.
\newblock In \emph{Proceedings of the 42nd {ACM} {SIGPLAN} {International}
  {Conference} on {Programming} {Language} {Design} and {Implementation}},
  {Pldi} 2021, pages 835--850, New York, NY, USA, June 2021. Association for
  Computing Machinery.
\newblock ISBN 978-1-4503-8391-2.

\bibitem[Entner and Hoyer(2010)]{enter2010causal}
D.~Entner and P.~O. Hoyer.
\newblock On causal discovery from time series data using {FCI}.
\newblock In P.~Myllym{\"a}ki, T.~Roos, and T.~Jaakkola, editors,
  \emph{Proceedings of the 5th European Workshop on Probabilistic Graphical
  Models}, pages 121--128, Helsinki, FI, 2010. Helsinki Institute for
  Information Technology HIIT.

\bibitem[Erichson et~al.(2020)Erichson, Mathelin, Yao, Brunton, Mahoney, and
  Kutz]{r5}
N.~B. Erichson, L.~Mathelin, Z.~Yao, S.~L. Brunton, M.~W. Mahoney, and J.~N.
  Kutz.
\newblock {Shallow neural networks for fluid flow reconstruction with limited
  sensors}.
\newblock \emph{Proc. R. Soc. Lond. A}, 476:\penalty0 20200097, 2020.

\bibitem[Evans and Rzhetsky(2010)]{evans2010machine}
J.~Evans and A.~Rzhetsky.
\newblock Machine science.
\newblock \emph{Science}, 329\penalty0 (5990):\penalty0 399--400, 2010.

\bibitem[Evans(1979)]{Evans79:DFT}
R.~Evans.
\newblock The nature of the liquid-vapour interface and other topics in the
  statistical mechanics of non-uniform, classical fluids.
\newblock \emph{Advances in Physics}, 28\penalty0 (2):\penalty0 143--200, 1979.

\bibitem[Eyring et~al.(2016)Eyring, Bony, Meehl, Senior, Stevens, Stouffer, and
  Taylor]{eyring2016overview}
V.~Eyring, S.~Bony, G.~A. Meehl, C.~A. Senior, B.~Stevens, R.~J. Stouffer, and
  K.~E. Taylor.
\newblock Overview of the {Coupled Model Intercomparison Project Phase 6
  (CMIP6)} experimental design and organization.
\newblock \emph{Geoscientific Model Development}, 9\penalty0 (5):\penalty0
  1937--1958, 2016.

\bibitem[Eyring et~al.(2019)Eyring, Cox, Flato, Gleckler, Abramowitz, Caldwell,
  Collins, Gier, Hall, Hoffman, et~al.]{eyring2019taking}
V.~Eyring, P.~M. Cox, G.~M. Flato, P.~J. Gleckler, G.~Abramowitz, P.~Caldwell,
  W.~D. Collins, B.~K. Gier, A.~D. Hall, F.~M. Hoffman, et~al.
\newblock Taking climate model evaluation to the next level.
\newblock \emph{Nature Climate Change}, 9\penalty0 (2):\penalty0 102--110,
  2019.

\bibitem[Eyring et~al.(2020)Eyring, Bock, Lauer, Righi, Schlund, Andela,
  Arnone, Bellprat, Br{\"o}tz, Caron, et~al.]{eyring2020earth}
V.~Eyring, L.~Bock, A.~Lauer, M.~Righi, M.~Schlund, B.~Andela, E.~Arnone,
  O.~Bellprat, B.~Br{\"o}tz, L.-P. Caron, et~al.
\newblock {Earth System Model Evaluation Tool (ESMValTool) v2. 0}--{A}n
  extended set of large-scale diagnostics for quasi-operational and
  comprehensive evaluation of earth system models in {CMIP}.
\newblock \emph{Geoscientific Model Development}, 13\penalty0 (7):\penalty0
  3383--3438, 2020.

\bibitem[Faes et~al.(2012)Faes, Erla, and Nollo]{faes2012measuring}
L.~Faes, S.~Erla, and G.~Nollo.
\newblock Measuring connectivity in linear multivariate processes: definitions,
  interpretation, and practical analysis.
\newblock \emph{Computational and mathematical methods in medicine}, 2012,
  2012.

\bibitem[Fajardo-Fontiveros et~al.(2023)Fajardo-Fontiveros, Reichardt,
  De~Los~R{\'\i}os, Duch, Sales-Pardo, and Guimer{\`a}]{fajardo2023fundamental}
O.~Fajardo-Fontiveros, I.~Reichardt, H.~R. De~Los~R{\'\i}os, J.~Duch,
  M.~Sales-Pardo, and R.~Guimer{\`a}.
\newblock Fundamental limits to learning closed-form mathematical models from
  data.
\newblock \emph{Nature Communications}, 14\penalty0 (1):\penalty0 1043, 2023.

\bibitem[Falkenhainer and Michalski(1986)]{falkenhainer1986}
B.~Falkenhainer and R.~Michalski.
\newblock The structure mapping engine: {A}lgorithm and examples.
\newblock \emph{Artificial Intelligence}, 32\penalty0 (1):\penalty0 1--63,
  1986.

\bibitem[Feigenbaum et~al.(1971)Feigenbaum, Buchanan, and
  Lederberg]{feigenbaum1971}
E.~Feigenbaum, B.~Buchanan, and J.~Lederberg.
\newblock {The DENDRAL Project}.
\newblock \emph{AI Magazine}, 2:\penalty0 37--46, 1971.

\bibitem[Feyerabend(1981)]{feyerabend1965problems}
P.~K. Feyerabend.
\newblock \emph{Problems of Empiricism}, volume~2.
\newblock Cambridge University Press, 1981.

\bibitem[Fiedor(2014)]{fiedor2014networks}
P.~Fiedor.
\newblock Networks in financial markets based on the mutual information rate.
\newblock \emph{Physical Review E}, 89\penalty0 (5):\penalty0 052801, 2014.

\bibitem[Field et~al.(1995)Field, Jackson, and Mooney]{field1995stomatal}
C.~B. Field, R.~B. Jackson, and H.~A. Mooney.
\newblock Stomatal responses to increased {CO2}: {I}mplications from the plant
  to the global scale.
\newblock \emph{Plant, Cell \& Environment}, 18\penalty0 (10):\penalty0
  1214--1225, 1995.

\bibitem[Fisher(1935)]{fisher1935design}
R.~A. Fisher.
\newblock \emph{The Design of Experiments}.
\newblock Hafner Press, 1935.

\bibitem[Forr{\'e} and Mooij(2018)]{forre2018constraint}
P.~Forr{\'e} and J.~M. Mooij.
\newblock Constraint-based causal discovery for non-linear structural causal
  models with cycles and latent confounders.
\newblock In A.~Globerson and R.~Silva, editors, \emph{Proceedings of the 34th
  Conference on Uncertainty in Artificial Intelligence (UAI-18)}. AUAI Press,
  2018.

\bibitem[Fortunato et~al.(2018)Fortunato, Bergstrom, B{\"o}rner, Evans,
  Helbing, Milojevi{\'c}, Petersen, Radicchi, Sinatra, Uzzi,
  et~al.]{fortunato2018science}
S.~Fortunato, C.~T. Bergstrom, K.~B{\"o}rner, J.~A. Evans, D.~Helbing,
  S.~Milojevi{\'c}, A.~M. Petersen, F.~Radicchi, R.~Sinatra, B.~Uzzi, et~al.
\newblock Science of science.
\newblock \emph{Science}, 359\penalty0 (6379):\penalty0 eaao0185, 2018.

\bibitem[Freeman(1977)]{freeman1977set}
L.~C. Freeman.
\newblock A set of measures of centrality based on betweenness.
\newblock \emph{Sociometry}, pages 35--41, 1977.

\bibitem[Friston(2005)]{fris05}
K.~Friston.
\newblock A theory of cortical responses.
\newblock \emph{Philosophical Transactions of the Royal Society B: Biological
  Sciences}, 360\penalty0 (1456):\penalty0 815--836, 2005.

\bibitem[Friston et~al.(2013)Friston, Moran, and Seth]{friston2013analysing}
K.~Friston, R.~Moran, and A.~K. Seth.
\newblock Analysing connectivity with {G}ranger causality and dynamic causal
  modelling.
\newblock \emph{Current opinion in neurobiology}, 23\penalty0 (2):\penalty0
  172--178, 2013.

\bibitem[Friston et~al.(2019)Friston, Preller, Mathys, Cagnan, Heinzle, Razi,
  and Zeidman]{fris19}
K.~Friston, K.~H. Preller, C.~Mathys, H.~Cagnan, J.~Heinzle, A.~Razi, and
  P.~Zeidman.
\newblock Dynamic causal modelling revisited.
\newblock \emph{NeuroImage}, 199:\penalty0 730--744, 2019.

\bibitem[Friston et~al.(2003)Friston, Harrison, and Penny]{friston2003dynamic}
K.~J. Friston, L.~Harrison, and W.~Penny.
\newblock Dynamic causal modelling.
\newblock \emph{Neuroimage}, 19\penalty0 (4):\penalty0 1273--1302, 2003.

\bibitem[Friston et~al.(2020)Friston, Parr, Zeidman, Razi, Flandin, Daunizeau,
  Hulme, Billig, Litvak, Moran, et~al.]{friston2020dynamic}
K.~J. Friston, T.~Parr, P.~Zeidman, A.~Razi, G.~Flandin, J.~Daunizeau, O.~J.
  Hulme, A.~J. Billig, V.~Litvak, R.~J. Moran, et~al.
\newblock Dynamic causal modelling of {COVID-19}.
\newblock \emph{Wellcome open research}, 5, 2020.

\bibitem[Friston et~al.(2022)Friston, Flandin, and Razi]{friston2022dynamic}
K.~J. Friston, G.~Flandin, and A.~Razi.
\newblock Dynamic causal modelling of c{OVID-19} and its mitigations.
\newblock \emph{Scientific reports}, 12\penalty0 (1):\penalty0 12419, 2022.

\bibitem[Fuentes et~al.(2021)Fuentes, Nayek, Gardner, Dervilis, Rogers, Worden,
  and Cross]{fuentes2021equation}
R.~Fuentes, R.~Nayek, P.~Gardner, N.~Dervilis, T.~Rogers, K.~Worden, and
  E.~Cross.
\newblock Equation discovery for nonlinear dynamical systems: {A} {B}ayesian
  viewpoint.
\newblock \emph{Mechanical Systems and Signal Processing}, 154:\penalty0
  107528, 2021.

\bibitem[Fukami et~al.(2020)Fukami, Nakamura, and Fukagata]{r10}
K.~Fukami, T.~Nakamura, and K.~Fukagata.
\newblock {Convolutional neural network based hierarchical autoencoder for
  nonlinear mode decomposition of fluid field data}.
\newblock \emph{Phys. Fluids}, 32:\penalty0 095110, 2020.

\bibitem[Gain and Shpitser(2018)]{gain2018structure}
A.~Gain and I.~Shpitser.
\newblock Structure learning under missing data.
\newblock In \emph{International conference on probabilistic graphical models},
  pages 121--132. Pmlr, 2018.

\bibitem[Galgali et~al.(2023)Galgali, Sahani, and Mante]{galg23}
A.~R. Galgali, M.~Sahani, and V.~Mante.
\newblock Residual dynamics resolves recurrent contributions to neural
  computation.
\newblock \emph{Nature Neuroscience}, 26\penalty0 (2):\penalty0 326--338, Feb.
  2023.

\bibitem[Gao et~al.(2022)Gao, Bhattacharjya, Nelson, Liu, and Yu]{gao22IDYNO}
T.~Gao, D.~Bhattacharjya, E.~Nelson, M.~Liu, and Y.~Yu.
\newblock {IDYNO}: {L}earning nonparametric {DAGs} from interventional dynamic
  data.
\newblock In K.~Chaudhuri, S.~Jegelka, L.~Song, C.~Szepesvari, G.~Niu, and
  S.~Sabato, editors, \emph{Proceedings of the 39th International Conference on
  Machine Learning}, volume 162 of \emph{Proceedings of Machine Learning
  Research}, pages 6988--7001. Pmlr, 17--23 Jul 2022.

\bibitem[Geiger et~al.(1990)Geiger, Verma, and Pearl]{geiger1990identifying}
D.~Geiger, T.~Verma, and J.~Pearl.
\newblock Identifying independence in {B}ayesian networks.
\newblock \emph{Networks}, 20\penalty0 (5):\penalty0 507--534, 1990.

\bibitem[Gel\ss et~al.(2019)Gel\ss, Klus, Eisert, and Sch\"utte]{curse}
P.~Gel\ss, S.~Klus, J.~Eisert, and C.~Sch\"utte.
\newblock Multidimensional approximation of nonlinear dynamical systems.
\newblock \emph{J. Comput. Nonlinear Dyn.}, 14:\penalty0 061006, 2019.

\bibitem[Genkin and Engel(2020)]{genk20}
M.~Genkin and T.~A. Engel.
\newblock Moving beyond generalization to accurate interpretation of flexible
  models.
\newblock \emph{Nat. Mach. Intell.}, 2\penalty0 (11):\penalty0 674--683, Nov.
  2020.

\bibitem[Genkin et~al.(2021)Genkin, Hughes, and Engel]{genk21}
M.~Genkin, O.~Hughes, and T.~A. Engel.
\newblock Learning non-stationary {Langevin} dynamics from stochastic
  observations of latent trajectories.
\newblock \emph{Nature Communications}, 12\penalty0 (1):\penalty0 5986, 2021.

\bibitem[Gerhardus(2021)]{gerhardus2021characterisation}
A.~Gerhardus.
\newblock Characterization of causal ancestral graphs for time series with
  latent confounders.
\newblock \emph{Preprint arXiv:2112.08417}, 2021.

\bibitem[Gerhardus and Runge(2020)]{gerhardus2020high}
A.~Gerhardus and J.~Runge.
\newblock High-recall causal discovery for autocorrelated time series with
  latent confounders.
\newblock \emph{Advances in Neural Information Processing Systems},
  33:\penalty0 12615--12625, 2020.

\bibitem[Gerstner et~al.(2014)Gerstner, Kistler, Naud, and Paninski]{gern14}
W.~Gerstner, W.~M. Kistler, R.~Naud, and L.~Paninski.
\newblock \emph{Neuronal Dynamics: {F}rom Single Neurons to Networks and Models
  of Cognition}.
\newblock Cambridge University Press, 2014.

\bibitem[Geweke(1982)]{geweke1982measurement}
J.~Geweke.
\newblock Measurement of linear dependence and feedback between multiple time
  series.
\newblock \emph{Journal of the American statistical association}, 77\penalty0
  (378):\penalty0 304--313, 1982.

\bibitem[Gillies(1998)]{gillies1996artificial}
D.~Gillies.
\newblock Artificial intelligence and scientific method.
\newblock \emph{Mind}, 107\penalty0 (428), 1998.

\bibitem[Glass et~al.(2013)Glass, Goodman, Hern{\'a}n, and
  Samet]{glass2013causal}
T.~A. Glass, S.~N. Goodman, M.~A. Hern{\'a}n, and J.~M. Samet.
\newblock Causal inference in public health.
\newblock \emph{Annual review of public health}, 34:\penalty0 61--75, 2013.

\bibitem[Glymour et~al.(2019)Glymour, Zhang, and Spirtes]{glymour2019review}
C.~Glymour, K.~Zhang, and P.~Spirtes.
\newblock Review of causal discovery methods based on graphical models.
\newblock \emph{Frontiers in Genetics}, 10, 2019.

\bibitem[G{\"o}del(1931)]{godel1931formal}
K.~G{\"o}del.
\newblock {\"Uber formal unentscheidbare S\"atze der Principia Mathematica und
  verwandter Systeme I}.
\newblock \emph{Monatshefte f{\"u}r mathematik und physik}, 38\penalty0
  (1):\penalty0 173--198, 1931.

\bibitem[Gokcen et~al.(2022)Gokcen, Jasper, Semedo, Zandvakili, Kohn, Machens,
  and Yu]{gokc22}
E.~Gokcen, A.~I. Jasper, J.~D. Semedo, A.~Zandvakili, A.~Kohn, C.~K. Machens,
  and B.~M. Yu.
\newblock Disentangling the flow of signals between populations of neurons.
\newblock \emph{Nat. Comput. Sci.}, 2\penalty0 (8):\penalty0 512--525, Aug.
  2022.

\bibitem[Gong et~al.(2023)Gong, Jennings, Zhang, and Pawlowski]{gong2023rhino}
W.~Gong, J.~Jennings, C.~Zhang, and N.~Pawlowski.
\newblock Rhino: {D}eep causal temporal relationship learning with
  history-dependent noise.
\newblock In \emph{The Eleventh International Conference on Learning
  Representations}, 2023.

\bibitem[Gonçalves et~al.(2020)Gonçalves, Lueckmann, Deistler, Nonnenmacher,
  Öcal, Bassetto, Chintaluri, Podlaski, Haddad, Vogels, Greenberg, and
  Macke]{gonc20}
P.~J. Gonçalves, J.-M. Lueckmann, M.~Deistler, M.~Nonnenmacher, K.~Öcal,
  G.~Bassetto, C.~Chintaluri, W.~F. Podlaski, S.~A. Haddad, T.~P. Vogels, D.~S.
  Greenberg, and J.~H. Macke.
\newblock Training deep neural density estimators to identify mechanistic
  models of neural dynamics.
\newblock \emph{eLife}, 9:\penalty0 e56261, Sept. 2020.

\bibitem[Goodwell et~al.(2020)Goodwell, Jiang, Ruddell, and
  Kumar]{goodwell2020debates}
A.~E. Goodwell, P.~Jiang, B.~L. Ruddell, and P.~Kumar.
\newblock Debates—{D}oes information theory provide a new paradigm for
  {E}arth science? {C}ausality, interaction, and feedback.
\newblock \emph{Water Resources Research}, 56\penalty0 (2):\penalty0
  e2019WR024940, 2020.

\bibitem[Gordon et~al.(1994)Gordon, Moore, and Carlson]{gordon1994}
A.~Gordon, A.~Moore, and A.~Carlson.
\newblock Using genetic algorithms to discover good representations.
\newblock \emph{Machine Learning}, 15\penalty0 (1):\penalty0 239--263, 1994.

\bibitem[Gozolchiani et~al.(2011)Gozolchiani, Havlin, and
  Yamasaki]{Gozolchiani2011}
A.~Gozolchiani, S.~Havlin, and K.~Yamasaki.
\newblock Emergence of {El Ni\~{n}o} as an autonomous component in the climate
  network.
\newblock \emph{Phys. Rev. Lett.}, 107\penalty0 (14):\penalty0 148501, 2011.

\bibitem[Gradu et~al.(2022)Gradu, Zrnic, Wang, and Jordan]{gradu2022valid}
P.~Gradu, T.~Zrnic, Y.~Wang, and M.~Jordan.
\newblock Valid inference after causal discovery.
\newblock In \emph{NeurIPS 2022 Workshop on Causality for Real-world Impact},
  2022.

\bibitem[Granger(1969)]{granger1969investigating}
C.~W.~J. Granger.
\newblock Investigating causal relations by econometric models and
  cross-spectral methods.
\newblock \emph{Econometrica: journal of the Econometric Society}, 37:\penalty0
  424--438, 1969.

\bibitem[Granger(1980)]{granger1980testing}
C.~W.~J. Granger.
\newblock Testing for causality: {A} personal viewpoint.
\newblock \emph{Journal of Economic Dynamics and control}, 2:\penalty0
  329--352, 1980.

\bibitem[Green et~al.(2019)Green, Seneviratne, Berg, Findell, Hagemann,
  Lawrence, and Gentine]{green2019large}
J.~K. Green, S.~I. Seneviratne, A.~M. Berg, K.~L. Findell, S.~Hagemann, D.~M.
  Lawrence, and P.~Gentine.
\newblock Large influence of soil moisture on long-term terrestrial carbon
  uptake.
\newblock \emph{Nature}, 565\penalty0 (7740):\penalty0 476--479, 2019.

\bibitem[Gretton et~al.(2007)Gretton, Fukumizu, Teo, Song, Sch{\"o}lkopf, and
  Smola]{gretton2007kernel}
A.~Gretton, K.~Fukumizu, C.~Teo, L.~Song, B.~Sch{\"o}lkopf, and A.~Smola.
\newblock A kernel statistical test of independence.
\newblock \emph{Advances in neural information processing systems}, 20, 2007.

\bibitem[Groun et~al.(2022)Groun, Villalba-Orero, Lara-Pezzi, Valero,
  Garicano-Mena, and Le~Clainche]{Blood}
N.~Groun, M.~Villalba-Orero, E.~Lara-Pezzi, E.~Valero, J.~Garicano-Mena, and
  S.~Le~Clainche.
\newblock Higher order dynamic mode decomposition: {F}rom fluid dynamics to
  heart disease analysis.
\newblock \emph{Preprint arXiv:2201.03030}, 2022.

\bibitem[Grunberg and Modigliani(1954)]{grunberg1954predictability}
E.~Grunberg and F.~Modigliani.
\newblock The predictability of social events.
\newblock \emph{Journal of Political Economy}, 62\penalty0 (6):\penalty0
  465--478, 1954.

\bibitem[Guan et~al.(2021)Guan, Brunton, and Novosselov]{sindy2}
Y.~Guan, S.~L. Brunton, and I.~Novosselov.
\newblock {Sparse nonlinear models of chaotic electroconvection}.
\newblock \emph{R. Soc. Open Sci.}, 8\penalty0 (8):\penalty0 202367, 2021.

\bibitem[Guastoni et~al.(2021)Guastoni, G\"uemes, Ianiro, Discetti, Schlatter,
  Azizpour, and Vinuesa]{guastoni2}
L.~Guastoni, A.~G\"uemes, A.~Ianiro, S.~Discetti, P.~Schlatter, H.~Azizpour,
  and R.~Vinuesa.
\newblock Convolutional-network models to predict wall-bounded turbulence from
  wall quantities.
\newblock \emph{J. Fluid Mech.}, 928:\penalty0 A27, 2021.

\bibitem[G\"uemes et~al.(2021)G\"uemes, Discetti, Ianiro, Sirmacek, Azizpour,
  and Vinuesa]{guemes_gans}
A.~G\"uemes, S.~Discetti, A.~Ianiro, B.~Sirmacek, H.~Azizpour, and R.~Vinuesa.
\newblock {From coarse wall measurements to turbulent velocity fields through
  deep learning}.
\newblock \emph{Phys. Fluids}, 33:\penalty0 075121, 2021.

\bibitem[Guimer{\`a} et~al.(2020)Guimer{\`a}, Reichardt, Aguilar-Mogas,
  Massucci, Miranda, Pallar{\`e}s, and Sales-Pardo]{guimera2020bayesian}
R.~Guimer{\`a}, I.~Reichardt, A.~Aguilar-Mogas, F.~A. Massucci, M.~Miranda,
  J.~Pallar{\`e}s, and M.~Sales-Pardo.
\newblock A {Bayesian} machine scientist to aid in the solution of challenging
  scientific problems.
\newblock \emph{Science advances}, 6\penalty0 (5):\penalty0 eaav6971, 2020.

\bibitem[Halpern(2016)]{halpern2016actual}
J.~Y. Halpern.
\newblock \emph{Actual causality}.
\newblock MiT Press, 2016.

\bibitem[Hannart et~al.(2016)Hannart, Pearl, Otto, Naveau, and
  Ghil]{Hannart2016}
A.~Hannart, J.~Pearl, F.~E. Otto, P.~Naveau, and M.~Ghil.
\newblock Causal counterfactual theory for the attribution of weather and
  climate-related events.
\newblock \emph{Bull. Am. Meteorol. Soc.}, 97\penalty0 (1):\penalty0 99--110,
  2016.

\bibitem[Hansen and Sokol(2014)]{hansen2014causal}
N.~Hansen and A.~Sokol.
\newblock Causal interpretation of stochastic differential equations.
\newblock \emph{Electronic Journal of Probability}, 19:\penalty0 1 -- 24, 2014.

\bibitem[Hastie et~al.(2015)Hastie, Tibshirani, and
  Wainwright]{hastie2015statistical}
T.~Hastie, R.~Tibshirani, and M.~Wainwright.
\newblock Statistical learning with sparsity.
\newblock \emph{Monographs on statistics and applied probability},
  143:\penalty0 143, 2015.

\bibitem[Haufe et~al.(2014)Haufe, Meinecke, Görgen, Dähne, Haynes, Blankertz,
  and Bießmann]{Haufe2014}
S.~Haufe, F.~Meinecke, K.~Görgen, S.~Dähne, J.-D. Haynes, B.~Blankertz, and
  F.~Bießmann.
\newblock On the interpretation of weight vectors of linear models in
  multivariate neuroimaging.
\newblock \emph{NeuroImage}, 87:\penalty0 96--110, 2014.

\bibitem[Heinze-Deml et~al.(2018{\natexlab{a}})Heinze-Deml, Maathuis, and
  Meinshausen]{heinze2018causal}
C.~Heinze-Deml, M.~H. Maathuis, and N.~Meinshausen.
\newblock Causal structure learning.
\newblock \emph{Annual Review of Statistics and Its Application}, 5:\penalty0
  371--391, 2018{\natexlab{a}}.

\bibitem[Heinze-Deml et~al.(2018{\natexlab{b}})Heinze-Deml, Peters, and
  Meinshausen]{heinze2018invariant}
C.~Heinze-Deml, J.~Peters, and N.~Meinshausen.
\newblock Invariant causal prediction for nonlinear models.
\newblock \emph{Journal of Causal Inference}, 6\penalty0 (2),
  2018{\natexlab{b}}.

\bibitem[Heisenberg(1925)]{Heisenberg1925}
W.~Heisenberg.
\newblock {\"Uber quantentheoretische Umdeutung kinematischer und mechanischer
  Beziehungen}.
\newblock \emph{Zeitschrift f\"ur Physik}, 33\penalty0 (3):\penalty0 879--893,
  1925.

\bibitem[Hempel(2001)]{hempel2001philosophy}
C.~G. Hempel.
\newblock \emph{The philosophy of {Carl G. Hempel}: {S}tudies in science,
  explanation, and rationality}.
\newblock Oxford University Press, 2001.

\bibitem[Hernan and Robins(2020)]{hernan2020causal}
M.~Hernan and J.~Robins.
\newblock \emph{Causal Inference: What if}.
\newblock Chapman \& Hill/CRC, 2020.

\bibitem[Hern{\'a}n(2018)]{hernan2018c}
M.~A. Hern{\'a}n.
\newblock The {C-word}: {S}cientific euphemisms do not improve causal inference
  from observational data.
\newblock \emph{American journal of public health}, 108\penalty0 (5):\penalty0
  616--619, 2018.

\bibitem[Hicks et~al.(1980)]{hicks1980causality}
J.~Hicks et~al.
\newblock \emph{Causality in Economics}.
\newblock Australian National University Press, 1980.

\bibitem[Hiemstra and Jones(1994)]{hiemstra1994testing}
C.~Hiemstra and J.~D. Jones.
\newblock Testing for linear and nonlinear {G}ranger causality in the stock
  price-volume relation.
\newblock \emph{The Journal of Finance}, 49\penalty0 (5):\penalty0 1639--1664,
  1994.

\bibitem[Hinton and Salakhutdinov(2006)]{r1}
G.~E. Hinton and R.~Salakhutdinov.
\newblock {Reducing the dimensionality of data with neural networks}.
\newblock \emph{Science}, 313:\penalty0 504--507, 2006.

\bibitem[Hoyer et~al.(2008)Hoyer, Janzing, Mooij, Peters, and
  Sch\"{o}lkopf]{hoyer2008nonlinear}
P.~Hoyer, D.~Janzing, J.~M. Mooij, J.~Peters, and B.~Sch\"{o}lkopf.
\newblock Nonlinear causal discovery with additive noise models.
\newblock In D.~Koller, D.~Schuurmans, Y.~Bengio, and L.~Bottou, editors,
  \emph{Advances in Neural Information Processing Systems}, volume~21. Curran
  Associates, Inc., 2008.

\bibitem[Huang et~al.(2015)Huang, Zhang, and
  Sch\"{o}lkopf]{huang2015identification}
B.~Huang, K.~Zhang, and B.~Sch\"{o}lkopf.
\newblock Identification of time-dependent causal model: {A} {Gaussian} process
  treatment.
\newblock In \emph{Proceedings of the 24th International Conference on
  Artificial Intelligence}, Ijcai'15, page 3561–3568. AAAI Press, 2015.
\newblock ISBN 9781577357384.

\bibitem[Huang et~al.(2019)Huang, Zhang, Gong, and Glymour]{pmlr-v97-huang19g}
B.~Huang, K.~Zhang, M.~Gong, and C.~Glymour.
\newblock {Causal Discovery and Forecasting in Nonstationary Environments with
  State-Space Models}.
\newblock In K.~Chaudhuri and R.~Salakhutdinov, editors, \emph{Proceedings of
  the 36th International Conference on Machine Learning}, volume~97 of
  \emph{Proceedings of Machine Learning Research}, pages 2901--2910. Pmlr,
  09--15 Jun 2019.

\bibitem[Huang et~al.(2020)Huang, Zhang, Zhang, Ramsey, Sanchez-Romero,
  Glymour, and Sch{\"o}lkopf]{huang2020causal}
B.~Huang, K.~Zhang, J.~Zhang, J.~Ramsey, R.~Sanchez-Romero, C.~Glymour, and
  B.~Sch{\"o}lkopf.
\newblock Causal discovery from heterogeneous/nonstationary data.
\newblock \emph{The Journal of Machine Learning Research}, 21\penalty0
  (1):\penalty0 3482--3534, 2020.

\bibitem[Huang and Valtorta(2006)]{huang2006pearls}
Y.~Huang and M.~Valtorta.
\newblock Pearl's calculus of intervention is complete.
\newblock In R.~Dechter and T.~Richardson, editors, \emph{Proceedings of the
  Twenty-Second Conference on Uncertainty in Artificial Intelligence}, Uai'06,
  page 217–224, Arlington, Virginia, USA, 2006. AUAI Press.
\newblock ISBN 0974903922.

\bibitem[Hunt et~al.(1998)Hunt, Wray, and Moin]{q_Ref}
J.~C.~R. Hunt, A.~A. Wray, and P.~Moin.
\newblock {Eddies, streams, and convergence zones in turbulent flows}.
\newblock \emph{Center for Turbulence Research (CTR) Proceedings of Summer
  Program}, 1998.

\bibitem[Hyman et~al.(2012)Hyman, Ma, Balaguer-Ballester, Durstewitz, and
  Seamans]{hym12}
J.~M. Hyman, L.~Ma, E.~Balaguer-Ballester, D.~Durstewitz, and J.~K. Seamans.
\newblock Contextual encoding by ensembles of medial prefrontal cortex neurons.
\newblock \emph{Proc. Natl. Acad. Sci. U. S. A.}, 109\penalty0 (13):\penalty0
  5086--5091, Mar. 2012.

\bibitem[Hyttinen et~al.(2012)Hyttinen, Eberhardt, and Hoyer]{hyttinen12a}
A.~Hyttinen, F.~Eberhardt, and P.~O. Hoyer.
\newblock Learning linear cyclic causal models with latent variables.
\newblock \emph{Journal of Machine Learning Research}, 13\penalty0
  (109):\penalty0 3387--3439, 2012.

\bibitem[Hyv{\"a}rinen et~al.(2004)Hyv{\"a}rinen, Karhunen, and Oja]{ICA2001}
A.~Hyv{\"a}rinen, J.~Karhunen, and E.~Oja.
\newblock \emph{Independent Component Analysis}.
\newblock Adaptive and Cognitive Dynamic Systems: Signal Processing, Learning,
  Communications and Control. Wiley, 2004.
\newblock ISBN 9780471464198.

\bibitem[Hyv{\"a}rinen et~al.(2008)Hyv{\"a}rinen, Shimizu, and
  Hoyer]{hyvarinen2008causal}
A.~Hyv{\"a}rinen, S.~Shimizu, and P.~O. Hoyer.
\newblock Causal modelling combining instantaneous and lagged effects: {An}
  identifiable model based on non-{Gaussianity}.
\newblock In \emph{Proceedings of the 25th international conference on Machine
  learning}, pages 424--431, 2008.

\bibitem[Hyv{\"a}rinen et~al.(2010)Hyv{\"a}rinen, Zhang, Shimizu, and
  Hoyer]{hyvarinen2010estimation}
A.~Hyv{\"a}rinen, K.~Zhang, S.~Shimizu, and P.~O. Hoyer.
\newblock Estimation of a structural vector autoregression model using
  non-{Gaussianity}.
\newblock \emph{Journal of Machine Learning Research}, 11\penalty0
  (May):\penalty0 1709--1731, 2010.

\bibitem[Imbens and Rubin(2015)]{imbens2015causal}
G.~W. Imbens and D.~B. Rubin.
\newblock \emph{Causal inference in statistics, social, and biomedical
  sciences}.
\newblock Cambridge University Press, 2015.

\bibitem[Jansen and Held(2014)]{jansen2014parameterizing}
M.~F. Jansen and I.~M. Held.
\newblock Parameterizing subgrid-scale eddy effects using energetically
  consistent backscatter.
\newblock \emph{Ocean Modelling}, 80:\penalty0 36--48, 2014.

\bibitem[Janzing(2007)]{janzing2007causally}
D.~Janzing.
\newblock On causally asymmetric versions of {Occam's Razor} and their relation
  to thermodynamics.
\newblock \emph{arXiv preprint arXiv:0708.3411}, 2007.

\bibitem[Jim\'enez(2012)]{jimenez_cascades}
J.~Jim\'enez.
\newblock Cascades in wall-bounded turbulence.
\newblock \emph{Annu. Rev. Fluid Mech.}, 44:\penalty0 27--45, 2012.

\bibitem[Jim\'enez(2016)]{fluxes}
J.~Jim\'enez.
\newblock Optimal fluxes and {R}eynolds stresses.
\newblock \emph{J. Fluid Mech.}, 809:\penalty0 585--600, 2016.

\bibitem[Jim\'enez(2018{\natexlab{a}})]{jimenez_2d}
J.~Jim\'enez.
\newblock Machine-aided turbulence theory.
\newblock \emph{J. Fluid Mech.}, 854:\penalty0 R1, 2018{\natexlab{a}}.

\bibitem[Jim\'enez(2018{\natexlab{b}})]{jimenez_perspective}
J.~Jim\'enez.
\newblock Coherent structures in wall-bounded turbulence.
\newblock \emph{J. Fluid Mech.}, 842:\penalty0 P1, 2018{\natexlab{b}}.

\bibitem[Johnson et~al.(2007)Johnson, Harris, and
  Williams]{Johnson2007BRAINSFitMI}
H.~Johnson, G.~Harris, and K.~Williams.
\newblock {BRAINSFit}: {M}utual information registrations of whole-brain {3D}
  images, using the {Insight Toolkit}.
\newblock \emph{The Insight Journal}, 2007.

\bibitem[{Johnson} et~al.(2018){Johnson}, {Laparra}, and
  {Camps-Valls}]{Emman2018}
J.~E. {Johnson}, V.~{Laparra}, and G.~{Camps-Valls}.
\newblock Disentangling derivatives, uncertainty and error in {G}aussian
  process models.
\newblock In \emph{IGARSS 2018 - 2018 IEEE International Geoscience and Remote
  Sensing Symposium}, pages 4051--4054, July 2018.

\bibitem[Johnson-Laird and Byrne(1991)]{johnson1991deduction}
P.~N. Johnson-Laird and R.~M. Byrne.
\newblock \emph{Deduction}.
\newblock Lawrence Erlbaum Associates, Inc, 1991.

\bibitem[Juang and Pappa(1985)]{era}
J.~Juang and R.~Pappa.
\newblock An eigensystem realization algorithm for modal parameter
  identification and model reduction.
\newblock \emph{J. Guid. Control Dyn.}, 8:\penalty0 620, 1985.

\bibitem[Jung et~al.(2010)Jung, Reichstein, Ciais, Seneviratne, Sheffield,
  Goulden, Bonan, Cescatti, Chen, De~Jeu, et~al.]{jung2010recent}
M.~Jung, M.~Reichstein, P.~Ciais, S.~I. Seneviratne, J.~Sheffield, M.~L.
  Goulden, G.~Bonan, A.~Cescatti, J.~Chen, R.~De~Jeu, et~al.
\newblock {Recent decline in the global land evapotranspiration trend due to
  limited moisture supply}.
\newblock \emph{Nature}, 467\penalty0 (7318):\penalty0 951--954, 2010.

\bibitem[Kaddour et~al.(2022)Kaddour, Lynch, Liu, Kusner, and
  Silva]{kaddour2022causal}
J.~Kaddour, A.~Lynch, Q.~Liu, M.~J. Kusner, and R.~Silva.
\newblock {Causal machine learning: A survey and open problems}.
\newblock \emph{arXiv preprint arXiv:2206.15475}, 2022.

\bibitem[Kaheman et~al.(2022)Kaheman, Brunton, and
  Kutz]{kaheman2022:SindyNoise}
K.~Kaheman, S.~L. Brunton, and J.~N. Kutz.
\newblock {Automatic differentiation to simultaneously identify nonlinear
  dynamics and extract noise probability distributions from data}.
\newblock \emph{Machine Learning: Science and Technology}, 3\penalty0
  (1):\penalty0 015031, Mar. 2022.
\newblock Publisher: IOP Publishing.

\bibitem[Kaiser et~al.(2020)Kaiser, Kutz, and Brunton]{kaiser2020data}
E.~Kaiser, J.~N. Kutz, and S.~L. Brunton.
\newblock {Data-driven approximations of dynamical systems operators for
  control}.
\newblock In \emph{The {Koopman} Operator in Systems and Control}, pages
  197--234. Springer, 2020.

\bibitem[Kaiser et~al.(2021{\natexlab{a}})Kaiser, Kutz, and Brunton]{kais21}
E.~Kaiser, J.~N. Kutz, and S.~L. Brunton.
\newblock {Data-driven discovery of Koopman eigenfunctions for control}.
\newblock \emph{Machine Learning: Science and Technology}, 2\penalty0
  (3):\penalty0 035023, June 2021{\natexlab{a}}.

\bibitem[Kaiser et~al.(2021{\natexlab{b}})Kaiser, Kutz, and
  Brunton]{kaiser2021data}
E.~Kaiser, J.~N. Kutz, and S.~L. Brunton.
\newblock {Data-driven discovery of Koopman eigenfunctions for control}.
\newblock \emph{Machine Learning: Science and Technology}, 2\penalty0
  (3):\penalty0 035023, 2021{\natexlab{b}}.

\bibitem[Kalisch and B{\"u}hlman(2007)]{kalisch2007estimating}
M.~Kalisch and P.~B{\"u}hlman.
\newblock {Estimating high-dimensional directed acyclic graphs with the
  PC-algorithm.}
\newblock \emph{Journal of Machine Learning Research}, 8\penalty0 (3), 2007.

\bibitem[Kallenberg(2005)]{kallenberg2005probabilistic}
O.~Kallenberg.
\newblock \emph{{Probabilistic symmetries and invariance principles}},
  volume~9.
\newblock Springer, 2005.

\bibitem[Kalnay et~al.(1996)Kalnay, Kanamitsu, Kistler, Collins, Deaven,
  Gandin, Iredell, Saha, White, Woollen, et~al.]{kalnay1996ncep}
E.~Kalnay, M.~Kanamitsu, R.~Kistler, W.~Collins, D.~Deaven, L.~Gandin,
  M.~Iredell, S.~Saha, G.~White, J.~Woollen, et~al.
\newblock {The NCEP/NCAR 40-year reanalysis project}.
\newblock \emph{Bulletin of the American meteorological Society}, 77\penalty0
  (3):\penalty0 437--472, 1996.

\bibitem[Kemeth et~al.(2022)Kemeth, Bertalan, Thiem, Dietrich, Moon, Laing, and
  Kevrekidis]{keme22}
F.~Kemeth, T.~Bertalan, T.~Thiem, F.~Dietrich, S.~Moon, C.~Laing, and
  Y.~Kevrekidis.
\newblock {Learning emergent partial differential equations in a learned
  emergent space}.
\newblock \emph{Nature Communications}, 13, 06 2022.

\bibitem[Keynes(2013)]{keynes2013treatise}
J.~M. Keynes.
\newblock \emph{{A treatise on probability}}.
\newblock Courier Corporation, 2013.

\bibitem[Khodkar et~al.(2019)Khodkar, Hassanzadeh, and Antoulas]{koopman}
M.~Khodkar, P.~Hassanzadeh, and A.~Antoulas.
\newblock {A Koopman-based framework for forecasting the spatiotemporal
  evolution of chaotic dynamics with nonlinearities modeled as exogenous
  forcings}.
\newblock \emph{Preprint arXiv:1909.00076}, 2019.

\bibitem[Kim et~al.(1971)Kim, Kline, and Reynolds]{kim_et_al}
H.~T. Kim, S.~J. Kline, and W.~C. Reynolds.
\newblock {The production of turbulence near a smooth wall in a turbulent
  boundary layer}.
\newblock \emph{J. Fluid Mech.}, 50:\penalty0 133--160, 1971.

\bibitem[Kim et~al.(2021)Kim, Lu, Mukherjee, Gilbert, Jing, Čeperić, and
  Soljačić]{Kim2021:EQL-L05}
S.~Kim, P.~Y. Lu, S.~Mukherjee, M.~Gilbert, L.~Jing, V.~Čeperić, and
  M.~Soljačić.
\newblock {Integration of Neural Network-Based Symbolic Regression in Deep
  Learning for Scientific Discovery}.
\newblock \emph{IEEE Transactions on Neural Networks and Learning Systems},
  32\penalty0 (9):\penalty0 4166--4177, 2021.

\bibitem[King et~al.(2004)King, Langley, and Simon]{king2004}
R.~King, P.~Langley, and H.~Simon.
\newblock {Automated Discovery in the Biological Sciences}.
\newblock \emph{AI Magazine}, 25\penalty0 (3):\penalty0 21--36, 2004.

\bibitem[King et~al.(2009)King, Rowland, Aubrey, Liakata, Markham, Soldatova,
  Whelan, Clare, Young, Sparkes, et~al.]{king2009robot}
R.~D. King, J.~Rowland, W.~Aubrey, M.~Liakata, M.~Markham, L.~N. Soldatova,
  K.~E. Whelan, A.~Clare, M.~Young, A.~Sparkes, et~al.
\newblock {The robot scientist Adam}.
\newblock \emph{Computer}, 42\penalty0 (8):\penalty0 46--54, 2009.

\bibitem[Klahr and Simon(1999)]{klahr1999studies}
D.~Klahr and H.~A. Simon.
\newblock {Studies of scientific discovery: Complementary approaches and
  convergent findings}.
\newblock \emph{Psychological Bulletin}, 125\penalty0 (5):\penalty0 524, 1999.

\bibitem[Kline et~al.(1967)Kline, Reynolds, Schraub, and
  Runstadler]{kline_et_al}
S.~J. Kline, W.~C. Reynolds, F.~A. Schraub, and P.~W. Runstadler.
\newblock {The structure of turbulent boundary layers}.
\newblock \emph{J. Fluid Mech.}, 30:\penalty0 741--773, 1967.

\bibitem[Klus et~al.(2016)Klus, Koltai, and Sch{\"u}tte]{KKS16}
S.~Klus, P.~Koltai, and C.~Sch{\"u}tte.
\newblock {On the numerical approximation of the Perron--Frobenius and Koopman
  operator}.
\newblock \emph{Journal of Computational Dynamics}, 3:\penalty0 51--79, 2016.

\bibitem[Klus et~al.(2018)Klus, N\"uske, Koltai, Wu, Kevrekidis, Sch\"utte, and
  No\'e]{KNKWKSN18}
S.~Klus, F.~N\"uske, P.~Koltai, H.~Wu, I.~Kevrekidis, C.~Sch\"utte, and
  F.~No\'e.
\newblock {Data-driven model reduction and transfer operator approximation}.
\newblock \emph{Journal of Nonlinear Science}, 1010:\penalty0 9437--7, 2018.

\bibitem[Klus et~al.(2020)Klus, Schuster, and Muandet]{Klus2020}
S.~Klus, I.~Schuster, and K.~Muandet.
\newblock Eigendecompositions of transfer operators in reproducing kernel
  {Hilbert} spaces.
\newblock \emph{Journal of Nonlinear Science}, 30\penalty0 (1):\penalty0
  283--315, 2020.

\bibitem[Kocabas(1991)]{kocabas1991}
A.~Kocabas.
\newblock {A Genetic Programming System for Automated Discovery in the Physical
  Sciences}.
\newblock \emph{Machine Learning}, 7\penalty0 (3-4):\penalty0 295--314, 1991.

\bibitem[Koch(2004)]{koch04}
C.~Koch.
\newblock \emph{{Biophysics of Computation: Information Processing in Single
  Neurons (Computational Neuroscience Series)}}.
\newblock Oxford University Press, Inc., Usa, 2004.
\newblock ISBN 0195181999.

\bibitem[Kokar(1986)]{kokar1986}
M.~Kokar.
\newblock Knowledge acquisition: {A} realization of new artificial
  intelligence.
\newblock \emph{Artificial Intelligence}, 32:\penalty0 251--290, 1986.

\bibitem[Kolmogorov(1963)]{kolmogorov1963tables}
A.~N. Kolmogorov.
\newblock {On tables of random numbers}.
\newblock \emph{Sankhy{\=a}: The Indian Journal of Statistics, Series A}, pages
  369--376, 1963.

\bibitem[Kommenda et~al.(2020)Kommenda, Burlacu, Kronberger, and
  Affenzeller]{kommenda2020:SRLS}
M.~Kommenda, B.~Burlacu, G.~Kronberger, and M.~Affenzeller.
\newblock {Parameter identification for symbolic regression using nonlinear
  least squares}.
\newblock \emph{Genetic Programming and Evolvable Machines}, 21\penalty0
  (3):\penalty0 471--501, 2020.

\bibitem[Koppe et~al.(2019)Koppe, Toutounji, Kirsch, Lis, and
  Durstewitz]{kopp19}
G.~Koppe, H.~Toutounji, P.~Kirsch, S.~Lis, and D.~Durstewitz.
\newblock {Identifying nonlinear dynamical systems via generative recurrent
  neural networks with applications to fMRI}.
\newblock \emph{PLOS Computational Biology}, 15\penalty0 (8):\penalty0 1--35,
  08 2019.

\bibitem[Koster et~al.(2004)Koster, Dirmeyer, Guo, Bonan, Chan, Cox, Gordon,
  Kanae, Kowalczyk, Lawrence, et~al.]{koster2004regions}
R.~D. Koster, P.~A. Dirmeyer, Z.~Guo, G.~Bonan, E.~Chan, P.~Cox, C.~Gordon,
  S.~Kanae, E.~Kowalczyk, D.~Lawrence, et~al.
\newblock {Regions of strong coupling between soil moisture and precipitation}.
\newblock \emph{Science}, 305\penalty0 (5687):\penalty0 1138--1140, 2004.

\bibitem[Koster et~al.(2006)Koster, Sud, Guo, Dirmeyer, Bonan, Oleson, Chan,
  Verseghy, Cox, Davies, et~al.]{koster2006glace}
R.~D. Koster, Y.~Sud, Z.~Guo, P.~A. Dirmeyer, G.~Bonan, K.~W. Oleson, E.~Chan,
  D.~Verseghy, P.~Cox, H.~Davies, et~al.
\newblock {GLACE: the global land--atmosphere coupling experiment. Part I:
  overview}.
\newblock \emph{Journal of Hydrometeorology}, 7\penalty0 (4):\penalty0
  590--610, 2006.

\bibitem[Kostic et~al.(2022)Kostic, Novelli, Maurer, Ciliberto, Rosasco, and
  Pontil]{Kostic2022}
V.~Kostic, P.~Novelli, A.~Maurer, C.~Ciliberto, L.~Rosasco, and M.~Pontil.
\newblock Learning dynamical systems via {Koopman} operator regression in
  reproducing kernel {Hilbert} spaces.
\newblock In \emph{NeurIPS 2022}, pages 1--9, 2022.

\bibitem[Kotz and Drouet(2001)]{kotz2001correlation}
S.~Kotz and D.~Drouet.
\newblock \emph{{Correlation and dependence}}.
\newblock World Scientific, 2001.

\bibitem[Koza et~al.(2001)Koza, Bennett, Andre, and Keane]{koza2001}
J.~Koza, F.~Bennett, D.~Andre, and M.~Keane.
\newblock {Nonlinear Genetic Programming: Automatic Discovery of Reusable
  Programs}.
\newblock \emph{Machine Learning}, 42\penalty0 (1):\penalty0 185--223, 2001.

\bibitem[Koza(1990)]{Koza1990:GP}
J.~R. Koza.
\newblock {Genetic Programming: A Paradigm for Genetically Breeding Populations
  of Computer Programs to Solve Problems}.
\newblock Technical report, Dept. of Computer Science, Stanford University,
  Stanford, CA, USA, 1990.

\bibitem[Koza(1992)]{gep}
J.~R. Koza.
\newblock Genetic programming: {O}n the programming of computers by means of
  natural selection.
\newblock \emph{MIT Press}, 1992.

\bibitem[Koza(1994)]{koza1994genetic}
J.~R. Koza.
\newblock {Genetic programming as a means for programming computers by natural
  selection}.
\newblock \emph{Statistics and computing}, 4\penalty0 (2):\penalty0 87--112,
  1994.

\bibitem[Kretschmer et~al.(2016)Kretschmer, Coumou, Donges, and
  Runge]{kretschmer2016using}
M.~Kretschmer, D.~Coumou, J.~F. Donges, and J.~Runge.
\newblock {Using causal effect networks to analyze different Arctic drivers of
  midlatitude winter circulation}.
\newblock \emph{Journal of climate}, 29\penalty0 (11):\penalty0 4069--4081,
  2016.

\bibitem[Kretschmer et~al.(2017)Kretschmer, Runge, and
  Coumou]{kretschmer2017early}
M.~Kretschmer, J.~Runge, and D.~Coumou.
\newblock {Early prediction of extreme stratospheric polar vortex states based
  on causal precursors}.
\newblock \emph{Geophysical research letters}, 44\penalty0 (16):\penalty0
  8592--8600, 2017.

\bibitem[Krich et~al.(2020)Krich, Runge, Miralles, Migliavacca, Perez-Priego,
  El-Madany, Carrara, and Mahecha]{krich2020estimating}
C.~Krich, J.~Runge, D.~G. Miralles, M.~Migliavacca, O.~Perez-Priego,
  T.~El-Madany, A.~Carrara, and M.~D. Mahecha.
\newblock {Estimating causal networks in biosphere--atmosphere interaction with
  the PCMCI approach}.
\newblock \emph{Biogeosciences}, 17\penalty0 (4):\penalty0 1033--1061, 2020.

\bibitem[Krich et~al.(2021)Krich, Migliavacca, Miralles, Kraemer, El-Madany,
  Reichstein, Runge, and Mahecha]{krich2021functional}
C.~Krich, M.~Migliavacca, D.~G. Miralles, G.~Kraemer, T.~S. El-Madany,
  M.~Reichstein, J.~Runge, and M.~D. Mahecha.
\newblock {Functional convergence of biosphere--atmosphere interactions in
  response to meteorological conditions}.
\newblock \emph{Biogeosciences}, 18\penalty0 (7):\penalty0 2379--2404, 2021.

\bibitem[Kronberger et~al.(2022)Kronberger, de~Franca, Burlacu, Haider, and
  Kommenda]{Kronberger2022:shape-constraintSR}
G.~Kronberger, F.~O. de~Franca, B.~Burlacu, C.~Haider, and M.~Kommenda.
\newblock Shape-constrained symbolic regression—{I}mproving extrapolation
  with prior knowledge.
\newblock \emph{Evolutionary Computation}, 30\penalty0 (1):\penalty0 75--98, 03
  2022.

\bibitem[Kuhn(1962)]{Kuhn1962}
T.~S. Kuhn.
\newblock \emph{{The Structure of Scientific Revolutions}}.
\newblock University of Chicago Press, 1962.

\bibitem[Kutz et~al.(2016)Kutz, Brunton, Brunton, and Proctor]{brunton_dmd}
J.~Kutz, S.~Brunton, B.~Brunton, and J.~Proctor.
\newblock Dynamic mode decomposition: {D}ata-driven modeling of complex
  systems.
\newblock \emph{Siam}, 2016.

\bibitem[Kwapie{\'n} and Dro{\.z}d{\.z}(2012)]{kwapien2012physical}
J.~Kwapie{\'n} and S.~Dro{\.z}d{\.z}.
\newblock {Physical approach to complex systems}.
\newblock \emph{Physics Reports}, 515\penalty0 (3-4):\penalty0 115--226, 2012.

\bibitem[Lachapelle et~al.(2020)Lachapelle, Brouillard, Deleu, and
  Lacoste-Julien]{Lachapelle2020Gradient-Based}
S.~Lachapelle, P.~Brouillard, T.~Deleu, and S.~Lacoste-Julien.
\newblock {Gradient-Based Neural DAG Learning}.
\newblock In \emph{International Conference on Learning Representations}, 2020.

\bibitem[Landahl and Landahlt(1975)]{Landahl}
M.~T. Landahl and M.~T. Landahlt.
\newblock {Wave breakdown and turbulence}.
\newblock \emph{SIAM J. Appl. Maths}, 28:\penalty0 735--756, 1975.

\bibitem[Langley(2019)]{langley2019scientific}
P.~Langley.
\newblock Scientific discovery, causal explanation, and process model
  induction.
\newblock \emph{Mind \& Society}, 18\penalty0 (1):\penalty0 43--56, 2019.

\bibitem[Langley et~al.(1987{\natexlab{a}})Langley, Simon, and
  Bradshaw]{Langley1987}
P.~Langley, H.~Simon, and G.~Bradshaw.
\newblock Scientific discovery: {C}omputational explorations of the creative
  process.
\newblock \emph{AI Magazine}, 8\penalty0 (3):\penalty0 30--44,
  1987{\natexlab{a}}.

\bibitem[Langley et~al.(1987{\natexlab{b}})Langley, Simon, Bradshaw, and
  Zytkow]{langley1987scientific}
P.~Langley, H.~A. Simon, G.~L. Bradshaw, and J.~M. Zytkow.
\newblock \emph{Scientific discovery: {C}omputational explorations of the
  creative processes}.
\newblock MIT press, 1987{\natexlab{b}}.

\bibitem[Langley et~al.(2002{\natexlab{a}})Langley, Simon, and
  Bradshaw]{langley2002a}
P.~Langley, H.~Simon, and G.~Bradshaw.
\newblock Automated discovery in the {Physical Sciences}.
\newblock \emph{AI Magazine}, 23\penalty0 (3):\penalty0 11-- 28,
  2002{\natexlab{a}}.

\bibitem[Langley et~al.(2002{\natexlab{b}})Langley, Simon, and
  Bradshaw]{langley2002b}
P.~Langley, H.~Simon, and G.~Bradshaw.
\newblock Scientific discovery and the future of {AI}.
\newblock \emph{AI Magazine}, 23\penalty0 (3):\penalty0 29--39,
  2002{\natexlab{b}}.

\bibitem[Lapish et~al.(2015)Lapish, Balaguer-Ballester, Seamans, Phillips, and
  Durstewitz]{lapbal15}
C.~C. Lapish, E.~Balaguer-Ballester, J.~K. Seamans, A.~G. Phillips, and
  D.~Durstewitz.
\newblock {Amphetamine Exerts Dose-Dependent Changes in Prefrontal Cortex
  Attractor Dynamics during Working Memory}.
\newblock \emph{Journal of Neuroscience}, 35\penalty0 (28):\penalty0
  10172--10187. EB--B and CCL contributed equally., 2015.

\bibitem[Lazpita et~al.(2022)Lazpita, Mart\'inez-S\'anchez, Corrochano, Hoyas,
  Le~Clainche, and Vinuesa]{lazpita_pof}
E.~Lazpita, A.~Mart\'inez-S\'anchez, A.~Corrochano, S.~Hoyas, S.~Le~Clainche,
  and R.~Vinuesa.
\newblock {On the generation and destruction mechanisms of arch vortices in
  urban fluid flows}.
\newblock \emph{Phys. Fluids}, 34:\penalty0 051702, 2022.

\bibitem[Le~Clainche and Vega(2017)]{Sole}
S.~Le~Clainche and J.~M. Vega.
\newblock {Higher Order Dynamic Mode Decomposition}.
\newblock \emph{SIAM J. Appl. Dyn. Syst.}, 16:\penalty0 882--925, 2017.

\bibitem[Lee and Zaki(2018)]{dns_data}
J.~Lee and T.~A. Zaki.
\newblock Detection algorithm for turbulent interfaces and large-scale
  structures in intermittent flows.
\newblock \emph{Comput. Fluids}, 175\penalty0 (1):\penalty0 142--158, 2018.

\bibitem[Lee and Carlberg(2020)]{r6}
K.~Lee and K.~T. Carlberg.
\newblock {Model reduction of dynamical systems on nonlinear manifolds using
  deep convolutional autoencoders}.
\newblock \emph{J. Comput. Phys.}, 404:\penalty0 108973, 2020.

\bibitem[LeRoy(2004)]{leroy2004causality}
S.~LeRoy.
\newblock \emph{{Causality in economics}}.
\newblock London School of Economics, Centre for Philosophy of Natural and
  Social Sciences, 2004.

\bibitem[Lin and Oettel(2019)]{Lin2019}
S.-C. Lin and M.~Oettel.
\newblock {A classical density functional from machine learning and a
  convolutional neural network}.
\newblock \emph{SciPost Phys.}, 6:\penalty0 025, 2019.

\bibitem[Lin et~al.(2020)Lin, Martius, and Oettel]{lin2020:DFTEQL}
S.-C. Lin, G.~Martius, and M.~Oettel.
\newblock {Analytical classical density functionals from an equation learning
  network}.
\newblock \emph{The Journal of Chemical Physics}, 152\penalty0 (2):\penalty0
  021102, 2020.

\bibitem[Lindley(2013)]{lindley2013understanding}
D.~V. Lindley.
\newblock \emph{{Understanding uncertainty}}.
\newblock John Wiley \& Sons, 2013.

\bibitem[Ljung and Glad(1994)]{ljung1994global}
L.~Ljung and T.~Glad.
\newblock {On global identifiability for arbitrary model parametrizations}.
\newblock \emph{Automatica}, 30\penalty0 (2):\penalty0 265--276, 1994.

\bibitem[Lohmann et~al.(2012)Lohmann, Erfurth, Müller, and Turner]{lohm12}
G.~Lohmann, K.~Erfurth, K.~Müller, and R.~Turner.
\newblock {Critical comments on dynamic causal modelling}.
\newblock \emph{NeuroImage}, 59\penalty0 (3):\penalty0 2322--2329, 2012.

\bibitem[Loiseau(2020)]{sindy1}
J.-C. Loiseau.
\newblock {Data-driven modeling of the chaotic thermal convection in an annular
  thermosyphon}.
\newblock \emph{Theor. Comput. Fluid Dyn.}, 34\penalty0 (4):\penalty0 339--365,
  2020.

\bibitem[Long et~al.(2019)Long, Lu, and Dong]{long2019:PDE-Net}
Z.~Long, Y.~Lu, and B.~Dong.
\newblock {PDE-Net 2.0: Learning PDEs from data with a numeric-symbolic hybrid
  deep network}.
\newblock \emph{Journal of Computational Physics}, 399:\penalty0 108925, Dec.
  2019.

\bibitem[Louizos et~al.(2018)Louizos, Welling, and Kingma]{louizos2018learning}
C.~Louizos, M.~Welling, and D.~P. Kingma.
\newblock {Learning Sparse Neural Networks through $L_0$ Regularization}.
\newblock In \emph{International Conference on Learning Representations}, 2018.

\bibitem[Lozano-Dur\'an and Jim\'enez(2014)]{adrian_tracking}
A.~Lozano-Dur\'an and J.~Jim\'enez.
\newblock Time-resolved evolution of coherent structures in turbulent channels:
  characterization of eddies and cascades.
\newblock \emph{J. Fluid Mech.}, 759:\penalty0 432--471, 2014.

\bibitem[Lozano-Dur\'an et~al.(2012)Lozano-Dur\'an, Flores, and
  Jim\'enez]{adrian}
A.~Lozano-Dur\'an, O.~Flores, and J.~Jim\'enez.
\newblock The three-dimensional structure of momentum transfer in turbulent
  channels.
\newblock \emph{J. Fluid Mech.}, 694:\penalty0 100--130, 2012.

\bibitem[Lozano-Dur\'an et~al.(2020)Lozano-Dur\'an, Bae, and
  Encinar]{lozano_causality}
A.~Lozano-Dur\'an, H.~J. Bae, and M.~P. Encinar.
\newblock {Causality of energy-containing eddies in wall turbulence}.
\newblock \emph{J. Fluid Mech.}, 882:\penalty0 A2, 2020.

\bibitem[Lu and Willmarth(1973)]{lu_willmarth}
S.~S. Lu and W.~W. Willmarth.
\newblock {Measurements of the structure of the Reynolds stress in a turbulent
  boundary layer}.
\newblock \emph{J. Fluid Mech.}, 60:\penalty0 481--511, 1973.

\bibitem[Ludescher et~al.(2021)Ludescher, Martin, Boers, Bunde, Ciemer, Fan,
  Havlin, Kretschmer, Kurths, Runge, et~al.]{ludescher2021network}
J.~Ludescher, M.~Martin, N.~Boers, A.~Bunde, C.~Ciemer, J.~Fan, S.~Havlin,
  M.~Kretschmer, J.~Kurths, J.~Runge, et~al.
\newblock {Network-based forecasting of climate phenomena}.
\newblock \emph{Proceedings of the National Academy of Sciences}, 118\penalty0
  (47):\penalty0 e1922872118, 2021.

\bibitem[Lueckmann et~al.(2017)Lueckmann, Gon\c{c}alves, Bassetto, \"{O}cal,
  Nonnenmacher, and Macke]{luec17}
J.-M. Lueckmann, P.~J. Gon\c{c}alves, G.~Bassetto, K.~\"{O}cal,
  M.~Nonnenmacher, and J.~H. Macke.
\newblock {Flexible Statistical Inference for Mechanistic Models of Neural
  Dynamics}.
\newblock In \emph{Proceedings of the 31st International Conference on Neural
  Information Processing Systems}, NIPS'17, page 1289–1299, Red Hook, NY,
  USA, 2017. Curran Associates Inc.
\newblock ISBN 9781510860964.

\bibitem[Lumley(1967)]{lumley}
J.~L. Lumley.
\newblock {The structure of inhomogeneous turbulence}.
\newblock \emph{Atmospheric turbulence and wave propagation, A. M. Yaglom and
  V. I. Tatarski (eds). Nauka, Moscow}, pages 166--178, 1967.

\bibitem[Lun-Chau et~al.(2022)Lun-Chau, Hu, Gonzalez, and Sejdinovic]{shap}
S.~Lun-Chau, R.~Hu, J.~Gonzalez, and D.~Sejdinovic.
\newblock {RKHS-SHAP}: {Shapley} values for kernel methods.
\newblock \emph{Preprint arXiv:2110.09167v2}, 2022.

\bibitem[Lungarella et~al.(2007)Lungarella, Pitti, and
  Kuniyoshi]{lungarella2007information}
M.~Lungarella, A.~Pitti, and Y.~Kuniyoshi.
\newblock {Information transfer at multiple scales}.
\newblock \emph{Physical Review E}, 76\penalty0 (5):\penalty0 056117, 2007.

\bibitem[Lusch et~al.(2018)Lusch, Kutz, and Brunton]{Lusch2018}
B.~Lusch, J.~N. Kutz, and S.~L. Brunton.
\newblock {Deep learning for universal linear embeddings of nonlinear
  dynamics}.
\newblock \emph{Nature Communications}, 9\penalty0 (1), 2018.

\bibitem[M.~Mooij and Claassen(2020)]{mooij2020constraint}
J.~M.~Mooij and T.~Claassen.
\newblock {Constraint-Based Causal Discovery using Partial Ancestral Graphs in
  the presence of Cycles}.
\newblock In J.~Peters and D.~Sontag, editors, \emph{Proceedings of the 36th
  Conference on Uncertainty in Artificial Intelligence (UAI)}, volume 124 of
  \emph{Proceedings of Machine Learning Research}, pages 1159--1168. Pmlr,
  03--06 Aug 2020.

\bibitem[Madani et~al.(2020)Madani, Parazoo, Kimball, Ballantyne, Reichle,
  Maneta, Saatchi, Palmer, Liu, and Tagesson]{madani2020recent}
N.~Madani, N.~C. Parazoo, J.~S. Kimball, A.~P. Ballantyne, R.~H. Reichle,
  M.~Maneta, S.~Saatchi, P.~I. Palmer, Z.~Liu, and T.~Tagesson.
\newblock {Recent amplified global gross primary productivity due to
  temperature increase is offset by reduced productivity due to water
  constraints}.
\newblock \emph{AGU Advances}, 1\penalty0 (4):\penalty0 e2020AV000180, 2020.

\bibitem[Malinsky and Spirtes(2018)]{malinsky2018causal}
D.~Malinsky and P.~Spirtes.
\newblock Causal structure learning from multivariate time series in settings
  with unmeasured confounding.
\newblock In T.~D. Le, K.~Zhang, E.~K{\i}c{\i}man, A.~Hyv\"{a}rinen, and
  L.~Liu, editors, \emph{Proceedings of 2018 ACM SIGKDD Workshop on Causal
  Disocvery}, volume~92 of \emph{Proceedings of Machine Learning Research},
  pages 23--47, London, UK, 20 Aug 2018. PMLR.

\bibitem[Maraun et~al.(2017)Maraun, Shepherd, Widmann, Zappa, Walton,
  Guti{\'e}rrez, Hagemann, Richter, Soares, Hall, et~al.]{maraun2017towards}
D.~Maraun, T.~G. Shepherd, M.~Widmann, G.~Zappa, D.~Walton, J.~M.
  Guti{\'e}rrez, S.~Hagemann, I.~Richter, P.~M. Soares, A.~Hall, et~al.
\newblock Towards process-informed bias correction of climate change
  simulations.
\newblock \emph{Nature Climate Change}, 7\penalty0 (11):\penalty0 764--773,
  2017.

\bibitem[Marinazzo et~al.(2008{\natexlab{a}})Marinazzo, Pellicoro, and
  Stramaglia]{marinazzo2008kernel}
D.~Marinazzo, M.~Pellicoro, and S.~Stramaglia.
\newblock Kernel-{Granger} causality and the analysis of dynamical networks.
\newblock \emph{Physical review E}, 77\penalty0 (5):\penalty0 056215,
  2008{\natexlab{a}}.

\bibitem[Marinazzo et~al.(2008{\natexlab{b}})Marinazzo, Pellicoro, and
  Stramaglia]{marinazzo2008kernelunivariate}
D.~Marinazzo, M.~Pellicoro, and S.~Stramaglia.
\newblock Kernel method for nonlinear {Granger} causality.
\newblock \emph{Phys. Rev. Lett.}, 100:\penalty0 144103, Apr.
  2008{\natexlab{b}}.

\bibitem[Marini and Singer(1988)]{marini1988causality}
M.~M. Marini and B.~Singer.
\newblock Causality in the social sciences.
\newblock \emph{Sociological methodology}, 18:\penalty0 347--409, 1988.

\bibitem[Marreiros et~al.(2010)Marreiros, Stephan, and
  Friston]{marreiros2010dynamic}
A.~C. Marreiros, K.~E. Stephan, and K.~J. Friston.
\newblock Dynamic causal modeling.
\newblock \emph{Scholarpedia}, 5\penalty0 (7):\penalty0 9568, 2010.

\bibitem[Marshall and Adcroft(2010)]{marshall2010parameterization}
D.~P. Marshall and A.~J. Adcroft.
\newblock Parameterization of ocean eddies: {P}otential vorticity mixing,
  energetics and arnold’s first stability theorem.
\newblock \emph{Ocean Modelling}, 32\penalty0 (3-4):\penalty0 188--204, 2010.

\bibitem[Mart\'inez-S\'anchez et~al.(2022)Mart\'inez-S\'anchez, L\'opez,
  Le~Clainche, Lozano-Dur\'an, Srivastava, and Vinuesa]{alvaro_causality}
A.~Mart\'inez-S\'anchez, E.~L\'opez, S.~Le~Clainche, A.~Lozano-Dur\'an,
  A.~Srivastava, and R.~Vinuesa.
\newblock Causality analysis of large-scale structures in the flow around a
  wall-mounted square cylinder.
\newblock \emph{Preprint arXiv:2209.15356}, 2022.

\bibitem[Mart\'inez-S\'anchez et~al.(2023)Mart\'inez-S\'anchez, Lazpita,
  Corrochano, Le~Clainche, Hoyas, and Vinuesa]{alvaro_urban}
A.~Mart\'inez-S\'anchez, E.~Lazpita, A.~Corrochano, S.~Le~Clainche, S.~Hoyas,
  and R.~Vinuesa.
\newblock Data-driven assessment of arch vortices in urban flows.
\newblock \emph{Int. J. Heat Fluid Flow, To Appear. Preprint
  arXiv:2202.01667v1}, 2023.

\bibitem[Martius and Lampert(2016)]{MartiusLampert2017:EQL}
G.~Martius and C.~H. Lampert.
\newblock Extrapolation and learning equations, 2016.
\newblock \url{https://arxiv.org/abs/1610.02995}.

\bibitem[Mattia and Del~Giudice(2002)]{matt02}
M.~Mattia and P.~Del~Giudice.
\newblock Population dynamics of interacting spiking neurons.
\newblock \emph{Phys. Rev. E}, 66:\penalty0 051917, Nov. 2002.

\bibitem[Mattia et~al.(2019)Mattia, Biggio, Galluzzi, and Storace]{matt19}
M.~Mattia, M.~Biggio, A.~Galluzzi, and M.~Storace.
\newblock Dimensional reduction in networks of non-{Markovian} spiking neurons:
  {E}quivalence of synaptic filtering and heterogeneous propagation delays.
\newblock \emph{PLOS Computational Biology}, 15\penalty0 (10):\penalty0 1--35,
  Oct. 2019.

\bibitem[May(1972)]{may1972will}
R.~M. May.
\newblock Will a large complex system be stable?
\newblock \emph{Nature}, 238\penalty0 (5364):\penalty0 413--414, 1972.

\bibitem[McConaghy(2011)]{McConaghy:2011:GPTP}
T.~McConaghy.
\newblock {FFX}: {F}ast, scalable, deterministic symbolic regression
  technology.
\newblock In \emph{Genetic Programming Theory and Practice IX}, Genetic and
  Evolutionary Computation, chapter~13, pages 235--260. Springer, Ann Arbor,
  USA, 2011.

\bibitem[McDavid et~al.(2019)McDavid, Gottardo, Simon, and
  Drton]{mcdavid2019graphical}
A.~McDavid, R.~Gottardo, N.~Simon, and M.~Drton.
\newblock Graphical models for zero-inflated single cell gene expression.
\newblock \emph{The annals of applied statistics}, 13\penalty0 (2):\penalty0
  848, 2019.

\bibitem[McGibbon and Pande(2015)]{MP15}
R.~T. McGibbon and V.~S. Pande.
\newblock {Variational cross-validation of slow dynamical modes in molecular
  kinetics}.
\newblock \emph{The Journal of Chemical Physics}, 142, 2015.

\bibitem[McKnight et~al.(2007)McKnight, McKnight, Sidani, and
  Figueredo]{mcknight2007missing}
P.~E. McKnight, K.~M. McKnight, S.~Sidani, and A.~J. Figueredo.
\newblock \emph{Missing data: {A} gentle introduction}.
\newblock Guilford Press, 2007.

\bibitem[Medaglia et~al.(2015)Medaglia, Lynall, and
  Bassett]{medaglia2015cognitive}
J.~D. Medaglia, M.-E. Lynall, and D.~S. Bassett.
\newblock Cognitive network neuroscience.
\newblock \emph{Journal of cognitive neuroscience}, 27\penalty0 (8):\penalty0
  1471--1491, 2015.

\bibitem[Meek(1995)]{meek1995causal}
C.~Meek.
\newblock Causal inference and causal explanation with background knowledge.
\newblock In \emph{Proceedings of the Eleventh Conference on Uncertainty in
  Artificial Intelligence}, Uai'95, page 403–410, San Francisco, CA, USA,
  1995. Morgan Kaufmann Publishers Inc.
\newblock ISBN 1558603859.

\bibitem[Meek(1997)]{Meek1997GraphicalMS}
C.~Meek.
\newblock \emph{Graphical models: {S}electing causal and statistical models}.
\newblock PhD thesis, Carnegie Mellon University., 1997.

\bibitem[Mehler and Kording(2018)]{Mehler2018}
D.~M.~A. Mehler and K.~P. Kording.
\newblock The lure of causal statements: {R}ampant mis-inference of causality
  in estimated connectivity, 2018.

\bibitem[Meneveau and Katz(2000)]{meneveau2000}
C.~Meneveau and J.~Katz.
\newblock {Scale-invariance and turbulence models for large-eddy simulation}.
\newblock \emph{Annual Review of Fluid Mechanics}, 32\penalty0 (1):\penalty0
  1--32, 2000.

\bibitem[Menzly et~al.(2004)Menzly, Santos, and
  Veronesi]{menzly2004understanding}
L.~Menzly, T.~Santos, and P.~Veronesi.
\newblock {Understanding predictability}.
\newblock \emph{Journal of Political Economy}, 112\penalty0 (1):\penalty0
  1--47, 2004.

\bibitem[Meyer-Lindenberg et~al.(2005)Meyer-Lindenberg, Olsen, Kohn, Brown,
  Egan, Weinberger, and Berman]{meye05}
A.~S. Meyer-Lindenberg, R.~K. Olsen, P.~D. Kohn, T.~Brown, M.~F. Egan, D.~R.
  Weinberger, and K.~F. Berman.
\newblock {Regionally Specific Disturbance of Dorsolateral
  Prefrontal–Hippocampal Functional Connectivity in Schizophrenia}.
\newblock \emph{Archives of General Psychiatry}, 62\penalty0 (4):\penalty0
  379--386, 04 2005.

\bibitem[Mezi{\'c}(2005)]{mezic2005spectral}
I.~Mezi{\'c}.
\newblock {Spectral properties of dynamical systems, model reduction and
  decompositions}.
\newblock \emph{Nonlinear Dynamics}, 41\penalty0 (1):\penalty0 309--325, 2005.

\bibitem[Milano and Koumoutsakos(2002)]{r4}
M.~Milano and P.~Koumoutsakos.
\newblock {Neural network modeling for near wall turbulent flow}.
\newblock \emph{J. Comput. Phys.}, 182:\penalty0 1--26, 2002.

\bibitem[Milly(1992)]{milly1992potential}
P.~Milly.
\newblock {Potential evaporation and soil moisture in general circulation
  models}.
\newblock \emph{Journal of climate}, 5\penalty0 (3):\penalty0 209--226, 1992.

\bibitem[Moehlis et~al.(2004)Moehlis, Faisst, and Eckhardt]{moehlis_et_al}
J.~Moehlis, H.~Faisst, and B.~Eckhardt.
\newblock {A low-dimensional model for turbulent shear flows}.
\newblock \emph{New J. Phys.}, 6:\penalty0 56, 2004.

\bibitem[Mogensen(2022)]{mogensen2022equality}
S.~W. Mogensen.
\newblock {Equality Constraints in Linear Hawkes Processes}.
\newblock In B.~Schölkopf, C.~Uhler, and K.~Zhang, editors, \emph{Proceedings
  of the First Conference on Causal Learning and Reasoning}, volume 177 of
  \emph{Proceedings of Machine Learning Research}, pages 576--593. Pmlr, 11--13
  Apr 2022.

\bibitem[Mogensen and Hansen(2020)]{mogensen2020markov}
S.~W. Mogensen and N.~R. Hansen.
\newblock {Markov equivalence of marginalized local independence graphs}.
\newblock \emph{The Annals of Statistics}, 48\penalty0 (1):\penalty0 539--559,
  2020.

\bibitem[Mogensen et~al.(2018)Mogensen, Malinsky, and Hansen]{mogensencausal}
S.~W. Mogensen, D.~Malinsky, and N.~R. Hansen.
\newblock {Causal Learning for Partially Observed Stochastic Dynamical
  Systems}.
\newblock In \emph{Thirty-Fourth Conference on Uncertainty in Artifical
  Intelligence}. AUAI Press Corvallis, Oregon, 2018.

\bibitem[Monnier et~al.(2010)Monnier, Neiswander, and Wark]{monnier_old}
B.~Monnier, B.~Neiswander, and C.~Wark.
\newblock Stereoscopic particle image velocimetry measurements in an urban type
  boundary layer: {I}nsight into flow regimes and incidence angle effect.
\newblock \emph{Boundary-Layer Meteorol.}, 135:\penalty0 243–--268, 2010.

\bibitem[M{\o}nster et~al.(2017)M{\o}nster, Fusaroli, Tyl{\'e}n, Roepstorff,
  and Sherson]{monster2017causal}
D.~M{\o}nster, R.~Fusaroli, K.~Tyl{\'e}n, A.~Roepstorff, and J.~F. Sherson.
\newblock Causal inference from noisy time-series data—{T}esting the
  convergent cross-mapping algorithm in the presence of noise and external
  influence.
\newblock \emph{Future Generation Computer Systems}, 73:\penalty0 52--62, 2017.

\bibitem[Montbri\'o et~al.(2015)Montbri\'o, Paz\'o, and Roxin]{mont15}
E.~Montbri\'o, D.~Paz\'o, and A.~Roxin.
\newblock Macroscopic description for networks of spiking neurons.
\newblock \emph{Phys. Rev. X}, 5:\penalty0 021028, June 2015.

\bibitem[Monti et~al.(2020)Monti, Khemakhem, and Hyvarinen]{autoReg}
R.~P. Monti, l.~Khemakhem, and A.~Hyvarinen.
\newblock Autoregressive flow-based causal discovery and inference.
\newblock \emph{arXiv preprint: 2007.09390}, 07 2020.

\bibitem[Mooij et~al.(2009)Mooij, Janzing, Peters, and
  Sch{\"o}lkopf]{mooij2009regression}
J.~Mooij, D.~Janzing, J.~Peters, and B.~Sch{\"o}lkopf.
\newblock {Regression by dependence minimization and its application to causal
  inference in additive noise models}.
\newblock In \emph{Proceedings of the 26th annual international conference on
  machine learning}, pages 745--752, 2009.

\bibitem[Mooij et~al.(2013{\natexlab{a}})Mooij, Janzing, and
  Sch\"{o}lkopf]{moij2013ordinary}
J.~M. Mooij, D.~Janzing, and B.~Sch\"{o}lkopf.
\newblock From ordinary differential equations to structural causal models:
  {The} deterministic case.
\newblock In \emph{Proceedings of the Twenty-Ninth Conference on Uncertainty in
  Artificial Intelligence}, Uai'13, page 440–448, Arlington, Virginia, USA,
  2013{\natexlab{a}}. AUAI Press.

\bibitem[Mooij et~al.(2013{\natexlab{b}})Mooij, Janzing, and
  Schölkopf]{mooij2013deterministic}
J.~M. Mooij, D.~Janzing, and B.~Schölkopf.
\newblock {From Ordinary Differential Equations to Structural Causal Models:
  the deterministic case}.
\newblock \emph{arXiv preprint arXiv:1304.7920}, 2013{\natexlab{b}}.

\bibitem[Mooij et~al.(2016)Mooij, Peters, Janzing, Zscheischler, and
  Sch{\"o}lkopf]{mooij2016distinguishing}
J.~M. Mooij, J.~Peters, D.~Janzing, J.~Zscheischler, and B.~Sch{\"o}lkopf.
\newblock {Distinguishing cause from effect using observational data: methods
  and benchmarks}.
\newblock \emph{The Journal of Machine Learning Research}, 17\penalty0
  (1):\penalty0 1103--1204, 2016.

\bibitem[Mooij et~al.(2020)Mooij, Magliacane, and Claassen]{mooij2020joint}
J.~M. Mooij, S.~Magliacane, and T.~Claassen.
\newblock {Joint Causal Inference from Multiple Contexts}.
\newblock \emph{Journal of Machine Learning Research}, 21\penalty0
  (99):\penalty0 1--108, 2020.

\bibitem[Moraffah et~al.(2021)Moraffah, Sheth, Karami, Bhattacharya, Wang,
  Tahir, Raglin, and Liu]{moraffah2021causal}
R.~Moraffah, P.~Sheth, M.~Karami, A.~Bhattacharya, Q.~Wang, A.~Tahir,
  A.~Raglin, and H.~Liu.
\newblock {Causal inference for time series analysis: Problems, methods and
  evaluation}.
\newblock \emph{Knowledge and Information Systems}, 63:\penalty0 3041--3085,
  2021.

\bibitem[Morgan and Winship(2015)]{morgan2015counterfactuals}
S.~L. Morgan and C.~Winship.
\newblock \emph{{Counterfactuals and causal inference}}.
\newblock Cambridge University Press, 2015.

\bibitem[Moulet(1992)]{moulet1992}
P.~Moulet.
\newblock {Learning Rules from Structured Data}.
\newblock \emph{Machine Learning}, 8\penalty0 (1):\penalty0 47--75, 1992.

\bibitem[Munz(2014)]{munz2014our}
P.~Munz.
\newblock \emph{{Our knowledge of the growth of knowledge: Popper or
  Wittgenstein?}}
\newblock Routledge, 2014.

\bibitem[Murata and Tanaka(1994)]{murata1994}
T.~Murata and K.~Tanaka.
\newblock {A Constructive Induction Algorithm Incorporating Prior Knowledge}.
\newblock \emph{Machine Learning}, 14\penalty0 (1):\penalty0 71--96, 1994.

\bibitem[Murata et~al.(2020)Murata, Fukami, and Fukagata]{r2}
T.~Murata, K.~Fukami, and K.~Fukagata.
\newblock {Nonlinear mode decomposition with convolutional neural networks for
  fluid dynamics}.
\newblock \emph{J. Fluid Mech.}, 882:\penalty0 A13, 2020.

\bibitem[Ness et~al.(2017)Ness, Sachs, Mallick, and Vitek]{ness2017bayesian}
R.~O. Ness, K.~Sachs, P.~Mallick, and O.~Vitek.
\newblock {A Bayesian active learning experimental design for inferring
  signaling networks}.
\newblock In \emph{Research in Computational Molecular Biology: 21st Annual
  International Conference, RECOMB 2017, Hong Kong, China, May 3-7, 2017,
  Proceedings 21}, pages 134--156. Springer, 2017.

\bibitem[Neumann et~al.(2023)Neumann, Horn, and Kühn]{neum23}
W.-J. Neumann, A.~Horn, and A.~A. Kühn.
\newblock {Insights and opportunities for deep brain stimulation as a brain
  circuit intervention}, 2023.

\bibitem[Newton(1833)]{newton1833philosophiae}
I.~Newton.
\newblock \emph{{Philosophiae naturalis principia mathematica}}, volume~1.
\newblock G. Brookman, 1833.

\bibitem[Ng et~al.(2020)Ng, Ghassami, and Zhang]{ng2020golem}
I.~Ng, A.~Ghassami, and K.~Zhang.
\newblock {On the Role of Sparsity and DAG Constraints for Learning Linear
  DAGs}.
\newblock In H.~Larochelle, M.~Ranzato, R.~Hadsell, M.~Balcan, and H.~Lin,
  editors, \emph{Advances in Neural Information Processing Systems}, volume~33,
  pages 17943--17954. Curran Associates, Inc., 2020.

\bibitem[Ng et~al.(2022)Ng, Lachapelle, Rosemary~Ke, Lacoste-Julien, and
  Zhang]{ng2022convergence}
I.~Ng, S.~Lachapelle, N.~Rosemary~Ke, S.~Lacoste-Julien, and K.~Zhang.
\newblock {On the Convergence of Continuous Constrained Optimization for
  Structure Learning}.
\newblock In G.~Camps-Valls, F.~J.~R. Ruiz, and I.~Valera, editors,
  \emph{Proceedings of The 25th International Conference on Artificial
  Intelligence and Statistics}, volume 151 of \emph{Proceedings of Machine
  Learning Research}, pages 8176--8198. Pmlr, 28--30 Mar 2022.

\bibitem[Nieh et~al.(2021)Nieh, Schottdorf, Freeman, Low, Lewallen, Koay,
  Pinto, Gauthier, Brody, and Tank]{nieh21}
E.~Nieh, M.~Schottdorf, N.~Freeman, R.~Low, S.~Lewallen, S.~Koay, L.~Pinto,
  J.~Gauthier, C.~Brody, and D.~Tank.
\newblock Geometry of abstract learned knowledge in the hippocampus.
\newblock \emph{Nature}, 595\penalty0 (7865):\penalty0 80--84, July 2021.
\newblock \doi{10.1038/s41586-021-03652-7}.

\bibitem[Niemeijer and de~Groot(2008)]{niemeijer2008framing}
D.~Niemeijer and R.~S. de~Groot.
\newblock {Framing environmental indicators: moving from causal chains to
  causal networks}.
\newblock \emph{Environment, development and sustainability}, 10\penalty0
  (1):\penalty0 89--106, 2008.

\bibitem[Noack et~al.(2003)Noack, Afanasiev, Morzynski, Tadmor, and
  Thiele]{noack_galerkin}
B.~R. Noack, K.~Afanasiev, M.~Morzynski, G.~Tadmor, and F.~Thiele.
\newblock {A hierarchy of low-dimensional models for the transient and
  post-transient cylinder wake}.
\newblock \emph{J. Fluid Mech.}, 497:\penalty0 335--363, 2003.

\bibitem[No{\'e} and N{\"u}ske(2013)]{NoNu13}
F.~No{\'e} and F.~N{\"u}ske.
\newblock {A variational approach to modeling slow processes in stochastic
  dynamical systems}.
\newblock \emph{Multiscale Modeling \& Simulation}, 11:\penalty0 635--655,
  2013.

\bibitem[Nordhausen and Langley(1990)]{nordhausen1990}
K.~Nordhausen and P.~Langley.
\newblock {Inverse Entailment and Progol}.
\newblock \emph{Machine Learning}, 5\penalty0 (1):\penalty0 25--38, 1990.

\bibitem[Nowack et~al.(2020)Nowack, Runge, Eyring, and Haigh]{nowack2020causal}
P.~Nowack, J.~Runge, V.~Eyring, and J.~D. Haigh.
\newblock {Causal networks for climate model evaluation and constrained
  projections}.
\newblock \emph{Nature communications}, 11\penalty0 (1):\penalty0 1--11, 2020.

\bibitem[N\"uske et~al.(2014)N\"uske, Keller, P\'erez-Hern\'andez, Mey, and
  No\'e]{NKPMN14}
F.~N\"uske, B.~G. Keller, G.~P\'erez-Hern\'andez, A.~S. J.~S. Mey, and
  F.~No\'e.
\newblock {Variational approach to molecular kinetics}.
\newblock \emph{Journal of Chemical Theory and Computation}, 10:\penalty0
  1739--1752, 2014.

\bibitem[Observatory(2020)]{mauna:CO2data}
M.~L. Observatory.
\newblock
  https://www.climate.gov/teaching/resources/atmospheric-co2-mauna-loa-observatory,
  2020.

\bibitem[Olver(1999)]{olver1999classical}
P.~J. Olver.
\newblock \emph{Classical Invariant Theory}.
\newblock London Mathematical Society Student Texts. Cambridge University
  Press, 1999.

\bibitem[Orr(1907)]{orr}
W.~M. Orr.
\newblock {The stability or instability of the steady motions of a perfect
  liquid and of a viscous liquid. Part II. A viscous liquid}.
\newblock \emph{Math. Proc. R. Irish Acad.}, 27:\penalty0 69--138, 1907.

\bibitem[Pamfil et~al.(2020)Pamfil, Sriwattanaworachai, Desai, Pilgerstorfer,
  Georgatzis, Beaumont, and Aragam]{pamfil2020dynotears}
R.~Pamfil, N.~Sriwattanaworachai, S.~Desai, P.~Pilgerstorfer, K.~Georgatzis,
  P.~Beaumont, and B.~Aragam.
\newblock {DYNOTEARS: Structure Learning from Time-Series Data}.
\newblock In S.~Chiappa and R.~Calandra, editors, \emph{Proceedings of the
  Twenty Third International Conference on Artificial Intelligence and
  Statistics}, volume 108 of \emph{Proceedings of Machine Learning Research},
  pages 1595--1605. Pmlr, 26--28 Aug 2020.

\bibitem[Pandarinath et~al.(2018)Pandarinath, O’Shea, Collins, Jozefowicz,
  Stavisky, Kao, Trautmann, Kaufman, Ryu, Hochberg, Henderson, Shenoy, Abbott,
  and Sussillo]{pand18}
C.~Pandarinath, D.~O’Shea, J.~Collins, R.~Jozefowicz, S.~Stavisky, J.~Kao,
  E.~Trautmann, M.~Kaufman, S.~Ryu, L.~Hochberg, J.~Henderson, K.~Shenoy,
  L.~Abbott, and D.~Sussillo.
\newblock Inferring single-trial neural population dynamics using sequential
  auto-encoders.
\newblock \emph{Nature Methods}, 15, 10 2018.
\newblock \doi{10.1038/s41592-018-0109-9}.

\bibitem[Pearl(1988)]{pearl1988probabilistic}
J.~Pearl.
\newblock \emph{{Probabilistic Reasoning in Intelligent Systems: Networks of
  Plausible Inference}}.
\newblock Morgan Kaufmann Publishers Inc., San Francisco, CA, USA, 1988.
\newblock ISBN 0934613737.

\bibitem[Pearl(1995)]{pearl1995causal}
J.~Pearl.
\newblock {Causal diagrams for empirical research}.
\newblock \emph{Biometrika}, 82\penalty0 (4):\penalty0 669--688, 1995.

\bibitem[Pearl(2009{\natexlab{a}})]{Pearl2000}
J.~Pearl.
\newblock \emph{{Causality: Models, Reasoning and Inference}}.
\newblock Cambridge University Press, New York, NY, USA, 2nd edition,
  2009{\natexlab{a}}.
\newblock ISBN 052189560x, 9780521895606.

\bibitem[Pearl(2009{\natexlab{b}})]{Pearl2009}
J.~Pearl.
\newblock {Causal inference in statistics: An overview}.
\newblock \emph{Stat. Surv.}, 3:\penalty0 96--146, 2009{\natexlab{b}}.

\bibitem[Pearl(2009{\natexlab{c}})]{pearl2009causality}
J.~Pearl.
\newblock \emph{{Causality: Models, Reasoning, and Inference}}.
\newblock Cambridge University Press, Cambridge, UK, 2nd edition,
  2009{\natexlab{c}}.

\bibitem[Pearl(2011)]{pearl2011statistics}
J.~Pearl.
\newblock {Statistics and Causality: Separated to Reunite—Commentary on Bryan
  Dowd's “Separated at Birth”}.
\newblock \emph{Health Services Research}, 46\penalty0 (2):\penalty0 421--429,
  2011.

\bibitem[Pearl and Mackenzie(2018)]{Pearl2018a}
J.~Pearl and D.~Mackenzie.
\newblock \emph{{The Book of Why: The New Science of Cause and Effect}}.
\newblock Basic books, New York, 2018.

\bibitem[Pearl et~al.(2016)Pearl, Glymour, and Jewell]{Pearl2016}
J.~Pearl, M.~Glymour, and N.~P. Jewell.
\newblock \emph{{Causal inference in statistics: A Primer}}.
\newblock John Wiley {\&} Sons, 2016.
\newblock ISBN 1935-7516.

\bibitem[Penny et~al.(2004)Penny, Stephan, Mechelli, and
  Friston]{penny2004comparing}
W.~D. Penny, K.~E. Stephan, A.~Mechelli, and K.~J. Friston.
\newblock {Comparing dynamic causal models}.
\newblock \emph{Neuroimage}, 22\penalty0 (3):\penalty0 1157--1172, 2004.

\bibitem[Penny et~al.(2010)Penny, Stephan, Daunizeau, Rosa, Friston, Schofield,
  and Leff]{penny2010comparing}
W.~D. Penny, K.~E. Stephan, J.~Daunizeau, M.~J. Rosa, K.~J. Friston, T.~M.
  Schofield, and A.~P. Leff.
\newblock {Comparing families of dynamic causal models}.
\newblock \emph{PLoS computational biology}, 6\penalty0 (3):\penalty0 e1000709,
  2010.

\bibitem[P\'erez-Suay and Camps-Valls(2019)]{PerezSuay19shsic}
A.~P\'erez-Suay and G.~Camps-Valls.
\newblock {Causal Inference in Geoscience and Remote Sensing from Observational
  Data}.
\newblock \emph{IEEE Transactions on Geoscience and Remote Sensing},
  57\penalty0 (3):\penalty0 1502--1513, 2019.

\bibitem[Peters et~al.(2011)Peters, Mooij, Janzing, and
  Sch\"{o}lkopf]{peters2011identifiability}
J.~Peters, J.~M. Mooij, D.~Janzing, and B.~Sch\"{o}lkopf.
\newblock {Identifiability of Causal Graphs Using Functional Models}.
\newblock In \emph{Proceedings of the Twenty-Seventh Conference on Uncertainty
  in Artificial Intelligence}, Uai'11, page 589–598, Arlington, Virginia,
  USA, 2011. AUAI Press.
\newblock ISBN 9780974903972.

\bibitem[Peters et~al.(2013)Peters, Janzing, and
  Sch\"{o}lkopf]{peters2013causal}
J.~Peters, D.~Janzing, and B.~Sch\"{o}lkopf.
\newblock {Causal Inference on Time Series using Restricted Structural Equation
  Models}.
\newblock In C.~Burges, L.~Bottou, M.~Welling, Z.~Ghahramani, and
  K.~Weinberger, editors, \emph{Advances in Neural Information Processing
  Systems}, volume~26. Curran Associates, Inc., 2013.

\bibitem[Peters et~al.(2016)Peters, B{\"u}hlmann, and
  Meinshausen]{peters2016causal}
J.~Peters, P.~B{\"u}hlmann, and N.~Meinshausen.
\newblock {Causal inference by using invariant prediction: identification and
  confidence intervals}.
\newblock \emph{Journal of the Royal Statistical Society: Series B (Statistical
  Methodology)}, 78\penalty0 (5):\penalty0 947--1012, 2016.

\bibitem[Peters et~al.(2017{\natexlab{a}})Peters, Janzing, and
  Sch\"olkopf]{Peters2017book}
J.~Peters, D.~Janzing, and B.~Sch\"olkopf.
\newblock \emph{{Elements of Causal Inference: Foundations and Learning
  Algorithms}}.
\newblock MIT Press, Cambridge, MA, USA, 2017{\natexlab{a}}.

\bibitem[Peters et~al.(2017{\natexlab{b}})Peters, Janzing, and
  Sch{\"o}lkopf]{peters2017elements}
J.~Peters, D.~Janzing, and B.~Sch{\"o}lkopf.
\newblock \emph{{Elements of causal inference: foundations and learning
  algorithms}}.
\newblock The MIT Press, 2017{\natexlab{b}}.

\bibitem[Peters et~al.(2022)Peters, Bauer, and Pfister]{peters2022causal}
J.~Peters, S.~Bauer, and N.~Pfister.
\newblock \emph{Causal Models for Dynamical Systems}, page 671–690.
\newblock Association for Computing Machinery, New York, NY, USA, 1 edition,
  2022.
\newblock ISBN 9781450395861.

\bibitem[Peters(2012)]{peters2012restricted}
J.~M. Peters.
\newblock \emph{{Restricted structural equation models for causal inference}}.
\newblock PhD thesis, ETH Zurich and MPI for Intelligent Systems, 2012.

\bibitem[Petersen et~al.(2021)Petersen, Larma, Mundhenk, Santiago, Kim, and
  Kim]{petersen2021deep}
B.~K. Petersen, M.~L. Larma, T.~N. Mundhenk, C.~P. Santiago, S.~K. Kim, and
  J.~T. Kim.
\newblock {Deep symbolic regression: Recovering mathematical expressions from
  data via risk-seeking policy gradients}.
\newblock In \emph{International Conference on Learning Representations}, 2021.

\bibitem[Petersen and Hansen(2021)]{Petersen2021testing}
L.~Petersen and N.~R. Hansen.
\newblock {Testing Conditional Independence via Quantile Regression Based
  Partial Copulas}.
\newblock \emph{Journal of Machine Learning Research}, 22\penalty0
  (70):\penalty0 1--47, 2021.

\bibitem[Pfister et~al.(2019)Pfister, B{\"u}hlmann, and
  Peters]{pfister2019invariant}
N.~Pfister, P.~B{\"u}hlmann, and J.~Peters.
\newblock {Invariant causal prediction for sequential data}.
\newblock \emph{Journal of the American Statistical Association}, 114\penalty0
  (527):\penalty0 1264--1276, 2019.

\bibitem[Pope(2000)]{pope}
S.~B. Pope.
\newblock {Turbulent Flows}.
\newblock \emph{Cambridge University Press}, 2000.

\bibitem[Popper(2005)]{popper2005logic}
K.~Popper.
\newblock \emph{{The logic of scientific discovery}}.
\newblock Routledge, 2005.

\bibitem[Potthast(2013)]{pott13}
R.~Potthast.
\newblock \emph{{Amari Model}}, pages 1--6.
\newblock Springer New York, New York, NY, 2013.
\newblock ISBN 978-1-4614-7320-6.

\bibitem[Praksova(2011)]{eureqa}
R.~Praksova.
\newblock Eureqa: {S}oftware review.
\newblock \emph{Genet. Program. Evol. M.}, 12\penalty0 (1):\penalty0 173--178,
  2011.

\bibitem[Press(1967)]{press1967compound}
S.~J. Press.
\newblock {A compound events model for security prices}.
\newblock \emph{Journal of business}, pages 317--335, 1967.

\bibitem[Pukrittayakamee et~al.(2009)Pukrittayakamee, Malshe, Hagan, Raff,
  Narulkar, Bukkapatnum, and Komanduri]{pukrittayakamee2009simultaneous}
A.~Pukrittayakamee, M.~Malshe, M.~Hagan, L.~Raff, R.~Narulkar, S.~Bukkapatnum,
  and R.~Komanduri.
\newblock {Simultaneous fitting of a potential-energy surface and its
  corresponding force fields using feedforward neural networks}.
\newblock \emph{The Journal of chemical physics}, 130\penalty0 (13):\penalty0
  134101, 2009.

\bibitem[Quade et~al.(2019)Quade, Gout, and Abel]{Glyph}
M.~Quade, J.~Gout, and M.~Abel.
\newblock Glyph: {S}ymbolic regression tools.
\newblock \emph{Journal of Open Research Software}, June 2019.

\bibitem[Rabinovich and Varona(2018)]{rabi18}
M.~I. Rabinovich and P.~Varona.
\newblock {Discrete Sequential Information Coding: Heteroclinic Cognitive
  Dynamics}.
\newblock \emph{Frontiers in Computational Neuroscience}, 12, 2018.

\bibitem[Rabinovich et~al.(2008)Rabinovich, Huerta, Varona, and
  Afraimovich]{rabi08}
M.~I. Rabinovich, R.~Huerta, P.~Varona, and V.~S. Afraimovich.
\newblock {Transient Cognitive Dynamics, Metastability, and Decision Making}.
\newblock \emph{PLOS Computational Biology}, 4\penalty0 (5):\penalty0 1--9, 05
  2008.

\bibitem[Raia(2008)]{raia2008causality}
F.~Raia.
\newblock {Causality in complex dynamic systems: A challenge in earth systems
  science education}.
\newblock \emph{Journal of Geoscience Education}, 56\penalty0 (1):\penalty0
  81--94, 2008.

\bibitem[Ramsey et~al.(2006)Ramsey, Spirtes, and Zhang]{ramsey2006adjacency}
J.~Ramsey, P.~Spirtes, and J.~Zhang.
\newblock {Adjacency-Faithfulness and Conservative Causal Inference}.
\newblock In \emph{Proceedings of the Twenty-Second Conference on Uncertainty
  in Artificial Intelligence}, Uai'06, page 401–408, Arlington, Virginia,
  USA, 2006. AUAI Press.
\newblock ISBN 0974903922.

\bibitem[Ramsey et~al.(2017)Ramsey, Glymour, Sanchez-Romero, and
  Glymour]{ramsey2017million}
J.~Ramsey, M.~Glymour, R.~Sanchez-Romero, and C.~Glymour.
\newblock {A million variables and more: the fast greedy equivalence search
  algorithm for learning high-dimensional graphical causal models, with an
  application to functional magnetic resonance images}.
\newblock \emph{International journal of data science and analytics},
  3\penalty0 (2):\penalty0 121--129, 2017.

\bibitem[Rasmussen and Williams(2006)]{rasmussen2006gaussian}
C.~E. Rasmussen and C.~K.~I. Williams.
\newblock \emph{{Gaussian Processes for Machine Learning}}.
\newblock The MIT Press, 2006.

\bibitem[Razi and Friston(2016)]{razi2016connected}
A.~Razi and K.~J. Friston.
\newblock {The connected brain: causality, models, and intrinsic dynamics}.
\newblock \emph{IEEE Signal Processing Magazine}, 33\penalty0 (3):\penalty0
  14--35, 2016.

\bibitem[Reichenbach(1991)]{reichenbach1991direction}
H.~Reichenbach.
\newblock \emph{{The direction of time}}, volume~65.
\newblock Univ of California Press, 1991.

\bibitem[Reid et~al.(2019)Reid, Headley, Mill, sanchez romero, Uddin,
  Marinazzo, Lurie, Valdés-Sosa, Hanson, Biswal, Calhoun, Poldrack, and
  Cole]{reid19}
A.~Reid, D.~Headley, R.~Mill, R.~sanchez romero, L.~Uddin, D.~Marinazzo,
  D.~Lurie, P.~Valdés-Sosa, S.~Hanson, B.~Biswal, V.~Calhoun, R.~Poldrack, and
  M.~Cole.
\newblock {Advancing functional connectivity research from association to
  causation}.
\newblock \emph{Nature Neuroscience}, 22:\penalty0 1--10, Oct. 2019.

\bibitem[Reisach et~al.(2021)Reisach, Seiler, and Weichwald]{reisach2021beware}
A.~G. Reisach, C.~Seiler, and S.~Weichwald.
\newblock {Beware of the Simulated DAG! Causal Discovery Benchmarks May Be Easy
  To Game}.
\newblock In \emph{{Advances in Neural Information Processing Systems 34
  (NeurIPS)}}, 2021.

\bibitem[Reitsma(2010)]{reitsma2010geoscience}
F.~Reitsma.
\newblock {Geoscience explanations: Identifying what is needed for generating
  scientific narratives from data models}.
\newblock \emph{Environmental Modelling \& Software}, 25\penalty0 (1):\penalty0
  93--99, 2010.

\bibitem[Reynolds(1895)]{reynolds}
O.~Reynolds.
\newblock {On the dynamical theory of incompressible viscous fluids and the
  determination of the criterion}.
\newblock \emph{Phil. Trans. R. Soc. A}, 186:\penalty0 123--164, 1895.

\bibitem[Richardson(1996)]{richardson1996discovery}
T.~Richardson.
\newblock {A Discovery Algorithm for Directed Cyclic Graphs}.
\newblock In \emph{Proceedings of the Twelfth International Conference on
  Uncertainty in Artificial Intelligence}, Uai'96, page 454–461, San
  Francisco, CA, USA, 1996. Morgan Kaufmann Publishers Inc.
\newblock ISBN 155860412x.

\bibitem[Richardson and Spirtes(2002)]{richardson2002ancestral}
T.~Richardson and P.~Spirtes.
\newblock {Ancestral Graph Markov Models}.
\newblock \emph{The Annals of Statistics}, 30\penalty0 (4):\penalty0 962--1030,
  2002.

\bibitem[Richiardi et~al.(2013)Richiardi, Achard, Bunke, and Van
  De~Ville]{richiardi2013machine}
J.~Richiardi, S.~Achard, H.~Bunke, and D.~Van De~Ville.
\newblock {Machine learning with brain graphs: predictive modeling approaches
  for functional imaging in systems neuroscience}.
\newblock \emph{IEEE Signal processing magazine}, 30\penalty0 (3):\penalty0
  58--70, 2013.

\bibitem[Rinaldo et~al.(2019)Rinaldo, Wasserman, and
  G’Sell]{rinaldo2019bootstrapping}
A.~Rinaldo, L.~Wasserman, and M.~G’Sell.
\newblock {Bootstrapping and sample splitting for high-dimensional,
  assumption-lean inference}.
\newblock \emph{The Annals of Statistics}, 47\penalty0 (6):\penalty0
  3438--3469, 2019.

\bibitem[Robins et~al.(2003)Robins, Scheines, Spirtes, and
  Wasserman]{robins2003uniform}
J.~M. Robins, R.~Scheines, P.~Spirtes, and L.~Wasserman.
\newblock {Uniform consistency in causal inference}.
\newblock \emph{Biometrika}, 90\penalty0 (3):\penalty0 491--515, 2003.

\bibitem[Robins and Hernan(2020)]{robins2020causal}
M.~Robins and M.~Hernan.
\newblock {Causal inference: what if}.
\newblock \emph{Found Agnostic Stat}, pages 235--281, 2020.

\bibitem[Rolinek et~al.(2019)Rolinek, Zietlow, and
  Martius]{RolinekZietlowMartius:VAERecPCA}
M.~Rolinek, D.~Zietlow, and G.~Martius.
\newblock {Variational Autoencoders Pursue PCA Directions (by Accident)}.
\newblock In \emph{Proceedings IEEE Conf. on Computer Vision and Pattern
  Recognition (CVPR)}, pages 12406--12415, June 2019.

\bibitem[Ross et~al.(2023)Ross, Li, Perezhogin, Fernandez-Granda, and
  Zanna]{Ross-et-al2022}
A.~S. Ross, Z.~Li, P.~Perezhogin, C.~Fernandez-Granda, and L.~Zanna.
\newblock {Benchmarking of machine learning ocean subgrid parameterizations in
  an idealized model}.
\newblock \emph{Journal of Advances in Modeling Earth Systems}, 15:\penalty0
  e2022MS003258, 2023.

\bibitem[Ross(2015)]{ross15}
L.~N. Ross.
\newblock {Dynamical Models and Explanation in Neuroscience}.
\newblock \emph{Philosophy of Science}, 82\penalty0 (1):\penalty0 32–54,
  2015.

\bibitem[Rowley et~al.(2009{\natexlab{a}})Rowley, Mezic, Bagheri, Schlatter,
  and Henningson]{rowley_et_al}
C.~Rowley, I.~Mezic, S.~Bagheri, P.~Schlatter, and D.~Henningson.
\newblock {Spectral analysis of nonlinear flows}.
\newblock \emph{J. Fluid Mech.}, 641:\penalty0 115--127, 2009{\natexlab{a}}.

\bibitem[Rowley and Dawson(2017)]{rowley}
C.~W. Rowley and S.~T. Dawson.
\newblock {Model reduction for flow analysis and control}.
\newblock \emph{Annu. Rev. Fluid Mech.}, 49:\penalty0 387--417, 2017.

\bibitem[Rowley et~al.(2009{\natexlab{b}})Rowley, Mezi{\'c}, Bagheri,
  Schlatter, and Henningson]{RMBSH09}
C.~W. Rowley, I.~Mezi{\'c}, S.~Bagheri, P.~Schlatter, and D.~S. Henningson.
\newblock {Spectral analysis of nonlinear flows}.
\newblock \emph{Journal of Fluid Mechanics}, 641:\penalty0 115--127,
  2009{\natexlab{b}}.

\bibitem[Rubenstein et~al.(2018)Rubenstein, Bongers, Bernhard, and
  Mooij]{rubenstain2018}
P.~K. Rubenstein, B.~Bongers, S.~Bernhard, and J.~M. Mooij.
\newblock {From Deterministic ODEs to Dynamic Structural Causal Models}.
\newblock In \emph{Thirty-Fourth Conference on Uncertainty in Artifical
  Intelligence}. AUAI Press Corvallis, Oregon, 2018.

\bibitem[Runge(2015)]{runge2015quantifying}
J.~Runge.
\newblock {Quantifying information transfer and mediation along causal pathways
  in complex systems}.
\newblock \emph{Physical Review E}, 92\penalty0 (6):\penalty0 062829, 2015.

\bibitem[Runge(2018)]{runge2018conditional}
J.~Runge.
\newblock {Conditional independence testing based on a nearest-neighbor
  estimator of conditional mutual information}.
\newblock In A.~Storkey and F.~Perez-Cruz, editors, \emph{International
  Conference on Artificial Intelligence and Statistics}, volume~84 of
  \emph{Proceedings of Machine Learning Research}, pages 938--947. Pmlr, 2018.

\bibitem[Runge(2020)]{runge2020discovering}
J.~Runge.
\newblock {Discovering contemporaneous and lagged causal relations in
  autocorrelated nonlinear time series datasets}.
\newblock In J.~Peters and D.~Sontag, editors, \emph{Proceedings of the 36th
  Conference on Uncertainty in Artificial Intelligence (UAI)}, volume 124 of
  \emph{Proceedings of Machine Learning Research}, pages 1388--1397. Pmlr,
  03--06 Aug 2020.

\bibitem[Runge(2021)]{runge2021necessary}
J.~Runge.
\newblock {Necessary and sufficient graphical conditions for optimal adjustment
  sets in causal graphical models with hidden variables}.
\newblock In M.~Ranzato, A.~Beygelzimer, Y.~Dauphin, P.~Liang, and J.~W.
  Vaughan, editors, \emph{Advances in Neural Information Processing Systems 34
  (NeurIPS 2021)}, 2021.

\bibitem[Runge et~al.(2012)Runge, Heitzig, Petoukhov, and
  Kurths]{runge2012escaping}
J.~Runge, J.~Heitzig, V.~Petoukhov, and J.~Kurths.
\newblock {Escaping the Curse of Dimensionality in Estimating Multivariate
  Transfer Entropy}.
\newblock \emph{Physical Review Letters}, 108:\penalty0 258701, June 2012.

\bibitem[Runge et~al.(2014)Runge, Petoukhov, and Kurths]{runge2014quantifying}
J.~Runge, V.~Petoukhov, and J.~Kurths.
\newblock {Quantifying the strength and delay of climatic interactions: The
  ambiguities of cross correlation and a novel measure based on graphical
  models}.
\newblock \emph{Journal of climate}, 27\penalty0 (2):\penalty0 720--739, 2014.

\bibitem[Runge et~al.(2015{\natexlab{a}})Runge, Donner, and
  Kurths]{runge2015optimal}
J.~Runge, R.~V. Donner, and J.~Kurths.
\newblock {Optimal model-free prediction from multivariate time series}.
\newblock \emph{Physical Review E}, 91\penalty0 (5):\penalty0 052909,
  2015{\natexlab{a}}.

\bibitem[Runge et~al.(2015{\natexlab{b}})Runge, Petoukhov, Donges, Hlinka,
  Jajcay, Vejmelka, Hartman, Marwan, Palu{\v{s}}, and
  Kurths]{runge2015identifying}
J.~Runge, V.~Petoukhov, J.~F. Donges, J.~Hlinka, N.~Jajcay, M.~Vejmelka,
  D.~Hartman, N.~Marwan, M.~Palu{\v{s}}, and J.~Kurths.
\newblock {Identifying causal gateways and mediators in complex spatio-temporal
  systems}.
\newblock \emph{Nature communications}, 6\penalty0 (1):\penalty0 1--10,
  2015{\natexlab{b}}.

\bibitem[Runge et~al.(2019{\natexlab{a}})Runge, Bathiany, Bollt, Camps-Valls,
  Coumou, Deyle, Glymour, Kretschmer, Mahecha, Mu{\~n}oz-Mar{\'\i},
  et~al.]{runge2019inferring}
J.~Runge, S.~Bathiany, E.~Bollt, G.~Camps-Valls, D.~Coumou, E.~Deyle,
  C.~Glymour, M.~Kretschmer, M.~D. Mahecha, J.~Mu{\~n}oz-Mar{\'\i}, et~al.
\newblock {Inferring causation from time series in Earth system sciences}.
\newblock \emph{Nature communications}, 10\penalty0 (1):\penalty0 1--13,
  2019{\natexlab{a}}.

\bibitem[Runge et~al.(2019{\natexlab{b}})Runge, Nowack, Kretschmer, Flaxman,
  and Sejdinovic]{runge2019detecting}
J.~Runge, P.~Nowack, M.~Kretschmer, S.~Flaxman, and D.~Sejdinovic.
\newblock {Detecting and quantifying causal associations in large nonlinear
  time series datasets}.
\newblock \emph{Science advances}, 5\penalty0 (11):\penalty0 eaau4996,
  2019{\natexlab{b}}.

\bibitem[Runge et~al.(2020)Runge, Tibau, Bruhns, Mu{\~n}oz-Mar{\'\i}, and
  Camps-Valls]{runge2020causality}
J.~Runge, X.-A. Tibau, M.~Bruhns, J.~Mu{\~n}oz-Mar{\'\i}, and G.~Camps-Valls.
\newblock {The causality for climate competition}.
\newblock In \emph{NeurIPS 2019 Competition and Demonstration Track}, pages
  110--120. Pmlr, 2020.

\bibitem[Runge et~al.(2023)Runge, Gerhardus, Varando, Eyring, and
  Camps-Valls]{runge2023causal}
J.~Runge, A.~Gerhardus, G.~Varando, V.~Eyring, and G.~Camps-Valls.
\newblock {Causal inference for time series}.
\newblock \emph{Nature Reviews Earth \& Environment}, 10:\penalty0 2553, 2023.

\bibitem[Russell(2021)]{russell2021human}
S.~Russell.
\newblock {Human-compatible artificial intelligence}.
\newblock \emph{Human-Like Machine Intelligence}, pages 3--23, 2021.

\bibitem[Russo(2010)]{russo2010causality}
F.~Russo.
\newblock \emph{{Causality and causal modelling in the social sciences}}.
\newblock Springer, 2010.

\bibitem[Rutten et~al.(2020)Rutten, Bernacchia, Sahani, and Hennequin]{rutt20}
V.~Rutten, A.~Bernacchia, M.~Sahani, and G.~Hennequin.
\newblock {Non-reversible Gaussian processes for identifying latent dynamical
  structure in neural data}.
\newblock In H.~Larochelle, M.~Ranzato, R.~Hadsell, M.~Balcan, and H.~Lin,
  editors, \emph{Advances in Neural Information Processing Systems}, volume~33,
  pages 9622--9632. Curran Associates, Inc., 2020.

\bibitem[Saggioro et~al.(2020)Saggioro, de~Wiljes, Kretschmer, and
  Runge]{saggioro2020reconstructing}
E.~Saggioro, J.~de~Wiljes, M.~Kretschmer, and J.~Runge.
\newblock {Reconstructing regime-dependent causal relationships from
  observational time series}.
\newblock \emph{Chaos: An Interdisciplinary Journal of Nonlinear Science},
  30\penalty0 (11):\penalty0 113115, 2020.

\bibitem[Sahoo et~al.(2018)Sahoo, Lampert, and
  Martius]{SahooLampertMartius2018:EQLDiv}
S.~S. Sahoo, C.~H. Lampert, and G.~Martius.
\newblock {Learning equations for extrapolation and control}.
\newblock In J.~Dy and A.~Krause, editors, \emph{Proc. 35th International
  Conference on Machine Learning, {ICML} 2018, Stockholm, Sweden}, volume~80,
  pages 4442--4450. {Pmlr}, 2018.

\bibitem[Salcedo-Sanz et~al.(2022)Salcedo-Sanz, Casillas-P{\'e}rez, Del~Ser,
  Casanova-Mateo, Cuadra, Piles, and Camps-Valls]{salcedo2022persistence}
S.~Salcedo-Sanz, D.~Casillas-P{\'e}rez, J.~Del~Ser, C.~Casanova-Mateo,
  L.~Cuadra, M.~Piles, and G.~Camps-Valls.
\newblock {Persistence in complex systems}.
\newblock \emph{Physics Reports}, 957:\penalty0 1--73, 2022.

\bibitem[Sani et~al.(2021)Sani, Abbaspourazad, Wong, Pesaran, and
  Shanechi]{sani21}
O.~Sani, H.~Abbaspourazad, Y.~Wong, B.~Pesaran, and M.~Shanechi.
\newblock Modeling behaviorally relevant neural dynamics enabled by
  preferential subspace identification.
\newblock \emph{Nature Neuroscience}, 24:\penalty0 140--149, 2021.
\newblock \doi{10.1038/s41593-020-00733-0}.

\bibitem[Schaffer(1990)]{schaffer1990}
C.~Schaffer.
\newblock {Constructing Explanations for Propositional Knowledge Bases}.
\newblock \emph{Machine Learning}, 4\penalty0 (4):\penalty0 321--353, 1990.

\bibitem[Schirner et~al.(2022)Schirner, Domide, Perdikis, Triebkorn,
  Stefanovski, Pai, Prodan, Valean, Palmer, Langford, Blickensdörfer, {van der
  Vlag}, Diaz-Pier, Peyser, Klijn, Pleiter, Nahm, Schmid, Woodman, Zehl,
  Fousek, Petkoski, Kusch, Hashemi, Marinazzo, Mangin, Flöel, Akintoye, Stahl,
  Cepic, Johnson, Deco, McIntosh, Hilgetag, Morgan, Schuller, Upton, McMurtrie,
  Dickscheid, Bjaalie, Amunts, Mersmann, Jirsa, and Ritter]{schi22}
M.~Schirner, L.~Domide, D.~Perdikis, P.~Triebkorn, L.~Stefanovski, R.~Pai,
  P.~Prodan, B.~Valean, J.~Palmer, C.~Langford, A.~Blickensdörfer, M.~{van der
  Vlag}, S.~Diaz-Pier, A.~Peyser, W.~Klijn, D.~Pleiter, A.~Nahm, O.~Schmid,
  M.~Woodman, L.~Zehl, J.~Fousek, S.~Petkoski, L.~Kusch, M.~Hashemi,
  D.~Marinazzo, J.-F. Mangin, A.~Flöel, S.~Akintoye, B.~C. Stahl, M.~Cepic,
  E.~Johnson, G.~Deco, A.~R. McIntosh, C.~C. Hilgetag, M.~Morgan, B.~Schuller,
  A.~Upton, C.~McMurtrie, T.~Dickscheid, J.~G. Bjaalie, K.~Amunts, J.~Mersmann,
  V.~Jirsa, and P.~Ritter.
\newblock {Brain simulation as a cloud service: The Virtual Brain on EBRAINS}.
\newblock \emph{NeuroImage}, 251:\penalty0 118973, 2022.

\bibitem[Schmekel et~al.(2022)Schmekel, Alc\'antara-\'Avila, Hoyas, and
  Vinuesa]{structure_pred}
D.~Schmekel, F.~Alc\'antara-\'Avila, S.~Hoyas, and R.~Vinuesa.
\newblock Predicting coherent turbulent structures via deep learning.
\newblock \emph{Front. Phys.}, 10:\penalty0 888832, 2022.

\bibitem[Schmid(2010)]{Schmid10}
P.~J. Schmid.
\newblock {Dynamic mode decomposition of numerical and experimental data}.
\newblock \emph{Journal of Fluid Mechanics}, 656:\penalty0 5--28, 2010.

\bibitem[Schmidt and Lipson(2009{\natexlab{a}})]{SchmidtLipson}
M.~Schmidt and H.~Lipson.
\newblock {Distilling Free-Form Natural Laws from Experimental Data}.
\newblock \emph{Science}, 324\penalty0 (5923):\penalty0 81--85,
  2009{\natexlab{a}}.

\bibitem[Schmidt and Lipson(2009{\natexlab{b}})]{schmidt2009}
M.~Schmidt and H.~Lipson.
\newblock {Distilling Free-Form Natural Laws from Experimental Data}.
\newblock \emph{Science}, 324\penalty0 (5923):\penalty0 81--85,
  2009{\natexlab{b}}.

\bibitem[Schmidt et~al.(2011)Schmidt, Vallabhajosyula, Jenkins, Hood, Soni,
  Wikswo, and Lipson]{schmidt2011automated}
M.~D. Schmidt, R.~R. Vallabhajosyula, J.~W. Jenkins, J.~E. Hood, A.~S. Soni,
  J.~P. Wikswo, and H.~Lipson.
\newblock {Automated refinement and inference of analytical models for
  metabolic networks}.
\newblock \emph{Physical biology}, 8\penalty0 (5):\penalty0 055011, 2011.

\bibitem[Schneider et~al.(2023)Schneider, Lee, and Mathis]{schn23}
S.~Schneider, J.~H. Lee, and M.~W. Mathis.
\newblock Learnable latent embeddings for joint behavioural and neural
  analysis.
\newblock \emph{Nature}, May 2023.
\newblock ISSN 1476-4687.
\newblock \doi{10.1038/s41586-023-06031-6}.
\newblock URL \url{https://doi.org/10.1038/s41586-023-06031-6}.

\bibitem[Sch{\"o}lkopf and Smola(2008)]{kernels2008}
B.~Sch{\"o}lkopf and A.~Smola.
\newblock \emph{{Learning with Kernels}}.
\newblock MIT Press, Cambridge, MA, USA, 2008.

\bibitem[Sch{\"o}lkopf et~al.(2021)Sch{\"o}lkopf, Locatello, Bauer, Ke,
  Kalchbrenner, Goyal, and Bengio]{scholkopf2021toward}
B.~Sch{\"o}lkopf, F.~Locatello, S.~Bauer, N.~R. Ke, N.~Kalchbrenner, A.~Goyal,
  and Y.~Bengio.
\newblock {Toward causal representation learning}.
\newblock \emph{Proceedings of the IEEE}, 109\penalty0 (5):\penalty0 612--634,
  2021.

\bibitem[Schreiber(2000)]{schreiber2000measuring}
T.~Schreiber.
\newblock {Measuring information transfer}.
\newblock \emph{Physical review letters}, 85\penalty0 (2):\penalty0 461, 2000.

\bibitem[Sch{\"u}tt et~al.(2017)Sch{\"u}tt, Arbabzadah, Chmiela, M{\"u}ller,
  and Tkatchenko]{schutt2017quantum}
K.~T. Sch{\"u}tt, F.~Arbabzadah, S.~Chmiela, K.~R. M{\"u}ller, and
  A.~Tkatchenko.
\newblock {Quantum-chemical insights from deep tensor neural networks}.
\newblock \emph{Nature communications}, 8\penalty0 (1):\penalty0 13890, 2017.

\bibitem[Schwantes and Pande(2015)]{SP15}
C.~R. Schwantes and V.~S. Pande.
\newblock {Modeling molecular kinetics with tICA and the kernel trick}.
\newblock \emph{Journal of Chemical Theory and Computation}, 11:\penalty0
  600--608, 2015.

\bibitem[Scott~Armstrong and Collopy(1993)]{scott1993causal}
J.~Scott~Armstrong and F.~Collopy.
\newblock {Causal forces: Structuring knowledge for time-series extrapolation}.
\newblock \emph{Journal of Forecasting}, 12\penalty0 (2):\penalty0 103--115,
  1993.

\bibitem[Sellars et~al.(1956)]{sellars1956empiricism}
W.~Sellars et~al.
\newblock {Empiricism and the Philosophy of Mind}.
\newblock \emph{Minnesota studies in the philosophy of science}, 1\penalty0
  (19):\penalty0 253--329, 1956.

\bibitem[Shah and Peters(2020)]{shah2020hardness}
R.~D. Shah and J.~Peters.
\newblock {The hardness of conditional independence testing and the generalised
  covariance measure}.
\newblock \emph{The Annals of Statistics}, 48\penalty0 (3):\penalty0 1514 --
  1538, 2020.

\bibitem[Shannon(1948)]{shannon1948mathematical}
C.~E. Shannon.
\newblock {A mathematical theory of communication}.
\newblock \emph{The Bell system technical journal}, 27\penalty0 (3):\penalty0
  379--423, 1948.

\bibitem[Shea et~al.(2021)Shea, Brunton, and Kutz]{shea_et_al}
D.~Shea, S.~Brunton, and J.~Kutz.
\newblock {SINDy-BVP}: Sparse identification of nonlinear dynamics for boundary
  value problems.
\newblock \emph{Phys. Rev. Res.}, 3:\penalty0 023255, 2021.

\bibitem[Sheikhattar et~al.(2018)Sheikhattar, Miran, Liu, Fritz, Shamma,
  Kanold, and Babadi]{shei18}
A.~Sheikhattar, S.~Miran, J.~Liu, J.~B. Fritz, S.~A. Shamma, P.~O. Kanold, and
  B.~Babadi.
\newblock Extracting neuronal functional network dynamics via adaptive
  {Granger} causality analysis.
\newblock \emph{Proceedings of the National Academy of Sciences}, 115\penalty0
  (17):\penalty0 E3869--e3878, 2018.

\bibitem[Shepherd(2019)]{shepherd2019storyline}
T.~G. Shepherd.
\newblock {Storyline approach to the construction of regional climate change
  information}.
\newblock \emph{Proceedings of the Royal Society A}, 475\penalty0
  (2225):\penalty0 20190013, 2019.

\bibitem[Shimizu et~al.(2006)Shimizu, Hoyer, Hyv\"{a}inen, and
  Kerminen]{shimizu2006linear}
S.~Shimizu, P.~O. Hoyer, A.~Hyv\"{a}inen, and A.~Kerminen.
\newblock {A Linear Non-Gaussian Acyclic Model for Causal Discovery}.
\newblock \emph{Journal of Machine Learning Research}, 7\penalty0
  (72):\penalty0 2003--2030, 2006.

\bibitem[Shimizu et~al.(2011)Shimizu, Inazumi, Sogawa, Hyv{\"a}rinen, Kawahara,
  Washio, Hoyer, and Bollen]{shimizu2011directlingam}
S.~Shimizu, T.~Inazumi, Y.~Sogawa, A.~Hyv{\"a}rinen, Y.~Kawahara, T.~Washio,
  P.~O. Hoyer, and K.~Bollen.
\newblock {DirectLiNGAM: A direct method for learning a linear non-Gaussian
  structural equation model}.
\newblock \emph{The Journal of Machine Learning Research}, 12:\penalty0
  1225--1248, 2011.

\bibitem[Shpitser and Pearl(2006)]{shpitser2006identification}
I.~Shpitser and J.~Pearl.
\newblock {Identification of Conditional Interventional Distributions}.
\newblock In R.~Dechter and T.~Richardson, editors, \emph{Proceedings of the
  Twenty-Second Conference on Uncertainty in Artificial Intelligence}, Uai'06,
  page 437–444, Arlington, Virginia, USA, 2006. AUAI Press.
\newblock ISBN 0974903922.

\bibitem[Shpitser and Pearl(2008)]{shpitser2008complete}
I.~Shpitser and J.~Pearl.
\newblock {Complete identification methods for the causal hierarchy}.
\newblock \emph{Journal of Machine Learning Research}, 9:\penalty0 1941--1979,
  2008.

\bibitem[Shrager and Langley(1990)]{shrager1990}
J.~Shrager and P.~Langley.
\newblock \emph{{Computational Models of Scientific Discovery and Theory
  Formation}}.
\newblock Morgan Kaufmann, 1990.

\bibitem[Siddiqi et~al.(2022)Siddiqi, Kording, Parvizi, and
  Fox]{siddiqi2022causal}
S.~H. Siddiqi, K.~P. Kording, J.~Parvizi, and M.~D. Fox.
\newblock {Causal mapping of human brain function}.
\newblock \emph{Nature reviews neuroscience}, 23\penalty0 (6):\penalty0
  361--375, 2022.

\bibitem[Simidjievski et~al.(2020)Simidjievski, Todorovski, Kocijan, and
  D{\v{z}}eroski]{simidjievski2020equation}
N.~Simidjievski, L.~Todorovski, J.~Kocijan, and S.~D{\v{z}}eroski.
\newblock {Equation discovery for nonlinear system identification}.
\newblock \emph{IEEE Access}, 8:\penalty0 29930--29943, 2020.

\bibitem[Simon et~al.(1989)]{simon1989scientist}
H.~A. Simon et~al.
\newblock {The scientist as problem solver}.
\newblock \emph{Complex information processing: The impact of Herbert A.
  Simon}, pages 375--398, 1989.

\bibitem[Spalart(2000)]{spalart}
P.~R. Spalart.
\newblock {Strategies for turbulence modelling and simulations}.
\newblock \emph{Int. J. Heat Fluid Flow}, 21:\penalty0 252--263, 2000.

\bibitem[Spirtes and Glymour(1991)]{spirtes1991algorithm}
P.~Spirtes and C.~Glymour.
\newblock {An algorithm for fast recovery of sparse causal graphs}.
\newblock \emph{Social science computer review}, 9\penalty0 (1):\penalty0
  62--72, 1991.

\bibitem[Spirtes et~al.(1995)Spirtes, Meek, and Richardson]{spirtes1995causal}
P.~Spirtes, C.~Meek, and T.~Richardson.
\newblock {Causal Inference in the Presence of Latent Variables and Selection
  Bias}.
\newblock In P.~Besnard and S.~Hanks, editors, \emph{Proceedings of the
  Eleventh Conference on Uncertainty in Artificial Intelligence}, Uai'95, page
  499–506, San Francisco, CA, USA, 1995. Morgan Kaufmann Publishers Inc.
\newblock ISBN 1558603859.

\bibitem[Spirtes et~al.(2000)Spirtes, Glymour, and Scheines]{Spirtes2000}
P.~Spirtes, C.~Glymour, and R.~Scheines.
\newblock \emph{{Causation, Prediction, and Search}}.
\newblock MIT Press, Boston, 2000.

\bibitem[Stegle et~al.(2010)Stegle, Janzing, Zhang, Mooij, and
  Sch\"{o}lkopf]{probaLatZ}
O.~Stegle, D.~Janzing, K.~Zhang, J.~M. Mooij, and B.~Sch\"{o}lkopf.
\newblock {Probabilistic latent variable models for distinguishing between
  cause and effect}.
\newblock In J.~Lafferty, C.~Williams, J.~Shawe-Taylor, R.~Zemel, and
  A.~Culotta, editors, \emph{Advances in Neural Information Processing
  Systems}, volume~23. Curran Associates, Inc., 2010.

\bibitem[Stephan et~al.(2010)Stephan, Penny, Moran, den Ouden, Daunizeau, and
  Friston]{stephan2010ten}
K.~E. Stephan, W.~D. Penny, R.~J. Moran, H.~E. den Ouden, J.~Daunizeau, and
  K.~J. Friston.
\newblock {Ten simple rules for dynamic causal modeling}.
\newblock \emph{Neuroimage}, 49\penalty0 (4):\penalty0 3099--3109, 2010.

\bibitem[Stephens(2022)]{gplearn}
T.~Stephens.
\newblock gplearn: {G}enetic programming in {P}ython with a scikit-learn
  inspired {API}.
\newblock https://gplearn.readthedocs.io/en/stable, 2022.

\bibitem[Sterman(1994)]{sterman1994learning}
J.~D. Sterman.
\newblock {Learning in and about complex systems}.
\newblock \emph{System dynamics review}, 10\penalty0 (2-3):\penalty0 291--330,
  1994.

\bibitem[Stocker et~al.(2013)Stocker, Qin, Plattner, Tignor, Allen, Boschung,
  Nauels, Xia, Bex, and Midgley]{stocker2013climate}
T.~F. Stocker, D.~Qin, G.~Plattner, M.~Tignor, S.~Allen, J.~Boschung,
  A.~Nauels, Y.~Xia, V.~Bex, and P.~Midgley.
\newblock {Climate change 2013: the physical science basis. Intergovernmental
  panel on climate change, working group I contribution to the IPCC fifth
  assessment report (AR5)}.
\newblock \emph{New York}, 2013.

\bibitem[Stokes and Purdon(2017)]{stokes2017study}
P.~A. Stokes and P.~L. Purdon.
\newblock A study of problems encountered in {Granger} causality analysis from
  a neuroscience perspective.
\newblock \emph{Proceedings of the national academy of sciences}, 114\penalty0
  (34):\penalty0 E7063--e7072, 2017.

\bibitem[Strobl(2019)]{strobl2019constraint}
E.~V. Strobl.
\newblock A constraint-based algorithm for causal discovery with cycles, latent
  variables and selection bias.
\newblock \emph{International Journal of Data Science and Analytics},
  8\penalty0 (1):\penalty0 33--56, 2019.

\bibitem[Strobl et~al.(2018)Strobl, Visweswaran, and Spirtes]{strobl2018fast}
E.~V. Strobl, S.~Visweswaran, and P.~L. Spirtes.
\newblock Fast causal inference with non-random missingness by test-wise
  deletion.
\newblock \emph{International journal of data science and analytics},
  6:\penalty0 47--62, 2018.

\bibitem[Sugihara et~al.(2012)Sugihara, May, Ye, Hsieh, Deyle, Fogarty, and
  Munch]{sugihara2012detecting}
G.~Sugihara, R.~May, H.~Ye, C.-h. Hsieh, E.~Deyle, M.~Fogarty, and S.~Munch.
\newblock Detecting causality in complex ecosystems.
\newblock \emph{science}, 338\penalty0 (6106):\penalty0 496--500, 2012.

\bibitem[Sugiyama and Kawanabe(2012)]{sugiyama2012machine}
M.~Sugiyama and M.~Kawanabe.
\newblock \emph{Machine learning in non-stationary environments: {I}ntroduction
  to covariate shift adaptation}.
\newblock MIT press, 2012.

\bibitem[Sun et~al.(2023)Sun, Schulte, Liu, and Poupart]{sun2023nts}
X.~Sun, O.~Schulte, G.~Liu, and P.~Poupart.
\newblock {NTS-NOTEARS}: {L}earning nonparametric {DBNs} with prior knowledge.
\newblock In \emph{The 26th International Conference on Artificial Intelligence
  and Statistics (AISTATS)}, 2023.

\bibitem[Suseela et~al.(2022)Suseela, Feng, and Mao]{suseela2022comparative}
S.~S. Suseela, Y.~Feng, and K.~Mao.
\newblock A comparative study on machine learning algorithms for knowledge
  discovery.
\newblock In \emph{2022 17th International Conference on Control, Automation,
  Robotics and Vision (ICARCV)}, pages 131--136. Ieee, 2022.

\bibitem[Swearingen and Blackwelder(1987)]{b1}
J.~D. Swearingen and R.~F. Blackwelder.
\newblock The growth and breakdown of streamwise vortices in the presence of a
  wall.
\newblock \emph{J. Fluid Mech.}, 182:\penalty0 255--290, 1987.

\bibitem[Tabas et~al.(2019)Tabas, Andermann, Schuberth, Riedel,
  Balaguer-Ballester, and Rupp]{taba19}
A.~Tabas, M.~Andermann, V.~Schuberth, H.~Riedel, E.~Balaguer-Ballester, and
  A.~Rupp.
\newblock Modeling and {MEG} evidence of early consonance processing in
  auditory cortex.
\newblock \emph{PLOS Computational Biology}, 15\penalty0 (2):\penalty0 1--28,
  02 2019.

\bibitem[Taira et~al.(2017)Taira, Brunton, Dawson, Rowley, Colonius, McKeon,
  Schmidt, Gordeyev, Theofilis, and Ukeiley]{taira}
K.~Taira, S.~L. Brunton, S.~Dawson, C.~W. Rowley, T.~Colonius, B.~J. McKeon,
  O.~T. Schmidt, S.~Gordeyev, V.~Theofilis, and L.~S. Ukeiley.
\newblock Modal analysis of fluid flows: {A}n overview.
\newblock \emph{Aiaa J.}, 55\penalty0 (12):\penalty0 4013--4041, 2017.

\bibitem[Takagi(2020)]{Takagi2020PrinciplesOM}
K.~Takagi.
\newblock Principles of mutual information maximization and energy minimization
  affect the activation patterns of large scale networks in the brain.
\newblock \emph{Frontiers in Computational Neuroscience}, 13, 2020.

\bibitem[Takeishi et~al.(2017)Takeishi, Kawahara, and
  Yairi]{takeishi2017learning}
N.~Takeishi, Y.~Kawahara, and T.~Yairi.
\newblock Learning koopman invariant subspaces for dynamic mode decomposition.
\newblock \emph{Advances in neural information processing systems}, 30, 2017.

\bibitem[Takens(1981)]{Takens81}
F.~Takens.
\newblock Detecting strange attractors in turbulence.
\newblock In D.~A. Rand and L.-S. Young, editors, \emph{Dynamical Systems and
  Turbulence, Warwick 1980}, volume 898 of \emph{Lecture Notes in Mathematics},
  pages 366--381. Springer, Berlin, 1981.
\newblock ISBN 978-3-540-11171-9.

\bibitem[Tarski and Tarski(1994)]{tarski1994introduction}
A.~Tarski and J.~Tarski.
\newblock \emph{Introduction to Logic and to the Methodology of the Deductive
  Sciences}.
\newblock Number~24. Oxford University Press on Demand, 1994.

\bibitem[Tennekes and Lumley(1972)]{tennekes}
H.~Tennekes and J.~L. Lumley.
\newblock A first course in turbulence.
\newblock \emph{MIT press}, 1972.

\bibitem[Thiem et~al.(2020)Thiem, Kooshkbaghi, Bertalan, Laing, and
  Kevrekidis]{thie20}
T.~N. Thiem, M.~Kooshkbaghi, T.~Bertalan, C.~R. Laing, and I.~G. Kevrekidis.
\newblock Emergent spaces for coupled oscillators.
\newblock \emph{Front. Comput. Neurosci.}, 14:\penalty0 36, May 2020.

\bibitem[Tibau et~al.(2022)Tibau, Reimers, Gerhardus, Denzler, Eyring, and
  Runge]{tibau2022spatiotemporal}
X.-A. Tibau, C.~Reimers, A.~Gerhardus, J.~Denzler, V.~Eyring, and J.~Runge.
\newblock A spatiotemporal stochastic climate model for benchmarking causal
  discovery methods for teleconnections.
\newblock \emph{Environmental Data Science}, 1:\penalty0 e12, 2022.

\bibitem[Tibshirani(1996)]{tibshirani96regression}
R.~Tibshirani.
\newblock Regression shrinkage and selection via the lasso.
\newblock \emph{Journal of the Royal Statistical Society (Series B)},
  58:\penalty0 267--288, 1996.

\bibitem[Tipping(2001)]{tipping2001sparse}
M.~E. Tipping.
\newblock Sparse {Bayesian} learning and the relevance vector machine.
\newblock \emph{Journal of machine learning research}, 1\penalty0
  (Jun):\penalty0 211--244, 2001.

\bibitem[Todd et~al.(2013)Todd, Nystrom, and Cohen]{Todd2013}
M.~T. Todd, L.~E. Nystrom, and J.~D. Cohen.
\newblock Confounds in multivariate pattern analysis: {T}heory and rule
  representation case study.
\newblock \emph{NeuroImage}, 77:\penalty0 157--165, 2013.

\bibitem[Tognoli and Kelso(2014)]{tognoli2014metastable}
E.~Tognoli and J.~S. Kelso.
\newblock The metastable brain.
\newblock \emph{Neuron}, 81\penalty0 (1):\penalty0 35--48, 2014.

\bibitem[Tu et~al.(2014)Tu, Rowley, Luchtenburg, Brunton, and Kutz]{TRLBK14}
J.~H. Tu, C.~W. Rowley, D.~M. Luchtenburg, S.~L. Brunton, and J.~N. Kutz.
\newblock On dynamic mode decomposition: {T}heory and applications.
\newblock \emph{Journal of Computational Dynamics}, 1, 2014.

\bibitem[Tu et~al.(2019)Tu, Zhang, Ackermann, Mohan, Kjellstr{\"o}m, and
  Zhang]{tu2019causal}
R.~Tu, C.~Zhang, P.~Ackermann, K.~Mohan, H.~Kjellstr{\"o}m, and K.~Zhang.
\newblock Causal discovery in the presence of missing data.
\newblock In \emph{The 22nd International Conference on Artificial Intelligence
  and Statistics}, pages 1762--1770. Pmlr, 2019.

\bibitem[Udrescu and Tegmark(2020)]{udrescu2020ai}
S.-M. Udrescu and M.~Tegmark.
\newblock {AI Feynman}: {A} physics-inspired method for symbolic regression.
\newblock \emph{Science Advances}, 6\penalty0 (16):\penalty0 eaay2631, 2020.

\bibitem[Udrescu et~al.(2020)Udrescu, Tan, Feng, Neto, Wu, and
  Tegmark]{Udrescu2020:AI-feynman2_0}
S.-M. Udrescu, A.~Tan, J.~Feng, O.~Neto, T.~Wu, and M.~Tegmark.
\newblock {AI Feynman 2.0}: {Pareto}-optimal symbolic regression exploiting
  graph modularity.
\newblock In H.~Larochelle, M.~Ranzato, R.~Hadsell, M.~Balcan, and H.~Lin,
  editors, \emph{Advances in Neural Information Processing Systems}, volume~33,
  pages 4860--4871. Curran Associates, Inc., 2020.

\bibitem[Uhler et~al.(2013)Uhler, Raskutti, B{\"u}hlmann, and
  Yu]{uhler2013geometry}
C.~Uhler, G.~Raskutti, P.~B{\"u}hlmann, and B.~Yu.
\newblock Geometry of the faithfulness assumption in causal inference.
\newblock \emph{The Annals of Statistics}, pages 436--463, 2013.

\bibitem[Ulam(1960)]{Ulam60}
S.~M. Ulam.
\newblock \emph{A Collection of Mathematical Problems}.
\newblock Interscience Publisher NY, 1960.

\bibitem[Van~der Maaten and Hinton(2008)]{van2008visualizing}
L.~Van~der Maaten and G.~Hinton.
\newblock Visualizing data using {t-SNE}.
\newblock \emph{Journal of machine learning research}, 9\penalty0 (11), 2008.

\bibitem[VanderWeele(2015)]{vanderweele2015explanation}
T.~VanderWeele.
\newblock \emph{Explanation in causal inference: {M}ethods for mediation and
  interaction}.
\newblock Oxford University Press, 2015.

\bibitem[Vapnik(1999)]{vapnik1999overview}
V.~N. Vapnik.
\newblock An overview of statistical learning theory.
\newblock \emph{IEEE transactions on neural networks}, 10\penalty0
  (5):\penalty0 988--999, 1999.

\bibitem[Varando and Hansen(2020)]{varando2020graphical}
G.~Varando and N.~R. Hansen.
\newblock Graphical continuous {Lyapunov} models.
\newblock In \emph{Conference on Uncertainty in Artificial Intelligence}, pages
  989--998. PMLR, 2020.

\bibitem[Vaswani et~al.(2017)Vaswani, Shazeer, Parmar, Uszkoreit, Jones, Gomez,
  Kaiser, and Polosukhin]{vaswani2017:transformers}
A.~Vaswani, N.~Shazeer, N.~Parmar, J.~Uszkoreit, L.~Jones, A.~N. Gomez,
  {\L{}}.~Kaiser, and I.~Polosukhin.
\newblock Attention is all you need.
\newblock In I.~Guyon, U.~V. Luxburg, S.~Bengio, H.~Wallach, R.~Fergus,
  S.~Vishwanathan, and R.~Garnett, editors, \emph{Advances in {Neural}
  {Information} {Processing} {Systems}}, volume~30. Curran Associates, Inc.,
  2017.

\bibitem[Verma and Pearl(1990{\natexlab{a}})]{verma1990causal}
T.~Verma and J.~Pearl.
\newblock Causal networks: {S}emantics and expressiveness.
\newblock In R.~D. Shachter, T.~S. Levitt, L.~N. Kanal, and J.~F. Lemmer,
  editors, \emph{Uncertainty in Artificial Intelligence}, volume~9 of
  \emph{Machine Intelligence and Pattern Recognition}, pages 69--76.
  North-Holland, 1990{\natexlab{a}}.

\bibitem[Verma and Pearl(1990{\natexlab{b}})]{verma1990equivalence}
T.~Verma and J.~Pearl.
\newblock Equivalence and synthesis of causal models.
\newblock In P.~P. Bonissone, M.~Henrion, L.~N. Kanal, and J.~F. Lemmer,
  editors, \emph{Proceedings of the Sixth Annual Conference on Uncertainty in
  Artificial Intelligence}, Uai '90, page 255–270, New York, NY, USA,
  1990{\natexlab{b}}. Elsevier Science Inc.
\newblock ISBN 0444892648.

\bibitem[Versteeg et~al.(2022)Versteeg, Mooij, and Zhang]{versteeg22a}
P.~Versteeg, J.~Mooij, and C.~Zhang.
\newblock Local constraint-based causal discovery under selection bias.
\newblock In B.~Schölkopf, C.~Uhler, and K.~Zhang, editors, \emph{Proceedings
  of the First Conference on Causal Learning and Reasoning}, volume 177 of
  \emph{Proceedings of Machine Learning Research}, pages 840--860. Pmlr, 11--13
  Apr 2022.

\bibitem[Vinuesa and Sirmacek(2021)]{vinuesa_interp}
R.~Vinuesa and B.~Sirmacek.
\newblock Interpretable deep-learning models to help achieve the {Sustainable
  Development Goals}.
\newblock \emph{Nat. Mach. Intell.}, 3:\penalty0 926, 2021.

\bibitem[Vinuesa et~al.(2018)Vinuesa, Schlatter, and Nagib]{vinuesa_duct}
R.~Vinuesa, P.~Schlatter, and H.~M. Nagib.
\newblock Secondary flow in turbulent ducts with increasing aspect ratio.
\newblock \emph{Phys. Rev. Fluids}, 3:\penalty0 054606, 2018.

\bibitem[Von~Storch and Zwiers(2002)]{von2002statistical}
H.~Von~Storch and F.~W. Zwiers.
\newblock \emph{{Statistical analysis in climate research}}.
\newblock Cambridge university press, 2002.

\bibitem[Wagner(1999)]{wagner1999causality}
A.~Wagner.
\newblock Causality in complex systems.
\newblock \emph{Biology and Philosophy}, 14:\penalty0 83--101, 1999.

\bibitem[Waleffe(1995)]{b2}
F.~Waleffe.
\newblock Hydrodynamic stability and turbulence: {B}eyond transients to a
  self-sustaining process.
\newblock \emph{Stud. Appl. Maths}, 95:\penalty0 319--343, 1995.

\bibitem[Walker(1923)]{Walker1923}
G.~T. Walker.
\newblock Correlation in seasonal variations of weather, {VIII}: {A}
  preliminary study of world weather.
\newblock \emph{Mem. Indian Meteorol. Dep.}, 24\penalty0 (4):\penalty0 75--131,
  1923.

\bibitem[Wallace et~al.(1972)Wallace, Eckelman, and Brodkey]{wallace_et_al}
J.~M. Wallace, H.~Eckelman, and R.~S. Brodkey.
\newblock The wall region in turbulent shear flow.
\newblock \emph{J. Fluid Mech.}, 54:\penalty0 39--48, 1972.

\bibitem[Waltz and Buchanan(2009)]{waltz2009automating}
D.~Waltz and B.~G. Buchanan.
\newblock Automating science.
\newblock \emph{Science}, 324\penalty0 (5923):\penalty0 43--44, 2009.

\bibitem[Wang et~al.(2012)Wang, Akhtar, Borggaard, and Iliescu]{wang_galerkin}
Z.~Wang, I.~Akhtar, J.~Borggaard, and T.~Iliescu.
\newblock Proper orthogonal decomposition closure models for turbulent flows:
  {A} numerical comparison.
\newblock \emph{Comput. Methods Appl. Mech. Eng.}, 237:\penalty0 10--26, 2012.

\bibitem[Warne(2000)]{warne2000causality}
A.~Warne.
\newblock Causality and regime inference in a {Markov} switching {VAR}.
\newblock Technical report, Sveriges Riksbank Working Paper Series, 2000.

\bibitem[Washio and Motoda(1997)]{washio1997}
T.~Washio and H.~Motoda.
\newblock Inductive inference of first-order rules with non-linear structures.
\newblock \emph{Machine Learning}, 27\penalty0 (2):\penalty0 153--172, 1997.

\bibitem[Watson and Crick(1953)]{Watson1953}
J.~D. Watson and F.~H. Crick.
\newblock Molecular structure of nucleic acids: {A} structure for deoxyribose
  nucleic acid.
\newblock \emph{Nature}, 171\penalty0 (4356), 1953.

\bibitem[Weatheritt and Sandberg(2016)]{sandberg1}
J.~Weatheritt and R.~D. Sandberg.
\newblock {A novel evolutionary algorithm applied to algebraic modifications of
  the RANS stress-strain relationship}.
\newblock \emph{J. Comput. Phys.}, 325:\penalty0 22--37, 2016.

\bibitem[Weatheritt and Sandberg(2017)]{sandberg2}
J.~Weatheritt and R.~D. Sandberg.
\newblock {The development of algebraic stress models using a novel
  evolutionary algorithm}.
\newblock \emph{Int. J. Heat Fluid Flow}, 68:\penalty0 298--318, 2017.

\bibitem[Weichwald and Peters(2021)]{weichwald2021causality}
S.~Weichwald and J.~Peters.
\newblock Causality in cognitive neuroscience: concepts, challenges, and
  distributional robustness.
\newblock \emph{Journal of Cognitive Neuroscience}, 33\penalty0 (2):\penalty0
  226--247, 2021.

\bibitem[Weichwald et~al.(2015)Weichwald, Meyer, Özdenizci, Schölkopf, Ball,
  and Grosse-Wentrup]{weic15}
S.~Weichwald, T.~Meyer, O.~Özdenizci, B.~Schölkopf, T.~Ball, and
  M.~Grosse-Wentrup.
\newblock Causal interpretation rules for encoding and decoding models in
  neuroimaging.
\newblock \emph{NeuroImage}, 110:\penalty0 48--59, 2015.

\bibitem[Werner et~al.(2021)Werner, Junginger, Hennig, and
  Martius]{WernerEtal2021:InformedEQL}
M.~Werner, A.~Junginger, P.~Hennig, and G.~Martius.
\newblock Informed equation learning, 2021.

\bibitem[Werner et~al.(2022)Werner, Junginger, Hennig, and
  Martius]{Werner2022:UncertaintyEQL}
M.~Werner, A.~Junginger, P.~Hennig, and G.~Martius.
\newblock Uncertainty in equation learning.
\newblock In \emph{Proceedings of the Genetic and Evolutionary Computation
  Conference Companion (GECCO)}, pages 2298--2305. Association for Computing
  Machinery, 2022.

\bibitem[Williams et~al.(2015{\natexlab{a}})Williams, Kevrekidis, and
  Rowley]{WKR15}
M.~O. Williams, I.~G. Kevrekidis, and C.~W. Rowley.
\newblock A data-driven approximation of the {Koopman} operator: {E}xtending
  dynamic mode decomposition.
\newblock \emph{Journal of Nonlinear Science}, 25:\penalty0 1307--1346,
  2015{\natexlab{a}}.

\bibitem[Williams et~al.(2015{\natexlab{b}})Williams, Rowley, and
  Kevrekidis]{WRK15}
M.~O. Williams, C.~W. Rowley, and I.~G. Kevrekidis.
\newblock A kernel-based method for data-driven {Koopman} spectral analysis.
\newblock \emph{Journal of Computational Dynamics}, 2:\penalty0 247--265,
  2015{\natexlab{b}}.

\bibitem[Wilson and Cowan(1972)]{wils72}
H.~R. Wilson and J.~D. Cowan.
\newblock Excitatory and inhibitory interactions in localized populations of
  model neurons.
\newblock \emph{Biophysical Journal}, 12:\penalty0 1--24, 1972.

\bibitem[Wilson and Cowan(2021)]{wils21}
H.~R. Wilson and J.~D. Cowan.
\newblock Evolution of the {Wilson-Cowan} equations.
\newblock \emph{Biol. Cybern.}, 115\penalty0 (6):\penalty0 643--653, Dec. 2021.

\bibitem[Wimsatt and Wimsatt(2007)]{wimsatt2007re}
W.~C. Wimsatt and W.~K. Wimsatt.
\newblock \emph{Re-engineering philosophy for limited beings: {P}iecewise
  approximations to reality}.
\newblock Harvard University Press, 2007.

\bibitem[Winship and Morgan(1999)]{winship1999estimation}
C.~Winship and S.~L. Morgan.
\newblock The estimation of causal effects from observational data.
\newblock \emph{Annual review of sociology}, pages 659--706, 1999.

\bibitem[Winter(2002)]{shapley}
E.~Winter.
\newblock The shapley value.
\newblock \emph{Handbook of Game Theory with Economic Applications},
  3:\penalty0 2025--2054, 2002.

\bibitem[Woolgar et~al.(2014)Woolgar, Golland, and Bode]{Woolgar2014}
A.~Woolgar, P.~Golland, and S.~Bode.
\newblock Coping with confounds in multivoxel pattern analysis: {W}hat should
  we do about reaction time differences? {A} comment on {Todd, Nystrom \& Cohen
  2013}.
\newblock \emph{NeuroImage}, 98:\penalty0 506--512, 2014.

\bibitem[Wootton(2015)]{wootton2015invention}
D.~Wootton.
\newblock \emph{The invention of science: {A} new history of the scientific
  revolution}.
\newblock Penguin UK, 2015.

\bibitem[Wu et~al.(2021)Wu, Brunton, and Revzen]{wu2021challenges}
Z.~Wu, S.~L. Brunton, and S.~Revzen.
\newblock Challenges in dynamic mode decomposition.
\newblock \emph{Journal of the Royal Society Interface}, 18\penalty0
  (185):\penalty0 20210686, 2021.

\bibitem[Ye et~al.(2015)Ye, Deyle, Gilarranz, and
  Sugihara]{ye2015distinguishing}
H.~Ye, E.~R. Deyle, L.~J. Gilarranz, and G.~Sugihara.
\newblock Distinguishing time-delayed causal interactions using convergent
  cross mapping.
\newblock \emph{Scientific reports}, 5\penalty0 (1):\penalty0 1--9, 2015.

\bibitem[Yousif et~al.(2022)Yousif, Zhang, Yu, Vinuesa, and Lim]{heechang}
M.~Z. Yousif, M.~Zhang, L.~Yu, R.~Vinuesa, and H.~Lim.
\newblock A transformer-based synthetic-inflow generator for
  spatially-developing turbulent boundary layers.
\newblock \emph{J. Fluid Mech., To Appear. Preprint arXiv:2206.01618}, 2022.

\bibitem[Yu et~al.(2009)Yu, Cunningham, Santhanam, Ryu, Shenoy, and
  Sahani]{yu09}
B.~M. Yu, J.~P. Cunningham, G.~Santhanam, S.~I. Ryu, K.~V. Shenoy, and
  M.~Sahani.
\newblock {Gaussian}-process factor analysis for low-dimensional single-trial
  analysis of neural population activity.
\newblock \emph{Journal of Neurophysiology}, 102\penalty0 (1):\penalty0
  614--635, 2009.
\newblock Pmid: 19357332.

\bibitem[Yu et~al.(2020)Yu, Drton, and Shojaie]{yu2020directed}
S.~Yu, M.~Drton, and A.~Shojaie.
\newblock Directed graphical models and causal discovery for zero-inflated
  data.
\newblock \emph{arXiv preprint arXiv:2004.04150}, 2020.

\bibitem[Yu et~al.(2021)Yu, Gao, Yin, and Ji]{Yu2021curl}
Y.~Yu, T.~Gao, N.~Yin, and Q.~Ji.
\newblock {DAGs with No Curl: An Efficient DAG Structure Learning Approach}.
\newblock In M.~Meila and T.~Zhang, editors, \emph{Proceedings of the 38th
  International Conference on Machine Learning}, volume 139 of
  \emph{Proceedings of Machine Learning Research}, pages 12156--12166. PMLR,
  18--24 Jul 2021.

\bibitem[Zanna and Bolton(2020)]{zanna2020data}
L.~Zanna and T.~Bolton.
\newblock Data-driven equation discovery of ocean mesoscale closures.
\newblock \emph{Geophysical Research Letters}, 47\penalty0 (17):\penalty0
  e2020GL088376, 2020.

\bibitem[Zhang(2008{\natexlab{a}})]{zhang2008causal}
J.~Zhang.
\newblock Causal reasoning with ancestral graphs.
\newblock \emph{Journal of Machine Learning Research}, 9\penalty0
  (47):\penalty0 1437--1474, 2008{\natexlab{a}}.

\bibitem[Zhang(2008{\natexlab{b}})]{zhang2008completeness}
J.~Zhang.
\newblock On the completeness of orientation rules for causal discovery in the
  presence of latent confounders and selection bias.
\newblock \emph{Artificial Intelligence}, 172\penalty0 (16):\penalty0
  1873--1896, 2008{\natexlab{b}}.

\bibitem[Zhang and Spirtes(2002)]{zhang2002strong}
J.~Zhang and P.~Spirtes.
\newblock Strong faithfulness and uniform consistency in causal inference.
\newblock In \emph{Proceedings of the Nineteenth Conference on Uncertainty in
  Artificial Intelligence}, Uai'03, page 632–639, San Francisco, CA, USA,
  2002. Morgan Kaufmann Publishers Inc.
\newblock ISBN 0127056645.

\bibitem[Zhang and Spirtes(2008)]{zhang2008detection}
J.~Zhang and P.~Spirtes.
\newblock Detection of unfaithfulness and robust causal inference.
\newblock \emph{Minds and Machines}, 18\penalty0 (2):\penalty0 239--271, 2008.

\bibitem[Zheng et~al.(2018)Zheng, Aragam, Ravikumar, and
  Xing]{zheng2018notears}
X.~Zheng, B.~Aragam, P.~K. Ravikumar, and E.~P. Xing.
\newblock {DAGs with NO TEARS}: {C}ontinuous optimization for structure
  learning.
\newblock In S.~Bengio, H.~Wallach, H.~Larochelle, K.~Grauman, N.~Cesa-Bianchi,
  and R.~Garnett, editors, \emph{Advances in Neural Information Processing
  Systems}, volume~31. Curran Associates, Inc., 2018.

\bibitem[Zheng et~al.(2020)Zheng, Dan, Aragam, Ravikumar, and
  Xing]{zheng2020learning}
X.~Zheng, C.~Dan, B.~Aragam, P.~Ravikumar, and E.~Xing.
\newblock Learning sparse nonparametric {DAGs}.
\newblock In S.~Chiappa and R.~Calandra, editors, \emph{Proceedings of the
  Twenty Third International Conference on Artificial Intelligence and
  Statistics}, volume 108 of \emph{Proceedings of Machine Learning Research},
  pages 3414--3425. Pmlr, 26--28 Aug 2020.

\bibitem[Zou et~al.(2012)Zou, Hastie, and Tibshirani]{spca}
H.~Zou, T.~Hastie, and R.~Tibshirani.
\newblock {Sparse principal component analysis}.
\newblock \emph{J. Comput. Graph. Stat.}, 15\penalty0 (1):\penalty0 265--286,
  2012.

\bibitem[Zscheischler et~al.(2018)Zscheischler, Westra, Van Den~Hurk,
  Seneviratne, Ward, Pitman, AghaKouchak, Bresch, Leonard, Wahl,
  et~al.]{zscheischler2018future}
J.~Zscheischler, S.~Westra, B.~J. Van Den~Hurk, S.~I. Seneviratne, P.~J. Ward,
  A.~Pitman, A.~AghaKouchak, D.~N. Bresch, M.~Leonard, T.~Wahl, et~al.
\newblock Future climate risk from compound events.
\newblock \emph{Nature Climate Change}, 8\penalty0 (6):\penalty0 469--477,
  2018.

\bibitem[Zscheischler et~al.(2020)Zscheischler, Martius, Westra, Bevacqua,
  Raymond, Horton, van~den Hurk, AghaKouchak, J{\'e}z{\'e}quel, Mahecha,
  et~al.]{zscheischler2020typology}
J.~Zscheischler, O.~Martius, S.~Westra, E.~Bevacqua, C.~Raymond, R.~M. Horton,
  B.~van~den Hurk, A.~AghaKouchak, A.~J{\'e}z{\'e}quel, M.~D. Mahecha, et~al.
\newblock A typology of compound weather and climate events.
\newblock \emph{Nature reviews earth \& environment}, 1\penalty0 (7):\penalty0
  333--347, 2020.

\bibitem[Żytkow et~al.(1990)Żytkow, Michalski, and Stepp]{zytkow1990}
J.~Żytkow, R.~Michalski, and R.~Stepp.
\newblock Representation and learning of categorical structures.
\newblock \emph{Machine Learning}, 5\penalty0 (1):\penalty0 7--48, 1990.

\end{thebibliography}
